\documentclass[%
reprint,raggedbottom, superscriptaddress, nofootinbib, amsmath,amssymb, aps,
prd, ]{revtex4-2}

\usepackage[normalem]{ulem}

\usepackage{graphicx}
\usepackage{amsmath}
\usepackage[table]{xcolor}
\usepackage{booktabs}
\usepackage{multirow}
\usepackage{mathrsfs}
\allowdisplaybreaks

\definecolor{darkpink}{RGB}{200,55,113}
\usepackage[
colorlinks=true, linkcolor=darkpink, citecolor=green, urlcolor=darkpink
]{hyperref}

\begin{document}

\title{Neutron-star multipole moments through seventh order in rotation}

\author{Carlos Conde-Ocazionez} \email{carlosc7@illinois.edu} \affiliation{The Grainger College of Engineering, Illinois Center for Advanced Studies of the Universe, Department of Physics, University of Illinois at Urbana-Champaign, Urbana, IL 61801, USA}

\author{Hao-Jui Kuan} \email{hjkuan@illinois.edu} \affiliation{The Grainger College of Engineering, Illinois Center for Advanced Studies of the Universe, Department of Physics, University of Illinois at Urbana-Champaign, Urbana, IL 61801, USA}

\author{Jaquelyn Noronha-Hostler\vspace{0.1cm}} \email{jnorhos@illinois.edu} \affiliation{The Grainger College of Engineering, Illinois Center for Advanced Studies of the Universe, Department of Physics, University of Illinois at Urbana-Champaign, Urbana, IL 61801, USA}

\author{Nicolás Yunes\vspace{0.1cm}} \email{nyunes@illinois.edu} \affiliation{The Grainger College of Engineering, Illinois Center for Advanced Studies of the Universe, Department of Physics, University of Illinois at Urbana-Champaign, Urbana, IL 61801, USA}

\begin{abstract}

We compute several mass ($M_0$, $M_2$, $M_4$ and $M_6$) and current multipole
moments ($S_1$, $S_3$, $S_5$ and $S_7$) of uniformly rotating neutron stars
through seventh order in spin frequency within the Hartle--Thorne approximation.
We obtain the moments by integrating the interior perturbation equations
numerically and matching to the analytic exterior solutions at the stellar
surface. We construct stellar sequences at $716\,\mathrm{Hz}$ for the SLy and
BSk21 equations of state, as representative examples of soft and stiff cases,
respectively. For stars with nonrotating masses of $1.0$, $1.5$, and
$2.0\,M_\odot$, we compare successive truncations at fixed central density. The
leading terms fall short of the sums through the highest computed order by about
$2$ to $13\%$ for the mass, $6$ to $29\%$ for the angular momentum, and $6$ to
$23\%$ for the magnitude of the mass quadrupole. Along both sequences,
rotational corrections at fixed spin frequency are larger for less compact
stars, and their size varies with the multipole moment considered. Successive
corrections shrink by a factor that scales as the square of the spin frequency
in units of the mass-shedding frequency, and thus, the expansion remains well
behaved almost to the mass-shedding limit. Near that limit, however, a given
accuracy requires more orders in the expansion, especially when computing the
higher multipoles. For the $1.0\,M_\odot$ stars at $716\,\mathrm{Hz}$, the
$\mathcal{O}(\Omega^6)$ contribution changes the mass monopole $M_{0}$ by less
than $1\%$. The corrections to the angular momentum $S_{1}$ and the mass
quadrupole $M_{2}$, at $\mathcal{O}(\Omega^5)$ and $\mathcal{O}(\Omega^6)$
respectively, are below $10\%$. The resulting multipole moments can be used to
test how rotation changes quasi-universal relations and to assess how much
higher-order corrections matter for X-ray pulse profiles.
\end{abstract}

\maketitle


\section{\label{sec:level1}Introduction}

Neutron stars are natural laboratories for cold, neutron-rich matter in beta
equilibrium at supranuclear densities, a regime terrestrial experiments do not
reach. We now observe them across the electromagnetic spectrum and through
gravitational waves. X-ray timing measurements from
NICER~\cite{gendreau2016neutron} determine masses and radii through the pulse
profiles of hot spots on the stellar
surface~\cite{Watts:2016uzu,Miller:2019cac,Bogdanov:2019qjb}, and the
forthcoming eXTP mission will sharpen these measurements
further~\cite{eXTP:2016rzs}. Gravitational-wave observations by the
LIGO-Virgo-KAGRA
network~\cite{LIGOScientific:2007fwp,LIGOScientific:2014pky,VIRGO:2012dcp,Aso:2013eba}
provide access to tidal deformabilities during the late inspiral of neutron-star
binaries. These X-ray and gravitational-wave observations have the potential to
place tight bounds on the equation of state (EoS) of dense matter inside neutron
stars~\cite{LIGOScientific:2018hze, Dittmann:2024mbo, MUSES:2023hyz}.

Rotational effects may play an important role in setting these bounds. A
significant fraction of the observed neutron stars rotates rapidly. Of the
currently known pulsars, about $16.3\%$ (692 pulsars) spin at frequencies $f
\geq 100\,\mathrm{Hz}$~\cite{Manchester:2004bp, DiSalvo:2020xhn}, and the
fastest observed pulsar spins at $716\,\mathrm{Hz}$~\cite{Hessels:2006ze}.
Despite such high spin frequencies, a neutron star can rotate only so fast
before it begins to shed matter from its equator. A uniformly rotating star
reaches its mass-shedding limit when its spin frequency $f_{\mathrm{sh}}$ equals
that of a test particle on a circular orbit at the equator of the rotating star.
The Newtonian Keplerian frequency $f_{\mathrm{K}} =
\sqrt{GM_{*}/R_{*}^{3}}/(2\pi)$ if often used as a reference scale for
$f_{\mathrm{sh}}$, where $M_{*}$ and $R_{*}$ are the mass and radius of the
nonrotating star. Fits to relativistic models of uniformly rotating stars with
the same gravitational mass give $f_{\mathrm{sh}} \approx
0.6\,f_{\mathrm{K}}$~\cite{Lattimer:2004pg,Haensel:2009wa}.
Rotation changes the macroscopic observables of the star, and the change grows
in magnitude as the spin frequency increases~\cite{Hartle:1967he,
Hartle:1968si}. The pulsars targeted by X-ray pulse-profile
modeling~\cite{Miller:2019cac,Riley:2019yda,Miller:2021qha,Riley:2021pdl,Choudhury:2024xbk}
rotate rapidly, and the neutron stars that LIGO-Virgo-KAGRA will detect may also
rotate rapidly. If so, the measured masses and radii, used to set bounds on the
EoS, may acquire non-negligible spin corrections, which may contaminate EoS
inferences.

X-ray and gravitational-wave observations probe the interior indirectly, through
the external gravitational field that the stellar matter generates. Rotation
changes that field as well, and thus, inferences about dense matter can be
affected by how well we model rotation effects in the gravitational field.
Far from its matter source distribution, a gravitational field can be described
systematically by a multipolar expansion. In such an expansion, the field is
written as a sum of terms falling off with different powers of the field-point
distance, each weighted by an angular function. The coefficients of that sum are
the multipole moments~\cite{Thorne:1980ru, Geroch:1970cd, Hansen:1974zz}, which
carry information about the interior of the source. Such a multipolar
decomposition of the gravitational field is analogous to electromagnetism, where
the electromagnetic field of an arbitrary charge and current distribution is
decomposed into electric and magnetic multipoles.

In general relativity, two families of multipoles arise in analogy to
electromagnetic multipoles. Rotation sets the stellar matter in motion about the
symmetry axis, and this motion flattens the star into a deformed oblate
shape~\cite{hartle1967, Bardeen:1970vja, chandrasekhar1987ellipsoidal}. Both the
motion and the shape leave an imprint on the exterior
spacetime~\cite{hartle1967, Bardeen:1970vja, chandrasekhar1987ellipsoidal}. The
shape is captured by the mass multipole moments $M_\ell$, which trace how matter
is distributed within the star, much as the electric multipoles trace the
distribution of charge. The motion is captured by the mass-current multipole
moments $S_\ell$, which trace how matter moves, much as the magnetic multipoles
trace the flow of current~\cite{Geroch:1970cd, Hansen:1974zz, Thorne:1980ru}.
Here, $\ell$ is the multipole degree, and we use the Geroch--Hansen
normalization for both families of moments~\cite{Geroch:1970cd, Hansen:1974zz}.
The multipoles of lowest order are the most familiar, $M_0$ being the total
mass, $S_1$ the spin angular momentum, and $M_2$ the mass quadrupole moment,
which for an isolated rotating star measures the rotational
flattening~\cite{Laarakkers:1997hb}. The current moments emerge in general
relativity as an independent family, and only the mass moments have a Newtonian
analogue~\cite{Stein:2013ofa, Thorne:1980ru}. Far from the star, therefore, the
gravitational field is fully described by these two families of multipoles, and
from them valuable information about the structure of matter can be determined.

How accurately the multipole moments must be known depends on how strongly they
affect the observables used to interpret the data. Psaltis and \"Ozel argued
that neglecting the spacetime quadrupole alters X-ray pulse profiles by
$1$--$5\%$ at spin frequencies near $600\,\mathrm{Hz}$~\cite{Psaltis:2013zja}.
Yagi et al.~estimated that higher-order spin corrections to the stellar
eccentricity modify the pulse profiles by $3$--$9\%$~\cite{Yagi:2014bxa}. These
estimates motivate testing higher-order rotation in the pulse-profile models
used to infer masses and radii with NICER and eXTP~\cite{Watts:2016uzu,
Miller:2019cac}, especially given the $5$--$10\%$ measurement accuracy needed to
distinguish between EoSs in the study of~\cite{Psaltis:2013fha}.
The multipole moments must be predicted at least as accurately as they can be
measured, and the calculation must include higher multipoles to describe the
exterior field accurately. Whether the higher moments contribute at a level that
observations can detect cannot be settled without computing them first. Meeting
these conditions means solving the Einstein equations for a rotating star to
high precision and to high multipolar order.

Two strategies are commonly used to solve the stationary Einstein equations for
rotating neutron stars. In the nonperturbative strategy, one solves the
equations numerically without expanding in rotation~\cite{Stergioulas:1994ea,
Bonazzola:1998qx, Ansorg:2001pe}. In the perturbative strategy, one solves the
equations order-by-order in rotation, as in the Hartle--Thorne
formalism~\cite{Hartle:1967he, Hartle:1968si}. The latter was originally
developed to second order in the spin frequency and was later extended to
third~\cite{Hartle:1973zza, Benhar:2005gi} and fourth order~\cite{Yagi:2014bxa}.
The nonperturbative strategy is well-suited to the modeling of rapidly rotating
or strongly deformed configurations, while the perturbative strategy is better
suited when dealing with stars that rotate slowly or moderately.

With that said, the two strategies have complementary strengths and weaknesses.
In a non-perturbative calculation, one does not need to rely on a small spin
expansion, but the higher multipoles are harder to extract because their
contributions to the exterior field fall off more rapidly with distance. This
difficulty is compounded at slow rotation, where the moments themselves enter at
progressively higher powers of the spin frequency and can become difficult to
distinguish from numerical error~\cite{Yagi:2014bxa}. Such difficulties result
in an inability to reliably extract multipole moments beyond the mass
hexadecapole $M_4$ at high precision and moderate spins. In the perturbative
strategy, by contrast, a slow-rotation expansion isolates the contribution at
each order in spin, giving us access to higher multipoles and to rotational
corrections to the lower moments. However, a perturbative expansion loses
accuracy as the spin grows and it contains uncontrolled remainders that must be
estimated to determine the accuracy of the predicted moments.

The purpose of this paper is to extend the perturbative, slow-rotation study of
neutron stars to high perturbative order, which serves three purposes: (i) it
refines the moments to the precision future observations will demand, (ii) it
gives access to the higher multipoles that the nonperturbative scheme struggles
to resolve, and (iii) it extends the range of spin frequencies over which the
expansion remains accurate. We presented the analytic part of this calculation
in Ref.~\cite{Conde-Ocazionez:2025agu}, hereafter referred to as Paper~I. In
that work, we treated both the fluid and the spacetime perturbatively, derived
the interior and exterior perturbation equations, and solved the exterior
equations in closed form up to a set of exterior integration constants. The mass
and current multipole moments then followed in terms of those constants,
extracted through two independent methods as a consistency check. Here we
complete that program numerically by integrating the interior equations and
fixing the exterior integration constants by matching the (numerical) interior
and (analytic) exterior solutions at the stellar surface. Through this matching,
we are able to compute the multipole moments of slowly rotating neutron stars to
seventh order in the spin frequency for the first time.

\subsection*{Executive Summary}\label{sec:executive}

The numerical calculation proceeds order by order in rotation. Through seventh
order in the spin-frequency expansion, the interior equations require up to the
third radial derivative of the energy density. We obtain these derivatives from
segmented Chebyshev fits to tabulated EoSs, differentiating the fits
analytically and using the background structure equations to reduce the order of
derivatives when possible. In the exterior, we recast the analytic solutions to
reduce numerical cancellation and use extended precision where needed. We carry
out the calculation for two tabulated EoSs, SLy~\cite{Douchin:2001sv} and
BSk21~\cite{Goriely:2010bm}, which represent relatively soft and stiff models,
respectively. For each EoS, we compute the mass moments $M_0$, $M_2$, $M_4$, and
$M_6$ and the current moments $S_1$, $S_3$, $S_5$, and $S_7$, including all
rotational corrections through seventh order in the spin frequency. We follow
these moments along sequences labeled by the nonrotating stellar mass, from
$0.7\,M_\odot$ to the maximum nonrotating mass for each EoS. For the most part,
we fix the spin frequency at $716\,\mathrm{Hz}$, the highest measured value for
a pulsar, because it provides the most demanding test of our slow-rotation
expansion. The order required to compute a given multipole moment to a specified
accuracy falls steeply as the spin frequency decreases, so more slowly rotating
stars will be well described by fewer terms in the expansion, as we also show in
this paper.

\begin{figure*}[ht]
\centering
\includegraphics[width=\textwidth]{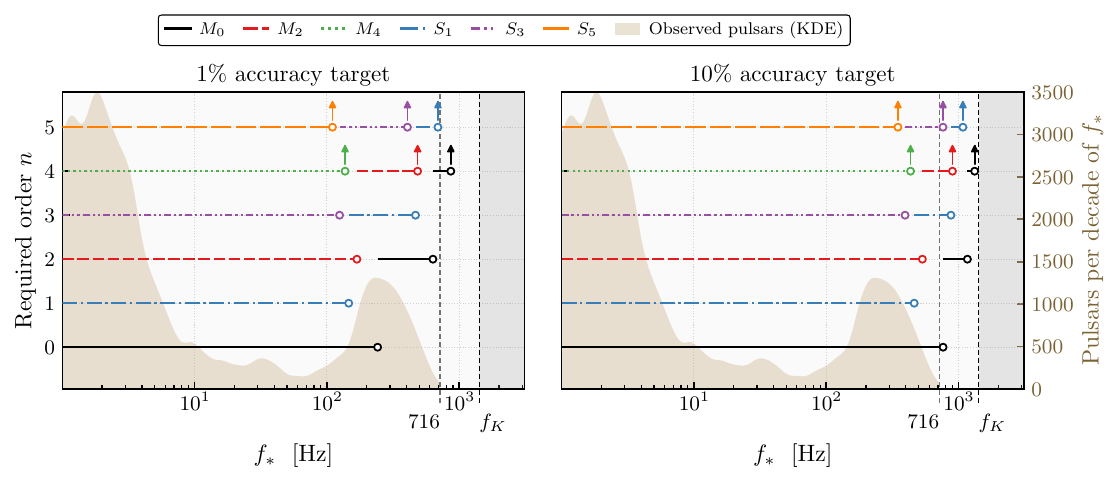}
\caption{Lowest order $n$ of the Hartle-Thorne expansion required to reach a given accuracy, as a function of spin frequency $f_{*}$, for an SLy star with a non-rotating mass of $1.0\,M_\odot$.
In the left and right panels, we retain enough terms in the spin expansion that adding the next correction changes the accumulated result by less than $1\%$ and $10\%$, respectively (see Eq.~\eqref{eq:truncation}). Each horizontal segment shows the required expansion order over the corresponding range of spin. The open circle marks the spin at which a higher order is needed to maintain the specified accuracy. An arrow indicates that the required order has risen beyond what seventh order can check. Dashed lines mark $716\,\mathrm{Hz}$ and the Newtonian mass-shedding frequency $f_K$, anything at higher frequency is shaded in gray.The background curve is a kernel density estimate of the measured pulsar spin frequencies, in pulsars per decade of $f_{*}$ (right vertical axis).}
\label{fig:required_order}
\end{figure*}

We find that corrections beyond leading order in spin can be relatively large
even for the lowest multipole moments. We first choose three central densities
that yield nonrotating stars with masses of $1.0$, $1.5$, and $2.0\,M_\odot$. At
each of these fixed central densities, we then include successively higher
orders in the spin expansion to determine how the rotational corrections change
the stellar properties. Compared to results that include all computed rotational
corrections, the leading-order calculation underestimate the full result by
about $2$--$13\%$ for the mass, $6$--$29\%$ for the angular momentum, and
$6$--$23\%$ for the magnitude of the mass quadrupole. The higher moments can
have larger fractional corrections. For example, in the $1.0\,M_\odot$ BSk21
model, the first rotational correction contributes about $25\%$ of the computed
mass hexadecapole and $34\%$ of the computed fifth current moment. Along both
sequences, the spin corrections are always largest for the least compact stars,
even for fixed spin frequency. In the cases where we can include several
corrections, we find a clear hierarchy in the terms, so successive terms
decrease in magnitude. How far the expansion needs to be carried forward depends
on both the compactness of the star, the moment of interest, and the precision
sought.

The spin frequency, however, also determines how far the expansion needs to be
carried out. The comparisons above held the spin frequency fixed at
$716\,\mathrm{Hz}$, but every term in the slow-rotation expansion grows as a
power of the spin, so the corrections at that frequency are the largest any
known star can have. At lower spin frequencies, therefore, the spin corrections
shrink quickly and the perturbative spin expansion needs far fewer orders. The
practical question is then the following: How many orders are required, for a
given accuracy, as the frequency grows gradually from slow rotation up to
break-up?

To answer this question, we estimate the truncation error of the slow-rotation
series from the first omitted term. Since successive contributions to a given
multipole differ by two powers of the angular spin frequency $\Omega$,
truncating at $\Omega^n$ leaves out a term proportional to $\Omega^{n+2}$ first.
Dividing the magnitude of that term by the magnitude of the sum og terms up to
the one proportional to $\Omega^n$ yields an estimate of the truncation error.

We can then ask what $n$ must be for this truncation error to fall below some
target accuracy, as a function of the spin frequency. This is precisely what
Fig.~\ref{fig:required_order} shows for two accuracy targets ($1\%$ on the left
and $10\%$ on the right) and for each multipole moment, using a $1.0\,M_\odot$
SLy neutron star. To read either panel, choose a multipole of interest and
identify its curve by its color and line style. Then, choose a spin frequency on
the horizontal axis and move vertically upward until you intersect that
multipole curve. The vertical coordinate $n$ gives the lowest power of $\Omega$
through which we must retain terms to meet the target accuracy according to our
error estimate. For instance, at $716\,\mathrm{Hz}$, the $1\%$ target for $M_0$
in the left panel and the $10\%$ target for $M_2$ in the right panel both
require terms through $\Omega^4$. This corresponds to fourth order terms in the
slow-rotation expansion relative to the non-rotating contribution to $M_0$ and
second order in the expansion relative to the leading $\Omega^2$ contribution to
$M_2$.

The truncation error of the expansion through $\Omega^n$ is here being estimated
from the first omitted term in the series, which is proportional to
$\Omega^{n+2}$. Our calculation includes mass terms through $\Omega^6$ and
current terms through $\Omega^7$, so we can check truncations through $\Omega^4$
for mass moments and $\Omega^5$ for current moments. These are also the leading
powers in $\Omega$ of $M_4$ and $S_5$, the highest multipoles shown in the
figure. Each arrow, therefore, marks the frequency at which the error estimate
for the last truncation we can check reaches the target. Above that frequency,
we would need to retain the next computed term, so checking the error after
including it would require extending the calculation by two more orders in
rotation. For the $1\%$ target, the test still gives an answer for $M_0$ at
$716\,\mathrm{Hz}$, while the other five curves reach their arrows below that
frequency. For the $10\%$ target at the same frequency, $M_0$ needs no
rotational correction, and $S_1$ needs only its $\Omega^3$ term, which is
second-order relative to its leading term proportional to $\Omega$. The curves
for $M_4$ and $S_5$ have already reached the limit that can be calculated at
seventh order in the expansion (indicated by the arrows). Since a moment of
degree $\ell$ starts at $\Omega^\ell$, the higher multipoles have fewer computed
corrections available for this test.

Our analysis also allows us to measure how fast the corrections in the
slow-rotation expansion shrink with expansion order.
At a fixed spin frequency, successive computed corrections decrease in
magnitude. For all stars considered here, the ratio of each correction to the
preceding one scales approximately as $\sim (f_{*}/f_K)^{2}$ where $f_K$ the
Newtonian mass shedding frequency. There is then a frequency above which the
next spin correction is no longer smaller than the previous one and the
expansion ceases to be mathematically well behaved. We find that this frequency
lies not far below $f_K$ for every star, as determined from consecutive pairs of
multipole corrections. At a rotation frequency of $716\,\mathrm{Hz}$, each star
therefore spins comfortably inside the range where the slow-rotation expansion
remains well behaved.

The $716\,\mathrm{Hz}$ frequency is the upper edge of the observed population.
Every other detected pulsar spins more slowly. The background of
Fig.~\ref{fig:required_order} shows a kernel density estimate (KDE) of the spin
frequencies measured to date~\cite{Manchester:2004bp,DiSalvo:2020xhn}. Its scale
is on the right vertical axis, in pulsars per decade of $f_{*}$, so the area
under the curve between two frequencies gives the number of pulsars in that
interval. The bulk of the pulsar population spins near $2\,\mathrm{Hz}$, where
the leading terms of the slow-rotation expansion already meet both accuracy
targets. Thus, for most neutron stars the expansion is accurate at its lowest
orders. However, the pulsar distribution is bimodal such that there is a second
peak centered at a few hundred Hz. For this second population of neutron stars,
these higher-order corrections do matter, and thus, they must be included when
calculated moments, as we do in this paper.

The results of our work rest on a careful and robust numerical implementation.
Both the numerical code we developed and the equations it solves are publicly
available: the numerical code in the \texttt{aQLIMR}
repository~\cite{conde2025aqlimr}, and the equations, together with the symbolic
material they are built from, in the \texttt{hartle-thorne-project}
repository~\cite{conde2025hartlethorne}. We checked the equations against their
published counterparts through fourth order. The second- and third-order
equations of Hartle and Thorne~\cite{Hartle:1967he, Hartle:1968si,
Hartle:1973zza} are recovered in full, the latter as revisited by Benhar et
al.~\cite{Benhar:2005gi}. At fourth order in spin, the equations of Yagi et
al.~\cite{Yagi:2014bxa} are recovered once the coordinate transformation to the
Hartle-Thorne frame is carried through consistently, as discussed in Paper~I.
Additionally, we have here recast all equations in a form suited to stable
numerical integration, and we tested the accuracy of the solutions at every
order in spin. With that foundation in place, a direct comparison against
non-perturbative, numerical codes in the regime where the two methods overlap
can be carried out, which is the subject of work now in progress.

\subsection*{Organization and units}\label{sec:organization}

The remainder of this paper is organized as follows. Sec.~\ref{sec:perturbation}
summarizes the perturbation scheme and the structure of the equations at each
order. Sec.~\ref{sec:exterior} presents the exterior solutions in the rescaled
form used here. Sec.~\ref{sec:matching} gives the matching conditions at the
stellar surface. Sec.~\ref{sec:numerical} describes the numerical
implementation, including the treatment of the higher derivatives of the energy
density. Sec.~\ref{sec:numerical_results} reports the results for single stars
and for stellar sequences. Sec.~\ref{sec:conclusions} discusses the findings and
their limitations. The appendices collect the asymptotic expansions, the source
terms, the analytic expressions for the moments, the exterior solutions, the
details of the differentiation algorithm, the complete profiles for one
configuration, the tabulated multipole moments with their spin-induced
corrections, and the exterior integration constants along both stellar
sequences.

Throughout this work, we adopt geometric units in which $G = 1 = c$, where $G$
is Newton's gravitational constant and $c$ is the speed of light. In these
units, all physical quantities are expressed in units of length $[L]$, and the
metric signature is $(-,+,+,+)$. In the numerical implementation, we
additionally adopt code units in which $M_\odot = 1$, and additional notation
follows Paper~I~\cite{Conde-Ocazionez:2025agu}.


\section{Perturbation Equations}
\label{sec:perturbation}

\begin{figure}[h]
    \centering
    \includegraphics[width=1.0\columnwidth]{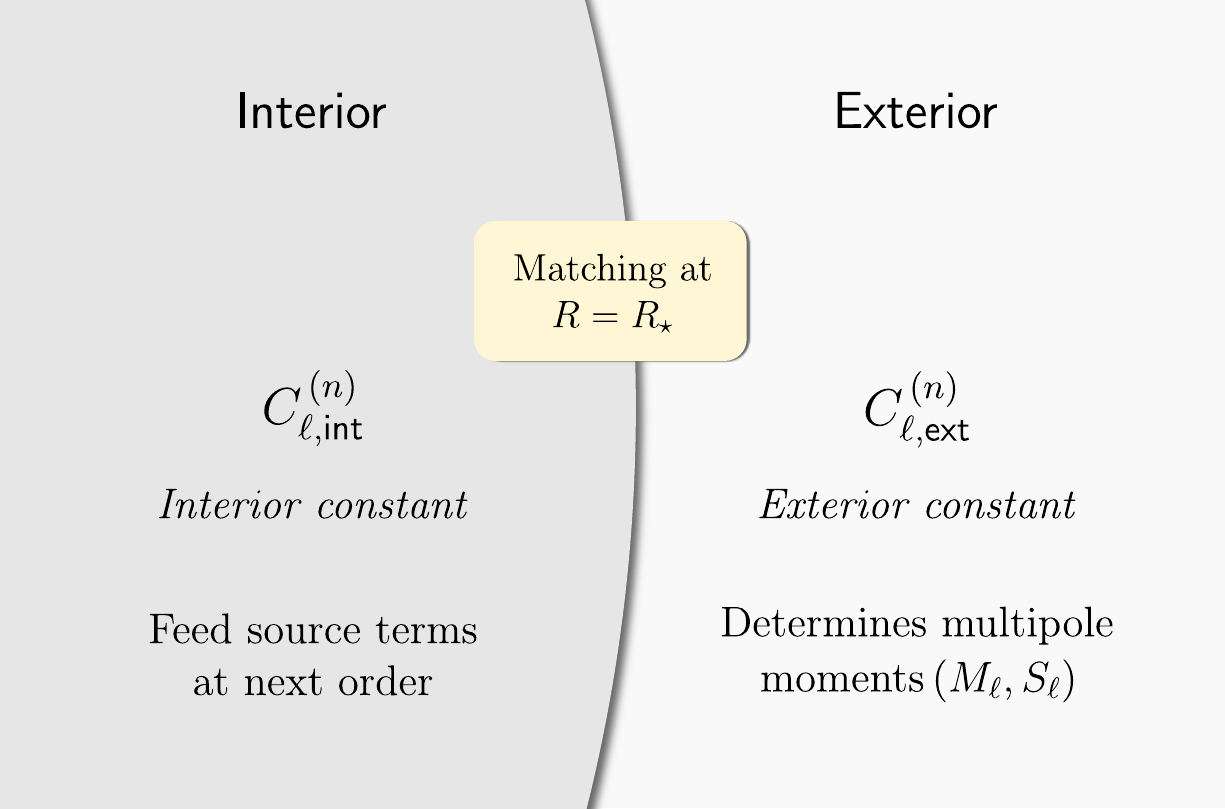}
\caption{The stellar interior and exterior regions separated by the stellar surface at $R = R_{*}$. The interior integration constant $C_{\ell,\mathsf{int}}^{(n)}$ feeds the source terms at the next perturbative order. The exterior integration constant $C_{\ell,\mathsf{ext}}^{(n)}$ determines the multipole moments $M_\ell$ and $S_\ell$ through the analytical expressions derived in Paper~I.}
    \label{fig:regions}
\end{figure}

We solve the perturbed Einstein equations by first dividing them into two
regions. In the stellar interior, matter is present and the fluid properties
enter through the EoS. Outside the star, in the exterior domain, one must solve
the vacuum-version of the interior equations. The exterior problem was solved
analytically in Paper~I, but the interior problem requires a numerical
treatment. The two regions meet at the stellar surface $R = R_*$, as illustrated
in Fig.~\ref{fig:regions}. Matching the interior and exterior solutions at the
surface fixes any integration constants of the interior and exterior solutions.

Determining the multipole moments $M_\ell$ and $S_\ell$ of a neutron star
requires fixing two integration constants for each harmonic mode $\ell$ at each
perturbative order. At a given order in spin frequency, the perturbation
equations for each mode reduce to a system of two coupled first-order ordinary
differential equations (ODEs) in the interior and another such system in the
exterior. The interior and exterior solutions, therefore, carry two integration
constants each. Regularity at the center removes one interior constant, and
asymptotic flatness removes one exterior constant. Two free constants remain,
$C^{(n)}_{\ell,\mathsf{int}}$ from the interior solution and
$C^{(n)}_{\ell,\mathsf{ext}}$ from the exterior solution, where the superscript
$n$ labels the order in spin frequency. Matching the interior and exterior
solutions at the stellar surface fixes both constants. The interior constant
obtained at one order enters the source terms at the next, which drives the
iterative process of the whole calculation. The exterior constants determine the
multipole moments through the analytic expressions of Paper~I.

We construct the interior solution sequentially, beginning with the nonrotating
background. Its TOV equations, together with Lindblom's
reformulation~\cite{lindblom1992determining}, are given in
Sec.~\ref{sec:background}. Sections~\ref{sec:interior_odd}
and~\ref{sec:interior_even} describe the interior equations in the odd-parity
and even-parity sectors, respectively, and Sec.~\ref{sec:source} closes with a
discussion of the source terms.

\subsection{Slow-rotation expansion}
\label{sec:slow_rotation}

\begin{figure}[t]
  \centering
  \includegraphics[width=\columnwidth]{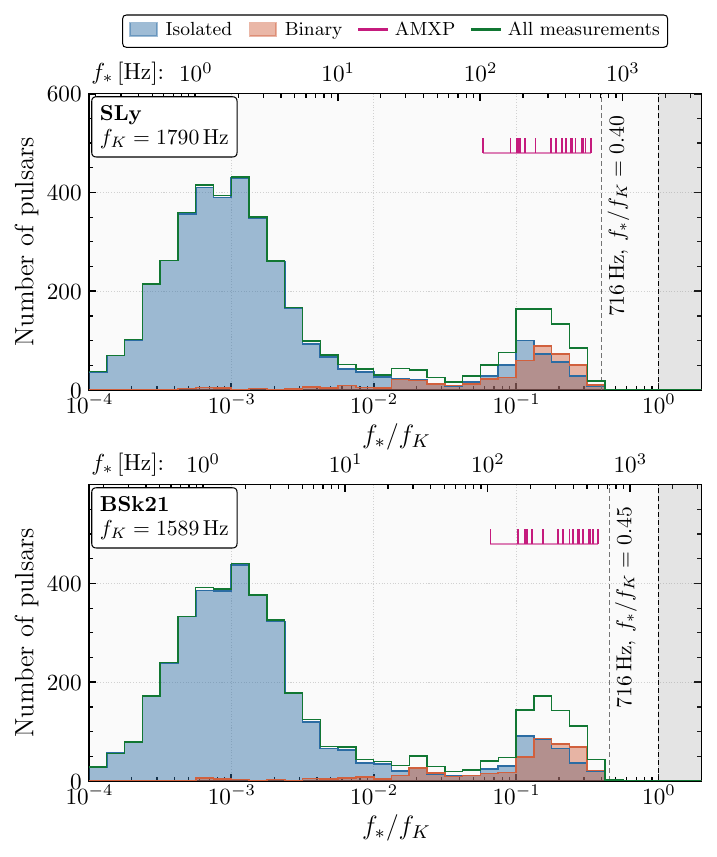}
    \caption{Distribution of measured pulsar spin frequencies $f_{*}$ in units of the Keplerian frequency $f_{\mathrm{K}}$ (lower axis) and in Hz (upper axis), for a nonrotating $1.5\,M_\odot$ star with SLy (top, $f_{\mathrm{K}} = 1790$~Hz) and BSk21 (bottom, $f_{\mathrm{K}} = 1589$~Hz). Histograms show isolated (blue) and binary (orange) pulsars~\cite{Manchester:2004bp}. The green line is the full sample, which adds the accreting millisecond X-ray pulsars (AMXP)~\cite{DiSalvo:2020xhn}, also marked as magenta ticks. The gray dashed line marks PSR~J1748$-$2446ad at $716$~Hz~\cite{Hessels:2006ze}, and the black dashed line marks $f_{*} = f_{\mathrm{K}}$. The mass-shedding limit lies near $0.6\,f_{\mathrm{K}}$.
    }
  \label{fig:pulsar_spin_fraction}
\end{figure}

The observed spin frequencies of pulsars show that the slow-rotation expansion
is well suited to neutron stars. Figure~\ref{fig:pulsar_spin_fraction} shows the
measured population, which includes isolated pulsars and pulsars in
binaries~\cite{Manchester:2004bp}, together with accreting millisecond X-ray
pulsars in low-mass X-ray binaries~\cite{DiSalvo:2020xhn}. The spin frequency
$f_{*}$ is given in Hz and in units of the Keplerian frequency $f_{\mathrm{K}}$
of a nonrotating $1.5\,M_\odot$ star, for SLy and BSk21. Most pulsars rotate
slowly, with $f_{*}/f_{\mathrm{K}} \sim 10^{-3}$. In that regime, the expansion
is accurate even if we keep only a few terms. Nonperturbative codes, such as
RNS~\cite{Stergioulas:1994ea}, by contrast, have difficulty extracting higher
$\ell$ multipole moments at such spins~\cite{Yagi:2014bxa}. A sizable fraction
of pulsars rotates rapidly, with $f_{*}/f_{\mathrm{K}}$ between $0.1$ and
$0.45$. For these stars, the ratio is no longer very small, yet it stays below
one half, which motivates extending the expansion to higher order. Higher-order
spin corrections will improve the accuracy of the description of these
rapidly-rotating stars, as discussed in the following sections. With enough
orders kept in the sum, the expansion remains accurate up to spin frequencies
close to the mass-shedding limit, as shown in Fig.~\ref{fig:required_order}. The
slow-rotation expansion can therefore cover the observed population.

We construct rotating neutron-star solutions perturbatively in $\epsilon$, a
dimensionless bookkeeping parameter that tracks powers of the angular spin
frequency $\Omega = 2\pi f_{*}$. The construction follows the Hartle--Thorne
formalism~\cite{Hartle:1967he, Hartle:1968si}, extended to
$\mathcal{O}(\Omega^{7})$ as detailed in Paper~I. Throughout this work, we
collect terms of the same order in $\epsilon$ and set $\epsilon=1$ after the
expansion is performed. In Schwarzschild-like coordinates $(t,r,\theta,\phi)$,
the metric of the rotating star takes the form
\begin{equation}
\begin{split}
ds^{2} &= -e^{\nu(r)}\bigl[1 + 2h(r,\theta)\bigr]dt^{2} \\
&\quad + e^{\lambda(r)}\left[1 + \frac{2m(r,\theta)}{r-2M(r)}\right]dr^{2} \\
&\quad + r^{2}\bigl[1 + 2k(r,\theta)\bigr] \\
&\quad \times \Big\{d\theta^{2} + \sin^{2}\theta\left[d\phi - \omega(r,\theta)\,dt \right]^{2}\Big\},
\end{split}
\label{eq:metric}
\end{equation}
where $\nu(r)$ and $M(r)$ are the background metric potential and enclosed
gravitational mass of the non rotating configuration. The metric function
$\lambda(r)$ is related to $M(r)$ through
\begin{equation}
e^{\lambda(r)} \equiv \left[1 - \frac{2M(r)}{r}\right]^{-1}.
\label{eq:lambda}
\end{equation}
The functions $h$, $k$, $m$, and $\omega$ encode the rotational perturbations of
the metric and are expanded jointly in $\epsilon$ and in spherical harmonics as
\begin{align}
h(r,\theta) &= \sum_{\substack{s = 2 \\ (\mathsf{even})}}^{6} \ \sum_{\substack{\ell = 0 \\ (\mathsf{even})}}^{s} \epsilon^{s}\, h^{(s)}_{\ell}(r)\,P_{\ell}(\cos\theta), \\[1ex]
k(r,\theta) &= \sum_{\substack{s = 2 \\ (\mathsf{even})}}^{6} \ \sum_{\substack{\ell = 2 \\ (\mathsf{even})}}^{s} \epsilon^{s}\, k^{(s)}_{\ell}(r)\,P_{\ell}(\cos\theta), \\[1ex]
m(r,\theta) &= \sum_{\substack{s = 2 \\ (\mathsf{even})}}^{6} \ \sum_{\substack{\ell = 0 \\ (\mathsf{even})}}^{s} \epsilon^{s}\, m^{(s)}_{\ell}(r)\,P_{\ell}(\cos\theta), \\[1ex]
\omega(r,\theta) &= \sum_{\substack{k = 1 \\ (\mathsf{odd})}}^{7} \ \sum_{\substack{\ell = 1 \\ (\mathsf{odd})}}^{k} \epsilon^{k}\, \omega^{(k)}_{\ell}(r)\,\frac{dP_{\ell}(\cos\theta)}{d\cos\theta} \, .
\end{align}
Axisymmetry leaves only the azimuthally-independent (i.e.~$m$--independnet)
harmonics. The labels $(\mathsf{even})$ and $(\mathsf{odd})$ underneath the
summation symbols indicate that each sum runs in even or odd steps. For the
even-parity functions $h$, $k$, and $m$, the order $s$ takes the values
$s=2,4,6$. At each order, the mode $\ell$ runs over $\ell=0,2,\ldots,s$ for $h$
and $m$, and over $\ell=2,4,\ldots,s$ for $k$. The $\ell=0$ term of $k$
contributes only a function of the radial coordinate, which shifts the areal
radius and can always be removed by a coordinate transformation. We therefore
set $k^{(s)}_{0}=0$ at each even order, as in
Paper~I~\cite{Conde-Ocazionez:2025agu}. For the odd-parity function $\omega$,
the order $k$ takes the values $k=1,3,5,7$, and at each order the mode $\ell$
runs over $\ell=1,3,\ldots,k$. The even-parity sums stop at $s=6$ because at odd
orders only the odd-parity function $\omega$ receives new contributions, so the
seventh-order calculation adds nothing to $h$, $k$ or $m$. The order and mode
content are summarized in Table~\ref{tab:modes}, where the general order $n$ can
be either $s$ or $k$ depending on whether the quantity is even-parity or
odd-parity, respectively.

\begin{table}[h]
\centering
\begin{tabular}{@{\hspace{0.7cm}} c @{\hspace{0.7cm}} c @{\hspace{0.7cm}} c @{\hspace{0.7cm}}}
\noalign{\hrule height 1.0pt}
\rule{0pt}{10pt} Order $n$ & Sector & Modes $\ell$ \\[3pt]
\noalign{\hrule height 0.7pt}
\\[-1ex]
$1$ & odd & $1$ \\[4pt]
$2$ & even & $0, 2$ \\[4pt]
$3$ & odd & $1, 3$ \\[4pt]
$4$ & even & $0, 2, 4$ \\[4pt]
$5$ & odd & $1, 3, 5$ \\[4pt]
$6$ & even & $0, 2, 4, 6$ \\[4pt]
$7$ & odd & $1, 3, 5, 7$ \\[4pt]
\noalign{\hrule height 0.7pt}
\end{tabular}
\caption{Nonvanishing harmonic modes $\ell$ at each perturbative order $n$ in the slow-rotation expansion up to $n=7$th order. Odd orders belong to the odd-parity sector and have non-vanishing odd-$\ell$ modes, while even orders belong to the even-parity sector and have non vanishing even-$\ell$ modes (including $\ell = 0$). At each order $n$, exactly one new mode is introduced, with maximum $\ell = n$. }
\label{tab:modes}
\end{table}

The matter source is a uniformly slowly rotating, barotropic, and unmagnetized
neutron star. Under these conditions, the stellar fluid is well described as a
perfect fluid, with stress-energy tensor
\begin{equation}
T_{\mu\nu} = (p+\varepsilon)\,u_{\mu}u_{\nu} + p\,g_{\mu\nu},
\end{equation}
where $p$ and $\varepsilon$ are the pressure and total mass-energy density of
the fluid, and $u^{\mu}$ is its four-velocity. Like the metric functions, $p$,
$\varepsilon$, and $u^{\mu}$ in general acquire rotational corrections
order-by-order in $\epsilon$. A valid perturbative expansion requires small
fractional changes in pressure and energy density under rotation. Adopting
Hartle's framework~\cite{Hartle:1967he}, we map the ellipsoidal contours of the
rotating state onto spherical shells of the static configuration. Each shell is
labeled by the radial coordinate $R$ of the nonrotating star, and the map is
given by the coordinate transformation
\begin{equation}
r(R,\Theta) = R + \xi(R,\Theta), \qquad \theta(R,\Theta) = \Theta,
\label{eq:coordinate_transformation}
\end{equation}
where $\xi(R,\Theta)$ is the radial displacement induced by the star's
rotational deformation. The mapping exists when the isobaric surfaces
(i.e.~surfaces of constant pressure) match isopycnal ones (i.e.~surfaces of
constant density). Since $\xi$ transforms as a scalar under rotations, it is
expanded in even powers of $\epsilon$ and decomposed into Legendre polynomials
as
\begin{equation}
\xi(R,\Theta) = \sum_{\substack{s = 2 \\ (\mathsf{even})}}^{6} \ \sum_{\substack{\ell = 0 \\ (\mathsf{even})}}^{s} \epsilon^{s}\, \xi^{(s)}_{\ell}(R)\,P_{\ell}(\cos\Theta),
\label{eq:xi_expansion}
\end{equation}
in direct analogy with $h$, $k$, and $m$ above. We refer to the resulting frame
with coordinates $(t,R,\Theta,\phi)$ as the Hartle-Thorne (\textsf{HT}) frame.

In the \textsf{HT} frame, some of the fields become functions of \textsf{HT}
radius only. By construction, the pressure and energy density carry no
rotational corrections along the mapped surfaces,
\begin{align}
p[r(R,\Theta),\theta] &= p(R) = p^{(0)}(R), \label{eq:pHT} \\[1ex]
\varepsilon[r(R,\Theta),\theta] &= \varepsilon(R) = \varepsilon^{(0)}(R), \label{eq:eHT}
\end{align}
so that $p = p(R)$ and $\varepsilon = \varepsilon(R)$ throughout this work. The
fluid's four-velocity, by contrast, does acquire nonzero spin corrections, since
rotation perturbs the fluid's motion along the axial flow. Rigid rotation with
angular velocity $\Omega$ fixes its form to
\begin{equation}
u^{\mu} = \bigl(u^{t},\,0,\,0,\,\epsilon\Omega\,u^{t}\bigr),
\end{equation}
with the remaining component $u^{t}$ determined by the normalization condition
$u^{\mu}u_{\mu} = -1$.

When numerically solving for the interior structure of the rotating neutron
star, we here work in the \textsf{HT} frame, which requires transforming the
metric in Eq.~\eqref{eq:metric} using the coordinate transformation of
Eq.~\eqref{eq:coordinate_transformation} along with the decomposition in
Eq.~\eqref{eq:xi_expansion}. With the metric and stress-energy tensor defined in
the \textsf{HT} frame, the Einstein equations are expanded as a power series in
$\epsilon$, yielding the perturbed field equations at each spin-frequency order
$n$. A complete discussion and derivation of this procedure can be found in
Paper I. In the following subsections, we extend Paper I by presenting the
equations in a suitable form for numerical integration.

\subsection{Background configuration}
\label{sec:background}

At $\mathcal{O}(\epsilon^{0})$, the metric in Eq.~\eqref{eq:metric} reduces to a
static, spherically symmetric line element. Using the $(t,t)$ and $(R,R)$
components of Einstein's equations, along with the radial component of the
Bianchi identity, we obtain the equations of stellar structure of the
non-rotating configuration,
\begin{align}
\frac{dM}{dR} &= 4\pi R^{2}\varepsilon, \label{eq:enclosed} \\[1ex]
\frac{d\nu}{dR} &= 2\,\frac{M + 4\pi R^{3}p}{R(R-2M)}, \label{eq:nu} \\[1ex]
\frac{dp}{dR} &= -(\varepsilon+p)\,\frac{M + 4\pi R^{3}p}{R(R-2M)}. \label{eq:TOV}
\end{align}
Equations~\eqref{eq:enclosed}--\eqref{eq:TOV}, together with a barotropic EoS
$p=p(\varepsilon)$, form a closed system of ordinary differential equations for
the metric functions $M$ and $\nu$, and the fluid variables $p$ and
$\varepsilon$. The solution inside the star is obtained by numerically
integrating these equations from the center outward until the stellar surface is
reached.

The global solution is obtained by matching the interior solution to the
exterior one at the stellar surface, $R=R_*$. This matching determines the
stellar radius $R_*$ and gravitational mass $M_*$, as follows. Beyond the
stellar surface, the pressure and energy density vanish, $p_{\mathsf{ext}}(R)=0$
and $\varepsilon_{\mathsf{ext}}(R)=0$, and
Eqs.~\eqref{eq:enclosed}--\eqref{eq:TOV} reduce to the vacuum Einstein
equations. By Birkhoff's theorem, their unique solution is the Schwarzschild
metric,
\begin{align}
M_{\mathsf{ext}}(R) &= M_*, \\[1ex]
\nu_{\mathsf{ext}}(R) &= \ln f(R),
\label{eq:nu_exterior}
\end{align}
with the Schwarzschild factor
\begin{equation}
f(R) = 1 - \frac{2M_*}{R}.
\end{equation}
Continuity of the metric functions at the stellar surface then gives the
matching conditions
\begin{align}
p(R_*) &= 0, \label{eq:p_match} \\[1ex]
M(R_*) &= M_*, \label{eq:M_match} \\[1ex]
\nu(R_*) &= \ln f(R_*). \label{eq:nu_match}
\end{align}
The stellar radius $R_*$ follows from the zero-pressure condition in
Eq.~\eqref{eq:p_match}, and the gravitational mass $M_*$ then follows from
Eq.~\eqref{eq:M_match} once $R_*$ is known.

In practice, the exact location where pressure and energy density vanish is not
known \textit{a priori}, and thus, a numerical integration step is extremely
unlikely to land precisely on $p=0$. A more convenient reformulation is obtained
by Lindblom~\cite{lindblom1992determining} using the dimensionless
\textit{pseudo-enthalpy}\footnote{At zero temperature, the baryon chemical
potential is also the thermodynamic enthalpy per baryon, $\mu_B=
(\varepsilon+p)/n_B$, which is not equal to $\mathsf{h}$. With $\mathsf{h}=0$ at
the stellar surface, integrating $d\mathsf{h}=d\mu_B/\mu_B$ gives
$\mathsf{h}=\ln[\mu_B/\mu_B(p=0)]$; we call this dimensionless logarithmic
variable the pseudo-enthalpy to differentiate it from $\mu_B$} $\mathsf{h}$,
defined through
\begin{equation}
d\mathsf{h} \equiv \frac{dp}{\varepsilon+p} = \frac{d\mu_B}{\mu_B}\,,
\label{eq:h}
\end{equation}
where $\mu_B \equiv (\varepsilon+p)/n_B$ is the baryon chemical potential and
$n_B$ the baryon number density at $T=0$. In terms of $\mathsf{h}$,
Eqs.~\eqref{eq:enclosed}--\eqref{eq:TOV} become
\begin{align}
\frac{dR}{d\mathsf{h}} &= -\frac{R(R-2M)}{M+4\pi R^{3}p}, \label{eq:Rh} \\[1ex]
\frac{dM}{d\mathsf{h}} &= -4\pi R^{2}\varepsilon\,\frac{R(R-2M)}{M+4\pi R^{3}p}, \label{eq:Mh} \\[1ex]
\frac{d\nu}{d\mathsf{h}} &= -2. \label{eq:nuh}
\end{align}
The set of Eqs.~\eqref{eq:h}, \eqref{eq:Rh}, and~\eqref{eq:Mh}, together with
the EoS $p=p(\varepsilon)$, form a closed system for the variables
$p(\mathsf{h})$, $\varepsilon(\mathsf{h})$, $R(\mathsf{h})$, and
$M(\mathsf{h})$. From the barotropic EoS $p=p(\varepsilon)$, we first compute
$\mathsf{h}(\varepsilon)$ through
\begin{align}
\mathsf{h}(\varepsilon) = \int_{0}^{\varepsilon} \frac{c_s^2(\varepsilon')}{\varepsilon'+p(\varepsilon')}\,d\varepsilon', \label{eq:hofe}
\end{align}
where the squared sound speed\footnote{For a multi-dimensional EoS, the
adiabatic sound speed must be defined by holding the entropy density per baryon
number density fixed. However, for the barotropic, one-dimensional case
considered here, the pressure only depends on the energy density.} is $c_s^2
\equiv dp/d\varepsilon$. Then, we invert the result to express $\varepsilon$ as
a function of $\mathsf{h}$, i.e., $\varepsilon = \varepsilon(\mathsf{h})$. The
pressure $p(\mathsf{h})$ is then computed from the EoS as $p(\mathsf{h}) =
p[\varepsilon(\mathsf{h})]$, or equivalently,
\begin{align}
\mathsf{h}(p) &= \int_{0}^{p} \frac{dp'}{\varepsilon(p')+p'}\,. \label{eq:hofp}
\end{align}
Integrating Eq.~\eqref{eq:nuh} immediately yields
\begin{equation}
\nu(\mathsf{h}) = -2\mathsf{h} + \nu_*,
\label{eq:nu_interior}
\end{equation}
where $\nu_*$ is an integration constant. From Eq.~\eqref{eq:hofp}, the stellar
surface is located exactly at $\mathsf{h}=0$, where $p=0$. Matching the interior
solution, Eq.~\eqref{eq:nu_interior}, to the exterior one,
Eq.~\eqref{eq:nu_exterior}, at $\mathsf{h}=0$ ($R=R_*$) then yields
\begin{equation}
\nu_* = \ln f(R_*),
\end{equation}
consistent with the matching condition already found in Eq.~\eqref{eq:nu_match}.

\subsubsection*{Boundary conditions}

Equations~\eqref{eq:Rh} and~\eqref{eq:Mh} become singular at the stellar center,
where $\mathsf{h} = \mathsf{h}_c \equiv \mathsf{h}(\varepsilon=\varepsilon_c)$,
given by Eq.~\eqref{eq:hofe}. To obtain regular boundary conditions, we perform
an asymptotic expansion of $R(\mathsf{h})$ and $M(\mathsf{h})$ about
$\mathsf{h}_c$, and initiate the numerical integration at a nearby initial value
$\mathsf{h}_i$, slightly displaced from the center (i.e. $h_i<h_c$ or in other
words $R(h_i)>R(h_c)$ where $R(h_c)=0$). The resulting boundary conditions at
$\mathsf{h}=\mathsf{h}_i$ are
\begin{align}
R(\mathsf{h}_i) &\simeq \left[\frac{3(\mathsf{h}_c-\mathsf{h}_i)}{2\pi(\varepsilon_c+3p_c)}\right]^{1/2} \equiv R_{i}, \label{eq:Rofh} \\[1ex]
M(\mathsf{h}_i) &\simeq \frac{4\pi}{3}\varepsilon_c R_{i}^{3}, \label{eq:Mofh}
\end{align}
where $R_{i}$ is a small radial displacement from the center used to initialize
the integration. Equivalently, inverting Eq.~\eqref{eq:Rofh} gives
$\mathsf{h}_i$ directly in terms of $R_{i}$,
\begin{equation}
\label{eq:hi}
\mathsf{h}_i = \mathsf{h}_c - \frac{2\pi}{3}R_{i}^{2}\left(\varepsilon_c+3p_c\right),
\end{equation}
where $p_c$ and $\varepsilon_c$ are the values of the pressure and the total
energy density at the center of the star. As established above, the stellar
surface corresponds to $\mathsf{h}=0$, where $p=0$ by Eq.~\eqref{eq:hofp}. The
integration therefore proceeds from $\mathsf{h}=\mathsf{h}_i$ to $\mathsf{h}=0$,
and the stellar radius and mass follow from
\begin{align}
R_* &= R(\mathsf{h}=0), \\[1ex]
M_* &= M(\mathsf{h}=0) \, .
\end{align}
One could, in principle carry out these asymptotic expansions to higher order in
$\mathsf{h}_c - \mathsf{h}_i$, or equivalently, in $R_i$, but we find that this
is not necessary if one chooses a sufficiently small initial $R_i$, as we
discuss in Sec.~\ref{sec:single_star}.

\subsection{Odd-parity sector}
\label{sec:interior_odd}

In the odd-parity sector, the perturbation equations govern the metric functions
associated with the rotational degrees of freedom of the stellar fluid. At first
order $n = 1$ and mode $\ell = 1$, the equations reduce to a single second-order
homogeneous ODE for the metric function $\varpi^{(1)}_{1} \equiv \Omega -
\omega^{(1)}_{1}$, the first-order angular velocity of the stellar fluid
relative to the local inertial frame~\cite{Hartle:1967he}. At higher odd orders
$n = k = 3, 5, 7$, the equations are second-order inhomogeneous ODEs for the
functions $\omega^{(k)}_{\ell}$, sourced by lower-order perturbation functions.
Any solution can then be decomposed into a particular part
$\omega^{(k),\mathsf{P}}_{\ell}$, driven entirely by the source term, and a
homogeneous part $\omega^{(k),\mathsf{H}}_{\ell}$, scaled by the interior
integration constant $C^{(k)}_{\ell, \mathsf{int}}$. Throughout, the
superscripts $\mathsf{H}$ and $\mathsf{P}$ denote the homogeneous and particular
parts, respectively. The general interior solution is then written as
\begin{equation}
\omega^{(k)}_{\ell}(R) = \omega^{(k),\mathsf{P}}_{\ell}(R) \, + \,  C^{(k)}_{\ell, \mathsf{int}}\,
\omega^{(k),\mathsf{H}}_{\ell}(R) \, ,
\label{eq:general_solution_odd}
\end{equation}
where $C^{(k)}_{\ell, \mathsf{int}}$ is uniquely determined by matching at the
stellar surface\footnote{Such a decomposition is general for any linear ODE. At
order $\mathcal{O}(\epsilon)$, the equation is homogeneous, so the particular
part vanishes identically and the solution takes the form $\varpi^{(1)}_1 =
C^{(1)}_{1,\mathsf{int}}\, \varpi^{(1),\mathsf{H}}_{1}$, where
$C^{(1)}_{1,\mathsf{int}}$ is an arbitrary scaling constant, reflecting the
scale invariance of the homogeneous equation.}. Each piece of the decomposition
satisfies its own ODE system: the homogeneous part is governed by the
source-free equations, while the particular part is driven by the source term.

For numerical convenience, let us reduce each second-order ODE to a system of
two, coupled, first-order ODEs. At $\mathcal{O}(\epsilon)$, we introduce the
auxiliary variable $\bar{\alpha}^{(1)}_{1} \equiv d\varpi^{(1)}_{1}/dR$, so that
the differential system is given by
\begin{align}
    \frac{d}{dR}
        \begin{bmatrix}
        \varpi^{(1)}_{1} \\[1ex]
        \bar{\alpha}^{(1)}_{1}
        \end{bmatrix}
        &=
        \begin{bmatrix}
        0 & 1 \\[1ex]
        X_{\bar{\alpha}^{(1)}_{1}} & Y_{\bar{\alpha}^{(1)}_{1}}
        \end{bmatrix}
        \begin{bmatrix}
        \varpi^{(1)}_{1} \\[1ex]
        \bar{\alpha}^{(1)}_{1}
        \end{bmatrix},
        \label{eq:varpi_alpha_system}
\end{align}
where
\begin{align}
X_{\bar{\alpha}^{(1)}_{1}} &= 16\pi (\varepsilon + p) e^{\lambda},
\\[1ex]
Y_{\bar{\alpha}^{(1)}_{1}} &= - 4 \frac{1-\pi R^{2}(\varepsilon + p)
e^{\lambda}}{R}.
\end{align}
Here all background quantities are evaluated in the Hartle--Thorne radial
coordinate $R$, with $\lambda$ given by Eq.~\eqref{eq:lambda}.

At higher odd orders $n=k = 3, 5, 7$, the particular solution is non-trivial and
thus we introduce $\alpha^{(k),\mathsf{H}}_{\ell} \equiv
d\omega^{(k),\mathsf{H}}_{\ell}/dR$ and $\alpha^{(k),\mathsf{P}}_{\ell} \equiv
d\omega^{(k),\mathsf{P}}_{\ell}/dR$ for each component. The homogeneous
differential system is
\begin{align}
    \frac{d}{dR}
        \begin{bmatrix}
        \omega^{(k),\mathsf{H}}_{\ell} \\[1ex]
        \alpha^{(k),\mathsf{H}}_{\ell}
        \end{bmatrix}
        &=
        \begin{bmatrix}
        0 & 1 \\[1ex]
        X_{\alpha^{(k)}_{\ell}} & Y_{\alpha^{(k)}_{\ell}}
        \end{bmatrix}
        \begin{bmatrix}
        \omega^{(k),\mathsf{H}}_{\ell} \\[1ex]
        \alpha^{(k),\mathsf{H}}_{\ell}
        \end{bmatrix},
        \label{eq:omega_homogeneous}
\end{align}
while the particular differential system is
\begin{align}
    \frac{d}{dR}
        \begin{bmatrix}
        \omega^{(k),\mathsf{P}}_{\ell} \\[1ex]
        \alpha^{(k),\mathsf{P}}_{\ell}
        \end{bmatrix}
        &=
        \begin{bmatrix}
        0 & 1 \\[1ex]
        X_{\alpha^{(k)}_{\ell}} & Y_{\alpha^{(k)}_{\ell}}
        \end{bmatrix}
        \begin{bmatrix}
        \omega^{(k),\mathsf{P}}_{\ell} \\[1ex]
        \alpha^{(k),\mathsf{P}}_{\ell}
        \end{bmatrix}
        +
        \begin{bmatrix}
        0 \\[1ex]
        \mathcal{S}_{\alpha^{(k)}_{\ell}}
        \end{bmatrix} \, .
    \label{eq:omega_particular}
\end{align}
The quantities $X_{\alpha^{(k)}_{\ell}}$ and $Y_{\alpha^{(k)}_{\ell}}$ are
coefficient functions that depend only on the non-rotating fluid and metric
functions,
\begin{align}
X_{\alpha^{(k)}_{\ell}} &= \left[ \frac{(\ell+2)(\ell-1)}{R^{2}} +
16\pi (\varepsilon + p) \right] e^{\lambda}, \\[1ex]
Y_{\alpha^{(k)}_{\ell}} &= Y_{\bar{\alpha}^{(1)}_{1}},
\end{align}
and $\mathcal{S}_{\alpha^{(k)}_{\ell}}$ denotes the source term driven by
lower-order perturbation functions. Note that for $\ell = 1$,
$X_{\alpha^{(k)}_{1}}$ reduces to $X_{\bar{\alpha}^{(1)}_{1}}$, and the
homogeneous system recovers the same structural form as the first-order
expression in Eq.~\eqref{eq:varpi_alpha_system}, which governs the mass-current
dipole mode of the star. The coefficient $Y_{\alpha^{(k)}_{\ell}}$ is the same
across all odd-parity modes and coincides with $Y_{\bar{\alpha}^{(1)}_{1}}$,
reflecting a common structural feature of the odd-parity sector. The structure
of the source terms is discussed in Sec.~\ref{sec:source}.

The complete set of metric perturbation functions that must be solved
numerically in this sector is
\begin{equation*}
\left\{ \varpi^{(1)}_{1},\, \omega^{(3)}_{1},\, \omega^{(3)}_{3},\,
\omega^{(5)}_{1},\, \omega^{(5)}_{3},\, \omega^{(5)}_{5},\,
\omega^{(7)}_{1},\, \omega^{(7)}_{3},\, \omega^{(7)}_{5},\,
\omega^{(7)}_{7} \right\}\,.
\end{equation*}
Including their derivatives $\bar{\alpha}^{(1)}_1$ and $\alpha^{(k)}_\ell$ and
split into homogeneous and particular components for orders $n>1$, the system
totals 38 unknown radial functions.

\subsubsection*{Boundary conditions}

To initiate the integration, boundary conditions must be specified at the
center, where regularity should be meticulously preserved. In
Paper~I~\cite{Conde-Ocazionez:2025agu}, we derived in detail a regular
asymptotic expansion of each ODE system around $R = 0$, so that one could
initialize the numerical integration by evaluating the resulting near-center
expansions at the starting radius $R_i$. We summarize the boundary conditions
required for both the homogeneous and particular systems of the odd-parity
sector below.

At order $\mathcal{O}(\epsilon)$, the asymptotic expansion around $R = 0$ yields
\begin{align}
\label{eq:wb11_asym}
\varpi^{(1)}_{1}(R) &= \varpi^{(1)}_{1,c} \left[ 1 \, + \, \dfrac{8}{5} \pi (p_{c} + \varepsilon_{c}) R^{2} \, + \, \mathcal{O}\left(R^{4}\right) \right] \, ,  \\[1ex]
\label{eq:ab11_asym}
\bar{\alpha}^{(1)}_{1}(R) &=  \dfrac{16}{5} \pi \varpi^{(1)}_{1,c}  (p_{c} + \varepsilon_{c}) R \, + \, \mathcal{O}\left(R^{3}\right) \, ,
\end{align}
where $\varpi^{(1)}_{1,c} \equiv \varpi^{(1)}_{1}(R=0)$ is the central value of
the metric function, and recall that $p_c \equiv p(R=0)$ and $\varepsilon_c
\equiv \varepsilon(R=0)$ denote the central pressure and energy density of the
non-rotating background configuration, respectively. Notice that although the
numerical integrations start at $R=R_i$, we have a global solution that extends
to the center of the star, $R=0$, throught he asymptotic expansions carried out.

For the homogeneous system at orders $n=k = 3, 5, 7$ and modes $\ell$ as listed
in Table~\ref{tab:modes}, the asymptotic expansion around $R = 0$ yields
\begin{align}
\label{eq:wkl_H_asym}
\omega^{(k),\mathsf{H}}_{\ell}(R) &= A^{(k), \mathsf{H}}_{\ell,\mathsf{int}} R^{ \ell-1} \left[ 1 \, + \,\mathcal{B}_{\ell} R^{2} \, + \, \mathcal{O}\left(R^{4}\right) \right]\, ,  \\[1ex] \nonumber
\alpha^{(k),\mathsf{H}}_{\ell}(R) &= A^{(k),\mathsf{H}}_{\ell,\mathsf{int}} R^{\ell-2} \Big[(\ell-1) \, \\
&+ \,  (\ell+1)\,\mathcal{B}_{\ell} R^{2} \,
+ \, \mathcal{O}\left(R^{4}\right) \Big] \, ,
\label{eq:akl_H_asym}
\end{align}
where $A^{(k),\mathsf{H}}_{\ell,\mathsf{int}}$ is a free asymptotic constant
that sets the overall amplitude of the homogeneous solution, and
$\mathcal{B}_{\ell}$ is a coefficient determined entirely by the central values
of the pressure and energy density, given as
\begin{align}
 \mathcal{B}_{\ell} := \pi \left[ \dfrac{2(\ell + 3)}{2\ell +3} p_{c} +  \dfrac{2(2\ell^{2} + 5\ell + 5)}{3(2\ell + 3)} \varepsilon_{c}  \right] \, .
\end{align}
Note that from Eq.~\eqref{eq:wkl_H_asym}, the $\ell=1$ asymptotic constants,
$A^{(k),\mathsf{H}}_{1,\mathsf{int}}$, are just the central value of the metric
function, $A^{(k),\mathsf{H}}_{1,\mathsf{int}} = \omega^{(k),\mathsf{H}}_{1}(0)
\equiv \omega^{(k),\mathsf{H}}_{1,c}$ for all $k$, in direct analogy with
$\varpi^{(1)}_{1,c}$ at first order.

For the particular system at orders $n = k = 3, 5, 7$, the asymptotic expansion
around $R = 0$ gives

\begin{align}
\label{eq:wkl_P_asym}
\nonumber
\omega^{(k),\mathsf{P}}_{\ell}(R) &= A^{(k),\mathsf{P}}_{\ell,\mathsf{int}} R^{\ell-1} \, \\[1ex]
&+ \, \Big[ \mathcal{B}_{\ell}\,A^{(k),\mathsf{P}}_{\ell,\mathsf{int}} \, + \, \widetilde{\mathcal{B}}^{(k)}_{\ell} \Big] R^{\ell+1} \, + \, \mathcal{O}\left(R^{\ell+3}\right) , \\[1ex] \nonumber
\alpha^{(k),\mathsf{P}}_{\ell}(R) &= (\ell-1)\,A^{(k),\mathsf{P}}_{\ell,\mathsf{int}} R^{\ell-2} \\
&+ \, (\ell+1)\Big[ \mathcal{B}_{\ell}\,A^{(k),\mathsf{P}}_{\ell,\mathsf{int}} \, + \, \widetilde{\mathcal{B}}^{(k)}_{\ell} \Big] R^{\ell} \, + \, \mathcal{O}\left(R^{\ell+2}\right) ,
\label{eq:akl_P_asym}
\end{align}
where $A^{(k),\mathsf{P}}_{\ell,\mathsf{int}}$ is a free asymptotic constant of
the particular solution. The particular solution is therefore \emph{affine} in
that constant: each coefficient beyond the leading one is the sum of a term
proportional to $A^{(k),\mathsf{P}}_{\ell,\mathsf{int}}$, with
$\mathcal{B}_{\ell}$ from the homogeneous expansion, and a source-led term
$\widetilde{\mathcal{B}}^{(k)}_{\ell}$. The asymptotic expansions from which
$\widetilde{\mathcal{B}}^{(k)}_{\ell}$ can be read off are given in full through
third order in Appendix~\ref{app:bc}.

The source coefficient $\widetilde{\mathcal{B}}^{(k)}_{\ell}$ depends not only
on the central values $\varepsilon_c$, $p_c$, and $\nu_c$, but also on the
asymptotic constants of all lower-order asymptotic solutions,
\begin{align}
 \widetilde{\mathcal{B}}^{(k)}_{\ell} = \widetilde{\mathcal{B}}^{(k)}_{\ell}\left( \varepsilon_{c}, p_{c}, \nu_{c} ; \left\{ A^{(n < k)}_{\ell', \mathsf{int}} \right\} \right) \, ,
\end{align}
with $\ell' \le n$. In the limit $\left\{ A^{(n < k)}_{\ell', \mathsf{int}}
\right\} \to 0$ the source terms vanish, $\widetilde{\mathcal{B}}^{(k)}_{\ell}
\to 0$, and the particular solution reduces to a multiple of the homogeneous
one, as it must when the sourced equation becomes homogeneous.

This dependence reflects the iterative structure of the problem. \emph{The
source terms at a fixed spin-frequency order are built from solutions at all
previous orders, and their behavior near the origin is inherited by the
particular solution through the asymptotic expansion.} Therefore, the system of
equations must be solved sequentially, order-by-order. Once all orders are
solved, the exterior integration constants $C^{(k)}_{\ell,\mathsf{ext}}$
determine the mass-current multipole moments $S_1, S_3, S_5, S_7$ and their
spin-induced corrections through the analytical expressions of Paper~I.

\subsection{Even-parity sector}
\label{sec:interior_even}

The even-parity sector consists of the corrections in mass $m^{(s)}_{\ell}$,
radial displacement $\xi^{(s)}_{\ell}$, angular metric deformation
$k^{(s)}_{\ell}$, and lapse function $h^{(s)}_{\ell}$, where $k^{(s)}_{\ell}$
does not have a monopole piece. The monopole ($\ell=0$) the spherically
symmetric response of the star to rotation, while the higher even modes
($\ell=2,4,\ldots$) describe its departure from spherical symmetry. The two
cases differ not only in their physical interpretation but also in the structure
of the perturbation equations, and are therefore treated separately.

The even-parity monopole is a special case for which a decomposition into
homogeneous and particular parts is not necessary, and the reason is the
following. The monopole perturbation consists of a response to rotation and a
homogeneous contribution corresponding to a shift along the TOV sequence, in
general. Fixing the central density eliminates the nontrivial homogeneous
solution, and thus the particular equation determines the solution uniquely.
Accordingly, the matching procedure for this mode follows more directly by
solving the full inhomogeneous system and imposing continuity of the solutions
at the stellar surface, as detailed in Paper~I and in Sec.~\ref{sec:matching}.

Mathematically, the differential system for the even-parity monopole case reads
\begin{align}
\frac{d}{dR}
\begin{bmatrix}
m^{(s)}_{0} \\[1ex]
\xi^{(s)}_{0} \\[1ex]
h^{(s)}_{0}
\end{bmatrix}
&=
\begin{bmatrix}
0 & Y_{m^{(s)}_{0}} & 0 \\[1ex]
X_{\xi^{(s)}_{0}} & Y_{\xi^{(s)}_{0}} & 0 \\[1ex]
X_{h^{(s)}_{0}} & Y_{h^{(s)}_{0}} & 0
\end{bmatrix}
\begin{bmatrix}
m^{(s)}_{0} \\[1ex]
\xi^{(s)}_{0} \\[1ex]
h^{(s)}_{0}
\end{bmatrix}
+
\begin{bmatrix}
\mathcal{S}_{m^{(s)}_{0}} \\[1ex]
\mathcal{S}_{\xi^{(s)}_{0}} \\[1ex]
\mathcal{S}_{h^{(s)}_{0}}
\end{bmatrix}
\label{eq:mxh_system}
\end{align}
where the coefficient functions are given by
\begin{align}
X_{\xi^{(s)}_{0}} &= -\dfrac{1+8 \pi p R^{2}}{M + 4 \pi p R^{3} }e^{\lambda} \, , \\[1ex]
X_{h^{(s)}_{0}} &= \dfrac{1 + 8 \pi p R^{2}}{R^{2}}e^{2 \lambda} \, , \\[1ex]
Y_{m^{(s)}_{0}}   &= -4 \pi R^{2} \dfrac{d\varepsilon}{dR} \, ,\\[1ex]
Y_{\xi^{(s)}_{0}} &=
\dfrac{2e^{\lambda}}{R^{2} \big(M+ 4\pi p R^{3} \big)} \Big[M(R+8\pi p R^{3})  \,
\nonumber \\
&\quad
- 2 \pi R^{4} \big(p + \varepsilon
+ 8 \pi p R^{2} \varepsilon \big) - M^{2} \Big] \, , \\
Y_{h^{(s)}_{0}} &= \dfrac{4 \pi  (M + 4 \pi p R^{3})(p +\varepsilon)}{R}e^{2\lambda} \, ,
\end{align}
and $\mathcal{S}_{m^{(s)}_{0}}$, $\mathcal{S}_{\xi^{(s)}_{0}}$,
$\mathcal{S}_{h^{(s)}_{0}}$, where $s = 2, 4, 6$, denote the source terms driven
by lower-order perturbation functions, which are discussed in
Sec.~\ref{sec:source} and App.~\ref{app:Source}, and are available in this
paper's accompanying repository. The complete set of metric perturbation
functions for the even-parity, $\ell = 0$ case is
\begin{equation*}
\left\{ m^{(2)}_{0},\, \xi^{(2)}_{0},\, h^{(2)}_{0},\, m^{(4)}_{0},\,
\xi^{(4)}_{0},\, h^{(4)}_{0},\, m^{(6)}_{0},\, \xi^{(6)}_{0},\,
h^{(6)}_{0} \right\},
\end{equation*}
comprising nine radial functions in total. The third column of the coefficient
matrix in Eq.~\eqref{eq:mxh_system} vanishes, so no equation in the system
depends on $h^{(s)}_{0}$. The equations for $m^{(s)}_{0}$ and $\xi^{(s)}_{0}$
form a self-contained pair, and $h^{(s)}_{0}$ follows from them by a single
quadrature, which can be used to decouple the equations, as discussed in
Sec.~\ref{sec:numerical}.

For the even-parity modes with $\ell \geq 2$, the functions $h^{(s)}_{\ell}$ and
$k^{(s)}_{\ell}$ satisfy a system of two coupled first-order ODEs, while
$m^{(s)}_{\ell}$ and $\xi^{(s)}_{\ell}$ are determined algebraically from
$h^{(s)}_{\ell}$ once the ODE system has been solved. In principle, such ODE
system is written in terms of $h^{(s)}_{\ell}$ and $k^{(s)}_{\ell}$, but this
formulation is not well-suited for numerical integration. To see this, note from
Paper~I~\cite{Conde-Ocazionez:2025agu} that the coupled ODE system for
$h^{(s)}_{\ell}$ and $k^{(s)}_{\ell}$ can be written in matrix form as
\begin{equation}
\mathbf{M}
\dfrac{d}{dR}
\begin{bmatrix}
h^{(s)}_{\ell} \\[1ex]
k^{(s)}_{\ell}
\end{bmatrix}
=
\begin{bmatrix}
F^{(s)}_{h,\ell} \\[1ex]
F^{(s)}_{k,\ell}
\end{bmatrix},
\label{eq:hk_matrix}
\end{equation}
where $\mathbf{M}$ is the $2 \times 2$ matrix
\begin{equation}
\mathbf{M} =
\begin{bmatrix}
1 & \dfrac{R - M + 4\pi p R^{3}}{R}\,e^{\lambda} \\[2ex]
1 & 1
\end{bmatrix},
\end{equation}
with
\begin{align}
F^{(s)}_{h,\ell} &= \left[\dfrac{\ell(\ell+1)}{2R} -
4\pi(\varepsilon + p)R\right]e^{\lambda} h^{(s)}_{\ell} \\[1ex] \nonumber
&+ \dfrac{(\ell+2)(\ell-1)}{2R}e^{\lambda} k^{(s)}_{\ell}
+ \dfrac{1+8\pi p R^{2}}{R^{2}}e^{2\lambda} m^{(s)}_{\ell}
+ S_{h^{(s)}_{\ell}}, \\[2ex] \nonumber
F^{(s)}_{k,\ell} &= \dfrac{R-3M-4\pi p R^{3}}{R^{2}}e^{\lambda}
h^{(s)}_{\ell} \\[1ex]
&+ \dfrac{R-M+4\pi p R^{3}}{R^{3}}e^{2\lambda} m^{(s)}_{\ell}
+ S_{k^{(s)}_{\ell}}.
\end{align}
Recall that $m^{(s)}_{\ell}$ in the above equations is not an independent
unknown, but rather, it is determined algebraically from $h^{(s)}_{\ell}$
through
\begin{align}
m^{(s)}_{\ell} &= -R\,e^{-\lambda}\,h^{(s)}_{\ell} +
\mathcal{S}_{m^{(s)}_{\ell}},
\label{eq:msl_int}
\end{align}
where $\mathcal{S}_{m^{(s)}_{\ell}}$ is a function driven by lower-order
perturbation functions. The system in Eq.~\eqref{eq:hk_matrix} is therefore a
true $2\times 2$ system in $(h^{(s)}_{\ell}, k^{(s)}_{\ell})$. However, the
determinant of $\mathbf{M}$,
\begin{equation}
\det \mathbf{M} = 1 - \frac{R - M + 4\pi p R^{3}}{R}\,e^{\lambda}= \mathcal{O}(R^{2}) \xrightarrow{R \to 0} 0,\,,
\label{eq:detM}
\end{equation}
is singular at the center of the star, because $M(R) = \frac{4}{3}\pi
\varepsilon_{c} R^{3} + \mathcal{O}(R^{5})$, $p(R) = p_c + \mathcal{O}(R^{2})$,
and $e^{\lambda} = 1 + \mathcal{O}(R^{2})$. Consequently, the inversion of
Eq.~\eqref{eq:hk_matrix} becomes progressively ill-conditioned near the origin,
making the $(h^{(s)}_{\ell}, k^{(s)}_{\ell})$ formulation numerically unstable
for integrations that start at the center.

To overcome this numerical instability, we follow Hartle \&
Thorne~\cite{Hartle:1967he, Hartle:1968si} and introduce the combination
\begin{equation}
v^{(s)}_{\ell} \equiv h^{(s)}_{\ell} + k^{(s)}_{\ell}\,.
\label{eq:v_definition}
\end{equation}
The ODE system for $(v^{(s)}_{\ell}, h^{(s)}_{\ell})$ instead of
$(h^{(s)}_{\ell}, k^{(s)}_{\ell})$ avoids the inversion of $\mathbf{M}$ near the
origin and is correspondingly stable for numerical integration. The matching
procedure for $v^{(s)}_{\ell}$ and $h^{(s)}_{\ell}$ is best performed by
decomposing the general solution into a homogeneous part and a particular part,
\begin{align}
\label{eq:vsl_int}
v^{(s)}_{\ell} &= C^{(s)}_{\ell,\mathsf{int}}\, v^{(s),\mathsf{H}}_{\ell}
+ v^{(s),\mathsf{P}}_{\ell}, \\[1ex]
h^{(s)}_{\ell} &= C^{(s)}_{\ell,\mathsf{int}}\, h^{(s),\mathsf{H}}_{\ell}
+ h^{(s),\mathsf{P}}_{\ell},
\label{eq:hsl_int}
\end{align}
where $C^{(s)}_{\ell,\mathsf{int}}$ is the interior integration constant
uniquely determined by matching at the stellar surface. The homogeneous ODE
system reads
\begin{align}
\frac{d}{dR}
\begin{bmatrix}
v^{(s),\mathsf{H}}_{\ell} \\[1ex]
h^{(s),\mathsf{H}}_{\ell}
\end{bmatrix}
&=
\begin{bmatrix}
0 & Y_{v^{(s)}_{\ell}} \\[1ex]
X_{h^{(s)}_{\ell}} & Y_{h^{(s)}_{\ell}}
\end{bmatrix}
\begin{bmatrix}
v^{(s),\mathsf{H}}_{\ell} \\[1ex]
h^{(s),\mathsf{H}}_{\ell}
\end{bmatrix}\,,
\label{eq:vh_homogeneous}
\end{align}
while the particular ODE system is given by
\begin{align}
\frac{d}{dR}
\begin{bmatrix}
v^{(s),\mathsf{P}}_{\ell} \\[1ex]
h^{(s),\mathsf{P}}_{\ell}
\end{bmatrix}
&=
\begin{bmatrix}
0 & Y_{v^{(s)}_{\ell}} \\[1ex]
X_{h^{(s)}_{\ell}} & Y_{h^{(s)}_{\ell}}
\end{bmatrix}
\begin{bmatrix}
v^{(s),\mathsf{P}}_{\ell} \\[1ex]
h^{(s),\mathsf{P}}_{\ell}
\end{bmatrix}
+
\begin{bmatrix}
\mathcal{S}_{v^{(s)}_{\ell}} \\[1ex]
\mathcal{S}_{h^{(s)}_{\ell}}
\end{bmatrix}\,.
\label{eq:vh_particular}
\end{align}
The coefficient functions are shared between them and are given by
\begin{align}
X_{h^{(s)}_{\ell}} &= -\dfrac{(\ell-1)(\ell+2)}{2\big(M + 4\pi R^{3}
p\big)}, \\[1ex]
Y_{v^{(s)}_{\ell}} &= -\dfrac{2\big(M + 4\pi R^{3} p\big)}{R^{2}}\,
e^{\lambda}, \\[1ex]
Y_{h^{(s)}_{\ell}} &= -\dfrac{2e^{\lambda}}{R^{2}\big(M + 4\pi R^{3}p
\big)} \Big\{ M\big[R + 4\pi R^{3}(3p+\varepsilon)\big] \nonumber \\
&\quad - 2\pi R^{4}\big(p - 8\pi R^{2}p^{2} + \varepsilon\big) -
M^{2}\Big\}\,,
\end{align}
while $\mathcal{S}_{v^{(s)}_{\ell}}$ and $\mathcal{S}_{h^{(s)}_{\ell}}$ denote
the source terms driven by lower-order perturbation functions (see
Sec.~\ref{sec:source} for their expressions). Appendix~\ref{app:Source} gives
their polynomial structure together with the explicit coefficients up to third
order, and the complete set is available in the accompanying repository. One can
check that the matrix in Eq.~\eqref{eq:vh_homogeneous} has non-zero determinant
at the stellar center.

The complete set of metric perturbation functions for modes $\ell \geq 2$ in the
even-parity sector is
\begin{align*}
\Big\{ &v^{(2)}_{2},\, h^{(2)}_{2},\, v^{(4)}_{2},\, h^{(4)}_{2},\,
v^{(4)}_{4},\, h^{(4)}_{4},\, \\
&v^{(6)}_{2},\, h^{(6)}_{2},\,
v^{(6)}_{4},\, h^{(6)}_{4},\, v^{(6)}_{6},\, h^{(6)}_{6} \Big\}\,,
\end{align*}
comprising twelve radial functions in total. Including their homogeneous and
particular parts, this gives 24 unknown functions governed by 24 first-order
ODEs. Once $v^{(s)}_{\ell}$ and $h^{(s)}_{\ell}$ are obtained, the remaining
metric perturbation functions,
\begin{align*}
\Big\{ &k^{(2)}_{2},\, m^{(2)}_{2},\, \xi^{(2)}_{2},\,
k^{(4)}_{2},\, m^{(4)}_{2},\, \xi^{(4)}_{2},\,
k^{(4)}_{4},\, m^{(4)}_{4},\, \xi^{(4)}_{4},\, \\
&k^{(6)}_{2},\, m^{(6)}_{2},\, \xi^{(6)}_{2},\,
k^{(6)}_{4},\, m^{(6)}_{4},\, \xi^{(6)}_{4},\,
k^{(6)}_{6},\, m^{(6)}_{6},\, \xi^{(6)}_{6} \Big\}\,,
\end{align*}
are determined algebraically through
\begin{align}
\label{eq:ksl_int}
k^{(s)}_{\ell} &= v^{(s)}_{\ell} - h^{(s)}_{\ell}, \\[1ex]
\xi^{(s)}_{\ell} &= -\dfrac{R^{2}\,e^{-\lambda}}{M +
4\pi p R^{3}}\,h^{(s)}_{\ell} +
\mathcal{S}_{\xi^{(s)}_{\ell}},
\label{eq:xisl_int}
\end{align}
and Eq.~\eqref{eq:msl_int}. The quantity $\mathcal{S}_{\xi^{(s)}_{\ell}}$ is a
function of lower-order perturbation solutions, described in
App.~\ref{app:Source} and provided in this paper's accompanying repository.

\subsubsection*{Boundary conditions}

As in the odd-parity sector, boundary conditions at the center $R = 0$ are
obtained from a regular asymptotic expansion of each ODE system around the
origin, derived in Paper~I; numerical integrations are then carried out by
evaluating these expressions at a small initial radius $R_i$. For the $\ell = 0$
mode, the asymptotic expansion around $R = 0$ yields
\begin{align}
\label{eq:ms0_asym}
m^{(s)}_{0}(R) &= R^{5} \left[ \mathcal{C}^{(s)}_{0} \, + \, \mathcal{O}(R^{2}) \right] \\[1ex]
\label{eq:xis0_asym}
\xi^{(s)}_{0}(R) &= R \left[ \mathcal{D}^{(s)}_{0} \, + \, \mathcal{E}^{(s)}_{0} R^{2} + \mathcal{O}(R^{4}) \right] \\[1ex]
h^{(s)}_{0}(R) &= h^{(s)}_{0,c} \,  + \,  R^{4}\left[ \mathcal{F}^{(s)}_{0} + \mathcal{O}(R^{2}) \right]
\label{eq:hs0_asym}
\end{align}
where $h^{(s)}_{0,c} \equiv h^{(s)}_{0}(R=0)$ is the central value of
$h^{(s)}_{0}$, and the coefficients $\mathcal{C}^{(s)}_{0}$,
$\mathcal{D}^{(s)}_{0}$, $\mathcal{E}^{(s)}_{0}$, and $\mathcal{F}^{(s)}_{0}$
depend on lower-order asymptotic constants (see Appendix~\ref{app:bc} for their
explicit expressions). For modes $\ell \geq 2$, the asymptotic expansion of the
homogeneous system around $R = 0$ gives
\begin{align}
\label{eq:vsl_H_asym}
v^{(s),\mathsf{H}}_{\ell}(R) &= A^{(s),\mathsf{H}}_{\ell,\mathsf{int}} R^{\ell+2}\left[ \mathcal{G}_{\ell} \, + \, \mathcal{O}(R^{2}) \right] \\[1ex]
h^{(s),\mathsf{H}}_{\ell}(R) &= A^{(s), \mathsf{H}}_{\ell,\mathsf{int}} R^{\ell}\left[1 \, + \, \mathcal{H}_{\ell} R^{2} \, + \, \mathcal{O} (R^{4}) \right]  \, .
\label{eq:hsl_H_asym}
\end{align}
where $A^{(s),\mathsf{H}}_{\ell,\mathsf{int}}$ sets the overall amplitude of the
homogeneous solution, and $\mathcal{G}_{\ell}$ and $\mathcal{H}_{\ell}$ are
coefficients determined entirely by the central values $p_c,\,\varepsilon_c$
through
\begin{align}
\mathcal{G}_{\ell} &:= -\frac{8\pi}{3(\ell+2)}\left(3p_c+\varepsilon_c\right) \, , \\[1ex]
\mathcal{H}_{\ell} &:= -\frac{2\pi}{2\ell+3}\left[(\ell+9)\,p_c+\frac{15-3\ell-2\ell^2}{3}\,\varepsilon_c+\frac{p_c+\varepsilon_c}{c^2_{s,c}}\right] \,,
\end{align}
where $c^2_{s,c} \equiv (\mathrm{d}p/\mathrm{d}\varepsilon)_{\varepsilon_c}$
denotes the central value of the sound speed squared. For the particular system,
the asymptotic expansion reads
\begin{align}
\label{eq:vsl_P_asym}
v^{(s),\mathsf{P}}_{\ell}(R) &= \Big[\mathcal{G}_{\ell}\,A^{(s),\mathsf{P}}_{\ell,\mathsf{int}} \, + \, \widetilde{\mathcal{G}}^{(s)}_{\ell}\Big] R^{\ell+2} \, + \, \mathcal{O}\left(R^{\ell+4}\right) \\[1ex]
\nonumber
h^{(s),\mathsf{P}}_{\ell}(R) &=A^{(s),\mathsf{P}}_{\ell,\mathsf{int}} R^{\ell} \, \\[1ex]
&+ \, \Big[\mathcal{H}_{\ell}\,A^{(s),\mathsf{P}}_{\ell,\mathsf{int}} \, + \, \widetilde{\mathcal{H}}^{(s)}_{\ell}\Big] R^{\ell+2} \, + \, \mathcal{O}\left(R^{\ell+4}\right)
\label{eq:hsl_P_asym}
\end{align}
so that the even-parity particular solution is likewise \emph{affine} in
$A^{(s),\mathsf{P}}_{\ell,\mathsf{int}}$, with $\mathcal{G}_{\ell}$ and
$\mathcal{H}_{\ell}$ the same central-value coefficients as in the homogeneous
case. The source coefficients $\widetilde{\mathcal{G}}^{(s)}_{\ell}$ and
$\widetilde{\mathcal{H}}^{(s)}_{\ell}$ depend not only on the central values
$\varepsilon_c$, $p_c$, and $\nu_c$, but also on the asymptotic constants of all
lower-order solutions,
\begin{align}
\widetilde{\mathcal{G}}^{(s)}_{\ell} &= \widetilde{\mathcal{G}}^{(s)}_{\ell}\left( \varepsilon_{c},
p_{c}, \nu_{c} ; \left\{ A^{(n < s)}_{\ell', \mathsf{int}} \right\}
\right), \\[1ex]
\widetilde{\mathcal{H}}^{(s)}_{\ell} &= \widetilde{\mathcal{H}}^{(s)}_{\ell}\left( \varepsilon_{c},
p_{c}, \nu_{c} ; \left\{ A^{(n < s)}_{\ell', \mathsf{int}} \right\}
\right),
\end{align}
with $\ell' \le n$. In the limit $\left\{ A^{(n < s)}_{\ell', \mathsf{int}}
\right\} \to 0$, the source terms vanish, $\widetilde{\mathcal{G}}^{(s)}_{\ell}$
and $\widetilde{\mathcal{H}}^{(s)}_{\ell} \to 0$, and the particular solution
reduces to a multiple of the homogeneous one. The asymptotic expansions from
which these coefficients can be read off are given in full through third order
in Appendix~\ref{app:bc}, with the higher orders available in the
\texttt{hartle-thorne-project} repository~\cite{conde2025hartlethorne}.

The even-parity sector follows the same sequential integration strategy as the
odd-parity one. The exterior integration constants $C^{(s)}_{\ell,\mathsf{ext}}$
extracted from the matching procedure at $R = R_*$, determine the mass multipole
moments $M_0, M_2, M_4, M_6$ and their spin-induced corrections through the
analytical expressions of Paper~I~\cite{Conde-Ocazionez:2025agu}. The parameter
space of these constants as well as the odd-parity ones, and how we calculate
them in practice, are discussed in Sec.~\ref{sec:numerical}.

\subsection{Source terms}
\label{sec:source}

At each perturbative order $n \geq 2$, the perturbation equations in both parity
sectors are driven by source terms built from lower-order solutions. A key
structural feature of the source terms emerges after applying a systematic
\textit{order reduction} procedure. Specifically, lower-order perturbation
derivatives in the source terms are eliminated by substituting the corresponding
lower-order relations, until all source terms are expressed purely in terms of
the perturbation functions themselves. One of the findings of Paper~I is that,
in the Hartle-Thorne coordinate frame, this repeated order reduction inevitably
introduces higher radial derivatives of the energy density profile; namely,
derivatives $d\varepsilon/dR$, $d^2\varepsilon/dR^2$, $d^3\varepsilon/dR^3$
appear progressively at higher perturbative orders. Up to first order in
rotation, none of these radial derivatives are required. From second order in
rotation onward, $d\varepsilon/dR$ is needed, from fourth order in rotation
onward $d^2\varepsilon/dR^2$ enters as well, and from sixth order onwards
$d^3\varepsilon/dR^3$ is also required. After order reduction, the source terms
therefore depend on $\varepsilon$ and its derivatives, multiplied by products of
lower-order perturbation functions.

The structure of the resulting expression can be further simplified by sorting
them with respect to the spin-frequency order counting. We find that organizing
source terms by spin-frequency order and factoring out parameter combinations
(such as terms that depend on $\varepsilon$, on $d\varepsilon/dR$, etc.) enables
a hierarchical grouping of lower-order perturbation contributions. This
algebraic restructuring slashes symbolic expression sizes by up to $91\%$
compared to those presented in Paper~I. The resulting forms are subsequently
optimized for numerical evaluation using common subexpression elimination.

This hierarchical framework suggests a strategy for treating models with
non-vanishing surface densities (e.g., quark stars), where source terms
involving $d\varepsilon/dR$, $d^2\varepsilon/dR^2$ and $d^3\varepsilon/dR^3$
become singular at the stellar surface. Isolating these contributions would
allow term-by-term integration by parts to derive surface junction conditions
(although products of high-order distributions are deferred to future work).
Detailed structural decompositions through third order are provided in
Appendix~\ref{app:Source}, with higher-order expressions hosted in the
\texttt{hartle-thorne-project} repository~\cite{conde2025hartlethorne}.


\section{Exterior Solutions}
\label{sec:exterior}

The general exterior solution at each order in rotation and for each harmonic
mode number also splits into homogeneous and particular parts. In the original
derivation of Paper~I, the homogeneous part is weighted by an integration
constant, whose dimension varies with both orders and mode number. In this work,
we rescale these constants by integer prefactors and powers of $M_*$ so that
they are dimensionless across all orders and mode number, thus standardizing
literature conventions. The particular part of the solution in Paper~I is laden
with large, high-precision integer coefficients and logarithmic terms $\ln f$,
where $f \equiv 1 - 2M_{*}/R$, so that direct evaluation causes severe
cancellation errors near the stellar surface. To circumvent this, the
expressions are algebraically restructured by factoring out $\ln f$ combinations
and simplifying coefficients (see Mathematica notebooks of the
\texttt{hartle-thorne-project} repository~\cite{conde2025hartlethorne}). Doing
so mitigates the precision issues described above, allowing us to carry out
matching in Sec.~\ref{sec:matching} and the extraction of the multipoles to high
precision.

\subsection{Rescaling of the exterior integration constants}

The exterior integration constants carry dimensions in the literature that
reflect their physical meaning at each order and mode number. As discussed in
Paper~I, the exterior solution at each order $n$ and mode number $\ell$ is
characterized by a constant $C^{(n)}_{\ell,\mathsf{ext}}$ that weights the
homogeneous part of the solution and is fixed by matching at $R = R_*$. These
constants are related to the multipole moments of the star and carry information
about the interior that is imprinted on the exterior field. At first order in
rotation, the constant is the spin angular momentum itself, with dimensions
$[L]^2$. At second order in rotation, $C^{(2)}_{0,\mathsf{ext}}$ is the first
correction to the mass monopole, with dimensions $[L]$, while
$C^{(2)}_{2,\mathsf{ext}}$ is dimensionless, since Hartle and Thorne wrote the
$\ell=2$ homogeneous solution in terms of the Legendre function $Q^2_2(\zeta)$
with $\zeta \equiv R/M_* - 1$. At third order in rotation,
$C^{(3)}_{1,\mathsf{ext}}$ carries dimensions of $[L]^2$ and
$C^{(3)}_{3,\mathsf{ext}}$ of $[L]^4$. The correspondence with the symbols of
earlier work is collected in Table~\ref{tab:legacy}. The pattern continues to
higher orders in rotation, where the dimensions of $C^{(n)}_{\ell,\mathsf{ext}}$
grow with $\ell$.
\begin{table}[t]
\setlength{\tabcolsep}{6pt}
\renewcommand{\arraystretch}{1.2}
\begin{tabular}{c c c c l}
\toprule[1.2pt]
$n$ & $\ell$ & $C^{(n)}_{\ell,\mathsf{ext}}$ & Symbol & Reference \\
\midrule
1 & 1 & $S$ & $J$ & Hartle~\cite{Hartle:1967he} \\
2 & 0 & $C^{(2)}_{0,\mathsf{ext}}$ & $\delta M$ & Hartle \& Thorne~\cite{Hartle:1968si} \\
2 & 2 & $C^{(2)}_{2,\mathsf{ext}}$ & $A$ & Hartle \& Thorne~\cite{Hartle:1968si} \\
3 & 1 & $C^{(3)}_{1,\mathsf{ext}}$ & $\delta J$ & Hartle~\cite{Hartle:1973zza}, Benhar et al.~\cite{Benhar:2005gi} \\
3 & 3 & $C^{(3)}_{3,\mathsf{ext}}$ & $D$ & Benhar et al.~\cite{Benhar:2005gi} \\
\bottomrule[1.2pt]
\end{tabular}
\caption{Correspondence between the exterior integration constants of Paper~I
and the symbols used for these quantities in earlier work. Dimensions in
geometrized units are $[L]^2$, $[L]$, $[L]^0$, $[L]^2$ and $[L]^4$
respectively.}
\label{tab:legacy}
\end{table}

These conventions for the constants of integration were inherited from the past
literature, rather than chosen. Hartle and
Thorne~\cite{Hartle:1967he,Hartle:1968si,Hartle:1973zza} fixed the dimensions of
$S$, $C^{(2)}_{0,\mathsf{ext}}$ and $C^{(3)}_{1,\mathsf{ext}}$, and Benhar et
al.~\cite{Benhar:2005gi} those of $C^{(3)}_{3,\mathsf{ext}}$. Yagi et
al.~\cite{Yagi:2014bxa} adopted dimensionless constants through fourth order,
but their exterior solutions were not used directly here, since the source terms
differ once the coordinate transformation to the Hartle-Thorne frame is carried
through consistently, as discussed in Paper~I. When the solutions were
propagated to higher orders, the dimensional conventions of Hartle-Thorne and
Benhar were kept. Working with constants of different dimensions across orders
and modes is numerically inconvenient, since their magnitudes can vary
substantially, and this can cause numerical error to accumulate.

\begin{table}[t]
\centering
\renewcommand{\arraystretch}{1.8}
\setlength{\tabcolsep}{4pt}
\small
\newcommand{\ctr}[1]{\rule[-5pt]{0pt}{14pt}#1}
\begin{tabular}{|c|c|r@{\ $=$\ }l|}
\hline
\textbf{Order} $(n)$ & \textbf{Mode} $\ell$ & \multicolumn{2}{c|}{\textbf{Rescaling relation}} \\ \hline
$(1)$ & $1$ & $\ctr{S}$ & $M_*^{2}\,j$ \\ \hline
 & $0$ & $\ctr{C^{\,(2)}_{0,\,\mathsf{ext}}}$ & $M_*\,\tilde{C}^{\,(2)}_{0,\,\mathsf{ext}}$ \\ \cline{2-4}
\multirow{-2}{*}{$(2)$} & $2$ & $\ctr{C^{\,(2)}_{2,\,\mathsf{ext}}}$ & $\tilde{C}^{\,(2)}_{2,\,\mathsf{ext}}$ \\ \hline
 & $1$ & $\ctr{C^{\,(3)}_{1,\,\mathsf{ext}}}$ & $M_*^{2}\,\tilde{C}^{\,(3)}_{1,\,\mathsf{ext}}$ \\ \cline{2-4}
\multirow{-2}{*}{$(3)$} & $3$ & $\ctr{C^{\,(3)}_{3,\,\mathsf{ext}}}$ & $M_*^{4}\,\tilde{C}^{\,(3)}_{3,\,\mathsf{ext}}$ \\ \hline
 & $0$ & $\ctr{C^{\,(4)}_{0,\,\mathsf{ext}}}$ & $M_*\,\tilde{C}^{\,(4)}_{0,\,\mathsf{ext}}$ \\ \cline{2-4}
 & $2$ & $\ctr{C^{\,(4)}_{2,\,\mathsf{ext}}}$ & $-8M_*^{5}\,\tilde{C}^{\,(4)}_{2,\,\mathsf{ext}}$ \\ \cline{2-4}
\multirow{-3}{*}{$(4)$} & $4$ & $\ctr{C^{\,(4)}_{4,\,\mathsf{ext}}}$ & $-192M_*^{9}\,\tilde{C}^{\,(4)}_{4,\,\mathsf{ext}}$ \\ \hline
 & $1$ & $\ctr{C^{\,(5)}_{1,\,\mathsf{ext}}}$ & $-6M_*^{2}\,\tilde{C}^{\,(5)}_{1,\,\mathsf{ext}}$ \\ \cline{2-4}
 & $3$ & $\ctr{C^{\,(5)}_{3,\,\mathsf{ext}}}$ & $-7M_*^{4}\,\tilde{C}^{\,(5)}_{3,\,\mathsf{ext}}$ \\ \cline{2-4}
\multirow{-3}{*}{$(5)$} & $5$ & $\ctr{C^{\,(5)}_{5,\,\mathsf{ext}}}$ & $11M_*^{6}\,\tilde{C}^{\,(5)}_{5,\,\mathsf{ext}}$ \\ \hline
 & $0$ & $\ctr{C^{\,(6)}_{0,\,\mathsf{ext}}}$ & $M_*\,\tilde{C}^{\,(6)}_{0,\,\mathsf{ext}}$ \\ \cline{2-4}
 & $2$ & $\ctr{C^{\,(6)}_{2,\,\mathsf{ext}}}$ & $-8M_*^{5}\,\tilde{C}^{\,(6)}_{2,\,\mathsf{ext}}$ \\ \cline{2-4}
 & $4$ & $\ctr{C^{\,(6)}_{4,\,\mathsf{ext}}}$ & $-192M_*^{9}\,\tilde{C}^{\,(6)}_{4,\,\mathsf{ext}}$ \\ \cline{2-4}
\multirow{-4}{*}{$(6)$} & $6$ & $\ctr{C^{\,(6)}_{6,\,\mathsf{ext}}}$ & $-1024M_*^{13}\,\tilde{C}^{\,(6)}_{6,\,\mathsf{ext}}$ \\ \hline
 & $1$ & $\ctr{C^{\,(7)}_{1,\,\mathsf{ext}}}$ & $-6M_*^{2}\,\tilde{C}^{\,(7)}_{1,\,\mathsf{ext}}$ \\ \cline{2-4}
 & $3$ & $\ctr{C^{\,(7)}_{3,\,\mathsf{ext}}}$ & $-7M_*^{4}\,\tilde{C}^{\,(7)}_{3,\,\mathsf{ext}}$ \\ \cline{2-4}
 & $5$ & $\ctr{C^{\,(7)}_{5,\,\mathsf{ext}}}$ & $11M_*^{6}\,\tilde{C}^{\,(7)}_{5,\,\mathsf{ext}}$ \\ \cline{2-4}
\multirow{-4}{*}{$(7)$} & $7$ & $\ctr{C^{\,(7)}_{7,\,\mathsf{ext}}}$ & $15M_*^{8}\,\tilde{C}^{\,(7)}_{7,\,\mathsf{ext}}$ \\ \hline
\end{tabular}
\caption{Rescaling of the exterior integration constants
$C^{\,(n)}_{\ell,\,\mathsf{ext}}$ for spin orders $n = 1$ through $n = 7$ and
their corresponding modes $\ell$. At each order and mode, the factor
$\sigma^{(n)}_{\ell}\,M_{*}^{q_{\ell}}$ is absorbed into the homogeneous
exterior solution, ensuring it carries the correct dimensions of the
corresponding metric function and the normalization of its closed-form
representation. The dimensionless constant
$\tilde{C}^{\,(n)}_{\ell,\,\mathsf{ext}}$ appears on the right of each
relation.}
\label{tab:rescaling_ext}
\end{table}

In the present work, the convention we choose is to standardize the constants
across all spin orders and mode numbers through rescalings. The homogeneous
exterior solutions are multiplied by a factor
$\sigma^{(n)}_{\ell}\,M_{*}^{q_{\ell}}$, where $q_\ell \in \mathbb{N}$ and
$\sigma^{(n)}_{\ell} \in \mathbb{Z}$, both of which can be read from
Table~\ref{tab:rescaling_ext}. Leaving aside the first-order constant $S$, the
prefactor $\sigma^{(n)}_{\ell}$ equals unity at the lowest order of each mode
with $\ell \le 3$, while it takes a fixed, $\ell$-dependent value at every
higher order; for $\ell \ge 4$ that $\ell$-dependent value applies already at
the first appearance, at $n=\ell$. We choose this prefactor so that the rescaled
homogeneous solution carries the normalization of a known closed-form special
function. The rescaling in $M_{*}^{q_{\ell}}$ gives both the homogeneous and
particular parts of the solution the same dimensions as the corresponding metric
function, which makes $\tilde{C}^{(n)}_{\ell,\mathsf{ext}}$ dimensionless for
all $n$ and $\ell$. No rescaling is needed for $C^{(2)}_{2,\mathsf{ext}}$, since
Hartle and Thorne's second-order solution is adopted directly and already uses a
dimensionless constant, while at first order the dimensionless form of $S$ is $j
\equiv S/M_*^2$. The analytic expressions for the multipoles $M_\ell$ and
$S_\ell$ from Paper~I are modified by the same factors, as shown in
Appendix~\ref{app:multipoles}. This is to be expected, since the moments are
read from the asymptotic behavior of the exterior metric and so inherit the
rescaling of its constants directly.

The rescaled form of the constants is adopted throughout. Paper~I derived the
even-parity homogeneous solutions for $\ell = 2, 4, 6$ in terms of Legendre
functions $Q^2_\ell(\zeta)$, generalizing Hartle and Thorne's second-order
result. In that form, the constants are dimensionless, just as
$C^{(2)}_{2,\mathsf{ext}}$ is. The calculation was nonetheless propagated to
higher orders with dimensionful constants. Here, every homogeneous solution is
recast with dimensionless constants according to Table~\ref{tab:rescaling_ext},
and the \texttt{hartle-thorne-project} repository~\cite{conde2025hartlethorne}
is updated to take that form. The odd-parity homogeneous solution admits a
similar closed form, once rescaled, as a hypergeometric function, an
identification made here for the first time. The following subsections give both
sectors in this rescaled form.

\subsection{Odd-parity sector}

The odd-parity sector is described by odd spin-frequency orders $k$ and the odd
modes $\ell$ available at each order, as shown in Table~\ref{tab:modes}. At
first order in rotation, the only non-vanishing mode is $\ell = 1$, and the
exterior perturbation equation for $\varpi^{(1)}_{1,\mathsf{ext}} \equiv \Omega
- \omega^{(1)}_{1,\mathsf{ext}}$, where $\Omega$ is the stellar angular
frequency, is homogeneous. In terms of the dimensionless spin angular momentum
$j \equiv S/M_*^2$, given in Table~\ref{tab:rescaling_ext}, the exterior
solutions are given by
\begin{align}
\varpi^{(1)}_{1,\mathsf{ext}}(R) &= \Omega - \dfrac{2j}{R^{3}}M_{*}^{2}, \\[1ex]
\bar{\alpha}^{(1)}_{1,\mathsf{ext}}(R) &= \dfrac{6j}{R^{4}}M^{2}_{*},
\end{align}
where $\bar{\alpha}^{(1)}_{1,\mathsf{ext}} \equiv
d\varpi^{(1)}_{1,\mathsf{ext}}/dR$. No rescaling of $\Omega$ is required, since
it does not appear explicitly in the exterior solutions at higher orders,
entering only through $j$.

At higher odd orders $k = 3, 5, 7$, the exterior equations are inhomogeneous,
sourced by the solutions of all previous orders. Similarly to the interior
problem, the general solution at each order $k$ and mode $\ell$ splits into a
homogeneous part $\omega^{(k),\mathsf{H}}_{\ell,\mathsf{ext}}$ and a particular
part $\omega^{(k),\mathsf{P}}_{\ell,\mathsf{ext}}$, so that
\begin{equation}
\omega^{(k)}_{\ell,\mathsf{ext}} =
\omega^{(k),\mathsf{P}}_{\ell,\mathsf{ext}} \, + \, \tilde{C}^{(k)}_{\ell,\mathsf{ext}}\,
\omega^{(k),\mathsf{H}}_{\ell,\mathsf{ext}},
\label{eq:exterior_odd_decomposition}
\end{equation}
where $\tilde{C}^{(k)}_{\ell,\mathsf{ext}}$ is the dimensionless exterior
integration constant to be determined from the matching procedure at $R = R_*$.
As anticipated, we restructure both the homogeneous and particular pieces of the
general solution given in Eq.~\eqref{eq:exterior_odd_decomposition}.

\subsubsection*{Homogeneous solutions}

The homogeneous piece is restructured through a rescaling alone, which brings it
into the compact form of a hypergeometric function. Recall from Paper~I that the
homogeneous vacuum exterior equation for each spin-frequency order $k$ and its
corresponding available mode $\ell$ reads
\begin{align}
\label{eq:woddExt}
\frac{d^{2}\omega^{(k) \, \mathsf{H}}_{\ell, \, \mathsf{ext}}}{dR^{2}}  +  \dfrac{4}{R}  \frac{d\omega^{(k) \, \mathsf{H}}_{\ell, \, \mathsf{ext}}}{dR}  -  \left[ \dfrac{(\ell+2)(\ell-1)}{R(R- 2M_{*})} \right] \omega^{(k) \, \mathsf{H}}_{\ell, \, \mathsf{ext}} = 0 \, .
\end{align}
We now show how to massage this equation into that of a hypergeometric function.

Using Hartle's notation, we introduce the following coordinate
\begin{equation}
\zeta \equiv \frac{R}{M_*} - 1.
\end{equation}
In the exterior, $R > R_{*}$ by definition, so $\infty > \zeta > R_*/M_* - 1$.
The background is a static sphere of isotropic perfect fluid whose density does
not increase outward, so Buchdahl's bound~\cite{Buchdahl:1959zz} applies and
gives $\mathcal{C}_* = M_*/R_* < 4/9$. It follows that $R > R_* > 2M_{*}$ and
therefore $\zeta > 1$ in general. Changing the ODE independent variable from the
Hartle radial coordinate $R$ to $\zeta$, we obtain
\begin{equation}
\dfrac{d^{2}\omega^{(k),\mathsf{H}}_{\ell,\mathsf{ext}}}{d\zeta^{2}} + \dfrac{4}{\zeta+1}\dfrac{d\omega^{(k),\mathsf{H}}_{\ell,\mathsf{ext}}}{d\zeta} -  \dfrac{(\ell+2)(\ell-1)}{\zeta^{2}-1} \omega^{(k),\mathsf{H}}_{\ell,\mathsf{ext}}=0 \, .
\end{equation}
This equation is closer to that of a hypergeometric function, but not quite
there yet.

We further transform this equation using the change of variable
\begin{equation}
x \equiv \frac{1-\zeta}{2} \, .
\end{equation}
The equation becomes
\begin{equation}
x(1-x)\dfrac{d^{2} \omega^{(k),\mathsf{H}}_{\ell,\mathsf{ext}}}{dx^{2}}  -  4x \dfrac{d\omega^{(k),\mathsf{H}}_{\ell,\mathsf{ext}}}{dx} + (\ell+2)(\ell-1)\omega^{(k),\mathsf{H}}_{\ell,\mathsf{ext}} = 0 \, ,
\label{eq:hyper_eq}
\end{equation}
which is a standard hypergeometric equation. Matching to the canonical form
$x(1-x)y'' + [c-(a+b+1)x]\,y' - ab\,y = 0$ gives $c = 0$, $a + b = 3$ and $ab =
-(\ell+2)(\ell-1)$, so that $a = \ell+2$ and $b = 1-\ell$.

The general solution of Eq.~\eqref{eq:hyper_eq} is $y(x)=y_A y_1(x)+y_B y_2(x)$,
where $y_{A,B}$ are integration constants. Since the boundary condition is
imposed at spatial infinity, we write the two solutions in a form adapted to
$x=\infty$~\cite[\S15.10]{DLMF},
\begin{align}
y_1 &= x^{-a}\,{}_2F_1\!\left(a,\ a-c+1;\ a-b+1;\ \tfrac{1}{x}\right), \\
y_2 &= x^{-b}\,{}_2F_1\!\left(b,\ b-c+1;\ b-a+1;\ \tfrac{1}{x}\right).
\end{align}
For the parameters above, the series in $y_1$ is regular at $1/x=0$, and the
series in $y_2$ terminates before its lower parameter can produce a vanishing
denominator. As $x\to-\infty$, the two solutions behave as $x^{-(\ell+2)}$ and
$x^{\ell-1}$, respectively, so they are linearly independent. At spatial
infinity, $y_1$ vanishes, while $y_2$ is constant for $\ell=1$ and grows for
$\ell=3,5,7$. Asymptotic flatness therefore sets $y_B=0$.

Up to normalization, the final solution of Eq.~\eqref{eq:hyper_eq} is then
\begin{equation}
\omega^{(k),\mathsf{H}}_{\ell, \mathsf{ext}}(x) = x^{-(\ell+2)} {}_{2}F_{1} \left(\ell+2,\ \ell+3;\ 2\ell+2;\ \frac{1}{x}\right) \, .
\end{equation}
Expressed in terms of the coordinate $\zeta$, and with a normalization constant
$C_\ell$ fixed to reproduce the solutions available in the literature up to
third order, the homogeneous solution and its derivative take the form
\begin{align}
\label{eq:hypergeometric_homogeneous}
\nonumber
\omega^{(k),\mathsf{H}}_{\ell, \mathsf{ext}}(\zeta) &= \dfrac{C_{\ell}}{M_{*}} (1-\zeta)^{-(\ell+2)} \\
& \quad \times {}_{2}F_{1} \left(\ell+2,\ \ell+3;\ 2\ell+2;\ \frac{2}{1-\zeta}\right) \, , \\[2ex]
\alpha^{(k),\mathsf{H}}_{\ell,\mathsf{ext}}(\zeta) &= \dfrac{1}{M_{*}} \dfrac{ d\omega^{(k),\mathsf{H}}_{\ell, \mathsf{ext}}(\zeta)}{d\zeta} \, ,
\end{align}
where
\begin{equation}
    C_{\ell} =
\begin{cases}
-2, & \ell = 1 \\
-1, & \ell = 3 \\
\phantom{-} 1, & \ell = 5 \\
\phantom{-}1, & \ell = 7
\end{cases}
\, .
\end{equation}
The rescaling introduced in Table~\ref{tab:rescaling_ext} for the odd
integration constants is precisely what brings the exterior homogeneous
solutions into the hypergeometric form of
Eq.~\eqref{eq:hypergeometric_homogeneous}. We provide more explicit expressions
in Appendix~\ref{app:Exterior_solutions}.

\subsubsection*{Particular solutions}

Unlike the homogeneous piece, the particular part of the exterior solution does
not reduce to a single closed-form special function common to all orders and
modes. At each spin order $k$, the vacuum exterior equation for
$\omega^{(k),\mathsf{P}}_{\ell,\mathsf{ext}}$ is inhomogeneous, with a source
term built from the exterior solutions of all lower orders. As a result, the
particular solution depends explicitly on both the spin order $k$ and the mode
number $\ell$, unlike the homogeneous piece. Compared to the original
expressions derived in Paper~I, we simplify these solutions by collecting terms
proportional to $\ln f$ and its powers, where recall that $f = 1-2M_*/R$. This
restructuring substantially reduces the size of the coefficients appearing in
each term, avoiding the many-digit integers discussed at the beginning of this
section.

Based on this structure, we express the particular solution for
$\omega^{(k),\mathsf{P}}_{\ell,\mathsf{ext}}$ and its derivative
$\alpha^{(k),\mathsf{P}}_{\ell,\mathsf{ext}}$ as a finite sum of coefficient
functions, each multiplying an increasing power of $\ln f$, for the allowed
$\ell$-modes at each spin-frequency order $k$ listed in Table~\ref{tab:modes}:
\begin{align}
\omega^{(3),\mathsf{P}}_{\ell,\mathsf{ext}} &= \frac{1}{M_*}\bigg[\mathcal{L}^{[0]}_{\omega^{(3)}_\ell} + \mathcal{L}^{[1]}_{\omega^{(3)}_\ell}(\ln f)\bigg], \\
\alpha^{(3),\mathsf{P}}_{\ell,\mathsf{ext}} &= \frac{1}{M_*^2}\bigg[\mathcal{L}^{[0]}_{\alpha^{(3)}_\ell} + \mathcal{L}^{[1]}_{\alpha^{(3)}_\ell}(\ln f)\bigg], \\
\omega^{(5),\mathsf{P}}_{\ell,\mathsf{ext}} &= \frac{1}{M_*}\bigg[\dfrac{1}{f}\mathcal{L}^{[0]}_{\omega^{(5)}_\ell} + \mathcal{L}^{[1]}_{\omega^{(5)}_\ell}(\ln f) + \mathcal{L}^{[2]}_{\omega^{(5)}_\ell}(\ln f)^2\bigg], \\
\alpha^{(5),\mathsf{P}}_{\ell,\mathsf{ext}} &= \frac{1}{M_*^2}\bigg[\dfrac{1}{f^2}\mathcal{L}^{[0]}_{\alpha^{(5)}_\ell} + \mathcal{L}^{[1]}_{\alpha^{(5)}_\ell}(\ln f) + \mathcal{L}^{[2]}_{\alpha^{(5)}_\ell}(\ln f)^2\bigg], \\
\omega^{(7),\mathsf{P}}_{\ell,\mathsf{ext}} &= \frac{1}{M_*}\bigg[\dfrac{1}{f^2}\mathcal{L}^{[0]}_{\omega^{(7)}_\ell} + \dfrac{1}{f}\mathcal{L}^{[1]}_{\omega^{(7)}_\ell}(\ln f) \nonumber \\
&\quad + \mathcal{L}^{[2]}_{\omega^{(7)}_\ell}(\ln f)^2 + \mathcal{L}^{[3]}_{\omega^{(7)}_\ell}(\ln f)^3\bigg], \\
\alpha^{(7),\mathsf{P}}_{\ell,\mathsf{ext}} &= \frac{1}{M_*^2}\bigg[\dfrac{1}{f^3}\mathcal{L}^{[0]}_{\alpha^{(7)}_\ell} + \dfrac{1}{f^2}\mathcal{L}^{[1]}_{\alpha^{(7)}_\ell}(\ln f) \nonumber \\
&\quad + \mathcal{L}^{[2]}_{\alpha^{(7)}_\ell}(\ln f)^2 + \mathcal{L}^{[3]}_{\alpha^{(7)}_\ell}(\ln f)^3\bigg].
\end{align}
The superscript $[m]$ in the $\mathcal{L}^{[m]}$ functions labels the term
proportional to $(\ln f)^m$, with $[0]$ denoting the term independent of $\ln
f$. These coefficient functions depend on the radius-dependent quantity
$\mathscr{C} \equiv M_{*}/R$ (not to be confused with the stellar compactness
$\mathscr{C}_* \equiv M_{*}/R_*$, which, for a given star, is just a number, and
not a function of $R$) and on the exterior integration constants of the lower
orders, and they have the schematic structure
\begin{equation}
\mathcal{L}^{[\cdot]} \sim \Big(\text{Laurent series in }
\mathscr{C}\Big) \times \Big(\text{products of }
\tilde{C}^{(k'<k)}_{\ell',\mathsf{ext}}\Big),
\label{eq:Lstructure}
\end{equation}
where the second factor collects the rescaled integration constants of the lower
orders $k' < k$ and modes $\ell' \le k'$. Within each coefficient function
$\mathcal{L}^{[\cdot]}$, the Laurent series in $\mathscr{C}$ is further recast
in Horner form, which we find better for numerical evaluation. Since both
$\mathscr{C}$ and each $\tilde{C}^{(k'<k)}_{\ell',\mathsf{ext}}$ are
dimensionless, every coefficient function
$\mathcal{L}^{[\cdot]}_{\omega^{(k)}_\ell}$ (and, similarly,
$\mathcal{L}^{[\cdot]}_{\alpha^{(k)}_\ell}$) is dimensionless as well. The
explicit $R$-dependence of the particular solution therefore enters through two
places: the factor $f = 1-2M_*/R$ appearing outside each $\mathcal{L}$, and
$\mathscr{C}$ appearing inside it. We provide the explicit coefficient
$\mathcal{L}$-functions through third order in
Appendix~\ref{app:Exterior_solutions}, while the higher-order expressions are
available in the \texttt{hartle-thorne-project}
repository~\cite{conde2025hartlethorne}.

\subsection{Even-parity sector}

The even-parity sector is described by even spin-frequency orders $s$ and the
even harmonic mode numbers $\ell$ available at each order, as shown in
Table~\ref{tab:modes}. This sector splits into two cases: $\ell = 0$, associated
with the monopole structure of the star, and $\ell \geq 2$, associated with its
mass multipolar deformations.

The $\ell = 0$ case is described by the metric perturbations
$m^{(s)}_{0,\mathsf{ext}}$ and $h^{(s)}_{0,\mathsf{ext}}$, which we decompose
into homogeneous and particular pieces as
\begin{align}
m^{(s)}_{0,\mathsf{ext}} &= m^{(s),\mathsf{P}}_{0,\mathsf{ext}} + \tilde{C}^{(s)}_{0,\mathsf{ext}}\,m^{(s),\mathsf{H}}_{0,\mathsf{ext}}, \\
h^{(s)}_{0,\mathsf{ext}} &= h^{(s),\mathsf{P}}_{0,\mathsf{ext}} + \tilde{C}^{(s)}_{0,\mathsf{ext}}\,h^{(s),\mathsf{H}}_{0,\mathsf{ext}}.
\label{eq:mh_ext}
\end{align}
The $\ell \geq 2$ case is described by the metric perturbations
$v^{(s)}_{\ell,\mathsf{ext}}$, $h^{(s)}_{\ell,\mathsf{ext}}$,
$k^{(s)}_{\ell,\mathsf{ext}}$, and $m^{(s)}_{\ell,\mathsf{ext}}$. The general
solution splits similarly as
\begin{align}
\label{eq:exterior_decomposition_v}
v^{(s)}_{\ell,\mathsf{ext}} &= v^{(s),\mathsf{P}}_{\ell,\mathsf{ext}} + \tilde{C}^{(s)}_{\ell,\mathsf{ext}}\,v^{(s),\mathsf{H}}_{\ell,\mathsf{ext}}, \\
\label{eq:exterior_decomposition_h}
h^{(s)}_{\ell,\mathsf{ext}} &= h^{(s),\mathsf{P}}_{\ell,\mathsf{ext}} + \tilde{C}^{(s)}_{\ell,\mathsf{ext}}\,h^{(s),\mathsf{H}}_{\ell,\mathsf{ext}}, \\
\label{eq:exterior_decomposition_k}
k^{(s)}_{\ell,\mathsf{ext}} &= k^{(s),\mathsf{P}}_{\ell,\mathsf{ext}} + \tilde{C}^{(s)}_{\ell,\mathsf{ext}}\,k^{(s),\mathsf{H}}_{\ell,\mathsf{ext}}, \\
m^{(s)}_{\ell,\mathsf{ext}} &= m^{(s),\mathsf{P}}_{\ell,\mathsf{ext}} + \tilde{C}^{(s)}_{\ell,\mathsf{ext}}\,m^{(s),\mathsf{H}}_{\ell,\mathsf{ext}}.
\label{eq:exterior_decomposition_m}
\end{align}
As in the odd-parity sector, we restructure both the homogeneous and particular
pieces of the general solution, given in
Eqs.~\eqref{eq:exterior_decomposition_v}-\eqref{eq:exterior_decomposition_m}.

\subsubsection*{Homogeneous solutions}

For the $\ell = 0$ case, Paper~I derived two vacuum ordinary differential
equations for $m^{(s)}_{0}$ and $h^{(s)}_{0}$. Their homogeneous solutions are
given by
\begin{align}
m^{(s),\mathsf{H}}_{0,\mathsf{ext}}(\zeta) &= M_*, \\
h^{(s),\mathsf{H}}_{0,\mathsf{ext}}(\zeta) &= \frac{1}{1-\zeta}.
\end{align}
For $\ell \geq 2$, $v^{(s)}_{\ell}$ and $h^{(s)}_{\ell}$ satisfy a coupled
system of vacuum ordinary differential equations. Once these are known,
$k^{(s)}_{\ell}$ follows algebraically from $h^{(s)}_{\ell}$ and
$v^{(s)}_{\ell}$, given in Eq.~\eqref{eq:ksl_int}, while $m^{(s)}_{\ell}$
follows algebraically from $h^{(s)}_{\ell}$, as shown in Eq.~\eqref{eq:msl_int}.
Although $k^{(s)}_{\ell}$ and $m^{(s)}_{\ell}$ also satisfy their own vacuum
equations, we use these algebraic expressions directly: within each, the piece
proportional to $\tilde{C}^{(s)}_{\ell,\mathsf{ext}}$ is identified as the
homogeneous solution, and the remainder as the particular solution, casting
$k^{(s)}_{\ell}$ and $m^{(s)}_{\ell}$ into the same decomposition used
throughout this work. The exterior homogeneous solutions are given by Legendre
functions of the second kind, which are the branch that is asymptotically flat
at spatial infinity,
\begin{align}
v^{(s),\mathsf{H}}_{\ell,\mathsf{ext}}(\zeta) &= \frac{2}{\sqrt{\zeta^{2}-1}}\,Q^{1}_{\ell}(\zeta), \\
h^{(s),\mathsf{H}}_{\ell,\mathsf{ext}}(\zeta) &= Q^{2}_{\ell}(\zeta), \\
k^{(s),\mathsf{H}}_{\ell,\mathsf{ext}}(\zeta) &= \frac{2}{\sqrt{\zeta^{2}-1}}\,Q^{1}_{\ell}(\zeta) - Q^{2}_{\ell}(\zeta), \\
m^{(s),\mathsf{H}}_{\ell,\mathsf{ext}}(\zeta) &= M_*(1-\zeta)\,Q^{2}_{\ell}(\zeta).
\end{align}
The explicit forms of $Q^{1}_{\ell}(\zeta)$ and $Q^{2}_{\ell}(\zeta)$ are given
in Paper~I~\cite{Conde-Ocazionez:2025agu}.

As in the interior problem, the two independent metric variables are
$h^{(s)}_{\ell,\mathsf{ext}}$ and $k^{(s)}_{\ell,\mathsf{ext}}$. However, the
interior problem is integrated in the variables $v^{(s)}_{\ell}$ and
$h^{(s)}_{\ell}$, as explained in Sec.~\ref{sec:interior_even}, so the matching
at the surface is performed in those same variables. The exterior solution is
therefore expressed in terms of $v^{(s)}_{\ell,\mathsf{ext}}$ and
$h^{(s)}_{\ell,\mathsf{ext}}$ as well. The remaining function,
$m^{(s)}_{\ell,\mathsf{ext}}$, follows from $h^{(s)}_{\ell,\mathsf{ext}}$.

\subsubsection*{Particular solutions}

As in the odd-parity sector, the particular piece depends on the spin-frequency
order through the metric perturbations of all lower orders. We decompose these
solutions following the same structure used for the odd-parity particular
solutions. For $\ell = 0$, the general structure of the particular solutions is
\begin{align}
m^{(2),\mathsf{P}}_{0,\mathsf{ext}} &= M_*\,\mathcal{L}^{[0]}_{m^{(2)}_0}, \\
h^{(2),\mathsf{P}}_{0,\mathsf{ext}} &= \dfrac{1}{f}\mathcal{L}^{[0]}_{h^{(2)}_0}, \\[1ex]
m^{(4),\mathsf{P}}_{0,\mathsf{ext}} &= M_*\bigg[\dfrac{1}{f}\mathcal{L}^{[0]}_{m^{(4)}_0} + \mathcal{L}^{[1]}_{m^{(4)}_0}(\ln f) + \mathcal{L}^{[2]}_{m^{(4)}_0}(\ln f)^2\bigg], \\
h^{(4),\mathsf{P}}_{0,\mathsf{ext}} &= \dfrac{1}{f^2}\mathcal{L}^{[0]}_{h^{(4)}_0} + \dfrac{1}{f}\mathcal{L}^{[1]}_{h^{(4)}_0}(\ln f) + \mathcal{L}^{[2]}_{h^{(4)}_0}(\ln f)^2, \\[1ex]
m^{(6),\mathsf{P}}_{0,\mathsf{ext}} &= M_*\bigg[\dfrac{1}{f^2}\mathcal{L}^{[0]}_{m^{(6)}_0} + \dfrac{1}{f}\mathcal{L}^{[1]}_{m^{(6)}_0}(\ln f) \nonumber \\
&\quad + \mathcal{L}^{[2]}_{m^{(6)}_0}(\ln f)^2 + \mathcal{L}^{[3]}_{m^{(6)}_0}(\ln f)^3\bigg], \\
h^{(6),\mathsf{P}}_{0,\mathsf{ext}} &= \dfrac{1}{f^3}\mathcal{L}^{[0]}_{h^{(6)}_0} + \dfrac{1}{f^2}\mathcal{L}^{[1]}_{h^{(6)}_0}(\ln f) \nonumber \\
&\quad + \dfrac{1}{f}\mathcal{L}^{[2]}_{h^{(6)}_0}(\ln f)^2 + \mathcal{L}^{[3]}_{h^{(6)}_0}(\ln f)^3,
\end{align}
where $\mathcal{L}^{[\cdot]}$ follows the same structure decomposition as in
Eq.~\eqref{eq:Lstructure}.

For $\ell \geq 2$, the same $f$- and $\ln f$-power pattern holds across all
modes $\ell$ available at a given order $s$, differing only in the coefficient
functions themselves. The general structure of the particular solutions is
\begin{align}
v^{(2),\mathsf{P}}_{\ell,\mathsf{ext}} &= \mathcal{L}^{[0]}_{v^{(2)}_\ell}, \\
h^{(2),\mathsf{P}}_{\ell,\mathsf{ext}} &= \mathcal{L}^{[0]}_{h^{(2)}_\ell}, \\
k^{(2),\mathsf{P}}_{\ell,\mathsf{ext}} &= \mathcal{L}^{[0]}_{k^{(2)}_\ell}, \\
m^{(2),\mathsf{P}}_{\ell,\mathsf{ext}} &= M_*\,\mathcal{L}^{[0]}_{m^{(2)}_\ell}, \\
v^{(4),\mathsf{P}}_{\ell,\mathsf{ext}} &= \dfrac{1}{f^2}\mathcal{L}^{[0]}_{v^{(4)}_\ell} + \dfrac{1}{f}\mathcal{L}^{[1]}_{v^{(4)}_\ell}(\ln f) + \mathcal{L}^{[2]}_{v^{(4)}_\ell}(\ln f)^2, \\
h^{(4),\mathsf{P}}_{\ell,\mathsf{ext}} &= \dfrac{1}{f^2}\mathcal{L}^{[0]}_{h^{(4)}_\ell} + \dfrac{1}{f}\mathcal{L}^{[1]}_{h^{(4)}_\ell}(\ln f) + \mathcal{L}^{[2]}_{h^{(4)}_\ell}(\ln f)^2, \\
k^{(4),\mathsf{P}}_{\ell,\mathsf{ext}} &= \dfrac{1}{f}\mathcal{L}^{[0]}_{k^{(4)}_\ell} + \mathcal{L}^{[1]}_{k^{(4)}_\ell}(\ln f) + \mathcal{L}^{[2]}_{k^{(4)}_\ell}(\ln f)^2, \\
m^{(4),\mathsf{P}}_{\ell,\mathsf{ext}} &= M_*\bigg[\dfrac{1}{f}\mathcal{L}^{[0]}_{m^{(4)}_\ell} + \mathcal{L}^{[1]}_{m^{(4)}_\ell}(\ln f) + \mathcal{L}^{[2]}_{m^{(4)}_\ell}(\ln f)^2\bigg], \\
v^{(6),\mathsf{P}}_{\ell,\mathsf{ext}} &= \dfrac{1}{f^3}\mathcal{L}^{[0]}_{v^{(6)}_\ell} + \dfrac{1}{f^2}\mathcal{L}^{[1]}_{v^{(6)}_\ell}(\ln f) \nonumber \\
&\quad + \dfrac{1}{f}\mathcal{L}^{[2]}_{v^{(6)}_\ell}(\ln f)^2 + \mathcal{L}^{[3]}_{v^{(6)}_\ell}(\ln f)^3, \\
h^{(6),\mathsf{P}}_{\ell,\mathsf{ext}} &= \dfrac{1}{f^3}\mathcal{L}^{[0]}_{h^{(6)}_\ell} + \dfrac{1}{f^2}\mathcal{L}^{[1]}_{h^{(6)}_\ell}(\ln f) \nonumber \\
&\quad + \dfrac{1}{f}\mathcal{L}^{[2]}_{h^{(6)}_\ell}(\ln f)^2 + \mathcal{L}^{[3]}_{h^{(6)}_\ell}(\ln f)^3, \\
k^{(6),\mathsf{P}}_{\ell,\mathsf{ext}} &= \dfrac{1}{f^2}\mathcal{L}^{[0]}_{k^{(6)}_\ell} + \dfrac{1}{f}\mathcal{L}^{[1]}_{k^{(6)}_\ell}(\ln f) \nonumber \\
&\quad + \mathcal{L}^{[2]}_{k^{(6)}_\ell}(\ln f)^2 + \mathcal{L}^{[3]}_{k^{(6)}_\ell}(\ln f)^3, \\
m^{(6),\mathsf{P}}_{\ell,\mathsf{ext}} &= M_*\bigg[\dfrac{1}{f^2}\mathcal{L}^{[0]}_{m^{(6)}_\ell} + \dfrac{1}{f}\mathcal{L}^{[1]}_{m^{(6)}_\ell}(\ln f) \nonumber \\
&\quad + \mathcal{L}^{[2]}_{m^{(6)}_\ell}(\ln f)^2 + \mathcal{L}^{[3]}_{m^{(6)}_\ell}(\ln f)^3\bigg],
\end{align}
where, as before, $\mathcal{L}^{[\cdot]}$ follows the decomposition of
Eq.~\eqref{eq:Lstructure}. Representative explicit coefficient
$\mathcal{L}$-functions are provided in Appendix~\ref{app:Exterior_solutions},
with the remaining higher-order expressions available in the
\texttt{hartle-thorne-project} repository~\cite{conde2025hartlethorne}.


\section{Matching Conditions}
\label{sec:matching}

As illustrated in Fig.~\ref{fig:regions} and discussed in detail in Paper~I, the
matching procedure at $R=R_*$ serves two purposes. First, it determines the
interior integration constant $C^{(n)}_{\ell,\mathsf{int}}$, which fixes the
interior solution and is required before the solution can be propagated to the
next order. Second, it determines the rescaled exterior integration constant
$\tilde{C}^{(n)}_{\ell,\mathsf{ext}}$, which completes the exterior solution and
determines the star's multipole moments $M_\ell$ and $S_\ell$. Together,
$C^{(n)}_{\ell,\mathsf{int}}$ and $\tilde{C}^{(n)}_{\ell,\mathsf{ext}}$ allow
the construction of the global spacetime geometry, interior and exterior, at
each spin order and harmonic mode number. In this section, we present the
explicit matching equations for both parity sectors, in the rescaled and
simplified form derived in Sec.~\ref{sec:exterior}.

\subsection{Odd-parity sector}

At first order in rotation, there is no source term and the exterior solution is
known in closed form, so continuity of $\varpi^{(1)}_{1}$ and its derivative
$\bar\alpha^{(1)}_{1}$ at $R=R_*$ determines the dimensionless leading spin
angular momentum $j$ and the stellar angular frequency $\Omega$ directly:
\begin{align}
j &= \frac{1}{6}\frac{R_*^{4}}{M_*^{2}}\,\bar\alpha^{(1)}_{1}(R_*), \\[1ex]
\Omega &= \varpi^{(1)}_{1}(R_*) + \frac{2jM_*^{2}}{R_*^{3}}.
\end{align}
At higher, odd, spin orders $k=3,5,7$, both the interior and exterior solutions
share the same homogeneous-plus-particular structure. Continuity of
$\omega^{(k)}_{\ell}$ and its derivative $\alpha^{(k)}_{\ell}$ is imposed at
$R=R_*$. This yields a linear $2\times2$ algebraic system for the two unknown
constants $C^{(k)}_{\ell,\mathsf{int}}$ and
$\tilde{C}^{(k)}_{\ell,\mathsf{ext}}$, whose solution is given by
\begin{align}
\label{eq:Codd_int}
C^{(k)}_{\ell,\mathsf{int}} =\ & \left.\frac{1}{\mathcal{W}^{(k)}_{\ell}}\Big\{ \left[\omega^{(k),\mathsf{P}}_{\ell,\mathsf{ext}} - \omega^{(k),\mathsf{P}}_{\ell,\mathsf{int}}\right] \alpha^{(k),\mathsf{H}}_{\ell,\mathsf{ext}} \right. \nonumber \\
& \left. + \ \omega^{(k),\mathsf{H}}_{\ell,\mathsf{ext}}\left[\alpha^{(k),\mathsf{P}}_{\ell,\mathsf{int}} - \alpha^{(k),\mathsf{P}}_{\ell,\mathsf{ext}}\right] \Big\}\right|_{R=R_*}, \\[2ex]
\label{eq:Codd_ext}
\tilde{C}^{(k)}_{\ell,\mathsf{ext}} =\ & \left.\frac{1}{\mathcal{W}^{(k)}_{\ell}}\Big\{ \left[\omega^{(k),\mathsf{P}}_{\ell,\mathsf{ext}} - \omega^{(k),\mathsf{P}}_{\ell,\mathsf{int}}\right] \alpha^{(k),\mathsf{H}}_{\ell,\mathsf{int}} \right. \nonumber \\
& \left. + \ \omega^{(k),\mathsf{H}}_{\ell,\mathsf{int}}\left[\alpha^{(k),\mathsf{P}}_{\ell,\mathsf{int}} - \alpha^{(k),\mathsf{P}}_{\ell,\mathsf{ext}}\right] \Big\}\right|_{R=R_*},
\end{align}
where the Wronskian $\mathcal{W}^{(k)}_{\ell}$ at the stellar surface is
\begin{equation}
\mathcal{W}^{(k)}_{\ell} \equiv \left.\Big[\omega^{(k),\mathsf{H}}_{\ell,\mathsf{int}}\,\alpha^{(k),\mathsf{H}}_{\ell,\mathsf{ext}} - \omega^{(k),\mathsf{H}}_{\ell,\mathsf{ext}}\,\alpha^{(k),\mathsf{H}}_{\ell,\mathsf{int}}\Big]\right|_{R=R_*}.
\end{equation}
The Wronskian does not vanish, since the interior and exterior homogeneous
solutions are linearly independent, one being selected by regularity at the
center and the other by asymptotic flatness at spatial infinity. Once
$C^{(k)}_{\ell,\mathsf{int}}$ is known, the interior solution is fully
determined and can be propagated to the next order. Once
$\tilde{C}^{(k)}_{\ell,\mathsf{ext}}$ is known, the exterior solution
$\omega^{(k)}_{\ell,\mathsf{ext}}$ is fully determined, and the multipole
moments $M_\ell$ and $S_\ell$ follow directly, as shown in
Appendix~\ref{app:multipoles}.

\subsection{Even-parity sector}

\subsubsection*{$\ell = 0$ case}

For the $\ell = 0$ case, the two constants are fixed one after the other, rather
than from a simultaneous pair of matching equations. The interior equation for
$h^{(s)}_{0,\mathsf{int}}$ is first order, and its right-hand side depends on
$m^{(s)}_{0,\mathsf{int}}$ and $\xi^{(s)}_{0}$, but not on
$h^{(s)}_{0,\mathsf{int}}$ itself. Its homogeneous solution is therefore simply
a constant, which we denote $h^{(s)}_{0,c}$. Unlike $m^{(s)}_{0,\mathsf{int}}$,
which vanishes at the center by regularity, this constant is not fixed by the
interior equation alone. Matching $m^{(s)}_{0}$ at the surface first determines
$\tilde{C}^{(s)}_{0,\mathsf{ext}}$, and $h^{(s)}_{0,c}$ then follows, as shown
below.

Continuity of $m^{(s)}_{0}$ at $R=R_*$ determines
$\tilde{C}^{(s)}_{0,\mathsf{ext}}$ directly, since the exterior homogeneous
solution $m^{(s),\mathsf{H}}_{0,\mathsf{ext}} = M_*$ is simply a constant:
\begin{equation}
\label{eq:Cs0ext}
\tilde{C}^{(s)}_{0,\mathsf{ext}} =  \dfrac{1}{M_*} \left. \left[ m^{(s)}_{0,\mathsf{int}} - m^{(s),\mathsf{P}}_{0,\mathsf{ext}} \right] \right|_{R=R_*}.
\end{equation}
With $\tilde{C}^{(s)}_{0,\mathsf{ext}}$ known, continuity of $h^{(s)}_{0}$ at
$R=R_*$ then fixes the free constant $h^{(s)}_{0,c}$:
\begin{equation}
\label{eq:hs0c}
h^{(s)}_{0,c} = \left.\left[h^{(s),\mathsf{P}}_{0,\mathsf{ext}} - h^{(s),\mathsf{P}}_{0,\mathsf{int}} + \tilde{C}^{(s)}_{0,\mathsf{ext}}\,h^{(s),\mathsf{H}}_{0,\mathsf{ext}}\right]\right|_{R=R_*}.
\end{equation}

\subsubsection*{$\ell \geq 2$ case}

For the $\ell \geq 2$ case, continuity of $v^{(s)}_{\ell}$ and $h^{(s)}_{\ell}$
is imposed at $R=R_*$. This yields a linear $2\times2$ algebraic system for the
two unknown constants $C^{(s)}_{\ell,\mathsf{int}}$ and
$\tilde{C}^{(s)}_{\ell,\mathsf{ext}}$, with solution
\begin{align}
\label{eq:Ceven_int}
C^{(s)}_{\ell,\mathsf{int}} =\ & \left.\frac{1}{\mathcal{W}^{(s)}_{\ell}}\Big\{ \big[v^{(s),\mathsf{P}}_{\ell,\mathsf{ext}} - v^{(s),\mathsf{P}}_{\ell,\mathsf{int}}\big]\,h^{(s),\mathsf{H}}_{\ell,\mathsf{ext}} \right. \nonumber \\
& \left. + \ v^{(s),\mathsf{H}}_{\ell,\mathsf{ext}}\big[h^{(s),\mathsf{P}}_{\ell,\mathsf{int}} - h^{(s),\mathsf{P}}_{\ell,\mathsf{ext}}\big] \Big\}\right|_{R=R_*}, \\[2ex]
\label{eq:Ceven_ext}
\tilde{C}^{(s)}_{\ell,\mathsf{ext}} =\ & \left.\frac{1}{\mathcal{W}^{(s)}_{\ell}}\Big\{ \big[v^{(s),\mathsf{P}}_{\ell,\mathsf{ext}} - v^{(s),\mathsf{P}}_{\ell,\mathsf{int}}\big]\,h^{(s),\mathsf{H}}_{\ell,\mathsf{int}} \right. \nonumber \\
& \left. + \ v^{(s),\mathsf{H}}_{\ell,\mathsf{int}}\big[h^{(s),\mathsf{P}}_{\ell,\mathsf{int}} - h^{(s),\mathsf{P}}_{\ell,\mathsf{ext}}\big] \Big\}\right|_{R=R_*},
\end{align}
where $\mathcal{W}^{(s)}_{\ell}$ is the corresponding Wronskian, again evaluated
at $R=R_*$:
\begin{equation}
\mathcal{W}^{(s)}_{\ell} \equiv \left.\Big[v^{(s),\mathsf{H}}_{\ell,\mathsf{int}}\,h^{(s),\mathsf{H}}_{\ell,\mathsf{ext}} - v^{(s),\mathsf{H}}_{\ell,\mathsf{ext}}\,h^{(s),\mathsf{H}}_{\ell,\mathsf{int}}\Big]\right|_{R=R_*}.
\end{equation}
Continuity of $k^{(s)}_{\ell}$ follows automatically, since
$k^{(s)}_{\ell}=v^{(s)}_{\ell}-h^{(s)}_{\ell}$ on both sides of the surface. The
mass function $m^{(s)}_{\ell}$ is instead reconstructed afterwards through
Eq.~\eqref{eq:msl_int}, but it enters neither the matching nor the moments,
which are built from $\tilde{C}^{(s)}_{\ell,\mathsf{ext}}$ alone.


\section{Numerical Implementation}
\label{sec:numerical}

Having established the perturbation equations, exterior solutions, and matching
conditions, we cast this formalism into a \texttt{c++} code to numerically
compute the multipole moments $M_\ell$ and $S_\ell$ and their spin-induced
corrections. After addressing the robust evaluation of high-order derivatives
($\mathrm{d}\varepsilon/\mathrm{d}R$ through
$\mathrm{d}^3\varepsilon/\mathrm{d}R^3$) in Sec.~\ref{sec:derivatives}, we
detail the order-by-order integration scheme (Sec.~\ref{sec:order_by_order}).
The interior equations are integrated and matched to the exterior solutions at
each step. The interior integration constants $C^{(n)}_{\ell,\mathsf{int}}$
propagate the calculation to the next order, while the exterior constants
$\tilde{C}^{(n)}_{\ell,\mathsf{ext}}$ determine the multipole moments and their
spin-induced corrections, using the expressions given in
Appendix~\ref{app:multipoles}.

\subsection{Higher derivatives of energy density profile}
\label{sec:derivatives}

The perturbation equations require higher radial derivatives of the background
energy density profile, $\varepsilon(R)$. As discussed in Paper~I, these
derivatives are related to $d^n c_s^2/d\varepsilon^n$, which in turn are set by
the susceptibilities of the EoS.\footnote{For a barotropic EoS, $c_s^2 =
n_B/(\chi_2^B \mu_B)$, where $n_B$ is the baryon density, $\mu_B$ is the baryon
chemical potential, and $\chi_n^B \equiv d^n p/d\mu_B^n$ is the $n$th-order
susceptibility, with $n_B = \chi_1^B$. Since $d\varepsilon/d\mu_B = \mu_B
\chi_2^B$, the derivatives $d^n c_s^2/d\varepsilon^n$ can be written in terms of
$\mu_B$ and the susceptibilities up to $\chi_{n+2}^B$.} These derivatives may
therefore carry important information relevant to the search for phase
transitions inside neutron stars. However, computing these derivatives
accurately is numerically challenging. Because susceptibilities can diverge when
phase transitions are present, we restrict this work to two EoSs with only
nucleonic degrees of freedom, which avoids the phase transitions that exotic
degrees of freedom may introduce. We leave the influence of phase transitions on
the multipole moments for a separate study.

It is possible to calculate susceptibilities within microscopic models, see
e.g.~\cite{Gholami:2025cfq,Cruz-Camacho:2024odu}, but these cannot be used
directly here, for three reasons. First, effective models of QCD do not directly
incorporate leptonic degrees of freedom, which are required to obtain the
$\beta$-equilibrium trajectory. Second, these models also require a crust EoS to
be attached, and the way it is attached can change $c_s^2$
significantly~\cite{ReinkePelicer:2025vuh}. Third, susceptibilities calculated
within microscopic models, typically in the grand canonical ensemble, are taken
at fixed chemical potentials. At $\beta$ equilibrium, the electric-charge
chemical potential varies with $\mu_B$, so the model's susceptibilities are not
the same as the derivatives along the $\beta$-equilibrium trajectory. The
derivatives must therefore be obtained numerically from the tabulated EoS.

Computing high-order derivatives from a tabulated EoS is challenging, even
without explicit phase transitions. Three common approaches either fail or
amplify noise. Differentiating piecewise polynomials, such as cubic splines,
yields discontinuous, staircase-like third derivatives. Finite-difference
schemes on nonuniform grids suffer from severe noise amplification that worsens
with grid refinement. Global EoS fits suppress noise, but at the cost of
smoothing out important local features, such as the core-crust transition.
Balancing local EoS fidelity with noise-suppressed high-order differentiation,
therefore, requires a dedicated approach, which we develop in this subsection.

\subsubsection*{Segmented differentiation method}

The strategy adopted in this work is based on a segmented fit of the curve
$p(\varepsilon)$. The main philosophy of this approach is that a new segment
boundary is introduced wherever the data exhibits a rapid change in curvature,
such as a smooth crossover between different regimes, or a kink discontinuity. A
kink may signal a phase transition or the junction between different
piecewise-polytropic pieces in some EoS-representations. A boundary is also
introduced where the interval has become too wide for a single fit. Each segment
is then described by an analytic curve, which can be differentiated in closed
form. The derivatives therefore do not compound the interpolation error that
repeated differencing of the table would introduce. The
$\varepsilon(R)$-derivatives needed by the source terms follow from these
derivatives through the chain rule and the structure equations. We summarize the
procedure in the five steps below, and Appendix~\ref{app:derivatives_algorithm}
develops each step in detail.

\vspace{0.2cm} \noindent \textbf{Step 1: }\textit{Choose the EoS to fit.} The
starting point of the procedure is a tabulated barotropic EoS, $p(\varepsilon)$,
and the fit is performed for $w\equiv\ln p$ against $u\equiv\ln\varepsilon$,
where the curve resembles a nearly straight line rather than the steeply varying
form it takes on a linear scale.

\vspace{0.2cm} \noindent \textbf{Step 2: }\textit{Choose a fitting
representation and acceptance criteria.} A candidate segment is a contiguous
block of $m$ rows of the EoS table, each row being a pair $(u_i, w_i)$. For each
candidate segment, we construct a least-squares fit to a chosen, smooth,
analytic function over the full range of $u$ spanned by those rows. Here we use
a smooth fit rather than interpolating the raw data, because the smooth fit
suppresses local irregularities in derivatives that can become dangerously
amplified at high order. We choose a truncated sum of Chebyshev polynomials as
our analytic function, although other functions could also be adopted. Since
Chebyshev polynomials are bounded by unity on the interval $[-1,1]$, each term
in the sum contributes at most the magnitude of its own coefficient on that
interval. Coefficients that become smaller in magnitude as the polynomial degree
increases therefore indicate that enough terms have been retained.

In practice, the domain of each candidate segment is first mapped affinely onto
$[-1,1]$, with $x$ the mapped variable, and the sum is truncated at degree
$\kappa=10$,
\begin{equation}
w_{\mathrm{fit}}(x) = \sum_{\kappa=0}^{10} a_\kappa\, T_\kappa(x),
\label{eq:cheb_expansion_main}
\end{equation}
where $T_\kappa$ is the Chebyshev polynomial of degree $\kappa$ and $a_\kappa$
are fitted coefficients. On short segments, the degree is reduced, so that the
number of free coefficients never exceeds half the number of rows, $m/2$, and
the fit cannot degenerate into an interpolation. A fit is accepted only if its
maximum $\ln p$ residual does not exceed $\tau = 10^{-7}$, that is,
$|w_{\mathrm{fit}}(x_i) - w_i| \le \tau$ at every row, where $x_i$ is the mapped
value of $u_i$ (see Appendix~\ref{app:derivatives_algorithm} for the affine
map). For fits at the full degree $\kappa=10$, the magnitudes of the last two
coefficients must also not exceed $10\tau$, that is, $|a_9| \le 10\tau$ and
$|a_{10}| \le 10\tau$. This coefficient test is not applied to reduced-degree
fits, whose highest-order coefficients can still carry the main shape of the
curve. For the tables considered here, tighter tolerances merely fragment the
table without improving accuracy.

\vspace{0.2cm} \noindent \textbf{Step 3: }\textit{Divide the domain into
segments.} Segment boundaries are determined automatically by the breakdown of
the acceptance criteria of Step 2. A polynomial of degree $\kappa=10$ smoothly
tracks continuous features over a finite interval, but can fail these criteria
as the interval is enlarged. This failure can occur near localized physical
non-smoothnesses, such as slope kinks associated with phase transitions, or in
smooth regions of high curvature. It can also occur simply because the interval
has become too wide for a single expansion of the chosen degree. The failure
therefore determines how far the current segment can extend while satisfying the
prescribed fitting criteria; it does not by itself identify a physical
interface.

Beginning at the lowest density, we progressively enlarge a candidate segment
and repeat the fit and acceptance tests of Step 2. When an enlarged candidate
fails, we retain the last accepted segment. The next segment begins at the last
accepted row or a subsequent row, depending on whether a valid fit can be
formed. This procedure repeats until the table is exhausted. A segment must
contain at least 22 rows, twice the eleven coefficients of a degree-ten fit, so
that the fit cannot degenerate into an interpolation. Only the final segment of
the table may be shorter. For the two EoSs used here, the shortest segment has
thirty rows. The rule is given in full in
Appendix~\ref{app:derivatives_algorithm}, together with the conditions under
which a table is refused as too coarsely sampled for the criteria to be
meaningful.

\vspace{0.2cm} \noindent \textbf{Step 4: }\textit{Classify each boundary.} A
segment boundary can mark a genuine feature in the EoS, such as a first-order
phase transition, or arise because a single degree-ten expansion no longer
satisfies the acceptance criteria over a longer span. We distinguish these cases
by evaluating both adjacent fits at the boundary and comparing their logarithmic
slopes $\Gamma = \mathrm{d}\ln p/\mathrm{d}\ln\varepsilon$. An absolute
difference in $\Gamma$ of at least $10^{-3}$ classifies the boundary as an
\textit{interface}. At an interface, discontinuities in the derivatives of
$\varepsilon$ must be accounted for when integrating the perturbation equations.
Otherwise, the boundary is treated as a \textit{knot}, a transition of the
representation with no physical content. The test reads the first derivative
alone and does not constrain the higher ones. A higher-order phase transition,
where the slope is continuous and only higher derivatives jump, would therefore
be classified as a knot. We tested the criterion on the seven-piece polytropic
representation of SLy from~\cite{Read:2008iy}, which has six known junctions.
The method recovers all six dividing densities to within $0.5\%$ and the slope
to within $0.2\%$. A first-order phase transition also produces a jump in
$\Gamma$, so it would be classified as an interface. We leave the validation of
the method on EoSs with other phase transitions for future work. For the tables
used in this work every boundary is a knot.

\vspace{0.2cm} \noindent \textbf{Step 5: }\textit{Compute the
$\varepsilon(R)$-derivatives.} Once the segments have been determined, the
accepted representations are differentiated analytically using exact Chebyshev
recurrence relations. These supply the derivatives of $w(u)$ (where recall that
$w = \ln(p)$ and $u = \ln(\varepsilon)$), which are converted into derivatives
of $p(\varepsilon)$. The source terms, however, require radial derivatives of
$\varepsilon$. Since the EoS is barotropic, these follow from the derivatives of
$p(\varepsilon)$ through the chain rule and the structure equations.\footnote{At
lowest order, for example, $c_s^2 = dp/d\varepsilon = (p/\varepsilon)\,dw/du$
and $d\varepsilon/dR =(dp/dR)/c_s^2$, with $dp/dR$ given by the TOV equation.}
The calculation proceeds in order, with each successive radial derivative
requiring only quantities already computed, so no tabulated or background
quantity is differentiated numerically. The recurrence relations, conversions
and order reduction are detailed in Appendix~\ref{app:derivatives_algorithm}.

\begin{figure*}[t]
\includegraphics[width=\textwidth]{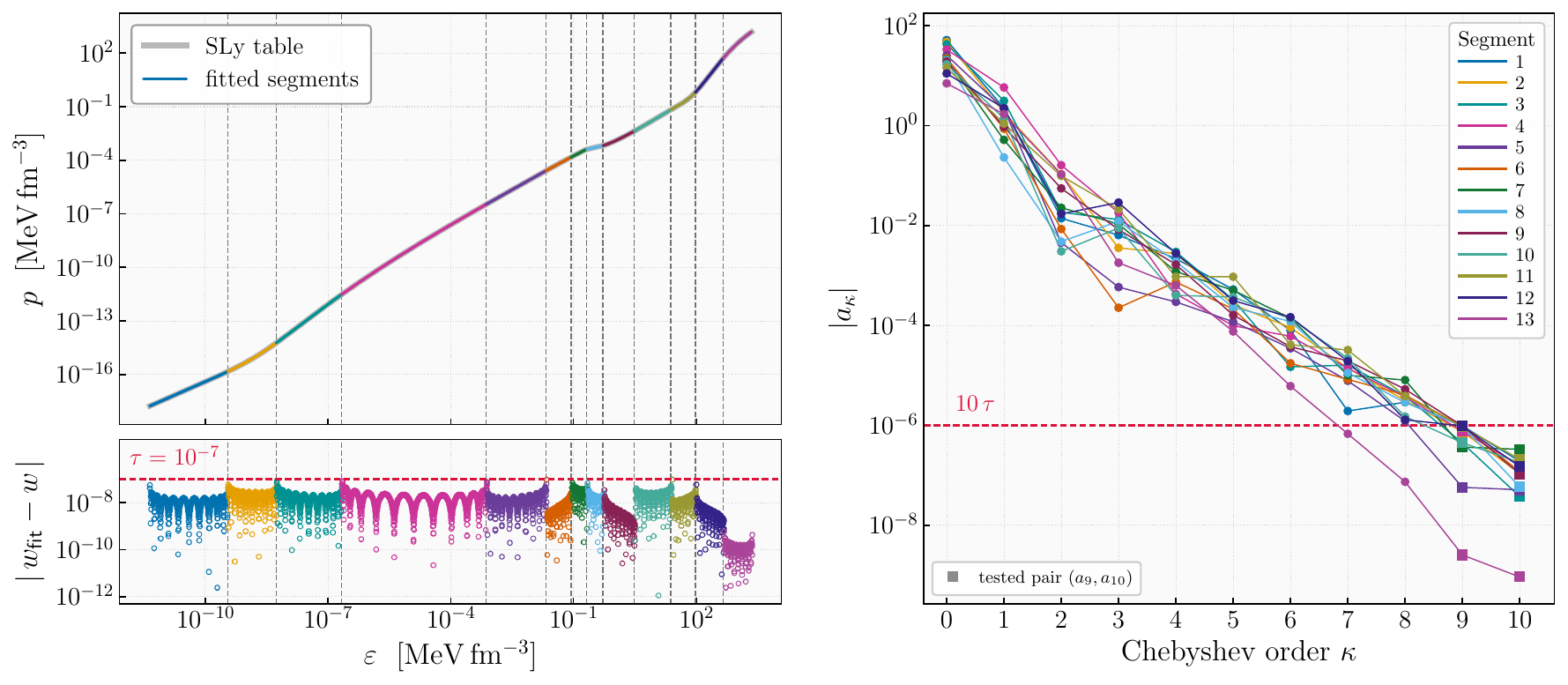}
\caption{The segmented Chebyshev representation of the SLy EoS used in this work. The fit is performed on $w=\ln p$ as a function of $u=\ln\varepsilon$. \textit{Left (upper):} the tabulated data and the fitted polynomial of each segment, with the boundaries located by the algorithm marked by vertical dashed lines. Thirteen segments are found. \textit{Left (lower):} the residual
$|w_{\mathrm{fit}}-w|$ at every tabulated point, measured against the tolerance $\tau=10^{-7}$ (horizontal dashed line). Every residual lies below $\tau$. \textit{Right:} the magnitudes of the Chebyshev coefficients against order
$\kappa$, one curve per segment, with colors matching the left panel. The squares mark the pair $(a_9,a_{10})$ read by the convergence criterion, and the dashed line is the bar $10\tau$ below which both must fall. Both criteria are
therefore satisfied on every segment.}
\label{fig:eos_fit}
\end{figure*}

The fit to the SLy EOS results in thirteen segments as shown in
Fig.~\ref{fig:eos_fit} and summarized in Table~\ref{table:eos_segments}. No
segment is short enough for the degree cap to act, so degree $\kappa=10$ is used
throughout. Both acceptance criteria hold everywhere. The largest residual over
all segments is $9.6\times10^{-8}$, below the tolerance $\tau=10^{-7}$, and the
lower left panel of Fig.~\ref{fig:eos_fit} shows that every tabulated point of
every segment lies below that bar. The coefficients (right panel) fall
geometrically with $\kappa$-order, consistent with a neglected tail below the
tolerance. The logarithmic slope of the fitted curve, $\Gamma=\mathrm{d}\ln
p/\mathrm{d}\ln\varepsilon$ is continuous across every boundary, so that all of
them are classified as \textit{knots}. The same procedure applied to BSk21
yields seventeen segments and behaves analogously. The coefficients of both
representations are provided with the code~\cite{conde2025aqlimr}, as one file
per EoS, so that the fits used here can be evaluated directly. Their format is
described in Appendix~\ref{app:derivatives_algorithm}.

\begin{table}[h!]
\setlength{\tabcolsep}{4pt}
\renewcommand{\arraystretch}{1.0}
\begin{tabular}{c r@{$,\;$}l c c r@{$\,$--$\,$}l}
\toprule
Seg. & \multicolumn{2}{c}{$\log_{10}\varepsilon$} & $\max|\Delta w|$
     & $\max|a_{9,10}|$ & \multicolumn{2}{c}{$\Gamma$} \\
\cmidrule(lr){2-3}\cmidrule(lr){6-7}
1 & $-11.4$ & $-9.4$ & $5.6{\times}10^{-8}$ & $9.8{\times}10^{-7}$ & $1.00$ & $1.08$ \\
2 & $-9.4$ & $-8.3$ & $8.2{\times}10^{-8}$ & $7.4{\times}10^{-7}$ & $1.08$ & $1.65$ \\
3 & $-8.3$ & $-6.7$ & $8.2{\times}10^{-8}$ & $4.4{\times}10^{-7}$ & $1.65$ & $1.62$ \\
4 & $-6.7$ & $-3.1$ & $8.5{\times}10^{-8}$ & $7.7{\times}10^{-7}$ & $1.62$ & $1.30$ \\
5 & $-3.1$ & $-1.7$ & $9.5{\times}10^{-8}$ & $5.8{\times}10^{-8}$ & $1.30$ & $1.27$ \\
6 & $-1.7$ & $-1.1$ & $9.5{\times}10^{-8}$ & $9.5{\times}10^{-7}$ & $1.27$ & $1.21$ \\
7 & $-1.1$ & $-0.7$ & $7.1{\times}10^{-8}$ & $3.7{\times}10^{-7}$ & $1.21$ & $0.74$ \\
8 & $-0.7$ & $-0.3$ & $6.4{\times}10^{-8}$ & $9.6{\times}10^{-7}$ & $0.74$ & $0.71$ \\
9 & $-0.3$ & $0.5$ & $3.5{\times}10^{-8}$ & $9.8{\times}10^{-7}$ & $0.71$ & $1.27$ \\
10 & $0.5$ & $1.4$ & $9.6{\times}10^{-8}$ & $4.5{\times}10^{-7}$ & $1.27$ & $1.29$ \\
11 & $1.4$ & $2.0$ & $9.6{\times}10^{-8}$ & $9.3{\times}10^{-7}$ & $1.29$ & $2.39$ \\
12 & $2.0$ & $2.7$ & $5.9{\times}10^{-8}$ & $9.7{\times}10^{-7}$ & $2.39$ & $2.67$ \\
13 & $2.7$ & $3.4$ & $5.9{\times}10^{-10}$ & $2.6{\times}10^{-9}$ & $2.67$ & $1.63$ \\
\bottomrule
\end{tabular}
\caption{Segmented Chebyshev representation of the SLy EoS, obtained by fitting $w=\ln p$ as a function of $u=\ln\varepsilon$, with $p$ and $\varepsilon$ in
$\mathrm{MeV\,fm^{-3}}$ so that both logarithms are taken of the numerical value in those units. For each segment the table lists the domain in $\log_{10}\varepsilon$, the largest residual $\max|\Delta w|=\max|w_{\mathrm{fit}}-w|$, the larger in magnitude of the last two Chebyshev
coefficients, and the logarithmic slope of the fitted curve, $\Gamma=\mathrm{d}\ln p/\mathrm{d}\ln\varepsilon$, evaluated analytically at each end of the segment. $\Gamma$ is dimensionless.}
\label{table:eos_segments}
\end{table}

\begin{figure}
\includegraphics[width=8cm]{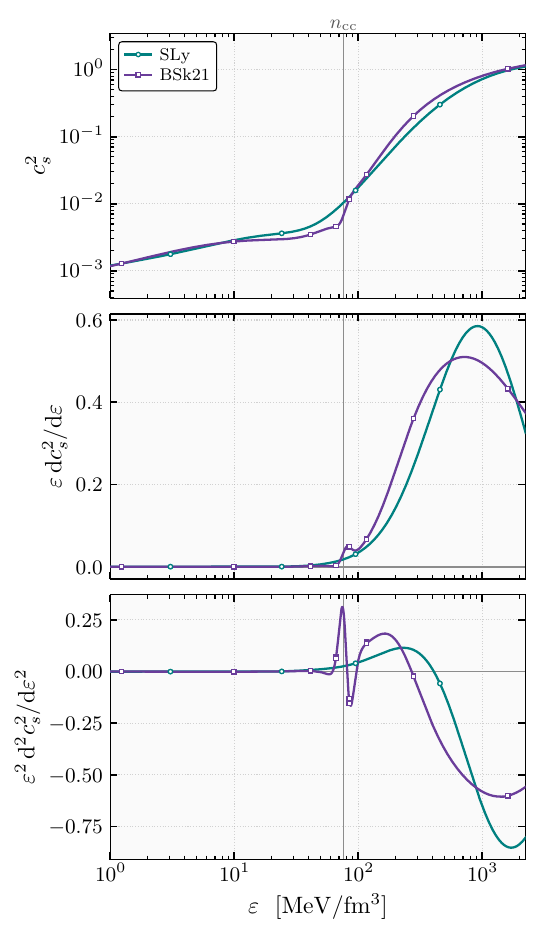}
\caption{The speed of sound squared, $c_s^2$ (top), and its first two derivatives with respect to the energy density $\varepsilon$, for SLy and BSk21. The derivatives are multiplied by powers of $\varepsilon$, so that $\varepsilon\,dc_s^2/d\varepsilon$ (middle) and $\varepsilon^2\,d^2c_s^2/d\varepsilon^2$ (bottom) are dimensionless. All curves follow analytically from the segmented Chebyshev fits. The vertical line marks the crust-core transition.}
\label{fig:cs2-derivatives}
\end{figure}

With the fits in hand, the derivatives of the EoS follow analytically, as
described in Step 5. The source terms contain the radial derivatives
$d\varepsilon/dR$, $d^2\varepsilon/dR^2$ and $d^3\varepsilon/dR^3$. Each
successive radial derivative brings in one more derivative of the sound speed,
namely $c_s^2$ for $d\varepsilon/dR$, $dc_s^2/d\varepsilon$ for
$d^2\varepsilon/dR^2$, and $d^2c_s^2/d\varepsilon^2$ for $d^3\varepsilon/dR^3$.
Figure~\ref{fig:cs2-derivatives} shows these three quantities for both EoSs. The
vertical line, labeled $n_{\mathrm{cc}}$, marks the crust-core transition, so
the crust lies to its left and the core to its right. The markers on each curve
indicate the knots, where one segment ends and the next begins. Both derivatives
are small in the crust and grow in the core, where the first derivative peaks
and the second changes sign. At the transition the two EoSs behave differently.
SLy varies smoothly, with a single knot nearby, while BSk21 changes rapidly in
all three panels, and the algorithm places several knots close together there.

\subsection{Numerical implementation order-by-order}
\label{sec:order_by_order}

We begin by solving the non-rotating background and then proceed through seventh
order in rotation, presenting at each order the harmonic modes that appear, the
systems they obey, and the multipole moments they determine.

\subsubsection*{Zeroth Order in Rotation}

The perturbation equations are integrated and matched at the stellar surface
with the same scheme at every order and for every mode. We therefore describe
the common elements once, in this subsection, together with the nonrotating
background. Each later subsection then describes only what is new at that order.

The non-rotating configuration begins with a tabulated EoS, specified by two
columns, $\varepsilon$ and $p$. For the two models considered here, the tables
are generated from the published analytic representations, that of Haensel and
Potekhin for SLy~\cite{Haensel:2004nu} and that of Potekhin et al.~for
BSk21~\cite{Potekhin:2013qqa}, and are provided with our numerical
code~\cite{conde2025aqlimr}. Sampling the tables from these analytic
representations lets us choose their resolution and numerical precision. The
same representations also give analytic derivatives, against which the
derivatives obtained in Sec.~\ref{sec:derivatives} can be checked\footnote{The
alternative of differentiating the global fit directly is not adopted, since it
would apply only to EoSs for which such a fit exists, and since the higher
derivatives of a global fit reflect its functional form rather than the
underlying data. The segmented procedure is therefore used uniformly.}. Since
the tabulated data are given only at discrete points, they are interpolated with
Steffen's method~\cite{Steffen1990}, yielding a continuous, monotonic barotropic
relation $p(\varepsilon)$. Thermodynamic stability requires that the
second-order susceptibility, $\chi_2^B \equiv d^2p/d\mu_B^2$, is positive, which
implies both that $p(\varepsilon)$ increases monotonically with $\varepsilon$
and that $c_s^2\geq 0$ always holds.

With the EoS in hand, a specific stellar model is selected by fixing the central
energy density $\varepsilon_c$. The corresponding central pseudo-enthalpy,
$\mathsf{h}_c$, follows by evaluating the integral in Eq.~\eqref{eq:hofe},
computed numerically with the adaptive quadrature routine
\texttt{gsl\_integration\_qag} from the GNU Scientific Library (GSL)~\cite{GSL}.
Since Eqs.~\eqref{eq:Rh} and~\eqref{eq:Mh} are singular exactly at the center,
the integration instead begins at a small initial radius $R_{i}$ away from it,
with the corresponding value $\mathsf{h}_i$ obtained from the near-center
expansion given in Eq.~\eqref{eq:hi}. The coupled system for $R(\mathsf{h})$ and
$M(\mathsf{h})$ is then integrated numerically from $\mathsf{h}=\mathsf{h}_i$ to
$\mathsf{h}=0$, using the GSL Runge-Kutta-Fehlberg (4,5) method
\texttt{gsl\_odeiv2\_step\_rkf45}. The relative and absolute tolerances are set
to $10^{-15}$, which forces the adaptive stepper to take small steps, producing
a dense set of radial points, about $\sim3300$ for SLy and $\sim3400$ for BSk21
at $M_* = 1.5\,M_\odot$. The integration terminates at $\mathsf{h}=0$, where the
stellar surface is reached exactly, and the values of $R$ and $M$ at that point
define the TOV radius and mass, $R_*$ and $M_*$, respectively. The radial points
recorded along this integration form the TOV $R$-grid on which every subsequent
order is evaluated, as described at the start of this subsection. With $R_*$ and
$M_*$ known, the metric function $\nu$ follows directly from
Eq.~\eqref{eq:nu_interior}. The integration constant $\nu_*$ is fixed by
Eq.~\eqref{eq:nu_match}, and a vector solution for $\nu$ is constructed by
evaluating this expression at each value of $\mathsf{h}$ already stored on the
TOV $R$-grid.

The profiles $M(R)$, $p(R)$, $\varepsilon(R)$, and $\nu(R)$ evaluated on the
$R$-grid are then interpolated as described at the start of this subsection.
From the background profiles, the derivatives $d\varepsilon/dR$,
$d^2\varepsilon/dR^2$, and $d^3\varepsilon/dR^3$ required by the source terms at
higher orders (Sec.~\ref{sec:source}) are computed on the same TOV-$R$ grid,
using the method described in Sec.~\ref{sec:derivatives}. In particular, the
perturbation system at every higher order is integrated between successive
reference-grid points using an adaptive Dormand–Prince 5(4) scheme
(\texttt{runge\_kutta\_dopri5} from the Boost.Odeint library~\cite{Boost}), with
adaptive steps confined to each interval\footnote{The absolute and relative
tolerances are set as follows. We choose a tolerance of $10^{-12}$, absolute and
relative, through fifth order. The sixth-order monopole and the four
seventh-order modes use $10^{-10}$ absolute and $10^{-8}$ relative tolerances.
The sixth-order $\ell=2$, $4$ and $6$ harmonic modes use a $10^{-6}$ tolerance
for both. The looser values at sixth and seventh order follow from the mechanism
of Sec.~\ref{sec:single_star}. There, the sources are assembled from the
solutions of all lower orders and involve large cancellations, so below a
certain step size the controller measures the noise of that assembly rather than
the error of the step. At $10^{-12}$, the sixth-order $\ell=2$, $4$ and $6$
integration no longer completes, since the controller shrinks the step below its
minimum in pursuit of an error it cannot reduce.}. Solutions are evaluated and
stored strictly at the grid points to ensure a uniform radial grid across all
orders. These stored solutions are then interpolated using cubic splines
(\texttt{gsl\_interp\_cspline}) to provide the smooth radial dependence required
by higher-order source terms. Evaluating all orders on this shared grid
eliminates cross-grid interpolation errors.

\subsubsection*{First Order in Rotation}

At first order, the metric function $\varpi^{(1)}_{1}$ and its derivative
$\bar\alpha^{(1)}_{1}$ satisfy the homogeneous system of
Eq.~\eqref{eq:varpi_alpha_system}. The rotating stellar model at first order is
uniquely determined by the parametric set $\{\varepsilon_c,
\varpi^{(1)}_{1,c}\}$. We obtain $\varpi^{(1)}_1$ and $\bar\alpha^{(1)}_1$ by
evaluating Eqs.~\eqref{eq:wb11_asym}--\eqref{eq:ab11_asym} at the initial radial
distance $R_i$, using a trial value
$\varpi^{(1)}_{1,c}=\varpi^{(1),\mathrm{trial}}_{1,c}$. Using these initial
conditions and the interpolated background profiles from zeroth order, the ODE
system of Eq.~\eqref{eq:varpi_alpha_system} is integrated outward until the
surface is reached at $R=R_*$. There, the trial solution is matched to the
exterior, giving the leading dimensionless spin angular momentum and the stellar
angular frequency,
\begin{align}
j^{\mathrm{trial}} &= \frac{1}{6}\frac{R_*^{4}}{M_*^{2}}\,\bar\alpha^{(1),\mathrm{trial}}_{1}(R_*), \label{eq:j_match} \\[1ex]
\Omega^{\mathrm{trial}} &= \varpi^{(1),\mathrm{trial}}_{1}(R_*) + \frac{2j^{\mathrm{trial}}M_*^{2}}{R_*^{3}}. \label{eq:Omega_match}
\end{align}
Since the spin of neutron stars is approximately bounded by its Keplerian
frequency, an arbitrarily large trial value can produce an
$\Omega^{\mathrm{trial}}$ that violates this limitation. For simplicity, we
adopt the Newtonian Keplerian frequency $\Omega_K=\sqrt{M_*/R_*^3}$ to check the
validity of $\Omega^{\mathrm{trial}}$. We use $\Omega_K$ as a reference scale
for rotation; the relativistic mass-shedding frequency generally differs from
this Newtonian estimate. In practice, the trial value
$\varpi^{(1),\mathrm{trial}}_{1,c}=1$ is adopted throughout this work and is
found to satisfy this condition for both the EoS models considered. We check the
target angular frequency $\Omega^{\mathrm{target}}\equiv\Omega_*=2\pi f_*$ in
the same way, with $f_*=716\,\mathrm{Hz}$~\cite{Hessels:2006ze}. Every star in
both sequences satisfies $\Omega_*<\Omega_K$, and the largest value of
$\Omega_*/\Omega_K$ is $0.64$.

Rather than iteratively searching for $\varpi^{(1),\mathrm{trial}}_{1,c}$ to
match $\Omega^{\mathrm{target}}$, we exploit the fact that
Eq.~\eqref{eq:varpi_alpha_system} is linear and homogeneous, so that any trial
solution can be rescaled to the target star in a single step via $C =
\Omega^{\mathrm{target}}/\Omega^{\mathrm{trial}}$. In particular, since
$j^{\mathrm{trial}}$ and $\Omega^{\mathrm{trial}}$ in
Eqs.~\eqref{eq:j_match}--\eqref{eq:Omega_match} are themselves linear in
$\varpi^{(1),\mathrm{trial}}_1$ and its derivative, this same rescaling carries
through directly to the matched quantities,
\begin{equation}
    C\,j^{\mathrm{trial}} = j^{\rm target}\,.
\end{equation}
Consequently, the normalized quantities $\hat{\varpi}^{(1)}_1 \equiv
\varpi^{(1)}_1/\Omega$ and $\hat{j} \equiv j/\Omega$ are strictly independent of
the trial choice, and hinge solely on the stellar structure ($\varepsilon_c$ and
the EoS). Hereafter, we drop the trial label and work directly with these
scale-invariant quantities, and the star's own angular frequency is trivially
$\Omega/\Omega=1$ under this normalization.

This normalization propagates naturally to all higher orders $n \ge 2$. Because
the source terms at order $n$ consist of products whose spin-frequency powers
sum to $n$, evaluating them with lower-order hatted variables automatically
factors out $\Omega^n$. The inhomogeneous perturbation equations and boundary
matching are thus solved entirely in terms of scale-invariant quantities,
$\hat{F}^{(n)} \equiv F^{(n)}/\Omega^n$. Evaluating the relations in
Appendix~\ref{app:multipoles} directly with the hatted quantities yields the
normalized multipole moments $\hat{M}_\ell$ and $\hat{S}_\ell$. In addition to
postponing the specification of $\Omega^{\mathrm{target}}$ to the final
post-processing step, working in the normalized units help maintain variables at
$\mathcal{O}(1)$, thereby avoiding the extremely wide range of unnormalized
solutions. For the results presented in this work, $\Omega^{\rm target} \equiv
\Omega_{*} = 2\pi f_{*}$, with $f_{*}=716\,\mathrm{Hz}$.
The resulting solutions $\hat\varpi^{(1)}_1(R)$ and
$\hat{\bar\alpha}^{(1)}_1(R)$ are integrated and stored as described at the
start of this subsection.

\subsubsection*{Second Order in Rotation}

At second order in the spin-frequency expansion, we solve two decoupled $\ell=0$
and $\ell=2$ modes independently (Sec.~\ref{sec:interior_even}). The $\ell=0$
mode yields the leading correction to the mass monopole, $\hat{M}^{(2)}_0$,
while the $\ell=2$ mode contributes to the mass quadrupole, $\hat{M}^{(2)}_2$.

\vspace{0.2cm} \noindent \textit{Mode $\ell=0$}. This sector is governed by
Eq.~\eqref{eq:mxh_system} for the normalized variables $\hat{m}^{(2)}_0$,
$\hat\xi^{(2)}_0$, and $\hat{h}^{(2)}_0$, with source terms constructed from the
interpolated lower-order profiles. The system is integrated over $R\in
[R_i,R_*]$ with boundary conditions given in
Eqs.~\eqref{eq:ms0_asym}--\eqref{eq:hs0_asym}. All central quantities entering
these boundary conditions are fixed by the background EoS at $\varepsilon_c$ and
the first-order solution $\hat\varpi^{(1)}_{1,c}$, except for the constant
$\hat{h}^{(2)}_{0,c}$, which is determined after matching at the stellar
surface\footnote{In principle the central value $\varpi^{(1)}_{1,c}$ is needed
as well. Because we work with the scaled variables, the initial condition is the
hatted combination $\hat\varpi^{(1)}_{1,c}=\varpi^{(1)}_{1,c}/\Omega$. This
carries $\Omega$, which itself follows from $\varepsilon_c$ through the
first-order matching.}. As mentioned above, the coupled subsystem for
$\{\hat{m}^{(2)}_0, \hat\xi^{(2)}_0\}$ is not sourced by $\hat{h}^{(2)}_0$ and
thus can be integrated on its own. At the stellar surface ($R=R_{*}$), matching
to the exterior solution via Eq.~\eqref{eq:Cs0ext} yields the integration
constant $\hat{\tilde{C}}^{(2)}_{0,\mathsf{ext}}$, from which the normalized
monopole correction $\hat{M}^{(2)}_0$ follows directly via Eq.~\eqref{eq:M20}.

Separately, the solution for $\hat{h}^{(2)}_0$ follows algebraically from
$\hat\xi^{(2)}_0$ through the $R$-component of the Bianchi identity, as shown in
Paper~I,
\begin{equation}
\hat{h}^{(2)}_0(R) = \hat{h}^{(2)}_{0,c} - \frac{M+4\pi p R^3}{R(R-2M)}\,\hat\xi^{(2)}_0 + \frac{1}{3}e^{-\nu}R^2\left(\hat\varpi^{(1)}_1\right)^2.
\label{eq:h20int}
\end{equation}
Here, $\hat{h}^{(2)}_{0,c}\equiv\hat{h}^{(2)}_0(R=0)$ is a shift-invariant
constant of integration, left undetermined by Eq.~\eqref{eq:h20int}. It is fixed
by imposing continuity between $\hat{h}^{(2)}_0$ and the exterior solution
$\hat{h}^{(2)}_{0,\mathsf{ext}}$ at $R=R_*$. Using Eq.~\eqref{eq:h20int} and the
hatted version of Eq.~\eqref{eq:hs0c}, this gives
\begin{align}
\hat{h}^{(2)}_{0,c} &= \frac{M_*^{4}\hat{j}^{2}}{R_*^{3}(R_*-2M_*)} - \frac{M_*\hat{\tilde{C}}^{(2)}_{0,\mathsf{ext}}}{R_*-2M_*} \nonumber \\
&\quad + \frac{M_*\hat\xi^{(2)}_0(R_*)}{R_*(R_*-2M_*)} - \frac{1}{3}\frac{R_*^{3}}{R_*-2M_*}\left(1-\frac{2M_*^{2}\hat{j}}{R_*^{3}}\right)^{2}.
\label{eq:h20c_hatted}
\end{align}
Alternatively, all three equations of Eq.~\eqref{eq:mxh_system} may be
integrated simultaneously, starting from a trial value
$\hat{h}^{(2),\mathrm{trial}}_{0,c}=0$, with the correct value recovered
afterward, from Eq.~\eqref{eq:h20c_hatted}. This alternative strategy is the one
adopted, and described in detail, for the $\ell=0$ modes at fourth and sixth
order.

\vspace{0.2cm} \noindent \textit{Mode $\ell=2$}. This mode is governed by the
coupled system for $\hat v^{(2)}_2$ and $\hat h^{(2)}_2$, decomposed into
homogeneous and particular problems as detailed in Sec.~\ref{sec:interior_even}.
The corresponding ODEs are given by the hatted versions of
Eqs.~\eqref{eq:vh_homogeneous} and~\eqref{eq:vh_particular}, subject to the
boundary conditions in Eqs.~\eqref{eq:vsl_H_asym}--\eqref{eq:hsl_H_asym}
and~\eqref{eq:vsl_P_asym}--\eqref{eq:hsl_P_asym}, respectively. These boundary
conditions depend on the central quantities fixed at lower orders: the
homogeneous conditions require only the background values ($\varepsilon_c$,
$p_c$, $c^2_{s,c}$, and $\nu_c$), whereas the particular conditions additionally
depend on the first-order asymptotic constant $\hat\varpi^{(1)}_{1,c}$.
Initializing both trial integration constants to unity
($\hat{A}^{(2),\mathsf{H},\mathrm{trial}}_{2,\mathsf{int}} =
\hat{A}^{(2),\mathsf{P},\mathrm{trial}}_{2,\mathsf{int}} = 1$), we integrate the
homogeneous and particular systems simultaneously from $R=R_i$ to $R=R_*$.

At the surface, applying the hatted matching conditions of
Eqs.~\eqref{eq:Ceven_int}--\eqref{eq:Ceven_ext} determines the interior and
exterior integration constants, $\hat{C}^{(2)}_{2,\mathsf{int}}$ and
$\hat{\tilde{C}}^{(2)}_{2,\mathsf{ext}}$. The exterior constant yields the
leading quadrupole moment contribution $\hat{M}^{(2)}_2$ via the hatted
Eq.~\eqref{eq:M22}, while the interior constant reconstructs the interior
solutions $\hat v^{(2)}_2$, $\hat h^{(2)}_2$, $\hat k^{(2)}_2$, $\hat
m^{(2)}_2$, and $\hat{\xi}^{(2)}_2$ through the hatted
Eqs.~\eqref{eq:vsl_int}--\eqref{eq:xisl_int}. Finally, the true value
$\hat{A}^{(2)}_{2,\mathsf{int}}$ is extracted from the matched solution by
equating the full interior solution for $\hat h^{(2)}_2$ (or $\hat v^{(2)}_2$)
at $R=R_i$ to its near-center expansion in Eq.~\eqref{eq:hsl_P_asym} and solving
for $\hat{A}^{(2)}_{2,\mathsf{int}}$.

\subsubsection*{Third Order in Rotation}

At third order, the equations decouple into independent $\ell=1$ and $\ell=3$
modes, both following the second-order solution pattern. In particular, the
normalized functions $\hat\omega^{(3)}_\ell$ and $\hat\alpha^{(3)}_\ell$ are
governed by the homogeneous and particular systems of
Eqs.~\eqref{eq:omega_homogeneous} and~\eqref{eq:omega_particular} for each mode
$\ell \in \{1,3\}$, subject to the boundary conditions in
Eqs.~\eqref{eq:wkl_H_asym}--\eqref{eq:akl_H_asym} and
Eqs.~\eqref{eq:wkl_P_asym}--\eqref{eq:akl_P_asym}, with source terms constructed
from the interpolated lower-order solutions. Initializing both the homogeneous
and particular trial constants to unity
($\hat{A}^{(3),\mathsf{H}}_{\ell,\mathsf{int}}\equiv\hat\omega^{(3),\mathsf{H},\mathrm{trial}}_{\ell,c}=1$
and
$\hat{A}^{(3),\mathsf{P}}_{\ell,\mathsf{int}}\equiv\hat\omega^{(3),\mathsf{P},\mathrm{trial}}_{\ell,c}=1$),
we integrate the coupled systems simultaneously from $R=R_i$ to $R=R_*$. At the
surface, the hatted matching conditions of
Eqs.~\eqref{eq:Codd_int}--\eqref{eq:Codd_ext} determine the integration
constants $\hat{C}^{(3)}_{\ell,\mathsf{int}}$ and
$\hat{\tilde{C}}^{(3)}_{\ell,\mathsf{ext}}$. From
$\hat{C}^{(3)}_{\ell,\mathsf{int}}$, we construct the interior solutions
$\hat\omega^{(3)}_\ell$ and $\hat\alpha^{(3)}_\ell$, and subsequently recover
the true asymptotic constant $\hat{A}^{(3)}_{\ell,\mathsf{int}} \equiv
\hat\omega^{(3)}_{\ell,c}$ by matching the solution at $R_i$ to the near-center
expansion of Eq.~\eqref{eq:wkl_P_asym}. The exterior constants
$\hat{\tilde{C}}^{(3)}_{\ell,\mathsf{ext}}$ yield the physical observables,
which differ in character between the two modes: for $\ell=1$, the hatted
Eq.~\eqref{eq:S31} gives the first correction to the spin angular momentum,
$\hat{S}^{(3)}_1$, whereas for $\ell=3$, the hatted Eq.~\eqref{eq:S33} provides
the leading contribution to the current octupole moment, $\hat{S}^{(3)}_3$.

\subsubsection*{Fourth Order in Rotation}

The equations at this order decouple into independent $\ell=0$, $2$, and $4$
modes, with all quantities again normalized by the appropriate power of $\Omega$
and lower-order central asymptotic constants (including reconstructed values)
held fixed in the boundary conditions. For the $\ell=0$ mode, the functions
$\hat{m}^{(4)}_0$, $\hat\xi^{(4)}_0$, and $\hat{h}^{(4)}_0$ are governed by
Eq.~\eqref{eq:mxh_system} subject to the boundary conditions in
Eqs.~\eqref{eq:ms0_asym}--\eqref{eq:hs0_asym}. Unlike at second order, no
closed-form relation analogous to Eq.~\eqref{eq:h20int} links $\hat{h}^{(4)}_0$
and $\hat\xi^{(4)}_0$. All three equations then need to be integrated
simultaneously from $R=R_i$ to $R=R_*$. Nevertheless, because $\hat{h}^{(4)}_0$
enters Eq.~\eqref{eq:mxh_system} only through its derivative (with the
right-hand side depending solely on $\hat{m}^{(4)}_0$ and $\hat\xi^{(4)}_0$),
the equation is shift-invariant, like $\nu$ is in Eq.~\eqref{eq:nu}, which
obviates any shooting method. Its general solution decomposes into the
homogeneous constant $\hat{h}^{(4)}_{0,c}$ plus a particular part
$\Delta\hat{h}^{(4)}_0\equiv\hat{h}^{(4)}_0-\hat{h}^{(4)}_{0,c}$, which we
integrate numerically using the trial initial condition
$\hat{h}^{(4),\mathrm{trial}}_{0,c}=0$. At the surface, matching
$\hat{m}^{(4)}_0$ via the hatted Eq.~\eqref{eq:Cs0ext} fixes the exterior
constant $\hat{\tilde{C}}^{(4)}_{0,\mathsf{ext}}$, which then determines both
the true central constant $\hat{h}^{(4)}_{0,c}$ via the hatted
Eq.~\eqref{eq:hs0c} and the second correction to the mass monopole,
$\hat{M}^{(4)}_0$, via the hatted Eq.~\eqref{eq:M40}.

For the $\ell \in \{2,4\}$ modes, the procedure mirrors the second-order
$\ell=2$ case but at fourth-order in rotation this time. We simultaneously
integrate the homogeneous and particular systems of
Eqs.~\eqref{eq:vh_homogeneous} and~\eqref{eq:vh_particular} for
$\hat{v}^{(4)}_\ell$ and $\hat{h}^{(4)}_\ell$, subject to the boundary
conditions of Eqs.~\eqref{eq:vsl_H_asym}--\eqref{eq:hsl_H_asym} and
Eqs.~\eqref{eq:vsl_P_asym}--\eqref{eq:hsl_P_asym}, using unit trial constants
($\hat{A}^{(4),\mathsf{H},\mathrm{trial}}_{\ell,\mathsf{int}}=\hat{A}^{(4),\mathsf{P},\mathrm{trial}}_{\ell,\mathsf{int}}=1$).
At the surface, the hatted matching conditions of
Eqs.~\eqref{eq:Ceven_int}--\eqref{eq:Ceven_ext} yield the constants
$\hat{C}^{(4)}_{\ell,\mathsf{int}}$ and
$\hat{\tilde{C}}^{(4)}_{\ell,\mathsf{ext}}$. From
$\hat{C}^{(4)}_{\ell,\mathsf{int}}$, we obtain the interior solutions
$\hat{v}^{(4)}_\ell$, $\hat{h}^{(4)}_\ell$, $\hat{k}^{(4)}_\ell$,
$\hat{m}^{(4)}_\ell$, and $\hat\xi^{(4)}_\ell$ via the hatted
Eqs.~\eqref{eq:vsl_int}--\eqref{eq:xisl_int}, respectively, and recover the true
asymptotic constant $\hat{A}^{(4)}_{\ell,\mathsf{int}}$ exactly as at second
order. Finally, $\hat{\tilde{C}}^{(4)}_{\ell,\mathsf{ext}}$ determines the
next-to-leading mass quadrupole correction $\hat{M}^{(4)}_2$ for $\ell=2$
[hatted Eq.~\eqref{eq:M42}] and the leading mass hexadecapole moment
$\hat{M}^{(4)}_4$ for $\ell=4$ [hatted Eq.~\eqref{eq:M44}].

\subsubsection*{Fifth through Seventh Order in Rotation}

The remaining three orders introduce no new procedures, as each mode follows the
established approach for its parity sector, i.e., odd-parity modes as at third
order and even-parity modes as at fourth. What changes is the number of modes,
which grows by one at each successive order of the same parity, following
Table~\ref{tab:modes}, alongside the extracted multipole moments. Throughout,
all quantities are normalized by the appropriate power of $\Omega$, with
lower-order perturbation functions building the source terms and their
reconstructed central asymptotic constants propagating into the boundary
conditions.

For the odd-parity fifth ($k=5$, $\ell=1,3,5$) and seventh ($k=7$,
$\ell=1,3,5,7$) orders, each mode is solved via the homogeneous and particular
systems of Eqs.~\eqref{eq:omega_homogeneous} and~\eqref{eq:omega_particular} and
matched through Eqs.~\eqref{eq:Codd_int}--\eqref{eq:Codd_ext}. At fifth order,
the exterior constants $\hat{\tilde{C}}^{(5)}_{\ell,\mathsf{ext}}$ yield the
second spin angular momentum correction $\hat{S}^{(5)}_1$, the next-to-leading
current octupole correction $\hat{S}^{(5)}_3$, and the first accessible leading
contribution to $S_5$, $\hat{S}^{(5)}_5$, via the hatted Eqs.~\eqref{eq:S51},
\eqref{eq:S53}, and~\eqref{eq:S55}, respectively. Similarly, at seventh order,
$\hat{\tilde{C}}^{(7)}_{\ell,\mathsf{ext}}$ yield the third spin correction
$\hat{S}^{(7)}_1$, the second current octupole correction $\hat{S}^{(7)}_3$, the
next-to-leading correction to $S_5$, $\hat{S}^{(7)}_5$, and the newly accessible
leading contribution to $S_7$, $\hat{S}^{(7)}_7$, via the hatted
Eqs.~\eqref{eq:S71}, \eqref{eq:S73}, \eqref{eq:S75}, and~\eqref{eq:S77},
respectively.

For the even-parity sixth order ($s=6$, $\ell=0,2,4,6$) case, the
shift-invariant $\ell=0$ system [Eq.~\eqref{eq:mxh_system} with
$\hat{h}^{(6),\mathrm{trial}}_{0,c}=0$, matched via Eqs.~\eqref{eq:Cs0ext}
and~\eqref{eq:hs0c}] and the $\ell=2,4,6$ modes
[Eqs.~\eqref{eq:vh_homogeneous}--\eqref{eq:vh_particular}, matched via
Eqs.~\eqref{eq:Ceven_int}--\eqref{eq:Ceven_ext}] are solved just as at fourth
order. The exterior constants $\hat{\tilde{C}}^{(6)}_{\ell,\mathsf{ext}}$ yield
the third mass monopole correction $\hat{M}^{(6)}_0$, the second quadrupole
correction $\hat{M}^{(6)}_2$, the next-to-leading hexadecapole correction
$\hat{M}^{(6)}_4$, and the newly accessible leading $M_6$ contribution,
$\hat{M}^{(6)}_6$, via the hatted Eqs.~\eqref{eq:M60}, \eqref{eq:M62},
\eqref{eq:M64}, and~\eqref{eq:M66}, respectively. At all three orders,
recovering the interior solutions and true asymptotic constants
$\hat{A}^{(n)}_{\ell,\mathsf{int}}$ as at lower orders completes the interior
solution up to $\mathcal{O}(\epsilon^7)$.


\section{Numerical Results}
\label{sec:numerical_results}

The preceding sections established how to cast the perturbed Einstein equations
into a form suitable for numerical integration, and detailed the order-by-order
implementation used to solve them. This section applies that implementation to
construct actual rotating neutron star models. We organize this section in two
parts.

Sec.~\ref{sec:single_star} constructs a single rotating neutron star, specified
by a fixed central energy density $\varepsilon_c$. Following the procedure of
Sec.~\ref{sec:order_by_order}, the near-center asymptotic constants are computed
order-by-order and held fixed as inputs to every higher order. From the
resulting solution, the full set of multipole moments, $M_0$, $S_1$, $M_2$,
$S_3$, $M_4$, $S_5$, $M_6$, and $S_7$, is obtained for a star rotating at $f_{*}
= 716$ Hz. This is repeated for both EoSs considered in this work, SLy and
BSk21.

Sec.~\ref{sec:sequences} extends this construction to entire stellar sequences.
Each star in a sequence is obtained by repeating the single-star procedure of
Sec.~\ref{sec:single_star} independently. For each chosen $\varepsilon_c$, the
near-center asymptotic constants are again computed order-by-order and held
fixed across all orders, exactly as in the single star case. Only the central
energy density $\varepsilon_c$ is varied across the sequence. The target
frequency, $f_{*} = 716$ Hz, is held fixed across every rotating equilibrium
solution presented in this work for both EoSs.

\subsection{Single-star solution}
\label{sec:single_star}

We consider six stellar configurations spanning two EoSs (SLy and BSk21) and
three gravitational TOV masses ($M_* = 1.0$, $1.5$, and $2.0\,M_\odot$), all
rotating at $f_* = 716\,\mathrm{Hz}$. Table~\ref{table:configurations} lists the
central energy density $\varepsilon_c$, the radius $R_*$ and the compactness
$\mathcal{C}_*$ of each nonrotating configuration. The rotating stars are
constructed at fixed central energy density. Each rotating star therefore has
the same $\varepsilon_c$ as its nonrotating configuration, and its total mass
$M_0$ differs from the nonrotating mass $M_*$ by spin-frequency corrections. The
last column gives the ratio $\Omega_*/\Omega_K$, where $\Omega_K =
\sqrt{M_*/R_*^3}$ is the Newtonian Keplerian frequency of the nonrotating
configuration. The ratio is larger for less compact stars, rising from
$0.30\text{--}0.38$ at $2.0\,M_\odot$ to above $0.5$ at $1.0\,M_\odot$. It
measures how close a star that spins at $f_* = 716\,\mathrm{Hz}$ and has
gravitational mass $M_*$ is to the mass-shedding limit\footnote{The
mass-shedding frequency $\Omega_{\mathrm{sh}}$ lies below $\Omega_K$ because
rotation enlarges the star. At mass shedding, the equatorial radius is about
$1.44\,R_*$ for a star of the same gravitational mass~\cite{Haensel:2009wa}, and
a particle orbiting at that radius has a lower frequency than at $R_*$.},
$\Omega_{\mathrm{sh}} \approx 0.6\,\Omega_K$. The rotating stars considered here
have a different mass, $M_0$, so for them the ratio is only an estimate, which
neglects the difference between $M_0$ and $M_*$.

We compute eight multipole moments up to $\mathcal{O}(\epsilon^7)$, viz.~four
mass moments ($M_0, M_2, M_4, M_6$) and four mass-current moments ($S_1, S_3,
S_5, S_7$), for each configuration. A moment of index $\ell$ enters at order
$\ell$ in the spin expansion and receives higher-order corrections from the same
mode $\ell$ at the orders listed in Table~\ref{tab:modes}. Consequently, $M_0$
and $S_1$ each acquire three corrections (e.g., $M^{(2)}_0, M^{(4)}_0,
M^{(6)}_0$), $M_2$ and $S_3$ acquire two, and $M_4$ and $S_5$ acquire one,
whereas $M_6$ and $S_7$ enter at sixth and seventh order and receive no further
corrections at $\mathcal{O}(\epsilon^7)$. Before presenting the moments, we
validate our numerical pipeline against three key factors: (i) the starting
radius of the near-center integration, (ii) the invariance of the exterior
constants $\tilde{C}^{(n)}_{\ell,\mathsf{ext}}$ under the choice of central
trial values, and (iii) the quality of the interior--exterior matching across
the stellar surface.

\begin{table}
\setlength{\tabcolsep}{6pt}
\renewcommand{\arraystretch}{1.1}
\begin{tabular}{l c c c c c}
\toprule[1.2pt]
EoS & $\varepsilon_c$ & $M_*$ & $R_*$ & $\mathcal{C}_*$ & $\Omega_*/\Omega_K$ \\
    & [MeV\,fm$^{-3}$] & [$M_\odot$] & [km] & [--] & [--] \\
\midrule
SLy   & $411.443$ & $1.0$ & $11.848$ & $0.1246$ & $0.504$ \\
BSk21 & $323.458$ & $1.0$ & $12.467$ & $0.1184$ & $0.544$ \\
\midrule
SLy   & $598.880$ & $1.5$ & $11.632$ & $0.1904$ & $0.400$ \\
BSk21 & $435.370$ & $1.5$ & $12.593$ & $0.1759$ & $0.451$ \\
\midrule
SLy   & $1148.581$ & $2.0$ & $10.621$ & $0.2781$ & $0.302$ \\
BSk21 & $644.660$ & $2.0$ & $12.295$ & $0.2402$ & $0.376$ \\
\bottomrule[1.2pt]
\end{tabular}
\caption{The six stellar configurations considered, each rotating at
$f_*=716\,\mathrm{Hz}$. The compactness is $\mathcal{C}_*=M_*/R_*$ and
$\Omega_K=\sqrt{M_*/R_*^{3}}$ is the Newtonian Keplerian frequency.}
\label{table:configurations}
\end{table}

\subsubsection*{Choice of the starting radius}

The perturbation equations have a regular singular point at the origin, so the
near-center expansions of Sec.~\ref{sec:perturbation} supply initial conditions
at a small but finite radius $R_i$. The choice of $R_i$ matters, since different
starting data propagate to the surface and into the multipole moments. One must
therefore decide how small $R_i$ should be.

One option is to choose $R_i$ to be absolutely tiny. After all, the expansions
of Sec.~\ref{sec:perturbation} are truncated after a finite number of terms, so
they are most accurate close to the origin and lose accuracy as $R$ grows.
Pushing the starting radius toward the center, however, is problematic. The
initial data are obtained by evaluating the near-center expansions at $R_i$, but
some radial functions vanish as powers of $R_i$, while others differ from finite
central values by corrections proportional to powers of $R_i$. Small values are
not intrinsically inaccurate in floating-point arithmetic, but small corrections
to nonzero central values can be lost through roundoff. For example, the
starting pseudo-enthalpy differs from its central value by a term of order
$R_i^2$. If $R_i$ is sufficiently small, these two values become numerically
indistinguishable, even though the initialization requires a finite displacement
from the center. Thus, the choice of $R_i$ must allow for the finite-radius
corrections required by the initialization to remain numerically resolved.

Starting closer to the center can also increase the computational cost. Small
$R_i$ requires integrating the equations in a region where inverse powers of $R$
and cancellations among terms can amplify numerical errors in the background and
lower-order solutions. At each step, the adaptive integrator constructs two
approximations of different order using the same evaluations of the right-hand
side of the equations. The adaptive integrator then takes their disagreement as
an estimate of the local integration error, shortening the step whenever that
estimate exceeds the tolerance. The right-hand side, however, is assembled from
interpolated background quantities and lower-order solutions, which themselves
carry numerical errors. Errors shared by both approximations can remain hidden
in their agreement, while errors that affect the approximations differently can
contaminate the local error estimate. When these errors are amplified near the
center, the estimated error may decrease more slowly as the step is shortened
than the controller expects. The integrator can then take increasingly small
steps without a corresponding improvement in the accuracy of the solution. The
choice of starting radius must therefore balance the accuracy of the near-center
expansions against numerical accuracy and computational cost.

\begin{figure}
\includegraphics[width=9cm]{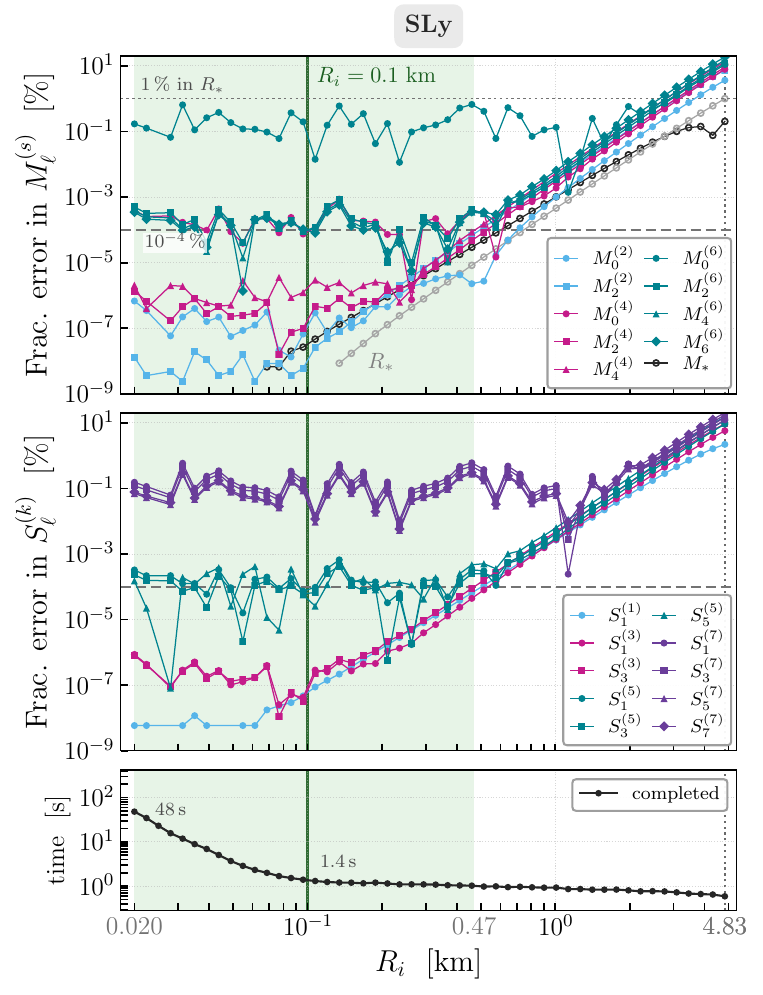}
\caption{Dependence of the multipole moments on the starting radius $R_i$, for
SLy at $M_{*}=1.5\,M_\odot$. \textit{Upper:} fractional error in the mass
moments, together with the shifts in $R_*$ and $M_*$ that accompany the change
in $R_i$. \textit{Middle:} the same for the current moments. Every quantity is
measured against the value obtained at $R_i=0.020\,\mathrm{km}$. The horizontal
dashed line marks $10^{-4}\%$, a fractional error of $10^{-6}$, and the dotted
line the level at which $R_*$ has shifted by one per cent. The shaded band is
the window of Table~\ref{table:rstart}, ending at $0.47\,\mathrm{km}$ where the
first curves cross the former level, and the vertical line marks the adopted
value $R_i= R_{\textrm{start}} = 0.1\,\mathrm{km}$. \textit{Lower:} runtime of
each run, falling from $48\,\mathrm{s}$ at $0.020\,\mathrm{km}$ to
$1.4\,\mathrm{s}$ at the adopted value. The corresponding figure for BSk21 is
not shown. It differs in the floors of the sixth-order correction to $M_0$ and
the seventh-order current terms, which sit higher there, and in the window edge,
as reported in the text. }
\label{fig:rstart}
\end{figure}

No analytic criterion settles the balance, so $R_i$ is fixed by direct numerical
experimentation. The star is the same in every run, so any variation among the
results is error introduced by $R_i$ alone, and what is sought is a range over
which that variation becomes small enough to be tolerable. The calculation is
repeated over fifty starting radii from $0.020\,\mathrm{km}$, which serves as
reference, up to the radius where $R_*$ has shifted by $1\%$,
$4.83\,\mathrm{km}$ for SLy and $5.65\,\mathrm{km}$ for BSk21, beyond which the
background is no longer the same star. Figure~\ref{fig:rstart} shows the result
for SLy at $1.5\,M_\odot$, and Table~\ref{table:rstart} collects the values for
both EoS.
Each curve shows two regimes, a floor at small $R_i$ set by the integration
itself and a power-law rise at large $R_i$ from the truncated expansions losing
accuracy away from the origin. The floor rises with spin-frequency order, since
each order inherits the error of the solutions that source it. The lowest curves
therefore track the run-to-run scatter of the integration itself, and the
highest is that of the full seven-order construction.

\begin{table}[h!]
\renewcommand{\arraystretch}{1.05}
\begin{tabular}{lcc}
\toprule
 & \textbf{SLy} & \textbf{BSk21} \\
\midrule
$\varepsilon_c$ [MeV\,fm$^{-3}$] & 598.880 & 435.370 \\
$M_*$ [$M_\odot$] & 1.5000 & 1.5000 \\
$R_*$ [km] & 11.632 & 12.593 \\
$\mathcal{C}_{*} = M_*/R_*$ & 0.1904 & 0.1759 \\
\addlinespace[2pt]
runs & 50 & 50 \\
$R_i$ swept [km] & 0.020--4.83 & 0.020--5.65 \\
\addlinespace[2pt]
window in $R_i$ [km] & $<0.47$ & $<0.52$ \\
set by & $S^{(3)}_{3}$ & $S^{(3)}_{3}$ \\
\addlinespace[2pt]
$R^{\mathsf{max}}_{i}$ [km] & 4.834 & 5.648 \\
\quad $\delta R_*/R_*$ & $1.0{\times}10^{-2}$ & $1.0{\times}10^{-2}$ \\
\quad $\delta M_*/M_*$ & $2.0{\times}10^{-3}$ & $5.6{\times}10^{-3}$ \\
\addlinespace[2pt]
runtime at $0.020$~km [s] & 48 & 114 \\
runtime at $0.1$~km [s] & 1.4 & 2.0 \\
\bottomrule
\end{tabular}
\caption{The sweep in starting radius, performed for the
configuration of each EoS with nonrotating TOV mass
$M_{*}=1.5\,M_\odot$. Fifty values of $R_i$ are sampled, up to
$R^{\mathsf{max}}_i$, the starting radius at which the
fractional shift of the TOV radius $R_{*}$ reaches one per cent,
$|\delta R_*/R_*| = 10^{-2}$, relative to the reference run at
$R_i=0.020\,\mathrm{km}$. The accompanying fractional shift
of the TOV mass is $\delta M_*/M_*$. Both shifts are numerical and are caused
only by the change in $R_i$. The window ends at the first crossing of a
fractional error of $10^{-6}$; the row beneath names the multipole that reaches
that level first, $S^{(3)}_{3}$ in both cases. Runtimes are for a single star.
}
\label{table:rstart}
\end{table}

The window of reliable starting radii ends where the expansion error first
overtakes a floor. Because the floors rise with order, only the lowest-order
terms can mark that edge. The leading terms of $S_1$, $M_2$, $S_3$ and $M_4$,
together with the lowest spin-induced corrections to $M_0$, $S_1$ and $M_2$,
start below a fractional error of $10^{-6}$, and everything else sits on a floor
above that level. The first to cross it is $S^{(3)}_3$ in both cases, at
$0.47\,\mathrm{km}$ for SLy and $0.52\,\mathrm{km}$ for BSk21. Inside the window
everything stays within $0.005\%$ of the reference except the sixth-order
correction to $M_0$, the seventh-order corrections to $S_1$, $S_3$ and $S_5$,
and $S_7$, the modes integrated at the $10^{-10}$ absolute and $10^{-8}$
relative tolerances of Sec.~\ref{sec:order_by_order}. In the median across the
window, their floors sit at $0.04$ to $0.08$ per cent for SLy and $1\%$ to $3\%$
for BSk21, with single runs reaching $0.5\%$ and $8.5\%$. This is scatter about
a flat mean rather than a drift with $R_i$, which the starting radius neither
creates nor can remove. Since all fifty runs share the integrator, tolerances
and EoS, this scatter is a lower bound on their numerical uncertainty. Because
these corrections are only a fraction of the moments they enter, the moments
themselves move less, in the median, by under $0.01\%$ for SLy and $0.24\%$ for
BSk21, and at worst by $0.04\%$ and $0.75\%$. The moment $S_7$ is the exception,
since it has no spin-induced corrections and inherits the full floor, in the
median about $0.04\%$ for SLy and $1\%$ for BSk21, and at worst $0.23\%$ and
$3.3\%$ per cent respectively. Within the window considered, accuracy does not
depend on $R_i$, so cost decides. The runtime falls from $48\,\mathrm{s}$ at
$0.020\,\mathrm{km}$ to $1.4\,\mathrm{s}$ at $0.1\,\mathrm{km}$ for SLy, and
from $114\,\mathrm{s}$ to $2.0\,\mathrm{s}$ for BSk21. Given this, all results
below use $R_i=0.1\,\mathrm{km}$, a factor of about five inside the window, so
the choice remains safe for stars of other masses and compactnesses whose window
has not been mapped.

\subsubsection*{Invariance under trial asymptotic constants}

The matched solution and the multipole moments are independent of the trial
values used to initialise the integration. At each order the general interior
solution is a particular solution plus a multiple of the homogeneous one. The
trial constants of Sec.~\ref{sec:order_by_order},
$A^{(n),\mathsf{H},\mathrm{trial}}_{\ell,\mathsf{int}}$ and
$A^{(n),\mathsf{P},\mathrm{trial}}_{\ell,\mathsf{int}}$, select which
representative of that family the integrator starts from. The claim that must be
checked is that neither choice leaves a trace in the observables. To keep the
notation compact, let $\mathbf{y}^{(n)}_{\ell}$ denote the pair of functions
integrated in either sector,
$\mathbf{y}^{(s)}_{\ell}=(v^{(s)}_\ell,\,h^{(s)}_\ell)$ for even parity and
$\mathbf{y}^{(k)}_{\ell}=(\omega^{(k)}_\ell,\,\alpha^{(k)}_\ell)$ for odd, so
that the interior solution reads
\begin{equation}
\mathbf{y}^{(n)}_{\ell,\mathsf{int}} = \mathbf{y}^{(n),\mathsf{P}}_{\ell,\mathsf{int}} + C^{(n)}_{\ell,\mathsf{int}}\,\mathbf{y}^{(n),\mathsf{H}}_{\ell,\mathsf{int}},
\label{eq:yint_general}
\end{equation}
with $C^{(n)}_{\ell,\mathsf{int}}$ and $\tilde{C}^{(n)}_{\ell,\mathsf{ext}}$
given by Eqs.~\eqref{eq:Ceven_int}--\eqref{eq:Ceven_ext} or their odd-parity
counterparts, and the Wronskian $\mathcal{W}^{(n)}_{\ell}$ built from the
homogeneous solutions alone. The near-center asymptotic expansions of
Appendix~\ref{app:bc} are linear in the trial constants, so that
$\mathbf{y}^{(n),\mathsf{H}}_{\ell,\mathsf{int}}\propto
A^{(n),\mathsf{H},\mathrm{trial}}_{\ell,\mathsf{int}}$ near $R_i$, while the
particular solution depends \emph{affinely} on its own constant, being the sum
of a multiple of the homogeneous series and a source-generated series
independent of it.

Consider first the homogeneous trial constant. The homogeneous system is linear
and its initial conditions are proportional to
$A^{(n),\mathsf{H},\mathrm{trial}}_{\ell,\mathsf{int}}$, so changing that
constant by a factor $\lambda$ rescales the whole solution,
$\mathbf{y}^{(n),\mathsf{H}}_{\ell,\mathsf{int}}\to\lambda\,\mathbf{y}^{(n),\mathsf{H}}_{\ell,\mathsf{int}}$.
The Wronskian then scales as
$\mathcal{W}^{(n)}_{\ell}\to\lambda\,\mathcal{W}^{(n)}_{\ell}$. The numerator of
$C^{(n)}_{\ell,\mathsf{int}}$ contains no interior homogeneous function and is
unchanged, while the numerator of $\tilde{C}^{(n)}_{\ell,\mathsf{ext}}$ is
linear in $\mathbf{y}^{(n),\mathsf{H}}_{\ell,\mathsf{int}}$ and picks up the
same factor $\lambda$ as the Wronskian. Hence,
\begin{equation}
C^{(n)}_{\ell,\mathsf{int}}\to\frac{C^{(n)}_{\ell,\mathsf{int}}}{\lambda},
\qquad
\tilde{C}^{(n)}_{\ell,\mathsf{ext}}\to\tilde{C}^{(n)}_{\ell,\mathsf{ext}}.
\end{equation}
The interior constant changes, but the solution built from it does not, since in
Eq.~\eqref{eq:yint_general} the factor $1/\lambda$ in
$C^{(n)}_{\ell,\mathsf{int}}$ cancels the factor $\lambda$ in
$\mathbf{y}^{(n),\mathsf{H}}_{\ell,\mathsf{int}}$.

Consider next the particular trial constant. Two particular solutions of the
same inhomogeneous system differ by a solution of the homogeneous one, since
their difference satisfies the homogeneous equation. Changing
$A^{(n),\mathsf{P},\mathrm{trial}}_{\ell,\mathsf{int}}$ therefore shifts the
particular solution by some multiple of the homogeneous one,
$\mathbf{y}^{(n),\mathsf{P}}_{\ell,\mathsf{int}}\to\mathbf{y}^{(n),\mathsf{P}}_{\ell,\mathsf{int}}+\mu\,\mathbf{y}^{(n),\mathsf{H}}_{\ell,\mathsf{int}}$.
The Wronskian is unaffected. Substituting the shift into the matching formulas,
the terms proportional to $\mu$ in the numerator of
$C^{(n)}_{\ell,\mathsf{int}}$ combine into $-\mu\,\mathcal{W}^{(n)}_{\ell}$,
while those in the numerator of $\tilde{C}^{(n)}_{\ell,\mathsf{ext}}$ cancel
identically. Hence,
\begin{equation}
C^{(n)}_{\ell,\mathsf{int}}\to C^{(n)}_{\ell,\mathsf{int}}-\mu,
\qquad
\tilde{C}^{(n)}_{\ell,\mathsf{ext}}\to\tilde{C}^{(n)}_{\ell,\mathsf{ext}}.
\end{equation}
Again the interior constant changes and the solution does not, since
$(\mathbf{y}^{(n),\mathsf{P}}_{\ell,\mathsf{int}}+\mu\,\mathbf{y}^{(n),\mathsf{H}}_{\ell,\mathsf{int}})+(C^{(n)}_{\ell,\mathsf{int}}-\mu)\,\mathbf{y}^{(n),\mathsf{H}}_{\ell,\mathsf{int}}$
reduces to Eq.~\eqref{eq:yint_general}.

In both cases the exterior constant is invariant, and with it every multipole
moment, since the moments are built from $\tilde{C}^{(n)}_{\ell,\mathsf{ext}}$
alone. The interior constant absorbs the change in trial value, but the matched
interior solution is invariant as well, and it is that solution, not the
constant, which sources the equations at the next order. The choice of trial
values therefore leaves no trace in the calculation. This was confirmed
numerically by repeating the integration with different trial values.

\subsubsection*{Matching at the stellar surface}
\begin{figure*}[t]
\includegraphics[width=2.05\columnwidth]{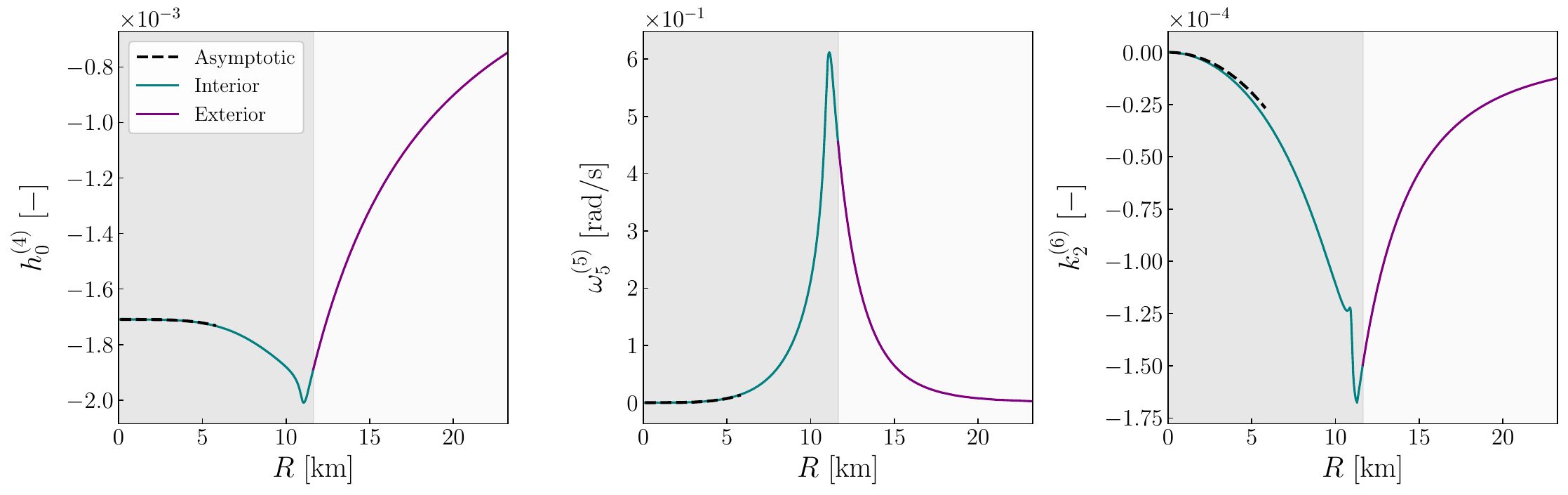}
\caption{Surface matching of the interior numerical and exterior analytic
solutions for SLy at $M_*=1.5\,M_\odot$, shown alongside the near-center
asymptotic expansion near the starting radius $R_i$. The shaded region denotes
the stellar interior ($R \leq R_* = 11.63\,\mathrm{km}$). Complete profiles up
to $\mathcal{O}(\epsilon^{7})$ are given in Appendix~\ref{app:profiles}.}
\label{fig:matching}
\end{figure*}
Because the multipole moments $(M_\ell, S_\ell)$ are constructed from the
exterior integration constants $\tilde{C}^{(n)}_{\ell,\mathsf{ext}}$, both the
interior constants $C^{(n)}_{\ell,\mathsf{int}}$ and exterior constants
$\tilde{C}^{(n)}_{\ell,\mathsf{ext}}$ must be fixed accurately by matching at
the stellar surface $R_*$. A mismatch in the matched variables displaces both
sets of constants: $C^{(n)}_{\ell,\mathsf{int}}$ reconstructs the metric
functions entering higher-order source terms, while
$\tilde{C}^{(n)}_{\ell,\mathsf{ext}}$ enters the multipole expressions
(Appendix~\ref{app:multipoles}). Both channels involve powers and products of
lower-order quantities, so such an error can grow with spin order.

Although the matching conditions are purely algebraic, numerical precision can
be degraded by lower-order error accumulation or three local effects: (i)
catastrophic cancellation in the interior when the homogeneous and particular
pieces—scaled by their central trial values—differ vastly at $R_*$; (ii)
analytic cancellation in the exterior solution at low compactness, familiar from
static tidal deformability
calculations~\cite{Postnikov:2010yn,VanOeveren:2017xkv}; and (iii) surface
density discontinuities, as in quark stars, which require supplementary junction
conditions~\cite{Reina:2014fga} (unlike the smooth-surface EoSs used here). In
practice, we verify the match dynamically at $R_*$ from second order onward
using Eqs.~\eqref{eq:Ceven_int}--\eqref{eq:Ceven_ext} and
\eqref{eq:Codd_int}--\eqref{eq:Codd_ext}, issuing a warning if the relative
difference exceeds $10^{-9}$; across all reported configurations, the maximum
relative mismatch in the matched variables, $(v^{(s)}_{\ell},h^{(s)}_{\ell})$,
$(\omega^{(k)}_{\ell},\alpha^{(k)}_{\ell})$, and $(m^{(s)}_{0},h^{(s)}_{0})$, is
$3.6\times10^{-11}$ (occurring at seventh order for $\ell=1$). The reconstructed
mass functions $m^{(n)}_{\ell}$ at $\ell\geq2$ show a larger surface residual,
reaching $1.1\times10^{-3}$ of the larger of the interior and exterior values in
$m^{(6)}_{2}$. This appears only for the lightest stars of each sequence, below
about $0.85\,M_\odot$ (and thus, for unrealistic neutron star masses), and it
arises from a strong cancellation between two terms in the exterior solution,
which amplifies small evaluation errors by four orders of magnitude.
Figure~\ref{fig:matching} illustrates this seamless transition at $R_i$ and
$R_*$ for three representative functions, with the complete set of metric,
pressure, and energy density profiles up to $\mathcal{O}(\epsilon^7)$ provided
in Appendix~\ref{app:profiles}.

\begin{figure}
\includegraphics[width=\columnwidth]{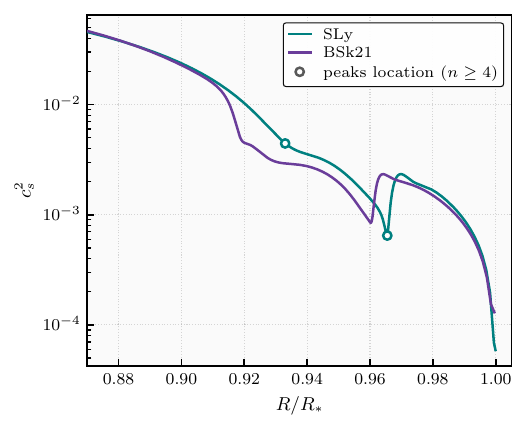}
\caption{Speed of sound squared $c_s^2$ through
the crust for SLy and BSk21 against fractional radius $R/R_*$. Markers on the
SLy curve indicate the two slope breaks where $n\geq4$ perturbation profiles
peak: the neutron-drip notch ($R/R_*=0.966$) and the crust--core knee
($R/R_*=0.933$). BSk21 exhibits analogous features (omitted in
Appendix~\ref{app:profiles} for clarity).}
\label{fig:cs2_crust}
\end{figure}

\begin{figure*}
\centering
\includegraphics[width=\textwidth]{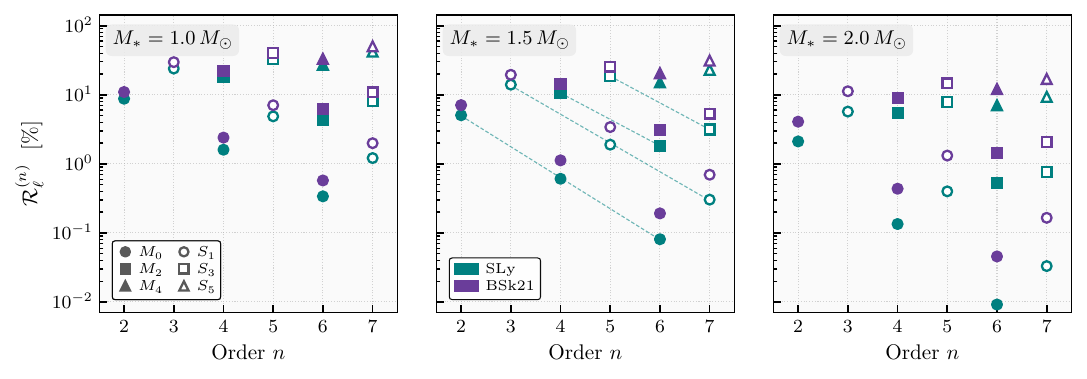}
\caption{The relative order-$n$ correction $\mathcal{R}^{(n)}_{\ell}$ of Eq.~\eqref{eq:truncation} against the spin-frequency order $n$, for the six configurations of Table~\ref{table:configurations}, each rotating at $\Omega_*=2\pi\times716\,\mathrm{Hz}$. The left panel shows $M_*=1.0\,M_\odot$, the middle $M_*=1.5\,M_\odot$, and the right panel $M_*=2.0\,M_\odot$ with SLy in teal and BSk21 in purple. Filled symbols denote the mass moments and open symbols the mass-current moments. As an example, the dashed lines in the middle panel show the straight-line fits for the four SLy moments that supply a slope. The lines only guide the eye, since the order $n$ takes integer values.}
\label{fig:truncation}
\end{figure*}

Notably, at order $n \geq 4$, the profiles in Appendix~\ref{app:profiles}
develop localized peaks a few percent inside $R_*$. As established in Step 5,
the radial density derivatives $d^2\varepsilon/dR^2$ and $d^3\varepsilon/dR^3$
(entering the sources from fourth and sixth order, respectively) depend on the
first and second derivatives of $c_s^2$. For our smooth EoSs, derivatives of
$c_s^2$ remain small across broad extrema, explaining why the prominent rebound
maximum at $R/R_* \simeq 0.97$ leaves no trace in the profiles—but become large
over narrow intervals where $c_s^2$ turns sharply. Consequently, the $n \geq 4$
profiles peak precisely at two sharp slope breaks in Fig.~\ref{fig:cs2_crust}:
the neutron-drip notch at $R/R_* = 0.966$ ($\varepsilon \simeq
0.32\,\mathrm{MeV/fm^3}$, $n_B \simeq 2\times 10^{-3}\,n_{\rm sat}$) and the
crust--core knee at $R/R_* = 0.933$ ($\varepsilon \simeq 40\,\mathrm{MeV/fm^3}$,
$n_B \simeq 0.26\,n_{\rm sat}$). Both points lie well within single polynomial
segments and are resolved by many grid points, so their positions follow the EoS
features, while their amplitudes, set by the second and third derivatives,
depend on the smoothness of the EoS representation.

\subsubsection*{Relative corrections of the slow-rotation expansion}

We come now to a question the calculation was built to answer: how large are the
spin-induced corrections, and how far must the expansion be carried before the
next order term ceases to matter? A natural measure to answer this question is
the size of each new contribution relative to the truncated sum accumulated
before it. Let us then define the fractional order-$n$ relative correction
\begin{equation}
\mathcal{R}^{(n)}_{\ell} = 100 \times
\frac{\bigl| \hat{P}^{(n)}_{\ell}\,\Omega_{*}^{n} \bigr|}
{\Bigl| \sum_{m=\ell,\,\ell+2,\,\ldots}^{\,n-2}
\hat{P}^{(m)}_{\ell}\,\Omega_{*}^{m} \Bigr|}\quad [\%]
\label{eq:truncation}
\end{equation}
for $P_\ell \in \{M_\ell, S_\ell\}$. A small $\mathcal{R}^{(n)}_{\ell}$ suggests
that the expansion has settled down (i.e., the term proportional to $\Omega_*^n$
changes the preceding partial sum only slightly\footnote{Spin-induced
corrections enter every two orders, so $n-2$ is the previous order contributing
to $P_\ell$.}). Read as an ``error,'' $\mathcal{R}^{(n)}_{\ell}$ estimates what
is lost by stopping the expansion just below order $n$. This error estimate,
however, does not bound the terms beyond $\Omega^{n}$, and the error of the sum
through the highest computed order would require the next term, which is not
available.

Figure~\ref{fig:truncation} shows $\mathcal{R}^{(n)}_{\ell}$ at $\Omega_* =
2\pi\times 716\,\mathrm{Hz}$ across all six configurations ($M_* = 1.0, 1.5,
2.0\,M_\odot$ from left to right) for all multipoles carrying beyond-leading
corrections, while excluding the leading-order-only moments (viz.~$M_6$ and
$S_7$). For every multipole, $\mathcal{R}^{(n)}_{\ell}$ decreases with $n$, so
each new contribution alters the accumulated sum by less than the one before it.
Note that $M_4$ and $S_5$ contribute only a single point each (at $n=6$ and
$7$); these next-to-leading-order corrections are the largest in every panel and
lack a subsequent order to judge whether they are diminishing.

Comparing across and within the panels of Fig.~\ref{fig:truncation} traces the
correlated effects of mass and EoS stiffness, both of which are governed by
compactness $\mathcal{C}_*$ and $\Omega_K$. At fixed $\Omega_*$, a less compact
star rotates closer to its $\Omega_K$ and undergoes stronger rotational
deformation. Indeed, moving from $1.0$ to $2.0\,M_\odot$ shifts every point
downward. We also see that $\mathcal{R}^{(n)}_{\ell}$ drops more steeply with
order $n$ as the mass increases, so a star of lower compactness carries larger
corrections and settles more slowly. Within each panel, the offset between the
two EoS reflects their difference in compactness. Because the stiffer BSk21 EoS
yields a larger radius at fixed mass (lowering $\mathcal{C}_*$ and $\Omega_K$),
its $\mathcal{R}^{(n)}_{\ell}$ values lie consistently above those of SLy, and
their separation widens from $1.0$ to $2.0\,M_\odot$ alongside the growing
difference in $\mathcal{C}_*$ ($0.006$ at $1.0\,M_\odot$ versus $0.038$ at
$2.0\,M_\odot$; Table~\ref{table:configurations}). In other words, the moments
of the neutron stars appear to be more sensitive to compactness as mass
increases such that we also see a stronger EoS dependence. The complete set of
multipole moments, relative corrections, and fractional contributions is
collected in Appendix~\ref{app:truncation}.

Moreover, lowering the spin frequency reduces every spin-induced correction. The
denominator of Eq.~\eqref{eq:truncation} is the partial sum through order $n-2$.
Because the corrections are small, that sum is set by its leading term, which is
of order $\Omega_*^{\ell}$. The denominator thus scales as $\Omega_*^{\ell}$,
while the numerator scales as $\Omega_*^{n}$. The relative correction therefore
obeys $\mathcal{R}^{(n)}_{\ell} \sim \Omega_*^{\,n-\ell}$. At
$71.6\,\mathrm{Hz}$, a tenth of the value used here, the first correction of
every moment falls by exactly $100$. The second corrections, which only $M_0$,
$M_2$, $S_1$ and $S_3$ carry, fall by between $7\times10^{3}$ and $10^{4}$, and
the third corrections, carried by $M_0$ and $S_1$ alone, fall by between
$7\times10^{5}$ and $10^{6}$. These factors stay below pure powers of ten
because the partial sum in the denominator itself grows with $\Omega_*$.
Redrawn at $71.6\,\mathrm{Hz}$, the points of Fig.~\ref{fig:truncation} would
not move down together. The first corrections would drop by two decades, the
second by nearly four and the third by nearly six. For each moment that carries
more than one correction, the line through its points would tilt down more
strongly with $n$, and the points would spread farther apart. The largest
relative correction falls from $50.3\%$ to $0.50\%$, and every entry lies below
one percent.

Figure~\ref{fig:truncation} also shows how fast the corrections fall off. Within
each panel the relative corrections of a given moment fall along a nearly
straight line on the logarithmic axis, and the lines of the different moments
run almost parallel to one another. The dashed lines in the middle panel show
them for SLy, as an example. A straight line on this scale means that each
correction is a fixed fraction of the one before it, and parallel lines mean
that the fraction is the same for every moment of a given star. Across stars of
different mass or EoS, however, the slopes clearly differ. The slopes of the
different moments versus the order steepen from $1.0$ to $2.0\,M_\odot$ and are
not the same for SLy and BSk21. The question is then the following: what sets
the slope, and does a single rule account for all six stars?

To quantify the fall-off, we define $m$ as the ratio of two consecutive relative
corrections of the same moment, $m = \mathcal{R}^{(n+2)}_{\ell} /
\mathcal{R}^{(n)}_{\ell}$. A smaller $m$ means a faster fall-off, and a steeper
line in Fig.~\ref{fig:truncation}. Four moments have more than one correction,
namely $M_0$, $S_1$, $M_2$ and $S_3$, so each star gives four values of $m$. For
$M_0$ and $S_1$, which have three corrections, we take $m$ from a straight-line
fit to their points. We then average the four values to obtain a single $m$ per
star.

At $716\,\mathrm{Hz}$, the four slopes within each star differ by $14$ to
$20\%$, and the averaged values are $m = 0.226$, $0.154$ and $0.084$ for the SLy
stars of $1.0$, $1.5$ and $2.0\,M_\odot$, and $m = 0.259$, $0.195$ and $0.132$
for the BSk21 stars with the same masses. The scaling $\mathcal{R}^{(n)}_{\ell}
\sim \Omega_{*}^{\,n-\ell}$ says that $m$ grows as $\Omega_{*}^{2}$, but all six
stars spin at the same $716\,\mathrm{Hz}$, so the spin frequency alone cannot be
what separates them. What differs is how close each star is to its own break-up
frequency. The last column of Table~\ref{table:configurations} gives
$\Omega_{*}/\Omega_K$ for the six stars, and the heavier the star, the smaller
the fraction. Dividing each $m$ by the square of its fraction brings all six to
$m/(\Omega_{*}/\Omega_K)^{2} = 0.92$, with a spread of $3\%$. A single rule
therefore accounts for all six stars. From one order to the next, the fractional
error drops by
\begin{equation}
    m \approx 0.9\,\left(\frac{\Omega_{*}}{\Omega_K}\right)^{2}=0.9  \frac{\Omega_{*}^2R_*^{3}}{M_*}
\end{equation}
for both EoS and all three masses.

The slope describes how fast the corrections shrink at a given spin frequency,
so we now ask up to what frequency they keep shrinking. That is, we want to know
where the next contribution in the series stops being smaller than the one
before it. The answer can be worked out pair by pair, because the contributions
$\hat{P}^{(n)}_{\ell}$ do not depend on the spin frequency. For two consecutive
contributions to the same moment, with $P_{\ell}$ standing for $M_{\ell}$ or
$S_{\ell}$ as in Eq.~\eqref{eq:truncation}, the ratio is
\begin{equation}
\frac{\bigl| \hat{P}^{(n+2)}_{\ell}\,\Omega_{*}^{\,n+2} \bigr|}
     {\bigl| \hat{P}^{(n)}_{\ell}\,\Omega_{*}^{\,n} \bigr|}
= \left| \frac{\hat{P}^{(n+2)}_{\ell}}{\hat{P}^{(n)}_{\ell}} \right| \Omega_{*}^{2}.
\label{eq:pair_ratio}
\end{equation}
Every power of the spin frequency cancels except the two that separate the
orders, and the ratio of contributions in front is a fixed number for a given
star. The pair therefore crosses equality at
\begin{equation}
f^{(n)}_{\ell} = \frac{1}{2\pi}
\left| \frac{\hat{P}^{(n)}_{\ell}}{\hat{P}^{(n+2)}_{\ell}} \right|^{1/2}.
\label{eq:pair_frequency}
\end{equation}
The frequency $f_{\ell}^{(n)}$ is then the spin frequency at which the
contributions at orders $n$ and $n+2$ have equal magnitude. At lower spin
frequencies, the higher-order contribution is smaller; at higher spin
frequencies, it is larger. A star has one such frequency for every pair of every
multipole moment, and the whole expansion stays ordered only up to the lowest of
them,
\begin{equation}
f_{\mathrm{lim}} = \min_{\ell,\,n}\, f^{(n)}_{\ell}.
\label{eq:flim}
\end{equation}

Table~\ref{table:ratio_test} lists $f_{\mathrm{lim}}$ for the six configurations
and the pair that sets it. Here, the spin order refers to the absolute power of
$\Omega$, not the order relative to the leading contribution to each moment. For
the four configurations with $1.0$ and $1.5\,M_\odot$, the threshold is set by
the pair formed by the leading $\Omega^5$ contribution to $S_5$ and its first
correction, proportional to $\Omega^7$. For the two configurations with
$2.0\,M_\odot$, it is set by the pair formed by the first and second corrections
to $S_3$ for SLy (proportional to $\Omega^5$ and $\Omega^7$) and to $M_2$ for
BSk21 (proportional to $\Omega^4$ and $\Omega^6$). Thus, the comparison for
$S_5$ tests its first correction against its leading term, whereas those for
$S_3$ and $M_2$ compare two beyond-leading corrections. These thresholds lie at
$77$ to $93\%$ of the Newtonian Keplerian frequency $f_K$. The spin frequency
used here, $716\,\mathrm{Hz}$, is $32$ to $71\%$ of the corresponding threshold,
with the widest margin for the heaviest stars.

\begin{table}
\setlength{\tabcolsep}{4pt}
\renewcommand{\arraystretch}{1.15}
\begin{tabular}{l c c c c}
\toprule
EoS & $M_{*}$ [$M_\odot$] & $f_{\mathrm{lim}}$ [Hz] & $f_{\mathrm{lim}}/f_{K}$ & Limiting pair \\
\midrule
SLy   & $1.0$ & $1102$ & $0.78$ & $S_{5}$, $5 \to 7$ \\
SLy   & $1.5$ & $1493$ & $0.83$ & $S_{5}$, $5 \to 7$ \\
SLy   & $2.0$ & $2213$ & $0.93$ & $S_{3}$, $5 \to 7$ \\
BSk21 & $1.0$ & $1010$ & $0.77$ & $S_{5}$, $5 \to 7$ \\
BSk21 & $1.5$ & $1281$ & $0.81$ & $S_{5}$, $5 \to 7$ \\
BSk21 & $2.0$ & $1719$ & $0.90$ & $M_{2}$, $4 \to 6$ \\
\bottomrule
\end{tabular}
\caption{The ordering limit $f_{\mathrm{lim}}$ of Eq.~\eqref{eq:flim} for the
six configurations of Table~\ref{table:configurations}, as an absolute frequency
and as a fraction of the Newtonian mass-shedding frequency $f_K = \Omega_K/2\pi$
of the nonrotating star. The last column gives the pair of contributions,
labeled by their orders in spin frequency, whose equality sets the limit.}
\label{table:ratio_test}
\end{table}

The slope rule provides a rough estimate of the frequency at which successive
relative corrections become comparable. Recall that $m$ is the ratio of two
consecutive relative corrections. As long as $m<1$, the next relative correction
is smaller than the current one, while when $m=1$, they are equal. Extrapolating
the empirical rule $m \approx 0.9\,(\Omega_{*}/\Omega_K)^2$ and setting $m=1$
gives $\Omega_{*}/\Omega_K \approx 0.9^{-1/2} \approx{} 1.05$, close to the
Newtonian Keplerian frequency scale. The pair test of Eq.~\eqref{eq:flim} gives
lower thresholds, at $0.77$ to $0.93$ of $f_K$, but one must remember that the
two tests compare different quantities. The pair test takes the minimum over all
computed pairs, so $f_{\mathrm{lim}}$ is set by the pair whose contributions
become equal at the lowest frequency. The slope rule, instead, describes the
average behavior of the relative corrections across four moments. Moreover, each
relative correction is normalized by a different partial sum. From
Eq.~\eqref{eq:truncation}, the ratio of two consecutive relative corrections is
\begin{equation}
m = \frac{\mathcal{R}^{(n+2)}_{\ell}}{\mathcal{R}^{(n)}_{\ell}}
= \left| \frac{\hat{P}^{(n+2)}_{\ell}}{\hat{P}^{(n)}_{\ell}} \right| \Omega_{*}^{2}
  \times
  \left| \frac{\sum_{k=\ell,\,\ell+2,\,\ldots}^{\,n-2}
               \hat{P}^{(k)}_{\ell}\,\Omega_{*}^{\,k}}
              {\sum_{k=\ell,\,\ell+2,\,\ldots}^{\,n}
               \hat{P}^{(k)}_{\ell}\,\Omega_{*}^{\,k}} \right|.
\label{eq:m_vs_pair}
\end{equation}
The first factor is the pair ratio of Eq.~\eqref{eq:pair_ratio}. The second
factor is smaller than one in our results, because the sum in its denominator
includes one more contribution. The ratio $m$ is therefore smaller than the pair
ratio, and it reaches one at a higher frequency. The two frequency estimates are
therefore comparable in scale and compatible with each other, although they are
not to be understood as independent determinations of the same threshold.

The analysis above shows that the computed contributions remain ordered up to
about $0.8$ to $0.9$ of the Newtonian mass-shedding frequency, and that they
shrink from one order to the next at a rate set by $(\Omega_{*}/\Omega_K)^{2}$.
The analysis, however, says nothing about convergence, which no finite number of
terms can establish. An asymptotic series behaves in the same way as a
convergent one up to the order where it does not, and the post-Newtonian
expansion, treated as a formal and probably asymptotic
series~\cite{Blanchet:2013haa}, is the standard example of truncations that
remain useful regardless of their convergence properties.

The $f_K$ used here is the Newtonian value for the nonrotating star. The actual
mass-shedding frequency of the rotating relativistic star is lower, by a factor
of ${\sim}0.6$ according to fits to numerical solutions~\cite{Lattimer:2004pg,
Haensel:2009wa}. Therefore, as measured against the relativistic mass shedding
frequency, the fractions in Table~\ref{table:ratio_test} would exceed one. In
other words, we expect (even considering differences in mass and the EoS) that
the fraction in our expansion verses $f_K$ will always be larger than the
relativistic mass-shedding limit. For the six configurations studied here, the
computed contributions appear to remain ordered over the whole range of spin
frequencies these stars can reach. If that holds more generally, the limit on
the expansion is not a frequency at which it breaks down, but rather, the number
of orders a given accuracy demands, which is what carrying it to seventh order
addresses.

A practical question remains: if one stops at the leading term, as is common
practice, how much of the moment is left out? Put another way, how different is
the leading-order prediction for a moment from the prediction obtained by
including all the corrections calculated in this paper? The answer is given by
the fraction of the total supplied by all corrections together,
\begin{align}
\eta^{\mathrm{lead}}_{\ell} &= 100 \times \frac{P_{\ell}^{\textrm{computed}} - P_{\ell}^{\textrm{leading}}}{P_{\ell}^{\textrm{computed}}}
\nonumber \\
&= 100 \times
\frac{\Bigl| \sum_{m=\ell+2,\,\ell+4,\,\ldots}^{n_{\mathrm{max}}}
\hat{P}^{(m)}_{\ell}\,\Omega_{*}^{m} \Bigr|}
{\Bigl| \sum_{m=\ell,\,\ell+2,\,\ldots}^{n_{\mathrm{max}}}
\hat{P}^{(m)}_{\ell}\,\Omega_{*}^{m} \Bigr|}\quad [\%],
\label{eq:lead}
\end{align}
collected in Table~\ref{table:lead}. Consistent with the $\Omega_*/\Omega_K$
scaling above, every entry falls with increasing mass and is larger for BSk21
than for SLy; values are also systematically larger for current moments ($S_1,
S_3, S_5$) than for mass moments ($M_0, M_2, M_4$), though the single-correction
entries for $M_4$ and $S_5$ should be read with the caution noted above.

\begin{table}
\setlength{\tabcolsep}{4.5pt}
\renewcommand{\arraystretch}{1.15}
\begin{tabular}{l c c c c c c}
\toprule[1.2pt]
 & \multicolumn{6}{c}{$\eta^{\mathrm{lead}}_{\ell}$ \ [\%]} \\
\cmidrule(lr){2-7}
 & \multicolumn{3}{c}{\textbf{SLy}} & \multicolumn{3}{c}{\textbf{BSk21}} \\
\cmidrule(lr){2-4}\cmidrule(lr){5-7}
$M_*$ \ [$M_\odot$] & $1.0$ & $1.5$ & $2.0$ & $1.0$ & $1.5$ & $2.0$ \\
\midrule
$M_{0}$ & $9.8$ & $5.5$ & $2.2$ & $12.5$ & $7.8$ & $4.4$ \\
$S_{1}$ & $24.0$ & $14.2$ & $5.8$ & $29.4$ & $19.6$ & $11.4$ \\
$M_{2}$ & $18.8$ & $11.2$ & $5.6$ & $22.9$ & $15.2$ & $9.7$ \\
$S_{3}$ & $30.3$ & $18.2$ & $8.0$ & $35.8$ & $24.4$ & $14.6$ \\
$M_{4}$ & $21.4$ & $13.1$ & $6.6$ & $24.9$ & $17.0$ & $10.8$ \\
$S_{5}$ & $29.7$ & $18.7$ & $8.5$ & $33.5$ & $23.8$ & $14.5$ \\
\bottomrule[1.2pt]
\end{tabular}
\caption{The leading-order discrepancy $\eta^{\mathrm{lead}}_{\ell}$, the
shortfall of the leading term of each multipole's spin series, of
Eq.~\eqref{eq:lead}, for the six configurations of
Table~\ref{table:configurations} at $f_*=716\,\mathrm{Hz}$. The entries for
$M_4$ and $S_5$ rest on a single correction each. $M_{6}$ and $S_{7}$ are
omitted, being computed at a single order and so carrying no correction at all.}
\label{table:lead}
\end{table}

\subsubsection*{Required order of the expansion}

The required order of our expansion is dependent on how precisely we can measure
these moments. Thus, let us briefly discus the precision of modern neutron star
mass measurements from electromagnetic radio observations and gravitational
waves. Radio observations can allow for extremely precise measurements of the
neutron star mass over long observational periods (e.g.~from relativistic radio
timing \cite{Kramer:2021jcw}); X-ray pulse-profile observations from NICER are
closer to the $10\%$ level \cite{Miller:2019cac,Kini:2026rjx}. Others
electromagnetic observations, such as through Shapiro-delay radio timing and
optical companion spectroscopy and light-curve modeling, fall in between these
two limits. Meanwhile, gravitational wave estimates for the LIGO-Virgo-KAGRA
fifth observing run, O5, suggest a $3-4\%$ level precision for the mass
\cite{Haberland:2026xvj} and sub-percent level from next-generation detectors
\cite{Smith:2021bqc,Huxford:2023qne,Guttman:2026fpc}. Thus, here we use
precision tests of $1\%$ and $10\%$ as reasonable bounds that reflect this range
of observational precision across different mass measurements.

Figure~\ref{fig:required_order} was introduced in Sec.~\ref{sec:level1} for an
SLy star of $1.0\,M_\odot$. Here, we provide the numbers behind this figure and
extend it to other masses and to the BSk21 EoS. The quantity to track is the
frequency at which each curve turns into an arrow, since, beyond that point, the
series can no longer verify if the truncation error of a given moment meets the
stated targeted accuracy. Table~\ref{table:required_order} lists these
frequencies for both EoSs at $1.0\,M_\odot$. For both EoSs, only the mass moment
$M_0$ stays within $1\%$ slightly beyond $716\,\mathrm{Hz}$, which requires
terms through $\mathcal{O}(\Omega^4)$. At the $10\%$ target, the picture is far
less demanding, and $M_0$ needs no rotational correction at all below
$765\,\mathrm{Hz}$ for SLy and below $710\,\mathrm{Hz}$ for BSk21.

\begin{table}[h!]
\setlength{\tabcolsep}{6pt}
\renewcommand{\arraystretch}{1.15}
\begin{tabular}{l c c c c}
\toprule
 & \multicolumn{2}{c}{\textbf{SLy}} & \multicolumn{2}{c}{\textbf{BSk21}} \\
\cmidrule(lr){2-3}\cmidrule(lr){4-5}
Moment & $1\%$ & $10\%$ & $1\%$ & $10\%$ \\
\midrule
$S_5$ & 110 &  349 & 101 &  319 \\
$M_4$ & 137 &  434 & 124 &  393 \\
$S_3$ & 407 &  763 & 372 &  699 \\
$M_2$ & 485 &  901 & 441 &  818 \\
$S_1$ & 691 & 1080 & 630 &  987 \\
$M_0$ & 865 & 1326 & 789 & 1212 \\
\bottomrule
\end{tabular}
\caption{Maximum spin frequency, in Hz, up to which each moment reaches the
stated accuracy when the Hartle-Thorne expansion is carried out to
$\mathcal{O}(\Omega^7)$, for $1.0\,M_\odot$ stars. The accuracy is estimated
from the fractional order-$n$ relative correction $\mathcal{R}^{(n)}_{\ell}$.
Above these frequencies, higher orders in rotation would be needed to establish
the target accuracy. The Keplerian mass-shedding frequency $f_{K}$ is
$1422\,\mathrm{Hz}$ for SLy and $1317\,\mathrm{Hz}$ for BSk21.}
\label{table:required_order}
\end{table}

These frequency thresholds also depend on the mass of the star. A heavier star
is more compact and deforms less, so every threshold moves to higher frequency
as the mass grows. For SLy, the frequency at which $M_0$ first needs a
rotational correction at the $1\%$ target rises from $205\,\mathrm{Hz}$ at
$0.70\,M_\odot$ to $596\,\mathrm{Hz}$ at $2.05\,M_\odot$. For $S_5$, the spin
frequency at which the $\Omega^7$ correction reaches $1\%$ of the leading
$\Omega^5$ term in magnitude rises from $92\,\mathrm{Hz}$ at $0.70\,M_\odot$ to
$278\,\mathrm{Hz}$ at $2.05\,M_\odot$. Above that frequency, we must retain the
$\Omega^7$ correction, and estimating the error of the resulting truncation
requires the $\Omega^9$ term. The same figure for other nonrotating masses, and
thus other compactnesses, is provided with the code~\cite{conde2025aqlimr},
together with an animation that follows it along the full mass sequence of each
EoS. In the animation, the staircases slide steadily to the right as the mass
increases, which shows directly that a more compact star needs fewer terms in
the spin expansion, at a given fixed spin frequency. For the SLy EoS, the
threshold frequencies increase monotonically with stellar mass, while for the
BSk21 EoS, they generally increase as well, although some thresholds determined
by the $\Omega^6$ and $\Omega^7$ corrections decrease locally by up to $2\%$ as
the mass increases. These small reversals are consistent with the sensitivity to
the third radial derivative of the energy density seen in
Fig.~\ref{fig:truncation_sequence}.

Finally, let us compare the two EoSs. At $1.0\,M_\odot$, for every multipole and
both accuracy targets, we need to include the next term in the expansion at a
lower spin frequency for BSk21 than for SLy. Much of this difference comes from
the Newtonian Keplerian frequency, which is $1317\,\mathrm{Hz}$ for the BSk21
EoS and $1422\,\mathrm{Hz}$ for the SLy EoS. At the same spin frequency, the
BSk21 star therefore has a larger $\Omega/\Omega_K$. When we express the segment
endpoints in terms of $\Omega/\Omega_K$, the thresholds for the two EoSs are
much closer to each other. The normalized endpoints differ by $1.1$ to $2.2\%$
at $1.0\,M_\odot$, $2.3$ to $3.4\%$ at $1.5\,M_\odot$, and $4.0$ to $7.5\%$ at
$2.0\,M_\odot$. For these two EoSs, much of the dependence of the required
truncation on the EoS is captured by the ratio $\Omega/\Omega_K$.

\subsection{Stellar sequences}
\label{sec:sequences}

The preceding subsection examined six representative stars at isolated masses.
The same analysis is now extended across the full mass range of each EoS by
varying the non-rotating central energy density continuously. Following the
procedure of Sec.~\ref{sec:order_by_order} at the same rotation rate $\Omega_*$,
we construct two sequences comprising 89 stars for SLy and 128 for BSk21,
spanning from $0.7\,M_\odot$ to the maximum mass (Table~\ref{table:sequences}).
Along each sequence, the compactness $\mathcal{C}_{*}$ varies by more than a
factor of three, and with it the rotation measure $\Omega_*/\Omega_K$, allowing
the multipole moments and their spin-induced corrections to be followed as
continuous trends. Three features of these sequences will be studied in turn in
this subsection. The smoothness of the sequences is established first, the
multipole moments are presented next, and the convergence of the expansion is
considered last.

\begin{table}
\setlength{\tabcolsep}{8pt}
\renewcommand{\arraystretch}{1.15}
\begin{tabular}{lcc}
\toprule[1.2pt]
 & \textbf{SLy} & \textbf{BSk21} \\
\midrule
stars in sequence & $89$ & $128$ \\
$\varepsilon_c$ [MeV\,fm$^{-3}$] & $325.88$--$1654.55$ & $268.52$--$1283.37$ \\
$M_{*}$ [$M_\odot$] & $0.7000$--$2.0477$ & $0.7000$--$2.2740$ \\
$R_{*}$ [km] & $9.922$--$11.946$ & $11.039$--$12.594$ \\
$\mathcal{C}_{*}$ & $0.0865$--$0.3047$ & $0.0838$--$0.3042$ \\
$\Omega_{*}/\Omega_{K}$ & $0.270$--$0.609$ & $0.300$--$0.640$ \\
\bottomrule[1.2pt]
\end{tabular}
\caption{The two stellar sequences, each evaluated at $f_*=716\,\mathrm{Hz}$.
Ranges are quoted over the whole sequence. }
\label{table:sequences}
\end{table}

\begin{figure*}
\includegraphics[width=2.05\columnwidth]{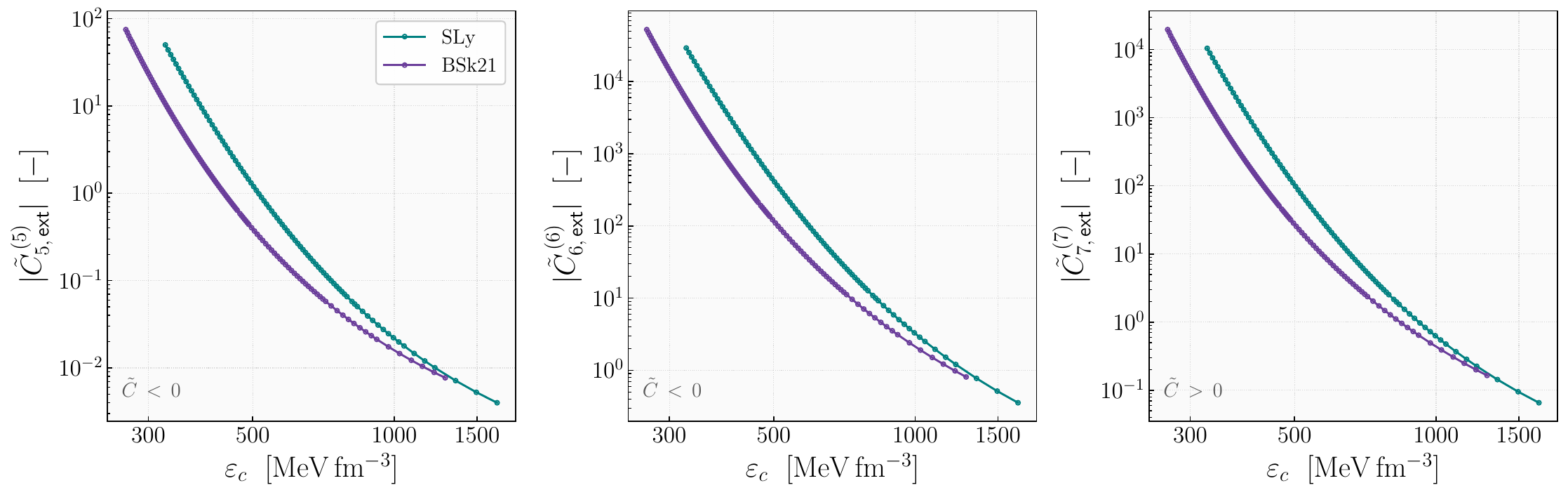}
\caption{Three representative exterior integration constants against the central
energy density, for both sequences. From left to right they are the leading
constants of $S_5$, $M_6$ and $S_7$, entering at fifth, sixth and seventh order
respectively. The magnitude is plotted, with the sign indicated in each panel.
SLy is shown in teal and BSk21 in purple. The remaining constants are given in
Appendix~\ref{app:exterior_constants}.}
\label{fig:exterior_constants_panel}
\end{figure*}

\subsubsection*{Smoothness of the sequences}

Because each sequence is generated by varying $\varepsilon_c$ alone—and the
exterior analytical solutions require care at low compactness
(Sec.~\ref{sec:single_star})—we must verify that the multipole moments vary
smoothly across the full mass range. Since the moments are built directly from
the exterior integration constants, we track the constants themselves. A small
change in $\varepsilon_c$ should produce a small change in each constant, just
as it does in the TOV mass, without numerical jumps.
Figure~\ref{fig:exterior_constants_panel} illustrates this for the leading
constants of $S_5$, $M_6$, and $S_7$, which belong to the three highest spin
orders and are therefore the most exposed to accumulated hierarchical error. As
$\varepsilon_c$ increases, each of these constants falls monotonically over
three to five decades, without changing sign and without kinks or jumps along
either sequence. The remaining even- and odd-sector constants (collected in
Appendix~\ref{app:exterior_constants}) vary just as smoothly, all maintaining
the same sign except $\tilde{C}^{(6)}_{0,\mathsf{ext}}$ and
$\tilde{C}^{(7)}_{3,\mathsf{ext}}$, which each pass smoothly through zero once.
The resulting multipole moments inherit this smoothness, which rules out abrupt
numerical artifacts in the trends examined below.

This smoothness originates directly from the underlying radial profiles, which
fix the constants via surface matching at $R_*$. To visualize this continuity,
the accompanying repository~\cite{conde2025aqlimr} provides animations of the
SLy profiles for each spin order and mode $\ell$, stepping through the sequence
star by star. As compactness increases and the stellar surface moves inward, the
interior and exterior solutions deform continuously and their match at $R_*$
slides without discontinuity.

\subsubsection*{Multipole moments along the sequences}

\begin{figure*}[t]
\includegraphics[width=\textwidth]{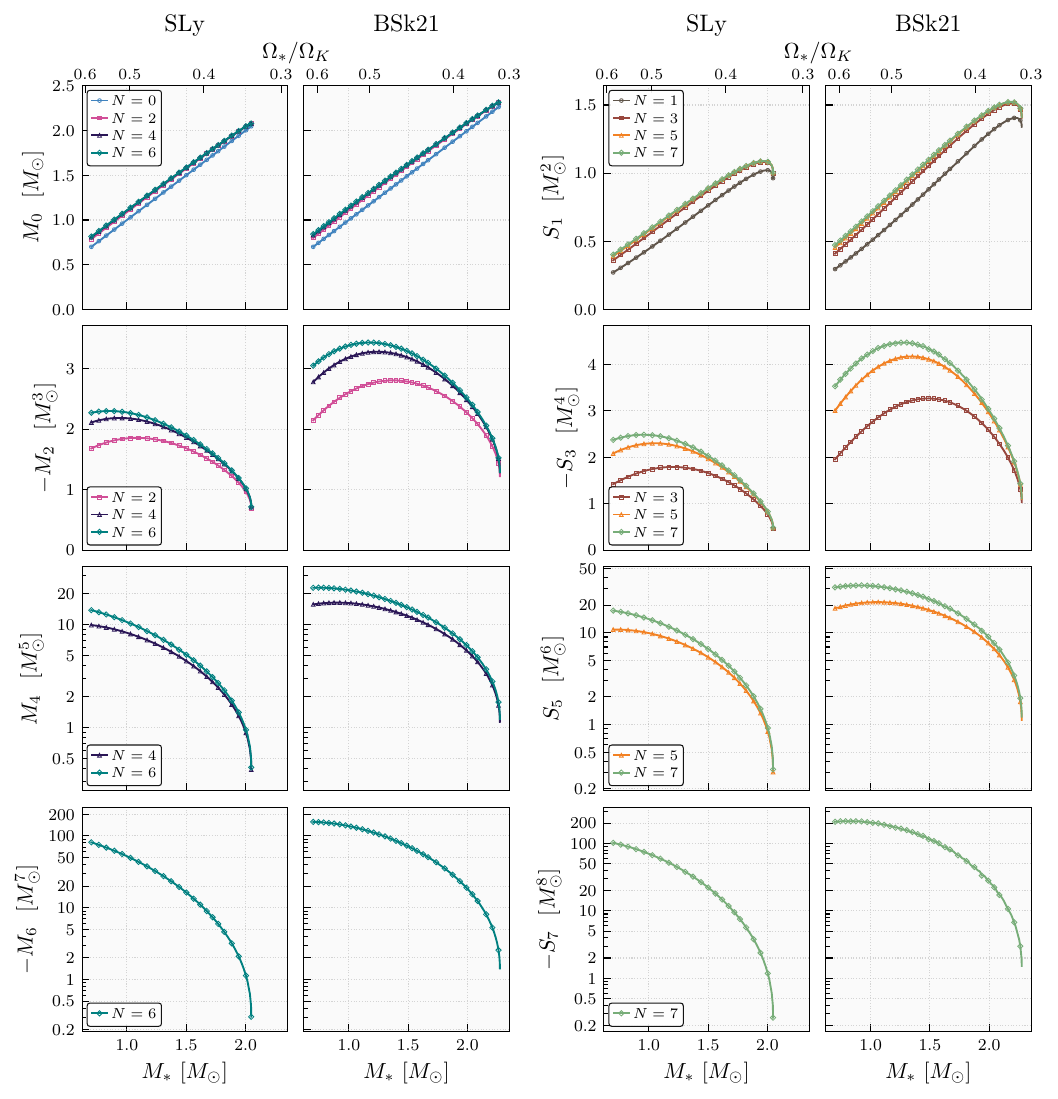}
\caption{The multipole moments $(M_{\ell}, S_{\ell})$ along both sequences, with
the expansion truncated at successively higher orders. Each row shows one
moment, the mass moments on the left and the current moments on the right, with
SLy and BSk21 in adjacent columns. Within a panel the curves are labeled by $N$,
the highest order retained in the sum of Eq.~\eqref{eq:cumulative}. The lower
axis gives the TOV stellar mass and the upper the rotation measure
$\Omega_*/\Omega_K$, which decreases as the mass increases. The moments $M_6$
and $S_7$ carry a single curve each, being computed only at leading order.}
\label{fig:cumulative}
\end{figure*}

Figure~\ref{fig:cumulative} presents each multipole moment along both sequences,
with the expansion truncated at successively higher orders. Every panel shows
one moment against the stellar TOV mass $M_{*}$. The curves within it are
labeled by $N$, the highest order retained in the sum,
\begin{equation}
P_{\ell}(N) = \sum_{n=\ell,\,\ell+2,\,\ldots}^{N}
\hat{P}^{(n)}_{\ell}\,\Omega_*^{n},
\label{eq:cumulative}
\end{equation}
where, as before, $P_\ell$ stands for either $M_\ell$ or $S_\ell$. A curve
labeled $N$ is thus the total value of the moment accumulated through that
order, and the gap between consecutive curves is the size of the correction
added between them. The upper axis carries the rotation measure
$\Omega_*/\Omega_K$, which decreases as the mass increases. Reading a panel from
left to right implies, therefore, reading it from the least to the most compact
configuration, and the vertical spread between the curves measures how much the
expansion is still changing at each point along the way.

The corrections matter most at the low-mass end of each sequence and least near
its maximum. In every panel the curves are widely separated on the left, where
the star is least compact and $\Omega_*/\Omega_K$ is largest, and draw together
towards the right as the mass increases. By the maximum mass they lie much more
closely together, so that truncating at a low order costs little there. The two
EoS behave in the same way, BSk21 tracing a curve of the same shape as SLy but
displaced upward, since at any given mass it yields the larger radius and
therefore the lower compactness. The ordering established for six stars in
Sec.~\ref{sec:single_star} thus holds continuously along the whole of both
sequences.

\begin{figure*}[t]
\centering
\includegraphics[width=0.98\textwidth]{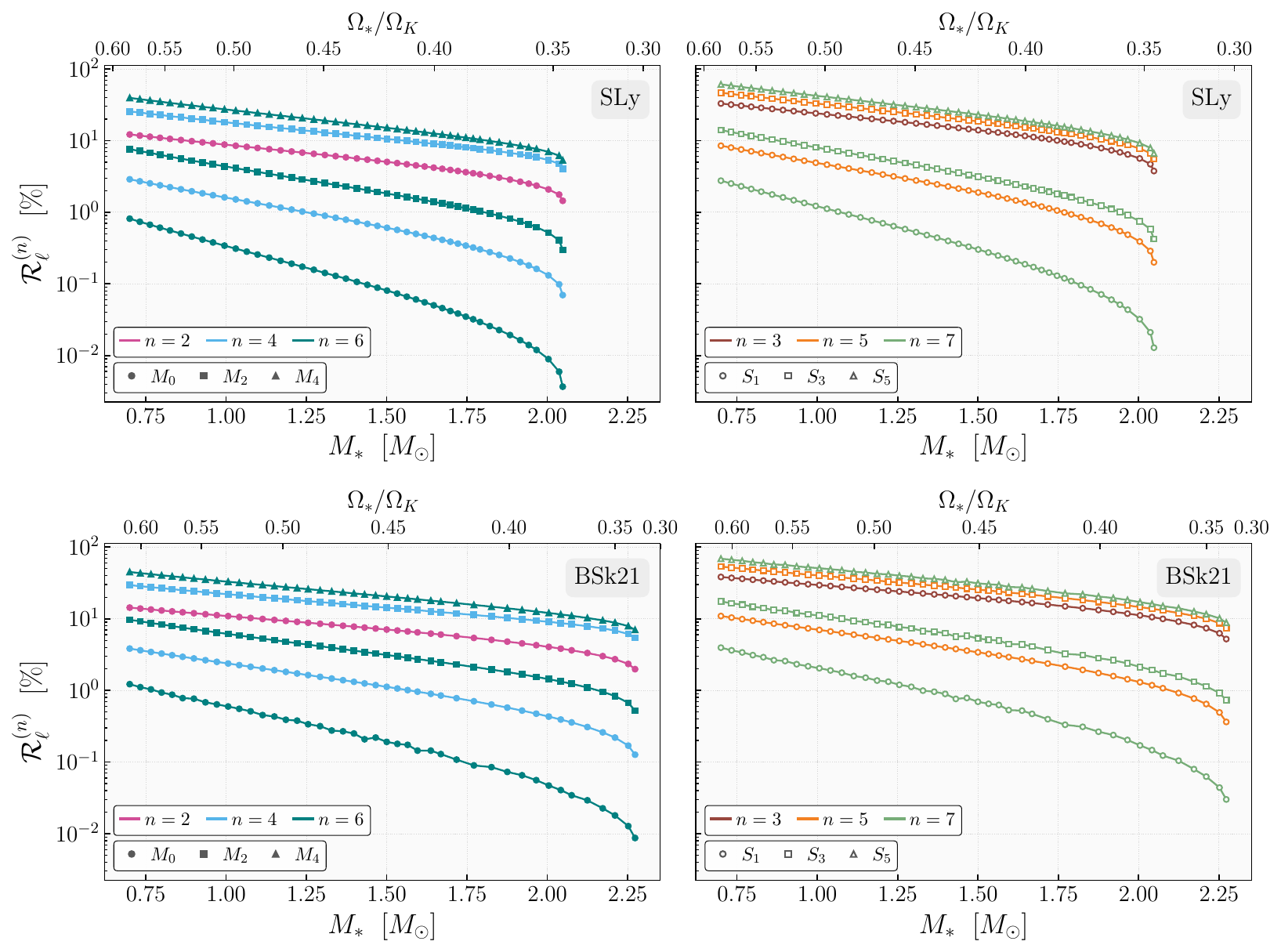}
\caption{The relative order-$n$ correction of Eq.~\eqref{eq:truncation} against
the stellar mass, for every multipole that carries a correction. The mass
moments appear in the left column and the current moments in the right, with SLy
in the upper row and BSk21 in the lower. Markers identify the multipole and
colors the order at which the correction enters, so that $M_0$ and $S_1$
contribute three curves each, $M_2$ and $S_3$ two, and $M_4$ and $S_5$ one.
Every star is evaluated at $f_*=716\,\mathrm{Hz}$. The compactness
$\mathcal{C}_*$ increases from left to right along the mass axis, and the
rotation measure $\Omega_*/\Omega_K$ decreases with it. The top axis of each
panel gives $\Omega_*/\Omega_K$ directly., which is drawn per row, since, at a
given mass, the two EoS differ in that ratio by up to a quarter. For clarity the
sequences are drawn with 45 of the 89 SLy stars and 43 of the 128 BSk21 stars.
The lowest BSk21 curves, $M_{0}$ at $n=6$ and $S_{1}$ at $n=7$, are slightly
wavy, while their SLy counterparts are not; these are the only orders whose
sources draw on $\mathrm{d}^3\varepsilon/\mathrm{d}R^3$, which is built from the
second derivative of the squared sound speed. The BSk21 EoS is smoother than the
SLy EoS in $c_s^2$ and its first derivative, but not in the second derivative,
which is what this waviness tracks.}
\label{fig:truncation_sequence}
\end{figure*}

The sign of each moment is preserved along both sequences. The sign is set by
the leading term in the expansion and is not overturned by any of the
corrections, at any mass or for either EoS. In Fig.~\ref{fig:cumulative}, $M_2$,
$S_3$, $M_6$ and $S_7$ are negative, and the other four moments are positive.
The signs of the three lowest moments, $M_0$, $S_1$ and $M_2$, have a direct
physical meaning. The mass monopole $M_0$ is positive everywhere, and rotation
adds to the gravitational mass rather than subtracting from it. The spin angular
momentum $S_1$ is positive for the chosen orientation of the rotation axis and
keeps a single sign throughout, the star rotating uniformly about a fixed axis.
The mass quadrupole is negative ($M_2<0$), as it must be for an equilibrium
figure that is flattened at the poles by its own rotation. The signs of the
higher moments describe finer details of the structure of the star. For both
EoSs they alternate in pairs, in the same pattern as the moments of a Kerr black
hole.\footnote{The multipole moments of a Kerr black hole satisfy $M_\ell + i
S_\ell = M_{0} (i a)^{\ell}$, where $M_{0}$ is the mass and $a = S_1/M_{0}$ is
the angular momentum per unit mass~\cite{Hansen:1974zz}. For $a>0$ the signs of
$M_0$, $S_1$, $M_2$, $S_3$, $M_4$, $S_5$, $M_6$ and $S_7$ are then
$+,+,-,-,+,+,-,-$.} None of this depends on the corrections staying small. Every
spin-induced correction found here carries the sign of the leading term of its
own multipole, and terms of one sign can only reinforce the total, whatever
their size. Nothing in the construction requires a priori that the corrections
share the sign of the leading term, and we record it as a property of the two
models studied here.

\subsubsection*{Convergence along the sequences}

Two complementary quantities track the convergence along the sequences. While
the fractional order-$n$ relative correction $\mathcal{R}^{(n)}_{\ell}$ of
Eq.~\eqref{eq:truncation} measures each correction against the partial sum
accumulated before it (testing whether the expansion is settling), the
fractional contribution $\mathcal{T}^{(n)}_{\ell}$ measures the same correction
against the completed moment,
\begin{equation}
\mathcal{T}^{(n)}_{\ell} = 100 \times
\frac{\bigl| \hat{P}^{(n)}_{\ell}\,\Omega_{*}^{n} \bigr|}
{\Bigl| \sum_{m=\ell,\,\ell+2,\,\ldots}^{n_{\mathrm{max}}}
\hat{P}^{(m)}_{\ell}\,\Omega_{*}^{m} \Bigr|}\quad [\%],
\label{eq:fractional_contribution}
\end{equation}
with $n_{\mathrm{max}}$ the highest order retained. Because
$\mathcal{T}^{(n)}_{\ell}$ sums to $100\%$ over all orders (up to sign
cancellations), it reveals how the moment is distributed among the orders that
build it. The three measures $\mathcal{R}^{(n)}_{\ell}$,
$\eta^{\mathrm{lead}}_{\ell}$, and $\mathcal{T}^{(n)}_{\ell}$, are closely
related and easily confused, so Table~\ref{table:diagnostics} presents them side
by side.

\begin{table}[h!]
\footnotesize
\setlength{\tabcolsep}{2pt}
\renewcommand{\arraystretch}{1.25}
\begin{tabular}{l l l}
\toprule
Symbol & Name & Ratio \\
\midrule
$\mathcal{R}^{(n)}_{\ell}$ & fractional relative correction & order $n$ / sum to $n{-}2$ \\
$\eta^{\mathrm{lead}}_{\ell}$ & leading-order discrepancy & corrections / total \\
$\mathcal{T}^{(n)}_{\ell}$ & fractional contribution & order $n$ / total \\
\bottomrule
\end{tabular}
\caption{The three diagnostics used in this section, defined in
Eqs.~\eqref{eq:truncation}, \eqref{eq:lead} and
\eqref{eq:fractional_contribution}. All are percentages and all carry the
multipole index $\ell$ and the spin order $n$. Here ``total'' means the sum
through the highest order computed, not the exact moment.}
\label{table:diagnostics}
\end{table}

Figure~\ref{fig:truncation_sequence} plots $\mathcal{R}^{(n)}_{\ell}$ against
stellar mass for all multipoles carrying corrections (mass moments on the left,
current moments on the right, for SLy in the upper row and BSk21 in the lower).
In every panel, $\mathcal{R}^{(n)}_{\ell}$ decreases with increasing mass,
dropping most steeply near the maximum mass. Because $\Omega_*/\Omega_K$ grows
from right to left along the mass axis, the corrections are largest where the
expansion parameter is largest, and at any fixed mass they remain systematically
larger for the stiffer BSk21 than for SLy. The behavior identified at three
isolated masses in Sec.~\ref{sec:single_star} therefore holds continuously
across the entire sequence.

\begin{figure*}[t]
\includegraphics[width=\textwidth]{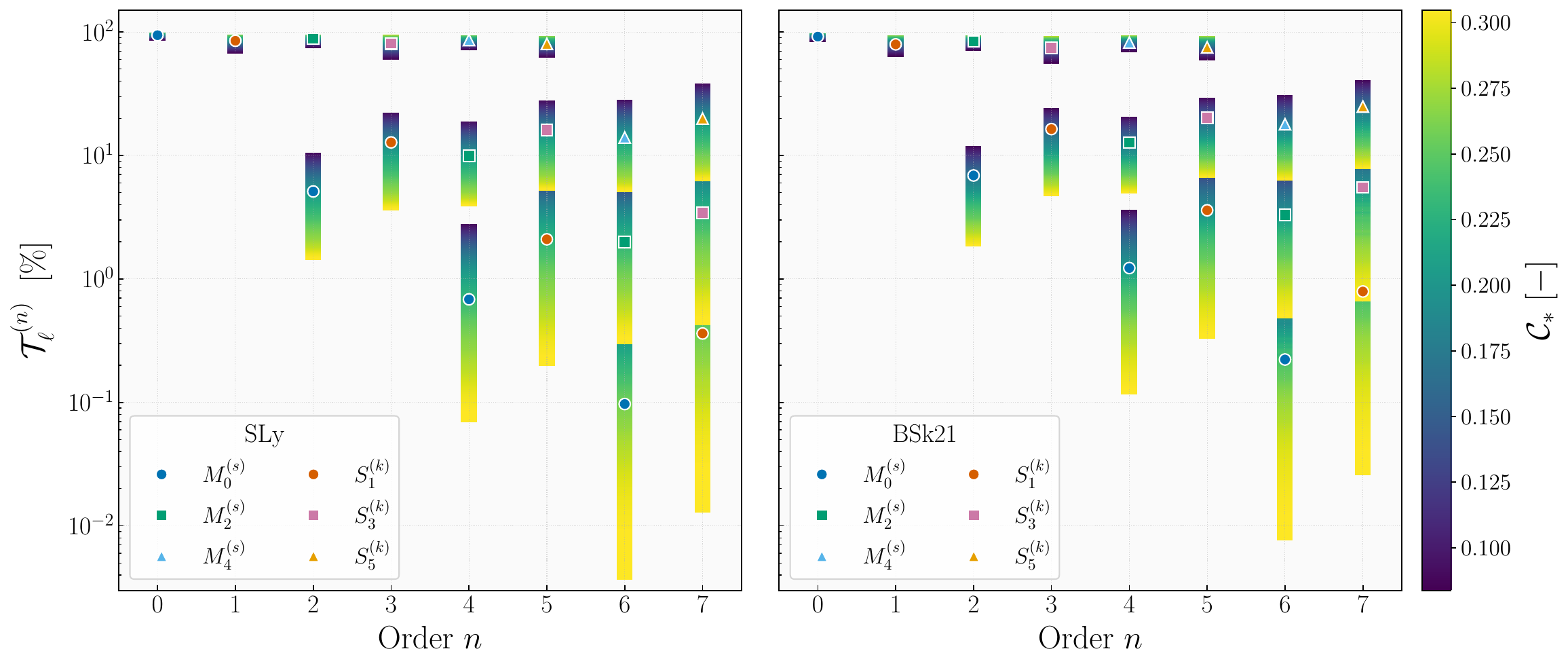}
\caption{Fractional contribution $\mathcal{T}^{(n)}_{\ell}$ of
Eq.~\eqref{eq:fractional_contribution} against spin-frequency order at
$f_*=716\,\mathrm{Hz}$, for SLy (left) and BSk21 (right). Each bar spans the
full range across the sequence (showing the variation between the least and most
compact stars) and is colored by compactness $\mathcal{C}_*$. Internal markers
identify the multipole (carrying a generic superscript $s$ for even mass orders
and $k$ for odd current orders) and indicate the sequence median.}
\label{fig:fractional_contribution}
\end{figure*}

Figure~\ref{fig:fractional_contribution} displays the corresponding distribution
of $\mathcal{T}^{(n)}_{\ell}$ across spin orders, where each vertical bar spans
the full sequence range colored by $\mathcal{C}_*$ and the internal marker
locates the median. The leading term supplies the bulk of every moment, reaching
nearly $100\%$ for the most compact stars, whereas at the light end the leading
share drops substantially—falling to $55\%$ for $S_3$ in BSk21, where
higher-order corrections account for nearly as much as the leading term itself.
At higher orders, $\mathcal{T}^{(n)}_{\ell}$ diminishes progressively (e.g.,
$|M^{(2)}_{0}|$ exceeds $|M^{(4)}_{0}|$, which in turn exceeds $|M^{(6)}_{0}|$).
Because the contributions sum to $100\%$, the compactness color gradient in the
leading bars is inverted relative to all correction bars. The most compact stars
have the smallest corrections and thus sit at the top of the leading bars and
the bottom of every correction bar, whereas the least compact stars do the
opposite. Consistent with the compactness scaling above, the correction bars of
BSk21 lie above those of SLy at every order, and its leading bars
correspondingly lie below.

Taken together, the full sequences confirm that the exterior constants vary
smoothly with $\varepsilon_c$, the sign of every moment is preserved, and
successive corrections diminish throughout. The correction magnitudes are
governed by $\Omega_*/\Omega_K$, with the difference between the two EoS
reflecting the compactness each implies. How far the expansion must be carried
therefore depends on both the stellar configuration and the multipole; at
$1.0\,M_\odot$, for instance, the leading term alone misses $10\%$ of $M_0$ but
$30\%$ of $S_3$. Near the maximum mass ($\Omega_*/\Omega_K \simeq 0.3$),
leading-order terms capture every moment that carries a correction to within a
few percent (below $8\%$, and below $2\%$ for $M_0$ via
$\eta^{\mathrm{lead}}_{\ell}$ of Eq.~\eqref{eq:lead}). At the light end near
$0.7\,M_\odot$ ($\Omega_*/\Omega_K \simeq 0.6$), however, the leading-order
discrepancy $\eta^{\mathrm{lead}}_{\ell}$ reaches $14\%$ for $M_0$, $26\%$ for
$M_2$, and $33\%$ for $S_1$ in SLy—and even higher in BSk21, where $S_3$ reaches
$44\%$. Across a sequence, the accuracy of a leading-order calculation thus
varies by roughly an order of magnitude, making the higher-order corrections
reported here essential at low masses.


\section{Discussion and Conclusions}
\label{sec:conclusions}

We have computed the multipole moments of rotating neutron stars through seventh
order in spin frequency within the Hartle--Thorne expansion. The higher-order
corrections turn out to be important not only for the higher multipoles, where
they are expected, but also for the mass and the angular momentum. The
calculation was carried out for the SLy and BSk21 EoS, chosen as representative
soft and stiff cases, for stars rotating at $716\,\mathrm{Hz}$. How much the
corrections matter depends on the star and on the multipole moment. At fixed
spin frequency, the corrections are largest for the least compact stars. The
corrections are also larger for the current moments than for the mass moments.
Finally, the first correction grows with the multipole degree, so it is largest
for the higher multipoles, which enter the expansion only at high orders in the
spin frequency. For instance, for a $1.0\,M_\odot$ neutron star, the leading
term alone omits $10\%$ (SLy) and $13\%$ (BSk21) of the mass, and $24\%$ (SLy)
and $29\%$ (BSk21) of the angular momentum. For a $2.0\,M_\odot$ star, these
fractions fall to $2\%$ (SLy) and $4\%$ (BSk21) for the mass, and to $6\%$ (SLy)
and $11\%$ (BSk21) for the angular momentum.

Figure~\ref{fig:required_order} places these results in the context of the
pulsars that have so far been observed. Most of them spin near $2\,\mathrm{Hz}$,
where the leading terms in the slow-rotation expansion are accurate enough.
About $16\%$ of the observed pulsars spin at $100\,\mathrm{Hz}$ or faster,
forming the second peak of the distribution. This group includes the pulsars
analyzed by NICER, which spin at about $170$ to
$350\,\mathrm{Hz}$~\cite{Miller:2019cac,Riley:2019yda,Miller:2021qha,Riley:2021pdl,Choudhury:2024xbk}.
The corrections shrink quickly as the spin frequency decreases, because each
spin correction carries two powers of $\Omega_*$. At $100\,\mathrm{Hz}$, the
first spin correction is below $1\%$ relative to the leading-order term for
every moment for all six configurations studied here. At $300\,\mathrm{Hz}$, the
first-order correction reaches $0.4$ to $2\%$ for the mass, $1$ to $5\%$ for the
angular momentum, $1$ to $4\%$ for the mass quadrupole, and it is larger for the
higher moments. These values are below the precision of current NICER radius
measurements, of about $10\%$~\cite{Miller:2019cac,Choudhury:2024xbk}, but they
could become relevant for future missions, such as eXTP~\cite{eXTP:2016rzs}.

The decrease of successive corrections supports the use of the truncated
expansion, but it does not by itself establish its accuracy. Four of the moments
that we calculated here carry two or three correction terms, and in all those
cases, each correction is smaller than the preious term by a factor that scales
as $(\Omega_{*}/\Omega_K)^{2}$.
This ordering of the corrections persists up to $77\%$ to $93\%$ of the
Newtonian Keplerian frequency $f_K$. Since the mass-shedding frequency lies
below $f_K$, at about $0.6\,f_K$~\cite{Lattimer:2004pg, Haensel:2009wa}, the
ordering in fact extends beyond the mass-shedding limit. What this shows is that
the computed terms behave as a well-ordered expansion should over the full range
of spin frequencies these stars can reach. For a $1.0\,M_\odot$ star, computing
the mass through $\mathcal{O}(\Omega^4)$ leaves an error below $1\%$ from the
$\mathcal{O}(\Omega^6)$ contribution up to $865\,\mathrm{Hz}$ for SLy and
$789\,\mathrm{Hz}$ for BSk21. Both frequencies lie above the $716\,\mathrm{Hz}$
spin frequency of the fastest observed pulsar. For calculations of the mass
along a stellar sequence, corrections through $\mathcal{O}(\Omega^4)$ are
therefore reasonable for such stars, given the accuracy of current astrophysical
and gravitational-wave observations, and the fact that heavier stars require
fewer terms in the expansion. The moments $M_6$ and $S_7$ are computed only at
leading order, so their first omitted terms are unavailable. The moments $M_4$
and $S_5$ carry a single correction each, which lets us estimate the error of
the leading term but not of the sum that includes that correction. Settling the
accuracy and the range of applicability of the expansion therefore requires an
independent benchmark. A comparison with nonperturbative solutions of rapidly
rotating stars is underway for exactly this purpose.

A broader set of EoS is needed to determine how general these results are. Two
EoS were used here, SLy as a soft case and BSk21 as a stiff one, and both gave
the same picture. In particular, for both EoSs, the required order depends on
the spin frequency mainly through $\Omega_{*}/\Omega_K$. Whether this holds for
EoS in general can only be settled by studying more of them. For EoS with a
smooth energy densityl, the calculation extends without change. For those with a
finite surface density, as in quark stars, or with jumps in the density or its
derivatives, as in piecewise polytropes or phase transitions, the high-order
equations need junction conditions at each jump before they can be integrated
across it. A caveat also applies within the present sequences. The corrections
scale with powers of $\Omega_{*}/\Omega_K$, which acts as the effective
expansion parameter, and so the expansion is to be trusted on the assumption
that this parameter is small. For the $2.0\,M_\odot$ stars at
$716\,\mathrm{Hz}$, this ratio ranges from $0.30$ to $0.38$. At the light end of
the sequences, near $0.7\,M_\odot$, the ratio reaches $0.61$ to $0.64$, and a
parameter that large does not guarantee that a truncation at seventh order is
accurate. Such a ratio is also comparable to the mass-shedding estimate of about
$0.6$, so these stars would rotate close to the limit they can sustain. For
those stars, the computed corrections still decrease from one order to the next,
and that is the only evidence that the truncation holds there. Stars this light
have not been observed, so the caveat is unlikely to matter in practice, but it
sets the lower mass at which the present results can be trusted.

Our results now make it possible to test how higher-order rotation effects
propagate into quasi-universal relations and the modeling of X-ray pulse
profiles. On the first front, quasi-universal relations link $M_2$ to the tidal
deformability $\Lambda$ and the moment of inertia $I$ in a manner largely
independent of the EoS~\cite{Yagi:2013awa, Chatziioannou:2018vzf}, and
approximate no-hair relations extend this universality to $S_3$ and
$M_4$~\cite{Yagi:2014bxa}. Three calculations are directly enabled by the
moments computed here. The I-Love-Q relations can be revisited with the
spin-induced corrections to the moment of inertia and to $M_2$ included. The
effective no-hair relations for $S_3$ and $M_4$ against $M_2$ can be revisited
in the same way. Finally, the three higher moments $S_5$, $M_6$ and $S_7$,
computed here for the first time, can be tested for effective no-hair relations
of their own. These tests bear on the use of such relations in interpreting
observations of neutron-star binaries. On the second front, pulse-profile models
need two inputs that the expansion also provides. One is the shape of the star,
in particular its equatorial radius and eccentricity, both of which are deformed
by rotation. The other is the exterior spacetime through which the light
propagates, now described more accurately by the higher multipole moments.
Whether the higher-order corrections shift an inferred mass or radius also
depends on the viewing geometry and on the pulse-profile model. Quantifying the
effect of the two inputs therefore requires computing both at seventh order and
incorporating them into a pulse-profile model. All of these applications remain
for future work. If the higher-order corrections shift the inferred mass or
equatorial radius by an amount comparable to the current measurement
uncertainty, they would need to be included in the model or accounted for as a
systematic error. Either way, knowing their size is a prerequisite for future
pulse-profile observations to constrain dense matter at the level the data will
allow.


\begin{acknowledgments}

We thank Kent Yagi for valuable discussions on integrating numerically the
perturbation equations. The authors are supported by the NSF under the MUSES
collaboration OAC-2103680. C.C., H.-J.K. and N.Y. acknowledge support from the
Simons Foundation through Award No. 896696, the Simons Foundation International
through Award No. SFI-MPS-BH-00012593-01, and the NSF through Grant No.
PHY-25-12423. J.N.H. acknowledges support from the US-DOE Nuclear Science Grant
No. DE-SC0023861.

\end{acknowledgments}


\section*{Data availability}
The implementation and the EoS inputs used in this work are publicly
available~\cite{conde2025aqlimr}, together with the Chebyshev coefficients of
the segmented fits of Sec.~\ref{sec:derivatives}; the equations it solves, with
the accompanying symbolic material, are available in the
\texttt{hartle-thorne-project} repository~\cite{conde2025hartlethorne}.


\bibliography{references/inspire, references/NOTinspire}


\appendix
\allowdisplaybreaks

\begin{widetext}


\section{Asymptotic Expansions}
\label{app:bc}

\medskip \noindent \textbf{The odd sector.} Each odd-parity system is integrated
as the pair $(\omega^{(k)}_\ell,\alpha^{(k)}_\ell)$ with $\alpha^{(k)}_\ell
\equiv \mathrm{d}\omega^{(k)}_\ell/\mathrm{d}R$, and likewise
$\bar\alpha^{(1)}_1\equiv\mathrm{d}\varpi^{(1)}_1/\mathrm{d}R$ at first order.
The expansion that starts $\alpha$ is therefore not listed separately: it is the
term-by-term derivative of the expansion listed for the corresponding $\omega$,
one order of $R$ shorter.

\medskip \noindent \textbf{The even sector.} For $\ell\neq 0$ each even-parity
system is integrated as the pair $(v^{(s)}_\ell,h^{(s)}_\ell)$, and both members
are listed below for the homogeneous and particular ODE systems. The remaining
metric function $k^{(s)}_\ell$ is not integrated and needs no initial condition
of its own, being recovered from $k^{(s)}_\ell=v^{(s)}_\ell-h^{(s)}_\ell$. For
$\ell=0$ the variables are instead $(m^{(s)}_0, \xi^{(s)}_0, h^{(s)}_0)$. At
second order only the pair $(m^{(2)}_0,\xi^{(2)}_0)$ is integrated, since
$h^{(2)}_0$ follows algebraically from the matching at the surface and so needs
no initial condition. From fourth order onward that is no longer the case, and
the asymptotic expansion for $h^{(s)}_0$ is required.

\medskip \noindent Listed below are only the expansions that start an integrator
through third order. The full set of asymptotic expansions is too long to
reproduce here and is available in the \texttt{hartle-thorne-project}
repository~\cite{conde2025hartlethorne}. Throughout, $c^2_{s,c} \equiv
(\mathrm{d}p/\mathrm{d}\varepsilon)_{\varepsilon_c}$ denotes the central sound
speed squared, as in the main text.

\begin{align}
M(R) &= \frac{4 \pi}{3}\,\varepsilon_c\,R^{3} + \mathcal{O}(R^{5}) \\[5pt]
p(R) &= p_c -\frac{2 \pi}{3}\,\left( \varepsilon_c + p_c\right)  \left( \varepsilon_c + 3 p_c\right) \,R^{2} + \mathcal{O}(R^{4}) \\[5pt]
\varepsilon(R) &= \varepsilon_c -\frac{2 \pi}{3 {c^2_{s,c}}}\,\left( \varepsilon_c + p_c\right)  \left( \varepsilon_c + 3 p_c\right) \,R^{2} + \mathcal{O}(R^{4}) \\[5pt]
\nu(R) &= \nu_c + \frac{4 \pi}{3}\,\left( \varepsilon_c + 3 p_c\right) \,R^{2} + \mathcal{O}(R^{4}) \\[5pt]
\varpi^{(1)}_{1}(R) &= {\varpi^{(1)}_{1,c}} + \frac{8 \pi}{5}\,{\varpi^{(1)}_{1,c}}\,\left( \varepsilon_c + p_c\right) \,R^{2} + \frac{8 \pi^{2}}{105 {c^2_{s,c}}}\,{\varpi^{(1)}_{1,c}}\,\left( \varepsilon_c + p_c\right)  \left( 33 {c^2_{s,c}} \varepsilon_c + 3 {c^2_{s,c}} p_c - 5 \varepsilon_c - 15 p_c\right) \,R^{4} + \mathcal{O}(R^{6}) \\[5pt]
m^{(2)}_{0}(R) &= \frac{4 \pi}{15 {c^2_{s,c}}}\,\left( {\varpi^{(1)}_{1,c}}\right) ^{2} e^{- \nu_c}\,\left( 2 {c^2_{s,c}} + 1\right)  \left( \varepsilon_c + p_c\right) \,R^{5} + \mathcal{O}(R^{7}) \\[5pt]
\xi^{(2)}_{0}(R) &= \frac{1}{4 \pi \left( \varepsilon_c + 3 p_c\right) }\,\left( {\varpi^{(1)}_{1,c}}\right) ^{2} e^{- \nu_c}\,R -\frac{1}{20 {c^2_{s,c}} \left( \varepsilon_c + 3 p_c\right) }\,\left( {\varpi^{(1)}_{1,c}}\right) ^{2} e^{- \nu_c}\,\left( {c^2_{s,c}} - 1\right)  \left( \varepsilon_c + p_c\right) \,R^{3} + \mathcal{O}(R^{5}) \\[5pt]
h^{(2),\mathsf{H}}_{2}(R) &= {A^{(2),\mathsf{H}}_{2,\mathsf{int}}}\,R^{2} -\frac{2 \pi}{21 {c^2_{s,c}}}\,{A^{(2),\mathsf{H}}_{2,\mathsf{int}}}\,\left( {c^2_{s,c}} \varepsilon_c + 33 {c^2_{s,c}} p_c + 3 \varepsilon_c + 3 p_c\right) \,R^{4} + \mathcal{O}(R^{6}) \\[5pt]
h^{(2),\mathsf{P}}_{2}(R) &= {A^{(2),\mathsf{P}}_{2,\mathsf{int}}}\,R^{2}
  \nonumber \\*
  &\quad -\frac{2 \pi}{21 {c^2_{s,c}}}\,\left[ {A^{(2),\mathsf{P}}_{2,\mathsf{int}}} \left( {c^2_{s,c}} \varepsilon_c + 33 {c^2_{s,c}} p_c + 3 \varepsilon_c + 3 p_c\right)  + \left( {\varpi^{(1)}_{1,c}}\right) ^{2} \left( 7 {c^2_{s,c}} + 1\right)  \left( \varepsilon_c + p_c\right)  e^{- \nu_c}\right] \,R^{4} + \mathcal{O}(R^{6}) \\[5pt]
v^{(2),\mathsf{H}}_{2}(R) &= -\frac{2 \pi}{3}\,{A^{(2),\mathsf{H}}_{2,\mathsf{int}}}\,\left( \varepsilon_c + 3 p_c\right) \,R^{4} + \mathcal{O}(R^{6}) \\[5pt]
v^{(2),\mathsf{P}}_{2}(R) &= \frac{2 \pi}{3}\,\left[ - {A^{(2),\mathsf{P}}_{2,\mathsf{int}}} \left( \varepsilon_c + 3 p_c\right)  + \left( {\varpi^{(1)}_{1,c}}\right) ^{2} \left( \varepsilon_c + p_c\right)  e^{- \nu_c}\right] \,R^{4} + \mathcal{O}(R^{6}) \\[5pt]
\omega^{(3),\mathsf{H}}_{1}(R) &= {\omega^{(3),\mathsf{H}}_{1,c}} + \frac{8 \pi}{5}\,{\omega^{(3),\mathsf{H}}_{1,c}}\,\left( \varepsilon_c + p_c\right) \,R^{2} + \frac{8 \pi^{2}}{105 {c^2_{s,c}}}\,{\omega^{(3),\mathsf{H}}_{1,c}}\,\left( \varepsilon_c + p_c\right)  \left( 33 {c^2_{s,c}} \varepsilon_c + 3 {c^2_{s,c}} p_c - 5 \varepsilon_c - 15 p_c\right) \,R^{4} + \mathcal{O}(R^{6}) \\[5pt]
\omega^{(3),\mathsf{P}}_{1}(R) &= {\omega^{(3),\mathsf{P}}_{1,c}} + \frac{8 \pi}{5}\,{\omega^{(3),\mathsf{P}}_{1,c}}\,\left( \varepsilon_c + p_c\right) \,R^{2}
  \nonumber \\*
  &\quad + \frac{4 \pi}{525 {c^2_{s,c}}}\,\biggl[ 10 \pi {\omega^{(3),\mathsf{P}}_{1,c}} \left( \varepsilon_c + p_c\right)  \left( 33 {c^2_{s,c}} \varepsilon_c + 3 {c^2_{s,c}} p_c - 5 \varepsilon_c - 15 p_c\right)  - 3 {A^{(2)}_{2,\mathsf{int}}} {\varpi^{(1)}_{1,c}} \left( 7 {c^2_{s,c}} + 5\right)  \left( \varepsilon_c + p_c\right)
  \nonumber \\*
  &\quad - 30 \left( {\varpi^{(1)}_{1,c}}\right) ^{3} \left( 3 {c^2_{s,c}} + 1\right)  \left( \varepsilon_c + p_c\right)  e^{- \nu_c}\biggr] \,R^{4} + \mathcal{O}(R^{6}) \\[5pt]
\omega^{(3),\mathsf{H}}_{3}(R) &= {A^{(3),\mathsf{H}}_{3,\mathsf{int}}}\,R^{2} + \frac{4 \pi}{27}\,{A^{(3),\mathsf{H}}_{3,\mathsf{int}}}\,\left( 19 \varepsilon_c + 9 p_c\right) \,R^{4} + \mathcal{O}(R^{6}) \\[5pt]
\omega^{(3),\mathsf{P}}_{3}(R) &= {A^{(3),\mathsf{P}}_{3,\mathsf{int}}}\,R^{2}
  \nonumber \\*
  &\quad + \frac{4 \pi}{675 {c^2_{s,c}}}\,\biggl[ 25 {A^{(3),\mathsf{P}}_{3,\mathsf{int}}} {c^2_{s,c}} \left( 19 \varepsilon_c + 9 p_c\right)  + 6 {A^{(2)}_{2,\mathsf{int}}} {\varpi^{(1)}_{1,c}} \left( 7 {c^2_{s,c}} + 5\right)  \left( \varepsilon_c + p_c\right)
  \nonumber \\*
  &\quad + 10 \left( {\varpi^{(1)}_{1,c}}\right) ^{3} \left( 3 {c^2_{s,c}} + 1\right)  \left( \varepsilon_c + p_c\right)  e^{- \nu_c}\biggr] \,R^{4} + \mathcal{O}(R^{6})
\end{align}


\section{Source Terms}
\label{app:Source}
\noindent Each perturbation equation solved in the interior is driven by a
source term built from the background star and from the solutions found at lower
orders. These sources are what tie the orders together, and they grow fast: by
fourth order a single coefficient runs to several hundred terms. Printing them
all is out of the question, so this appendix includes explicit expressions only
up to third order, and all the other source terms can be found explicitly in the
\texttt{hartle-thorne-project} repository~\cite{conde2025hartlethorne}.

\medskip \noindent \textbf{Structure.} Every source term is a polynomial in the
energy density $\varepsilon$ and its radial derivatives. Everything else sits in
the coefficients, which depend on the background and on the lower-order
solutions. The same letters are reused in every equation, so each coefficient
carries as a subscript the function whose equation it belongs to. Higher orders
in the spin lengthen the polynomial, bringing in higher powers of $\varepsilon$
and higher derivatives of it.

\begin{align}
\mathcal{S}_{m^{(2)}_{0}} &= a_{m^{(2)}_{0}} + b_{m^{(2)}_{0}} \varepsilon \\[1ex]
\mathcal{S}_{\xi^{(2)}_{0}} &= a_{\xi^{(2)}_{0}} \\[1ex]
\mathcal{S}_{v^{(2)}_{2}} &= a_{v^{(2)}_{2}} + b_{v^{(2)}_{2}}\varepsilon \\[1ex]
\mathcal{S}_{h^{(2)}_{2}} &= a_{h^{(2)}_{2}} + b_{h^{(2)}_{2}}\varepsilon \\[1ex]
\mathcal{S}_{\alpha^{(3)}_{\ell}} &= a_{\alpha^{(3)}_{\ell}} + b_{\alpha^{(3)}_{\ell}}\varepsilon + c_{\alpha^{(3)}_{\ell}}\varepsilon^{2} + A_{\alpha^{(3)}_{\ell}} \dfrac{d \varepsilon}{dR} \hspace{0.25cm} (\ell=1,3)  \\[1ex]
\mathcal{S}_{m^{(4)}_{0}} &= a_{m^{(4)}_{0}} + b_{m^{(4)}_{0}} \varepsilon + c_{m^{(4)}_{0}} \varepsilon^{2}
+ A_{m^{(4)}_{0}} \dfrac{d \varepsilon}{dR} + B_{m^{(4)}_{0}} \varepsilon \dfrac{d \varepsilon}{dR} + C_{m^{(4)}_{0}} \dfrac{d^{2} \varepsilon}{dR^{2}} \\[1ex]
\mathcal{S}_{\xi^{(4)}_{0}} &= a_{\xi^{(4)}_{0}} + b_{\xi^{(4)}_{0}} \varepsilon + c_{\xi^{(4)}_{0}} \varepsilon^{2}
+ A_{\xi^{(4)}_{0}} \dfrac{d \varepsilon}{dR}  \\[1ex]
\mathcal{S}_{h^{(4)}_{0}} &= a_{h^{(4)}_{0}} + b_{h^{(4)}_{0}} \varepsilon + c_{h^{(4)}_{0}} \varepsilon^{2}
+ A_{h^{(4)}_{0}} \dfrac{d \varepsilon}{dR}  \\[1ex]
\mathcal{S}_{v^{(4)}_{\ell}} &= a_{v^{(4)}_{\ell}} + b_{v^{(4)}_{\ell}} \varepsilon + c_{v^{(4)}_{\ell}} \varepsilon^{2}
+ A_{v^{(4)}_{\ell}} \dfrac{d \varepsilon}{dR} \hspace{0.25cm} (\ell=2,4) \\[1ex]
\mathcal{S}_{h^{(4)}_{\ell}} &= a_{h^{(4)}_{\ell}} + b_{h^{(4)}_{\ell}} \varepsilon + c_{h^{(4)}_{\ell}} \varepsilon^{2}
+ A_{h^{(4)}_{\ell}} \dfrac{d \varepsilon}{dR} \hspace{0.25cm} (\ell=2,4) \\[1ex]
\mathcal{S}_{\alpha^{(5)}_{\ell}} &= a_{\alpha^{(5)}_{\ell}} + b_{\alpha^{(5)}_{\ell}} \varepsilon + c_{\alpha^{(5)}_{\ell}} \varepsilon^{2} + d_{\alpha^{(5)}_{\ell}} \varepsilon^{3}
+ A_{\alpha^{(5)}_{\ell}} \dfrac{d \varepsilon}{dR} + B_{\alpha^{(5)}_{\ell}} \varepsilon \dfrac{d \varepsilon}{dR} + C_{\alpha^{(5)}_{\ell}} \dfrac{d^{2} \varepsilon}{dR^{2}} \hspace{0.25cm} (\ell=1,3,5) \\[1ex]
\mathcal{S}_{m^{(6)}_{0}} &= a_{m^{(6)}_{0}} + b_{m^{(6)}_{0}} \varepsilon + c_{m^{(6)}_{0}} \varepsilon^{2} + d_{m^{(6)}_{0}} \varepsilon^{3} \nonumber \\
&+ A_{m^{(6)}_{0}} \dfrac{d \varepsilon}{dR} + B_{m^{(6)}_{0}} \varepsilon \dfrac{d \varepsilon}{dR} + C_{m^{(6)}_{0}} \dfrac{d^{2} \varepsilon}{dR^{2}} + D_{m^{(6)}_{0}} \left( \dfrac{d\varepsilon}{dR} \right)^{2} + E_{m^{(6)}_{0}} \varepsilon^{2}\dfrac{d\varepsilon}{dR} + F_{m^{(6)}_{0}} \varepsilon \dfrac{d^{2}\varepsilon}{dR^{2}} + G_{m^{(6)}_{0}} \dfrac{d^{3}\varepsilon}{dR^{3}} \\[1ex]
\mathcal{S}_{\xi^{(6)}_{0}} &= a_{\xi^{(6)}_{0}} + b_{\xi^{(6)}_{0}} \varepsilon + c_{\xi^{(6)}_{0}} \varepsilon^{2} + d_{\xi^{(6)}_{0}} \varepsilon^{3}
+ A_{\xi^{(6)}_{0}} \dfrac{d \varepsilon}{dR} + B_{\xi^{(6)}_{0}} \varepsilon \dfrac{d \varepsilon}{dR} + C_{\xi^{(6)}_{0}} \dfrac{d^{2} \varepsilon}{dR^{2}}  \\[1ex]
\mathcal{S}_{h^{(6)}_{0}} &= a_{h^{(6)}_{0}} + b_{h^{(6)}_{0}} \varepsilon + c_{h^{(6)}_{0}} \varepsilon^{2} + d_{h^{(6)}_{0}} \varepsilon^{3} + A_{h^{(6)}_{0}} \dfrac{d \varepsilon}{dR} + B_{h^{(6)}_{0}} \varepsilon \dfrac{d \varepsilon}{dR} + C_{h^{(6)}_{0}} \dfrac{d^{2} \varepsilon}{dR^{2}} \\[1ex]
\mathcal{S}_{v^{(6)}_{\ell}} &= a_{v^{(6)}_{\ell}} + b_{v^{(6)}_{\ell}} \varepsilon + c_{v^{(6)}_{\ell}} \varepsilon^{2} + d_{v^{(6)}_{\ell}} \varepsilon^{3}
+ A_{v^{(6)}_{\ell}} \dfrac{d \varepsilon}{dR} + B_{v^{(6)}_{\ell}} \varepsilon \dfrac{d \varepsilon}{dR} + C_{v^{(6)}_{\ell}} \dfrac{d^{2} \varepsilon}{dR^{2}} \hspace{0.25cm} (\ell=2,4,6)\\[1ex]
\mathcal{S}_{h^{(6)}_{\ell}} &= a_{h^{(6)}_{\ell}} + b_{h^{(6)}_{\ell}} \varepsilon + c_{h^{(6)}_{\ell}} \varepsilon^{2} + d_{h^{(6)}_{\ell}} \varepsilon^{3}
+ A_{h^{(6)}_{\ell}} \dfrac{d \varepsilon}{dR} + B_{h^{(6)}_{\ell}} \varepsilon \dfrac{d \varepsilon}{dR} + C_{h^{(6)}_{\ell}} \dfrac{d^{2} \varepsilon}{dR^{2}} \hspace{0.25cm} (\ell=2,4,6) \\[1ex]
\mathcal{S}_{\alpha^{(7)}_{\ell}} &= a_{\alpha^{(7)}_{\ell}} + b_{\alpha^{(7)}_{\ell}} \varepsilon + c_{\alpha^{(7)}_{\ell}} \varepsilon^{2} + d_{\alpha^{(7)}_{\ell}} \varepsilon^{3} + e_{\alpha^{(7)}_{\ell}} \varepsilon^{4} \nonumber \\
&\quad + A_{\alpha^{(7)}_{\ell}} \dfrac{d \varepsilon}{dR} + B_{\alpha^{(7)}_{\ell}} \varepsilon \dfrac{d \varepsilon}{dR} + C_{\alpha^{(7)}_{\ell}} \dfrac{d^{2} \varepsilon}{dR^{2}} + D_{\alpha^{(7)}_{\ell}} \left( \dfrac{d\varepsilon}{dR} \right)^{2} + E_{\alpha^{(7)}_{\ell}} \varepsilon^{2}\dfrac{d\varepsilon}{dR} + F_{\alpha^{(7)}_{\ell}} \varepsilon \dfrac{d^{2}\varepsilon}{dR^{2}} + G_{\alpha^{(7)}_{\ell}} \dfrac{d^{3}\varepsilon}{dR^{3}}  \hspace{0.25cm} (\ell=1,3,5,7)
\end{align}
\textbf{Explicit form.} We give the coefficient functions named above explicitly
up to third order.
\begin{align}
a_{m^{(2)}_{0}} &= \frac{1}{12}\,R^{3} e^{- \nu}\,\left[ - \left( {\bar{\alpha}^{(1)}_{1}}\right) ^{2} \left( 2 M - R\right)  + 32 R p \pi \left( {\varpi^{(1)}_{1}}\right) ^{2}\right]  \\[6pt]
b_{m^{(2)}_{0}} &= \frac{8 \pi}{3}\,R^{4} \left( {\varpi^{(1)}_{1}}\right) ^{2} e^{- \nu} \\[6pt]
a_{\xi^{(2)}_{0}} &= \frac{1}{12 \left( M + 4 R^{3} p \pi\right) }\,R^{2} e^{- \nu}\,\left[ - 8 \left( {\varpi^{(1)}_{1}}\right) ^{2} \left( 3 M + 4 R^{3} p \pi - R\right)  + R {\bar{\alpha}^{(1)}_{1}} \left( 2 M - R\right)  \left( - R {\bar{\alpha}^{(1)}_{1}} - 8 {\varpi^{(1)}_{1}}\right) \right]  \\[6pt]
a_{v^{(2)}_{2}} &= \frac{1}{6 \left( 2 M - R\right) }\,R^{2} e^{- \nu}\,\left[ - 16 R p \pi \left( {\varpi^{(1)}_{1}}\right) ^{2} + \left( {\bar{\alpha}^{(1)}_{1}}\right) ^{2} \left( 2 M - R\right) \right]  \left( - M + 4 R^{3} p \pi + R\right)  \\[6pt]
b_{v^{(2)}_{2}} &= \frac{8 \pi}{3 \left( 2 M - R\right) }\,R^{3} \left( {\varpi^{(1)}_{1}}\right) ^{2} e^{- \nu}\,\left( M - 4 R^{3} p \pi - R\right)  \\[6pt]
a_{h^{(2)}_{2}} &= \frac{1}{12 \left( M + 4 R^{3} p \pi\right)  \left( 2 M - R\right) }\,R^{2} e^{- \nu}\,\biggl[ \left( {\bar{\alpha}^{(1)}_{1}}\right) ^{2} \left( 2 M - R\right)  \left( 2 M^{2} + 16 M R^{3} p \pi + 2 M R + 32 R^{6} p^{2} \pi^{2} - R^{2}\right)
  \nonumber \\*
  &\quad - 16 R p \pi \left( {\varpi^{(1)}_{1}}\right) ^{2} \left( 2 M^{2} + 16 M R^{3} p \pi - 2 M R + 32 R^{6} p^{2} \pi^{2} + R^{2}\right) \biggr]  \\[6pt]
b_{h^{(2)}_{2}} &= -\frac{4 \pi}{3 \left( M + 4 R^{3} p \pi\right)  \left( 2 M - R\right) }\,R^{3} \left( {\varpi^{(1)}_{1}}\right) ^{2} e^{- \nu}\,\left( 2 M^{2} + 16 M R^{3} p \pi - 2 M R + 32 R^{6} p^{2} \pi^{2} + R^{2}\right)  \\[6pt]
a_{\alpha^{(3)}_{1}} &= \frac{1}{15 R \left( M + 4 R^{3} p \pi\right)  \left( 2 M - R\right) ^{2}}\,\biggl\{ 60 R p \pi \left( M + 4 R^{3} p \pi\right)  \left( - R {\bar{\alpha}^{(1)}_{1}} - 4 {\varpi^{(1)}_{1}}\right) \,\biggl[ 2 R {m^{(2)}_{0}} + {\xi^{(2)}_{0}} \left( M + 4 R^{3} p \pi\right) \biggr]
  \nonumber \\*
  &\quad + 24 {\bar{\alpha}^{(1)}_{1}} \left( 2 M - R\right) ^{2} \left\{ R \left[ 2 R^{4} {\bar{\alpha}^{(1)}_{1}} p \pi {\varpi^{(1)}_{1}} \left( M + 4 R^{3} p \pi\right)  e^{- \nu} - {k^{(2)}_{2}}\right]  + {h^{(2)}_{2}} \left( - M + 2 R^{3} p \pi - R\right) \right\}
  \nonumber \\*
  &\quad + R^{3} \left( {\bar{\alpha}^{(1)}_{1}}\right) ^{3} \left( 2 M - R\right) ^{3} \left( 4 M + 16 R^{3} p \pi + R\right)  e^{- \nu}
  \nonumber \\*
  &\quad + 16 R^{2} p \pi {\varpi^{(1)}_{1}} \left( M + 4 R^{3} p \pi\right)  \left( 2 M - R\right)  \left[ - R^{2} \left( {\varpi^{(1)}_{1}}\right) ^{2} \left( 16 R^{2} p \pi - 13\right)  e^{- \nu} + 9 {h^{(2)}_{2}}\right]
  \nonumber \\*
  &\quad + 8 R^{4} {\bar{\alpha}^{(1)}_{1}} p \pi \left( {\varpi^{(1)}_{1}}\right) ^{2} \left( 2 M - R\right)  \left( 3 M^{2} + 24 M R^{3} p \pi + 9 M R + 48 R^{6} p^{2} \pi^{2} + 20 R^{4} p \pi - 2 R^{2}\right)  e^{- \nu}\biggr\}  \\[6pt]
b_{\alpha^{(3)}_{1}} &= \frac{4 \pi}{15 \left( M + 4 R^{3} p \pi\right)  \left( 2 M - R\right) ^{2}}\,\biggl\{ 15 \left( M + 4 R^{3} p \pi\right)  \left( - R {\bar{\alpha}^{(1)}_{1}} - 4 {\varpi^{(1)}_{1}}\right)  \left[ 2 R {m^{(2)}_{0}} + {\xi^{(2)}_{0}} \left( M + 4 R^{3} p \pi\right) \right]
  \nonumber \\*
  &\quad + 12 R^{2} {\bar{\alpha}^{(1)}_{1}} {h^{(2)}_{2}} \left( 2 M - R\right) ^{2}
  \nonumber \\*
  &\quad + 4 R {\varpi^{(1)}_{1}} \left( M + 4 R^{3} p \pi\right)  \left( 2 M - R\right)  \left[ 3 R^{3} \left( {\bar{\alpha}^{(1)}_{1}}\right) ^{2} \left( 2 M - R\right)  e^{- \nu} - R^{2} \left( {\varpi^{(1)}_{1}}\right) ^{2} \left( 32 R^{2} p \pi - 13\right)  e^{- \nu} + 9 {h^{(2)}_{2}}\right]
  \nonumber \\*
  &\quad + 2 R^{3} {\bar{\alpha}^{(1)}_{1}} \left( {\varpi^{(1)}_{1}}\right) ^{2} \left( 2 M - R\right)  \left( 3 M^{2} + 24 M R^{3} p \pi + 9 M R + 48 R^{6} p^{2} \pi^{2} + 20 R^{4} p \pi - 2 R^{2}\right)  e^{- \nu}\biggr\}  \\[6pt]
c_{\alpha^{(3)}_{1}} &= -\frac{256 \pi^{2}}{15 \left( 2 M - R\right) }\,R^{5} \left( {\varpi^{(1)}_{1}}\right) ^{3} e^{- \nu} \\[6pt]
A_{\alpha^{(3)}_{1}} &= \frac{4 \pi}{15 \left( M + 4 R^{3} p \pi\right)  \left( 2 M - R\right) }\,R\,\biggl\{ 15 {\xi^{(2)}_{0}} \left( M + 4 R^{3} p \pi\right)  \left( - R {\bar{\alpha}^{(1)}_{1}} - 4 {\varpi^{(1)}_{1}}\right)
  \nonumber \\*
  &\quad + 2 R {\varpi^{(1)}_{1}} \left( 2 M - R\right)  \left\{ R^{2} {\varpi^{(1)}_{1}} \left[ {\bar{\alpha}^{(1)}_{1}} \left( M + 4 R^{3} p \pi\right)  + 2 {\varpi^{(1)}_{1}}\right]  e^{- \nu} + 6 {h^{(2)}_{2}}\right\} \biggr\}  \\[6pt]
a_{\alpha^{(3)}_{3}} &= \frac{1}{15 R \left( M + 4 R^{3} p \pi\right)  \left( 2 M - R\right) }\,\biggl\{ 24 {\bar{\alpha}^{(1)}_{1}} \left( 2 M - R\right)  \left[ R {k^{(2)}_{2}} - {h^{(2)}_{2}} \left( - M + 2 R^{3} p \pi - R\right) \right]
  \nonumber \\*
  &\quad - R^{3} \left( {\bar{\alpha}^{(1)}_{1}}\right) ^{3} \left( 2 M - R\right) ^{2} \left( 4 M + 16 R^{3} p \pi + R\right)  e^{- \nu}
  \nonumber \\*
  &\quad + 16 R^{2} p \pi {\varpi^{(1)}_{1}} \left( M + 4 R^{3} p \pi\right)  \left[ - 3 R^{3} \left( {\bar{\alpha}^{(1)}_{1}}\right) ^{2} \left( 2 M - R\right)  e^{- \nu} + R^{2} \left( {\varpi^{(1)}_{1}}\right) ^{2} \left( 16 R^{2} p \pi - 3\right)  e^{- \nu} - 9 {h^{(2)}_{2}}\right]
  \nonumber \\*
  &\quad - 8 R^{4} {\bar{\alpha}^{(1)}_{1}} p \pi \left( {\varpi^{(1)}_{1}}\right) ^{2} \left( 3 M^{2} + 24 M R^{3} p \pi + 4 M R + 48 R^{6} p^{2} \pi^{2} - 2 R^{2}\right)  e^{- \nu}\biggr\}  \\[6pt]
b_{\alpha^{(3)}_{3}} &= \frac{8 \pi}{15 \left( M + 4 R^{3} p \pi\right)  \left( 2 M - R\right) }\,R\,\biggl\{ - R {\bar{\alpha}^{(1)}_{1}}\,\biggl[ R \left( {\varpi^{(1)}_{1}}\right) ^{2} e^{- \nu}\,\biggl( 3 M^{2} + 24 M R^{3} p \pi + 4 M R + 48 R^{6} p^{2} \pi^{2} - 2 R^{2}\biggr)
  \nonumber \\*
  &\quad + 6 {h^{(2)}_{2}} \left( 2 M - R\right) \biggr]
  \nonumber \\*
  &\quad + 2 {\varpi^{(1)}_{1}} \left( M + 4 R^{3} p \pi\right)  \left[ - 3 R^{3} \left( {\bar{\alpha}^{(1)}_{1}}\right) ^{2} \left( 2 M - R\right)  e^{- \nu} + R^{2} \left( {\varpi^{(1)}_{1}}\right) ^{2} \left( 32 R^{2} p \pi - 3\right)  e^{- \nu} - 9 {h^{(2)}_{2}}\right] \biggr\}  \\[6pt]
c_{\alpha^{(3)}_{3}} &= \frac{256 \pi^{2}}{15 \left( 2 M - R\right) }\,R^{5} \left( {\varpi^{(1)}_{1}}\right) ^{3} e^{- \nu} \\[6pt]
A_{\alpha^{(3)}_{3}} &= -\frac{8 \pi}{15 \left( M + 4 R^{3} p \pi\right) }\,R^{2} {\varpi^{(1)}_{1}}\,\left\{ 6 {h^{(2)}_{2}} + R^{2} {\varpi^{(1)}_{1}} \left[ {\bar{\alpha}^{(1)}_{1}} \left( M + 4 R^{3} p \pi\right)  + 2 {\varpi^{(1)}_{1}}\right]  e^{- \nu}\right\}
\end{align}


\section{Multipole Moments}
\label{app:multipoles}

This appendix collects the analytic expressions for the multipole moments in
terms of the rescaled exterior integration constants of Sec.~\ref{sec:exterior}.
Each moment is given order-by-order in the spin. The expression labeled
$P^{(n)}_{\ell}$ is the contribution of order $n$ to the moment $P_\ell$, where
$P$ stands for $M$ or $S$. It depends on the constants
$\tilde{C}^{(n')}_{\ell',\mathsf{ext}}$ of all orders $n' \le n$ and modes
$\ell' \le n'$. The full moment is the sum of its contributions from the order
at which it first appears through seventh order, so that $M_0$ collects four
terms, $M_2$ three, $M_4$ two and $M_6$ one, and the current moments follow the
same pattern from $S_1$ to $S_7$. The expressions are written in terms of the
unhatted, physical constants. Evaluating them with the normalized constants of
Sec.~\ref{sec:numerical} returns the normalized moment, and the physical moment
at a given spin frequency then follows from Eq.~\eqref{eq:cumulative}, which
restores the appropriate power of $\Omega_*$ to each term.
\begin{align}
\label{eq:M00}
M^{(0)}_{0} &= M_* \\
\label{eq:M20}
M^{(2)}_{0} &= M_*\, \tilde{C}^{(2)}_{0,\textsf{ext}} \\
\label{eq:M40}
M^{(4)}_{0} &= M_*\left(12\left(\tilde{C}^{(2)}_{2,\textsf{ext}}\right)^{2} + \tilde{C}^{(4)}_{0,\textsf{ext}}\right) \\
\label{eq:M60}
M^{(6)}_{0} &= -\frac{1}{8960}\Bigg\{ M_*\Bigg[ -256\left(52799\, j^2 - 2436\, \tilde{C}^{(2)}_{0,\textsf{ext}}\right)\left(\tilde{C}^{(2)}_{2,\textsf{ext}}\right)^{2} + 866304\left(\tilde{C}^{(2)}_{2,\textsf{ext}}\right)^{3} \notag\\
&\quad - 2240\, \tilde{C}^{(2)}_{2,\textsf{ext}}\left(2038\, j^4 + 96\, j^2 \tilde{C}^{(2)}_{0,\textsf{ext}} - 48\, j\, \tilde{C}^{(3)}_{1,\textsf{ext}} + 3057\, j\, \tilde{C}^{(3)}_{3,\textsf{ext}} + 96\, \tilde{C}^{(4)}_{2,\textsf{ext}}\right) \notag\\
&\quad + 35\left(140\, j^6 + 420\, j^3 \tilde{C}^{(3)}_{3,\textsf{ext}} + 315\left(\tilde{C}^{(3)}_{3,\textsf{ext}}\right)^{2} - 256\, \tilde{C}^{(6)}_{0,\textsf{ext}}\right)\Bigg]\Bigg\} \\
\label{eq:S11}
S^{(1)}_{1} &= j\, M_*^2 \\
\label{eq:S31}
S^{(3)}_{1} &= M_*^2\, \tilde{C}^{(3)}_{1,\textsf{ext}} \\
\label{eq:S51}
S^{(5)}_{1} &= \frac{1}{350}\, M_*^2 \Bigg[ 5896\, j\left(\tilde{C}^{(2)}_{2,\textsf{ext}}\right)^{2} - 840\, \tilde{C}^{(2)}_{0,\textsf{ext}}\left(j^3 - 4j\, \tilde{C}^{(2)}_{2,\textsf{ext}}\right) + \tilde{C}^{(2)}_{2,\textsf{ext}}\left(41942\, j^3 - 840\, \tilde{C}^{(3)}_{1,\textsf{ext}} - 105\, \tilde{C}^{(3)}_{3,\textsf{ext}}\right) \notag\\
&\quad + 70\left(168\, j^5 + 6j^2 \tilde{C}^{(3)}_{1,\textsf{ext}} + 252\, j^2 \tilde{C}^{(3)}_{3,\textsf{ext}} - 12j\, \tilde{C}^{(4)}_{2,\textsf{ext}} + 5\, \tilde{C}^{(5)}_{1,\textsf{ext}}\right)\Bigg] \\
\label{eq:S71}
S^{(7)}_{1} &= -\frac{1}{2016000}\, M_*^2 \Bigg\{ -1403566080\, j\left(\tilde{C}^{(2)}_{2,\textsf{ext}}\right)^{3} - 24192000\left(\tilde{C}^{(2)}_{0,\textsf{ext}}\right)^{2}\left(j^3 - 2j\, \tilde{C}^{(2)}_{2,\textsf{ext}}\right) \notag\\
&\quad + 64\left(\tilde{C}^{(2)}_{2,\textsf{ext}}\right)^{2}\left(709072558\, j^3 - 698400\, \tilde{C}^{(3)}_{1,\textsf{ext}} - 5287275\, \tilde{C}^{(3)}_{3,\textsf{ext}}\right) \notag\\
&\quad + 23040\, \tilde{C}^{(2)}_{0,\textsf{ext}}\Bigg[-45930\, j\left(\tilde{C}^{(2)}_{2,\textsf{ext}}\right)^{2} + \tilde{C}^{(2)}_{2,\textsf{ext}}\left(31823\, j^3 - 840\, \tilde{C}^{(3)}_{1,\textsf{ext}} - 27930\, \tilde{C}^{(3)}_{3,\textsf{ext}}\right) \notag\\
&\qquad + 5j\left(89\, j^4 + 210\, j\, \tilde{C}^{(3)}_{1,\textsf{ext}} + 3360\, j\, \tilde{C}^{(3)}_{3,\textsf{ext}} - 168\, \tilde{C}^{(4)}_{2,\textsf{ext}}\right)\Bigg] \notag\\
&\quad + 240\, \tilde{C}^{(2)}_{2,\textsf{ext}}\left(407750266\, j^5 - 3444624\, j^2 \tilde{C}^{(3)}_{1,\textsf{ext}} - 9042313\, j^2 \tilde{C}^{(3)}_{3,\textsf{ext}} - 80640\, j\, \tilde{C}^{(4)}_{0,\textsf{ext}} \right. \notag\\
&\qquad \left. - 327744\, j\, \tilde{C}^{(4)}_{2,\textsf{ext}} + 370240\, j\, \tilde{C}^{(4)}_{4,\textsf{ext}} + 20160\, \tilde{C}^{(5)}_{1,\textsf{ext}} + 2520\, \tilde{C}^{(5)}_{3,\textsf{ext}}\right) \notag\\
&\quad + 5\Bigg[-725760\, j\left(\tilde{C}^{(3)}_{1,\textsf{ext}}\right)^{2} + 27203085\, j\left(\tilde{C}^{(3)}_{3,\textsf{ext}}\right)^{2} - 2880\, \tilde{C}^{(3)}_{1,\textsf{ext}}\left(10226\, j^4 + 14133\, j\, \tilde{C}^{(3)}_{3,\textsf{ext}} - 336\, \tilde{C}^{(4)}_{2,\textsf{ext}}\right) \notag\\
&\qquad + 420\, \tilde{C}^{(3)}_{3,\textsf{ext}}\left(9267195\, j^4 + 288\, \tilde{C}^{(4)}_{2,\textsf{ext}} + 8800\, \tilde{C}^{(4)}_{4,\textsf{ext}}\right) \notag\\
&\quad + 4\left(818415069\, j^7 + 241920\, j^3 \tilde{C}^{(4)}_{0,\textsf{ext}} - 19143360\, j^3 \tilde{C}^{(4)}_{2,\textsf{ext}} - 222266560\, j^3 \tilde{C}^{(4)}_{4,\textsf{ext}} \right. \notag\\
&\qquad \left. - 120960\, j^2 \tilde{C}^{(5)}_{1,\textsf{ext}} - 5080320\, j^2 \tilde{C}^{(5)}_{3,\textsf{ext}} + 241920\, j\, \tilde{C}^{(6)}_{2,\textsf{ext}} - 100800\, \tilde{C}^{(7)}_{1,\textsf{ext}}\right)\Bigg]\Bigg\} \\
\label{eq:M22}
M^{(2)}_{2} &= -\frac{1}{5}\, M_*^3 \left(5 j^2 + 8\, \tilde{C}^{(2)}_{2,\textsf{ext}}\right) \\
\label{eq:M42}
M^{(4)}_{2} &= -\frac{1}{175}\, M_*^3 \Bigg[ 6576\, j^2 \tilde{C}^{(2)}_{2,\textsf{ext}} - 2340\left(\tilde{C}^{(2)}_{2,\textsf{ext}}\right)^{2} + 35\, \tilde{C}^{(2)}_{0,\textsf{ext}}\left(3j^2 + 8\, \tilde{C}^{(2)}_{2,\textsf{ext}}\right) \notag\\
&\quad + 35\left(78j^4 + 6j\, \tilde{C}^{(3)}_{1,\textsf{ext}} + 117j\, \tilde{C}^{(3)}_{3,\textsf{ext}} + 8\, \tilde{C}^{(4)}_{2,\textsf{ext}}\right)\Bigg] \\
\label{eq:M62}
M^{(6)}_{2} &= -\frac{1}{504000}\, M_*^3 \Bigg\{ 504000\, j^2 \left(\tilde{C}^{(2)}_{0,\textsf{ext}}\right)^{2} + 14530022144\, j^2 \left(\tilde{C}^{(2)}_{2,\textsf{ext}}\right)^{2} - 73144320\left(\tilde{C}^{(2)}_{2,\textsf{ext}}\right)^{3} \notag\\
&\quad + 1440\, \tilde{C}^{(2)}_{0,\textsf{ext}}\Bigg[7550\, j^4 - 29568\, j^2 \tilde{C}^{(2)}_{2,\textsf{ext}} + 6120\left(\tilde{C}^{(2)}_{2,\textsf{ext}}\right)^{2} + 140\, j\, \tilde{C}^{(3)}_{1,\textsf{ext}} - 8295\, j\, \tilde{C}^{(3)}_{3,\textsf{ext}} + 560\, \tilde{C}^{(4)}_{2,\textsf{ext}}\Bigg] \notag\\
&\quad + 1280\, \tilde{C}^{(2)}_{2,\textsf{ext}}\Bigg[47061\, j\, \tilde{C}^{(3)}_{1,\textsf{ext}} - 274323\, j\, \tilde{C}^{(3)}_{3,\textsf{ext}} + 2\left(8555513\, j^4 + 315\, \tilde{C}^{(4)}_{0,\textsf{ext}} - 5265\, \tilde{C}^{(4)}_{2,\textsf{ext}} + 5025\, \tilde{C}^{(4)}_{4,\textsf{ext}}\right)\Bigg] \notag\\
&\quad + 5\Bigg[60480\left(\tilde{C}^{(3)}_{1,\textsf{ext}}\right)^{2} + 524237700\, j^3 \tilde{C}^{(3)}_{3,\textsf{ext}} - 4037355\left(\tilde{C}^{(3)}_{3,\textsf{ext}}\right)^{2} + 8640\, \tilde{C}^{(3)}_{1,\textsf{ext}}\left(328j^3 + 273\, \tilde{C}^{(3)}_{3,\textsf{ext}}\right) \notag\\
&\qquad + 4\left(138942529\, j^6 + 15120\, j^2 \tilde{C}^{(4)}_{0,\textsf{ext}} + 1728000\, j^2 \tilde{C}^{(4)}_{2,\textsf{ext}} - 65962560\, j^2 \tilde{C}^{(4)}_{4,\textsf{ext}} \right. \notag\\
&\qquad\quad \left. + 30240\, j\, \tilde{C}^{(5)}_{1,\textsf{ext}} + 589680\, j\, \tilde{C}^{(5)}_{3,\textsf{ext}} + 40320\, \tilde{C}^{(6)}_{2,\textsf{ext}}\right)\Bigg]\Bigg\} \\
\label{eq:S33}
S^{(3)}_{3} &= \frac{3}{2}\, M_*^4\, \tilde{C}^{(3)}_{3,\textsf{ext}} \\
\label{eq:S53}
S^{(5)}_{3} &= -\frac{1}{8400}\, M_*^4 \Bigg[ -2447424\, j\left(\tilde{C}^{(2)}_{2,\textsf{ext}}\right)^{2} - 4\, \tilde{C}^{(2)}_{2,\textsf{ext}}\left(154682\, j^3 + 4560\, \tilde{C}^{(3)}_{1,\textsf{ext}} - 11655\, \tilde{C}^{(3)}_{3,\textsf{ext}}\right) \notag\\
&\quad - 240\, \tilde{C}^{(2)}_{0,\textsf{ext}}\left(146\, j^3 + 100\, j\, \tilde{C}^{(2)}_{2,\textsf{ext}} + 105\, \tilde{C}^{(3)}_{3,\textsf{ext}}\right) \notag\\
&\quad + 5\left(83218\, j^5 + 1824\, j^2 \tilde{C}^{(3)}_{1,\textsf{ext}} + 170583\, j^2 \tilde{C}^{(3)}_{3,\textsf{ext}} - 3648\, j\, \tilde{C}^{(4)}_{2,\textsf{ext}} + 39360\, j\, \tilde{C}^{(4)}_{4,\textsf{ext}} - 2520\, \tilde{C}^{(5)}_{3,\textsf{ext}}\right)\Bigg] \\
\label{eq:S73}
S^{(7)}_{3} &= \frac{1}{25872000}\, M_*^4 \Bigg\{ 6918179840\, j\left(\tilde{C}^{(2)}_{2,\textsf{ext}}\right)^{3} - 3326400\left(\tilde{C}^{(2)}_{0,\textsf{ext}}\right)^{2}\left(62\, j^3 + 248\, j\, \tilde{C}^{(2)}_{2,\textsf{ext}} + 105\, \tilde{C}^{(3)}_{3,\textsf{ext}}\right) \notag\\
&\quad + \left(\tilde{C}^{(2)}_{2,\textsf{ext}}\right)^{2}\left(-22815573755104\, j^3 + 7195372800\, \tilde{C}^{(3)}_{1,\textsf{ext}} + 6371038800\, \tilde{C}^{(3)}_{3,\textsf{ext}}\right) \notag\\
&\quad + 6160\, \tilde{C}^{(2)}_{0,\textsf{ext}}\Bigg[-3057168\, j\left(\tilde{C}^{(2)}_{2,\textsf{ext}}\right)^{2} - 90\, \tilde{C}^{(2)}_{2,\textsf{ext}}\left(157844\, j^3 - 336\, \tilde{C}^{(3)}_{1,\textsf{ext}} - 147\, \tilde{C}^{(3)}_{3,\textsf{ext}}\right) \notag\\
&\qquad + 5\left(274172\, j^5 + 7488\, j^2 \tilde{C}^{(3)}_{1,\textsf{ext}} + 274071\, j^2 \tilde{C}^{(3)}_{3,\textsf{ext}} + 6048\, j\, \tilde{C}^{(4)}_{2,\textsf{ext}} + 5600\, j\, \tilde{C}^{(4)}_{4,\textsf{ext}} + 2520\, \tilde{C}^{(5)}_{3,\textsf{ext}}\right)\Bigg] \notag\\
&\quad - 40\, \tilde{C}^{(2)}_{2,\textsf{ext}}\Bigg[134628032\, j^2 \tilde{C}^{(3)}_{1,\textsf{ext}} - 26400624131\, j^2 \tilde{C}^{(3)}_{3,\textsf{ext}} \notag\\
&\qquad + 12\left(48557113187\, j^5 - 154000\, j\, \tilde{C}^{(4)}_{0,\textsf{ext}} - 30694664\, j\, \tilde{C}^{(4)}_{2,\textsf{ext}} - 2422840\, j\, \tilde{C}^{(4)}_{4,\textsf{ext}} \right. \notag\\
&\qquad\quad \left. - 117040\, \tilde{C}^{(5)}_{1,\textsf{ext}} + 299145\, \tilde{C}^{(5)}_{3,\textsf{ext}} - 46200\, \tilde{C}^{(5)}_{5,\textsf{ext}}\right)\Bigg] \notag\\
&\quad + 35\Bigg[-1203840\, j\left(\tilde{C}^{(3)}_{1,\textsf{ext}}\right)^{2} + 33952590\, j\left(\tilde{C}^{(3)}_{3,\textsf{ext}}\right)^{2} \notag\\
&\qquad + 5\, \tilde{C}^{(3)}_{3,\textsf{ext}}\left(17292547049\, j^4 + 443520\, \tilde{C}^{(4)}_{0,\textsf{ext}} - 820512\, \tilde{C}^{(4)}_{2,\textsf{ext}} - 1410080\, \tilde{C}^{(4)}_{4,\textsf{ext}}\right) \notag\\
&\qquad - 2640\, \tilde{C}^{(3)}_{1,\textsf{ext}}\left(59447\, j^4 + 56084\, j\, \tilde{C}^{(3)}_{3,\textsf{ext}} - 608\, \tilde{C}^{(4)}_{2,\textsf{ext}} + 6560\, \tilde{C}^{(4)}_{4,\textsf{ext}}\right) \notag\\
&\qquad + 2\left(401893607\, j^7 + 1541760\, j^3 \tilde{C}^{(4)}_{0,\textsf{ext}} - 26139520\, j^3 \tilde{C}^{(4)}_{2,\textsf{ext}} + 34465853120\, j^3 \tilde{C}^{(4)}_{4,\textsf{ext}} - 401280\, j^2 \tilde{C}^{(5)}_{1,\textsf{ext}} \right. \notag\\
&\qquad\quad \left. - 37528260\, j^2 \tilde{C}^{(5)}_{3,\textsf{ext}} - 4246539000\, j^2 \tilde{C}^{(5)}_{5,\textsf{ext}} + 802560\, j\, \tilde{C}^{(6)}_{2,\textsf{ext}} - 8659200\, j\, \tilde{C}^{(6)}_{4,\textsf{ext}} + 554400\, \tilde{C}^{(7)}_{3,\textsf{ext}}\right)\Bigg]\Bigg\} \\
\label{eq:M44}
M^{(4)}_{4} &= \frac{1}{735}\, M_*^5 \left(749\, j^4 + 9428\, j^2 \tilde{C}^{(2)}_{2,\textsf{ext}} + 9432\left(\tilde{C}^{(2)}_{2,\textsf{ext}}\right)^{2} - 630\, j\, \tilde{C}^{(3)}_{3,\textsf{ext}} - 560\, \tilde{C}^{(4)}_{4,\textsf{ext}}\right) \\
\label{eq:M64}
M^{(6)}_{4} &= -\frac{1}{646800}\, M_*^5 \Bigg\{ 3221666688\, j^2 \left(\tilde{C}^{(2)}_{2,\textsf{ext}}\right)^{2} + 172983808\left(\tilde{C}^{(2)}_{2,\textsf{ext}}\right)^{3} + 653373420\, j^3 \tilde{C}^{(3)}_{3,\textsf{ext}} - 959175\left(\tilde{C}^{(3)}_{3,\textsf{ext}}\right)^{2} \notag\\
&\quad + 1760\, \tilde{C}^{(3)}_{1,\textsf{ext}}\left(631\, j^3 + 315\, \tilde{C}^{(3)}_{3,\textsf{ext}}\right) - 7494080\, j^2 \tilde{C}^{(4)}_{2,\textsf{ext}} \notag\\
&\quad + 16\, \tilde{C}^{(2)}_{2,\textsf{ext}}\left(265631383\, j^4 - 468160\, j\, \tilde{C}^{(3)}_{1,\textsf{ext}} + 1940085\, j\, \tilde{C}^{(3)}_{3,\textsf{ext}} - 1037520\, \tilde{C}^{(4)}_{2,\textsf{ext}} - 794480\, \tilde{C}^{(4)}_{4,\textsf{ext}}\right) \notag\\
&\quad - 1320\, \tilde{C}^{(2)}_{0,\textsf{ext}}\left(6200\, j^4 - 51400\, j^2 \tilde{C}^{(2)}_{2,\textsf{ext}} + 5456\left(\tilde{C}^{(2)}_{2,\textsf{ext}}\right)^{2} - 245\, j\, \tilde{C}^{(3)}_{3,\textsf{ext}} - 1120\, \tilde{C}^{(4)}_{4,\textsf{ext}}\right) \notag\\
&\quad + 100\left(5712643\, j^6 - 2174448\, j^2 \tilde{C}^{(4)}_{4,\textsf{ext}} + 5544\, j\, \tilde{C}^{(5)}_{3,\textsf{ext}} - 831600\, j\, \tilde{C}^{(5)}_{5,\textsf{ext}} + 4928\, \tilde{C}^{(6)}_{4,\textsf{ext}}\right)\Bigg\} \\
\label{eq:S55}
S^{(5)}_{5} &= \frac{1}{242550}\, M_*^6 \Bigg[ -25955952\, j\left(\tilde{C}^{(2)}_{2,\textsf{ext}}\right)^{2} - 20\, \tilde{C}^{(2)}_{2,\textsf{ext}}\left(1134733\, j^3 + 11907\, \tilde{C}^{(3)}_{3,\textsf{ext}}\right) \notag\\
&\quad + 35\left(42964\, j^5 + 159585\, j^2 \tilde{C}^{(3)}_{3,\textsf{ext}} + 81840\, j\, \tilde{C}^{(4)}_{4,\textsf{ext}} - 17325\, \tilde{C}^{(5)}_{5,\textsf{ext}}\right)\Bigg] \\
\label{eq:S75}
S^{(7)}_{5} &= -\frac{1}{189189000}\, M_*^6 \Bigg\{ 2986919018304\, j\left(\tilde{C}^{(2)}_{2,\textsf{ext}}\right)^{3} \notag\\
&\quad + 80\left(\tilde{C}^{(2)}_{2,\textsf{ext}}\right)^{2}\left(3413815477067\, j^3 + 256431708\, \tilde{C}^{(3)}_{1,\textsf{ext}} - 275401224\, \tilde{C}^{(3)}_{3,\textsf{ext}}\right) \notag\\
&\quad + 10\, \tilde{C}^{(2)}_{2,\textsf{ext}}\Bigg[2641861560\, j^2 \tilde{C}^{(3)}_{1,\textsf{ext}} - 133672119441\, j^2 \tilde{C}^{(3)}_{3,\textsf{ext}} \notag\\
&\qquad + 4\left(15604610281087\, j^5 + 1019004480\, j\, \tilde{C}^{(4)}_{2,\textsf{ext}} - 9595334640\, j\, \tilde{C}^{(4)}_{4,\textsf{ext}} + 4643730\, \tilde{C}^{(5)}_{3,\textsf{ext}} - 64500975\, \tilde{C}^{(5)}_{5,\textsf{ext}}\right)\Bigg] \notag\\
&\quad + 1950\, \tilde{C}^{(2)}_{0,\textsf{ext}}\Bigg[14746978\, j^5 + 6392256\, j\left(\tilde{C}^{(2)}_{2,\textsf{ext}}\right)^{2} + 1336461\, j^2 \tilde{C}^{(3)}_{3,\textsf{ext}} \notag\\
&\qquad - 4\, \tilde{C}^{(2)}_{2,\textsf{ext}}\left(39356216\, j^3 + 483777\, \tilde{C}^{(3)}_{3,\textsf{ext}}\right) - 3193120\, j\, \tilde{C}^{(4)}_{4,\textsf{ext}} + 970200\, \tilde{C}^{(5)}_{5,\textsf{ext}}\Bigg] \notag\\
&\quad + 25\Bigg[4990587952264\, j^7 - 1035039348\, j\left(\tilde{C}^{(3)}_{3,\textsf{ext}}\right)^{2} + 455801424\, j^3 \tilde{C}^{(4)}_{2,\textsf{ext}} - 861541152240\, j^3 \tilde{C}^{(4)}_{4,\textsf{ext}} \notag\\
&\qquad - 936\, \tilde{C}^{(3)}_{1,\textsf{ext}}\left(494107\, j^4 + 376334\, j\, \tilde{C}^{(3)}_{3,\textsf{ext}} + 95480\, \tilde{C}^{(4)}_{4,\textsf{ext}}\right) \notag\\
&\qquad + 84\, \tilde{C}^{(3)}_{3,\textsf{ext}}\left(77188419345\, j^4 + 88452\, \tilde{C}^{(4)}_{2,\textsf{ext}} + 443800\, \tilde{C}^{(4)}_{4,\textsf{ext}}\right) \notag\\
&\qquad - 174266820\, j^2 \tilde{C}^{(5)}_{3,\textsf{ext}} - 6750010575\, j^2 \tilde{C}^{(5)}_{5,\textsf{ext}} - 89369280\, j\, \tilde{C}^{(6)}_{4,\textsf{ext}} + 327539520\, j\, \tilde{C}^{(6)}_{6,\textsf{ext}} + 18918900\, \tilde{C}^{(7)}_{5,\textsf{ext}}\Bigg]\Bigg\} \\
\label{eq:M66}
M^{(6)}_{6} &= -\frac{1}{52026975}\, M_*^7 \Bigg\{ 610983177032\, j^2 \left(\tilde{C}^{(2)}_{2,\textsf{ext}}\right)^{2} + 9380111808\left(\tilde{C}^{(2)}_{2,\textsf{ext}}\right)^{3} \notag\\
&\quad + 4\, \tilde{C}^{(2)}_{2,\textsf{ext}}\left(320295972428\, j^4 - 1132730025\, j\, \tilde{C}^{(3)}_{3,\textsf{ext}} - 255081120\, \tilde{C}^{(4)}_{4,\textsf{ext}}\right) \notag\\
&\quad + 35\left(6858291931\, j^6 + 8633163360\, j^3 \tilde{C}^{(3)}_{3,\textsf{ext}} + 1142190\left(\tilde{C}^{(3)}_{3,\textsf{ext}}\right)^{2} \right. \notag\\
&\qquad \left. - 1414367040\, j^2 \tilde{C}^{(4)}_{4,\textsf{ext}} + 5665275\, j\, \tilde{C}^{(5)}_{5,\textsf{ext}} + 443520\, \tilde{C}^{(6)}_{6,\textsf{ext}}\right)\Bigg\} \\
\label{eq:S77}
S^{(7)}_{7} &= \frac{1}{13527013500}\, M_*^8 \Bigg\{ 68178172927488\, j\left(\tilde{C}^{(2)}_{2,\textsf{ext}}\right)^{3} + 80\left(\tilde{C}^{(2)}_{2,\textsf{ext}}\right)^{2}\left(103297051034396\, j^3 - 846137313\, \tilde{C}^{(3)}_{3,\textsf{ext}}\right) \notag\\
&\quad + 70\, \tilde{C}^{(2)}_{2,\textsf{ext}}\left(277485112706914\, j^5 + 885505392195\, j^2 \tilde{C}^{(3)}_{3,\textsf{ext}} - 143406140160\, j\, \tilde{C}^{(4)}_{4,\textsf{ext}} + 1022630400\, \tilde{C}^{(5)}_{5,\textsf{ext}}\right) \notag\\
&\quad + 35\Bigg[-51947998200\, j\left(\tilde{C}^{(3)}_{3,\textsf{ext}}\right)^{2} + 105\, \tilde{C}^{(3)}_{3,\textsf{ext}}\left(1456654219141\, j^4 + 13340640\, \tilde{C}^{(4)}_{4,\textsf{ext}}\right) \notag\\
&\qquad + 2\left(57563546858893\, j^7 - 8756129027520\, j^3 \tilde{C}^{(4)}_{4,\textsf{ext}} - 609753704175\, j^2 \tilde{C}^{(5)}_{5,\textsf{ext}} \right. \notag\\
&\qquad\quad \left. + 3405346560\, j\, \tilde{C}^{(6)}_{6,\textsf{ext}} - 676350675\, \tilde{C}^{(7)}_{7,\textsf{ext}}\right)\Bigg]\Bigg\}
\end{align}


\section{Exterior Solutions}
\label{app:Exterior_solutions}

The homogeneous solutions are written in the radial variable $\zeta = R/M_{*} -
1$, the particular solutions in the compactness $\mathscr{C} \equiv M_{*}/R =
(1+\zeta)^{-1}$. The homogeneous solutions are listed for every multipole and so
cover all orders. The particular solutions are written out only through third
order, in full, with the coefficient functions $\mathcal{L}^{[\,\cdot\,]}$ of
Sec.~\ref{sec:exterior} already substituted. For the remaining orders the
$\mathcal{L}^{[\,\cdot\,]}$ themselves are available in the
\texttt{hartle-thorne-project} repository~\cite{conde2025hartlethorne}.

\medskip \noindent \textit{Odd-parity homogeneous solutions.}
\begin{align}
\omega^{(k),\mathsf{H}}_{1,\mathsf{ext}}(\zeta) &= \frac{1}{M_{*}}\, \frac{2}{(1+\zeta)^3} \\
\omega^{(k),\mathsf{H}}_{3,\mathsf{ext}}(\zeta) &= \frac{1}{M_{*}}\left[\frac{7}{32(1+\zeta)^3}\left(-16 - 65\zeta - 15\zeta^2 + 75\zeta^3 + 45\zeta^4\right) - \frac{105}{64}\left(1 - 4\zeta + 3\zeta^2\right)\ln\left(\frac{\zeta + 1}{\zeta - 1}\right)\right] \\
\omega^{(k),\mathsf{H}}_{5,\mathsf{ext}}(\zeta) &= \frac{1}{M_{*}}\left[-\frac{11}{64(1+\zeta)^3}\left(64 + 791\zeta + 105\zeta^2 - 2940\zeta^3 - 1680\zeta^4 + 2205\zeta^5 + 1575\zeta^6\right) \right. \notag\\
&\quad \left. + \frac{1155}{128}\left(-1 + 4\zeta + 6\zeta^2 - 24\zeta^3 + 15\zeta^4\right)\ln\left(\frac{\zeta + 1}{\zeta - 1}\right)\right] \\
\omega^{(k),\mathsf{H}}_{7,\mathsf{ext}}(\zeta) &= \frac{1}{M_{*}}\Bigg[\frac{143}{8192(1+\zeta)^3}\left(2048 + 50571\zeta + 4725\zeta^2 - 385371\zeta^3 - 212373\zeta^4 \right. \notag\\
&\quad \left. + 738045\zeta^5 + 518595\zeta^6 - 405405\zeta^7 - 315315\zeta^8\right) \notag\\
&\quad + \frac{45045}{16384}\left(15 - 60\zeta - 285\zeta^2 + 880\zeta^3 + 165\zeta^4 - 1716\zeta^5 + 1001\zeta^6\right)\ln\left(\frac{\zeta + 1}{\zeta - 1}\right)\Bigg] \\
\alpha^{(k),\mathsf{H}}_{1,\mathsf{ext}}(\zeta) &= -\frac{1}{M_{*}^2}\, \frac{6}{(1+\zeta)^4} \\
\alpha^{(k),\mathsf{H}}_{3,\mathsf{ext}}(\zeta) &= \frac{1}{M_{*}^2}\left[\frac{7}{16(1+\zeta)^4}\left(-16 + 50\zeta + 165\zeta^2 + 150\zeta^3 + 45\zeta^4\right) - \frac{105}{32}(-2 + 3\zeta)\ln\left(\frac{\zeta + 1}{\zeta - 1}\right)\right] \\
\alpha^{(k),\mathsf{H}}_{5,\mathsf{ext}}(\zeta) &= \frac{1}{M_{*}^2}\left[-\frac{11}{16(1+\zeta)^4}\left(176 - 343\zeta - 2625\zeta^2 - 2205\zeta^3 + 2730\zeta^4 + 4410\zeta^5 + 1575\zeta^6\right) \right. \notag\\
&\quad \left. + \frac{1155}{32}\left(1 + 3\zeta - 18\zeta^2 + 15\zeta^3\right)\ln\left(\frac{\zeta + 1}{\zeta - 1}\right)\right] \\
\alpha^{(k),\mathsf{H}}_{7,\mathsf{ext}}(\zeta) &= \frac{1}{M_{*}^2}\Bigg[-\frac{429}{4096(1+\zeta)^4}\left(-8192 + 15282\zeta + 215523\zeta^2 + 170982\zeta^3 - 649467\zeta^4 \right. \notag\\
&\quad \left. - 893970\zeta^5 + 229845\zeta^6 + 810810\zeta^7 + 315315\zeta^8\right) \notag\\
&\quad + \frac{135135}{8192}\left(-10 - 95\zeta + 440\zeta^2 + 110\zeta^3 - 1430\zeta^4 + 1001\zeta^5\right)\ln\left(\frac{\zeta + 1}{\zeta - 1}\right)\Bigg]\end{align}

\noindent \textit{Even-parity homogeneous solutions.}
\begin{align}
h^{(s),\mathsf{H}}_{0,\mathsf{ext}}(\zeta) &= -\frac{1}{\zeta - 1} \\[6pt]
m^{(s),\mathsf{H}}_{0,\mathsf{ext}}(\zeta) &= M_{*} \\[6pt]
h^{(s),\mathsf{H}}_{2,\mathsf{ext}}(\zeta) &= \biggl[-\frac{1}{\left(\zeta - 1\right) \left(\zeta + 1\right)}\,\zeta \left(3 \zeta^{2} - 5\right) + \frac{3}{2}\,\left(\zeta - 1\right) \left(\zeta + 1\right)\,\ln\!\left(\frac{\zeta+1}{\zeta-1}\right)\biggr] \\[6pt]
k^{(s),\mathsf{H}}_{2,\mathsf{ext}}(\zeta) &= \biggl[\frac{1}{\zeta + 1}\,\left(3 \zeta^{2} + 9 \zeta + 4\right) - \frac{3}{2}\,\left(\zeta^{2} + 2 \zeta - 1\right)\,\ln\!\left(\frac{\zeta+1}{\zeta-1}\right)\biggr] \\[6pt]
v^{(s),\mathsf{H}}_{2,\mathsf{ext}}(\zeta) &= \biggl[\frac{2}{\left(\zeta - 1\right) \left(\zeta + 1\right)}\,\left(3 \zeta^{2} - 2\right) - 3\,\zeta\,\ln\!\left(\frac{\zeta+1}{\zeta-1}\right)\biggr] \\[6pt]
m^{(s),\mathsf{H}}_{2,\mathsf{ext}}(\zeta) &= \biggl[\frac{1}{\zeta + 1}\,M_{*} \zeta \left(3 \zeta^{2} - 5\right) - \frac{3}{2}\,M_{*} \left(\zeta - 1\right)^{2} \left(\zeta + 1\right)\,\ln\!\left(\frac{\zeta+1}{\zeta-1}\right)\biggr] \\[6pt]
h^{(s),\mathsf{H}}_{4,\mathsf{ext}}(\zeta) &= \biggl[-\frac{1}{2 \left(\zeta - 1\right) \left(\zeta + 1\right)}\,\zeta \left(105 \zeta^{4} - 190 \zeta^{2} + 81\right)
  \nonumber \\
  &\quad + \frac{15}{4}\,\left(\zeta - 1\right) \left(\zeta + 1\right) \left(7 \zeta^{2} - 1\right)\,\ln\!\left(\frac{\zeta+1}{\zeta-1}\right)\biggr] \\[6pt]
k^{(s),\mathsf{H}}_{4,\mathsf{ext}}(\zeta) &= \biggl[-\frac{1}{6 \left(\zeta + 1\right)}\,\left(- 315 \zeta^{4} - 525 \zeta^{3} + 45 \zeta^{2} + 275 \zeta + 32\right)
  \nonumber \\
  &\quad - \frac{5}{4}\,\left(21 \zeta^{4} + 14 \zeta^{3} - 24 \zeta^{2} - 6 \zeta + 3\right)\,\ln\!\left(\frac{\zeta+1}{\zeta-1}\right)\biggr] \\[6pt]
v^{(s),\mathsf{H}}_{4,\mathsf{ext}}(\zeta) &= \biggl[\frac{1}{3 \left(\zeta - 1\right) \left(\zeta + 1\right)}\,\left(105 \zeta^{4} - 115 \zeta^{2} + 16\right) - \frac{5}{2}\,\zeta \left(7 \zeta^{2} - 3\right)\,\ln\!\left(\frac{\zeta+1}{\zeta-1}\right)\biggr] \\[6pt]
m^{(s),\mathsf{H}}_{4,\mathsf{ext}}(\zeta) &= \biggl[\frac{1}{2 \left(\zeta + 1\right)}\,M_{*} \zeta \left(105 \zeta^{4} - 190 \zeta^{2} + 81\right)
  \nonumber \\
  &\quad - \frac{15}{4}\,M_{*} \left(\zeta - 1\right)^{2} \left(\zeta + 1\right) \left(7 \zeta^{2} - 1\right)\,\ln\!\left(\frac{\zeta+1}{\zeta-1}\right)\biggr] \\[6pt]
h^{(s),\mathsf{H}}_{6,\mathsf{ext}}(\zeta) &= \biggl[-\frac{1}{8 \left(\zeta - 1\right) \left(\zeta + 1\right)}\,\zeta \left(3465 \zeta^{6} - 7665 \zeta^{4} + 5103 \zeta^{2} - 919\right)
  \nonumber \\
  &\quad + \frac{105}{16}\,\left(\zeta - 1\right) \left(\zeta + 1\right) \left(33 \zeta^{4} - 18 \zeta^{2} + 1\right)\,\ln\!\left(\frac{\zeta+1}{\zeta-1}\right)\biggr] \\[6pt]
k^{(s),\mathsf{H}}_{6,\mathsf{ext}}(\zeta) &= \biggl[\frac{1}{40 \left(\zeta + 1\right)}\,\left(17325 \zeta^{6} + 24255 \zeta^{5} - 14070 \zeta^{4} - 24990 \zeta^{3} + 525 \zeta^{2} + 4851 \zeta + 256\right)
  \nonumber \\
  &\quad - \frac{21}{16}\,\left(165 \zeta^{6} + 66 \zeta^{5} - 255 \zeta^{4} - 60 \zeta^{3} + 95 \zeta^{2} + 10 \zeta - 5\right)\,\ln\!\left(\frac{\zeta+1}{\zeta-1}\right)\biggr] \\[6pt]
v^{(s),\mathsf{H}}_{6,\mathsf{ext}}(\zeta) &= \biggl[\frac{1}{20 \left(\zeta - 1\right) \left(\zeta + 1\right)}\,\left(3465 \zeta^{6} - 5460 \zeta^{4} + 2163 \zeta^{2} - 128\right)
  \nonumber \\
  &\quad - \frac{21}{8}\,\zeta \left(33 \zeta^{4} - 30 \zeta^{2} + 5\right)\,\ln\!\left(\frac{\zeta+1}{\zeta-1}\right)\biggr] \\[6pt]
m^{(s),\mathsf{H}}_{6,\mathsf{ext}}(\zeta) &= \biggl[\frac{1}{8 \left(\zeta + 1\right)}\,M_{*} \zeta \left(3465 \zeta^{6} - 7665 \zeta^{4} + 5103 \zeta^{2} - 919\right)
  \nonumber \\
  &\quad - \frac{105}{16}\,M_{*} \left(\zeta - 1\right)^{2} \left(\zeta + 1\right) \left(33 \zeta^{4} - 18 \zeta^{2} + 1\right)\,\ln\!\left(\frac{\zeta+1}{\zeta-1}\right)\biggr]
\end{align}
\medskip \noindent \textit{Particular solutions up to third order}
\begin{align}
h^{(2),\mathsf{P}}_{0,\mathsf{ext}} &= - \frac{\mathscr{C}^{4} j^{2}}{2 \mathscr{C} - 1} \\[6pt]
m^{(2),\mathsf{P}}_{0,\mathsf{ext}} &= - \mathscr{C}^{3} M_{*} j^{2} \\[6pt]
h^{(2),\mathsf{P}}_{2,\mathsf{ext}} &= \mathscr{C}^{3} j^{2} \left(\mathscr{C} + 1\right) \\[6pt]
k^{(2),\mathsf{P}}_{2,\mathsf{ext}} &= - \mathscr{C}^{3} j^{2} \left(2 \mathscr{C} + 1\right) \\[6pt]
v^{(2),\mathsf{P}}_{2,\mathsf{ext}} &= - \mathscr{C}^{4} j^{2} \\[6pt]
m^{(2),\mathsf{P}}_{2,\mathsf{ext}} &= - \mathscr{C}^{2} M_{*} j^{2} \left(2 \mathscr{C} - 1\right) \left(5 \mathscr{C} - 1\right) \\[6pt]
\omega^{(3),\mathsf{P}}_{1,\mathsf{ext}} &= \frac{1}{M_{*}}\,\biggl[- \frac{4 \mathscr{C}^{6} j^{3} \left(3 \mathscr{C} + 1\right)}{5} - \frac{2 \mathscr{C} {\tilde{C}^{(2)}_{2}} j \left(6 \mathscr{C}^{3} - 16 \mathscr{C}^{2} - 21 \mathscr{C} + 15\right)}{5}
  \nonumber \\
  &\quad + \Bigl(- \frac{3 {\tilde{C}^{(2)}_{2}} j \left(2 \mathscr{C} - 1\right) \left(4 \mathscr{C}^{2} + 2 \mathscr{C} - 5\right)}{5}\Bigr)\ln\!\left(1 - 2\mathscr{C}\right)\biggr] \\[6pt]
\omega^{(3),\mathsf{P}}_{3,\mathsf{ext}} &= \frac{1}{M_{*}}\,\biggl[\frac{j^{3} \left(576 \mathscr{C}^{8} + 32 \mathscr{C}^{7} - 160 \mathscr{C}^{6} + 140 \mathscr{C}^{4} + 350 \mathscr{C}^{3} + 1050 \mathscr{C}^{2} - 3675 \mathscr{C} + 1575\right)}{240 \mathscr{C}}
  \nonumber \\
  &\quad + \frac{{\tilde{C}^{(2)}_{2}} j \left(36 \mathscr{C}^{5} + 52 \mathscr{C}^{4} + 34 \mathscr{C}^{3} + 210 \mathscr{C}^{2} - 735 \mathscr{C} + 315\right)}{15 \mathscr{C}}
  \nonumber \\
  &\quad + \Bigl(\frac{35 j^{3} \left(2 \mathscr{C} - 1\right) \left(4 \mathscr{C} - 3\right)}{32 \mathscr{C}^{2}}
  \nonumber \\
  &\quad + \frac{{\tilde{C}^{(2)}_{2}} j \left(2 \mathscr{C} - 1\right) \left(4 \mathscr{C}^{4} + 12 \mathscr{C}^{3} + 140 \mathscr{C} - 105\right)}{10 \mathscr{C}^{2}}\Bigr)\ln\!\left(1 - 2\mathscr{C}\right)\biggr] \\[6pt]
\alpha^{(3),\mathsf{P}}_{1,\mathsf{ext}} &= \frac{1}{M_{*}^{2}}\,\biggl[\frac{12 \mathscr{C}^{7} j^{3} \left(7 \mathscr{C} + 2\right)}{5} + \frac{24 \mathscr{C}^{3} {\tilde{C}^{(2)}_{2}} j \left(2 \mathscr{C}^{2} - 3 \mathscr{C} - 3\right)}{5}
  \nonumber \\
  &\quad + \Bigl(\frac{36 \mathscr{C}^{2} {\tilde{C}^{(2)}_{2}} j \left(2 \mathscr{C}^{2} - 1\right)}{5}\Bigr)\ln\!\left(1 - 2\mathscr{C}\right)\biggr] \\[6pt]
\alpha^{(3),\mathsf{P}}_{3,\mathsf{ext}} &= \frac{1}{M_{*}^{2}}\,\biggl[- \frac{j^{3} \left(2016 \mathscr{C}^{8} + 96 \mathscr{C}^{7} - 400 \mathscr{C}^{6} + 210 \mathscr{C}^{4} + 350 \mathscr{C}^{3} + 525 \mathscr{C}^{2} + 1050 \mathscr{C} - 1575\right)}{120}
  \nonumber \\
  &\quad - \frac{2 {\tilde{C}^{(2)}_{2}} j \left(72 \mathscr{C}^{5} + 84 \mathscr{C}^{4} + 52 \mathscr{C}^{3} + 105 \mathscr{C}^{2} + 210 \mathscr{C} - 315\right)}{15}
  \nonumber \\
  &\quad + \Bigl(- \frac{35 j^{3} \left(5 \mathscr{C} - 3\right)}{16 \mathscr{C}}
  \nonumber \\
  &\quad - \frac{{\tilde{C}^{(2)}_{2}} j \left(12 \mathscr{C}^{5} + 20 \mathscr{C}^{4} - 6 \mathscr{C}^{3} + 175 \mathscr{C} - 105\right)}{5 \mathscr{C}}\Bigr)\ln\!\left(1 - 2\mathscr{C}\right)\biggr]
\end{align}

\end{widetext}


\section{Implementation of the Segmented Differentiation Algorithm}
\label{app:derivatives_algorithm}

This appendix supplies the mathematics behind the five steps of
Sec.~\ref{sec:derivatives}, in the same order. Throughout,
$u\equiv\ln\varepsilon$ and $w\equiv\ln p$ are the logarithmic variables of Step
1, a prime denotes a derivative with respect to the radial coordinate $R$, and
$\Gamma\equiv dw/du$ is the logarithmic slope of the EoS.

\subsection{Step 1: preparing the table}

Before any fit is attempted the table is cleaned. Rows with non-positive
$\varepsilon$ or $p$ are removed, since their logarithms are undefined. Repeated
values of $\varepsilon$, which occur when a published table prints a junction
row twice, are collapsed to a single entry, since the mapping of Step 2 requires
the abscissae to be strictly increasing. After cleaning, the table consists of
$N$ rows $(u_i, w_i)$ with $u_1 < u_2 < \cdots < u_N$, and these rows are the
only input the algorithm uses.

\subsection{Step 2: the fit inside one segment}

\vspace{0.2cm} \noindent \textit{The affine map.} A segment spans an interval
$[u_a, u_b]$ in $\ln\varepsilon$, where $u_a$ and $u_b$ are the first and last
rows of the segment. Chebyshev polynomials are defined on $[-1,1]$, so each row
is first carried onto that interval by the linear change of variable
\begin{equation}
x(u) = \frac{2u - (u_b + u_a)}{u_b - u_a},
\label{eq:affine_map}
\end{equation}
which sends $u_a$ to $x=-1$ and $u_b$ to $x=+1$. Because the map is linear,
differentiating with respect to $u$ and differentiating with respect to $x$
differ only by a constant factor, given in Eq.~\eqref{eq:jacobian} below.

\vspace{0.2cm} \noindent \textit{The expansion.} On the mapped interval the
fitting function is the truncated Chebyshev series
\begin{equation}
w_{\mathrm{fit}}(x) = \sum_{\kappa=0}^{\kappa_{\mathrm{eff}}} a_\kappa\, T_\kappa(x),
\label{eq:cheb_expansion}
\end{equation}
where $T_\kappa$ is the Chebyshev polynomial of degree $\kappa$, generated by
$T_0=1$, $T_1=x$ and $T_{\kappa+1}=2xT_\kappa - T_{\kappa-1}$, and $a_\kappa$
are the coefficients to be determined. The number of terms is set by
$\kappa_{\mathrm{eff}}$, explained below.

\vspace{0.2cm} \noindent \textit{The coefficients used in this work.} The
coefficients $a_\kappa$ obtained for the two EoSs considered here are provided
with the code~\cite{conde2025aqlimr}, as one file per EoS listing the segment
index, the order $\kappa$, the coefficient $a_\kappa$, and the density limits of
the segment. Together with Eqs.~\eqref{eq:affine_map}
and~\eqref{eq:cheb_expansion} they reproduce $p(\varepsilon)$ without rerunning
the code, with $\varepsilon$ and $p$ in $\mathrm{MeV\,fm^{-3}}$ as in
Table~\ref{table:eos_segments}.

\vspace{0.2cm} \noindent \textit{The least-squares problem.} The coefficients
are chosen to minimize the sum of squared deviations from the $m$ rows in the
segment,
\begin{equation}
\chi^2 = \sum_{i=1}^{m}\Big[w_i - \sum_{\kappa=0}^{\kappa_{\mathrm{eff}}} a_\kappa\,T_\kappa(x_i)\Big]^2,
\label{eq:chi2}
\end{equation}
where $x_i = x(u_i)$ is the mapped abscissa of row $i$. Since the unknowns
$a_\kappa$ enter linearly, this is an ordinary linear least-squares problem,
which we solve by singular-value decomposition with the routine
\texttt{gsl\_multifit\_linear} of the GNU Scientific Library~\cite{GSL}.

\vspace{0.2cm} \noindent \textit{The degree cap.} The nominal degree of the
expansion is $\kappa=10$. On a short segment, however, a degree-ten polynomial
has eleven free coefficients, and if the segment has only a dozen rows the fit
will pass almost exactly through every one of them. That is an interpolation,
not a fit, and it reproduces the row-level noise the fit is meant to smooth
away. To prevent this, the degree of the polynomial is reduced on short segments
according to
\begin{equation}
\kappa_{\mathrm{eff}} = \max\Big\{1,\ \min\Big[\kappa,\ \tfrac{1}{2}(m-2)\Big]\Big\},
\label{eq:degree_cap}
\end{equation}
so that the number of coefficients, $\kappa_{\mathrm{eff}}+1$, never exceeds
half the number of rows. The floor of unity guarantees that even the shortest
segment is represented by at least a straight line in the log-log plane, which
is a polytrope $p\propto\varepsilon^{\Gamma}$.

\vspace{0.2cm} \noindent \textit{The acceptance criteria.} A fit is accepted
only if it passes two tests. The first is the residual test,
\begin{equation}
\max_{i}\big|w_{\mathrm{fit}}(x_i) - w_i\big| \le \tau, \qquad \tau = 10^{-7},
\label{eq:residual_criterion}
\end{equation}
where the maximum runs over every row of the segment. A maximum is used rather
than an average because a single badly missed row is exactly what signals that a
kink has entered the segment, and an average would hide it. Since an error in
$\ln p$ is a relative error in $p$, Eq.~\eqref{eq:residual_criterion} says the
fit reproduces the tabulated pressure to one part in $10^{7}$ at every row.

The second test is convergence:
\begin{equation}
\max\big(|a_{\kappa-1}|,\ |a_{\kappa}|\big) \le 10\,\tau.
\label{eq:plateau_criterion}
\end{equation}
The residual test looks only at the rows, and a polynomial can pass through
every row while oscillating between them. Such oscillation is what the
highest-order coefficients carry. Because $|T_\kappa(x)|\le1$ on $[-1,1]$, the
size of $a_\kappa$ bounds the contribution of its term, so small top
coefficients mean the retained high terms are not producing a large oscillation.

These tests constrain the fit, not its derivatives, since differentiation
amplifies each coefficient by a factor that grows with $\kappa$. The derivatives
are checked instead against the analytic representations the tables were sampled
from. Across the whole density range the fitted first, second and third
derivatives reproduce the exact ones to medians of $10^{-7}$, $3\times10^{-6}$
and $5\times10^{-5}$, with departures confined to the segment boundaries and the
outermost rows. The first derivative stays within $10^{-4}$ everywhere. Nothing
ties the derivatives of adjacent segments together, so the third derivative is
most inaccurate at the junctions, by up to $43\%$ of the exact value. Outside
the outermost rows, that is fewer than one row in two hundred. What these
departures do to the high-order moments is not bounded. Two coefficients are
read rather than one because a single coefficient can be small by accident,
passing through zero as it changes sign. The test is applied only when the fit
runs at the full degree $\kappa$, since on a capped fit the top coefficients
still carry the main shape of the curve and do not measure convergence.

\vspace{0.2cm} \noindent \textit{Analytic differentiation.} The derivative of a
Chebyshev series is another Chebyshev series, of one degree lower. Given the
coefficients $a_\kappa$ of Eq.~\eqref{eq:cheb_expansion}, the coefficients
$b_\kappa$ of $dw_{\mathrm{fit}}/dx$ follow from the exact
recurrence~\cite{NumericalRecipes}
\begin{align}
b_{\kappa_{\mathrm{eff}}-1} &= 2\kappa_{\mathrm{eff}}\, a_{\kappa_{\mathrm{eff}}}, \nonumber\\
b_{\kappa_{\mathrm{eff}}-2} &= 2(\kappa_{\mathrm{eff}}-1)\, a_{\kappa_{\mathrm{eff}}-1}, \label{eq:cheb_derivative}\\
b_\kappa &= b_{\kappa+2} + 2(\kappa+1)\, a_{\kappa+1}, \qquad \kappa = \kappa_{\mathrm{eff}}-3,\ldots,0, \nonumber
\end{align}
after which $b_0$ is halved to account for the normalization of $T_0$. Applying
the recurrence to the $b_\kappa$ gives the second derivative, and once more the
third. All three coefficient sets are computed once per segment and thereafter
only evaluated. Because the map of Eq.~\eqref{eq:affine_map} is linear, each
derivative with respect to $u$ carries a constant factor,
\begin{equation}
\frac{d^n w}{du^n} = \Big(\frac{2}{u_b - u_a}\Big)^{n}\frac{d^n w}{dx^n},
\label{eq:jacobian}
\end{equation}
and we write $w_n \equiv d^n w/du^n$, so that $w_1 = \Gamma$. The three numbers
$w_1$, $w_2$, $w_3$ at each evaluation point are the entire output of the fit
that Step 5 consumes.

\subsection{Step 3: growing the segments}

\vspace{0.2cm} \noindent \textit{The growing rule.} Let $i_a$ denote the first
index row of a candidate segment and $i_b$ its last, so that the segment covers
rows $i_a, i_a+1, \ldots, i_b$ and spans the densities $u_a = u_{i_a}$ to $u_b =
u_{i_b}$ used in the map of Eq.~\eqref{eq:affine_map}. A range of rows $[i_a,
i_b]$ is said to \textit{pass} when the fit of Step 2 built on those rows
satisfies both Eqs.~\eqref{eq:residual_criterion}
and~\eqref{eq:plateau_criterion}. Starting from row $i_a$, the segment is
extended one row at a time for as long as the range keeps passing, and its last
row is the furthest one for which it does,
\begin{equation}
i_b(i_a) = \max\big\{\, i \;:\; [i_a, j]\ \text{passes for all } j \leq i \,\big\}.
\label{eq:reach}
\end{equation}
The condition applies to every range up to $i$, so growth stops at the first
failure.
A segment must contain at least
\begin{equation}
m_{\min} = 2(\kappa+1) = 22
\label{eq:mmin}
\end{equation}
rows, unless it reaches the end of the table, where a shorter final segment is
admitted. That is the smallest number of rows at which the degree cap of
Eq.~\eqref{eq:degree_cap} releases, since $\kappa_{\mathrm{eff}} =
\min\bigl(\kappa, \lfloor (m-2)/2 \rfloor\bigr)$ first reaches $\kappa$ when $m
= 2\kappa + 2$. Imposing it guarantees that every grown segment is fitted at the
full degree, so that the plateau criterion, which inspects the last two
coefficients of a degree-$\kappa$ expansion, is active on all of them. When no
range starting at $i_a$ of at least the minimum length passes, $i_b(i_a)$ is
undefined and no segment can be grown from that row.

In practice the search for $i_b$ does not test every row in turn. The range is
advanced in strides of eight, and when a stride fails the rows inside it are
tested backwards. This returns the same $i_b$ as Eq.~\eqref{eq:reach} for a
fraction of the fits, as long as a range that has failed does not pass again
further out. For the EoSs used here, the minimum length never binds, the
shortest accepted segment running to thirty rows.

\vspace{0.2cm} \noindent \textit{Where the next segment begins.} Suppose a
segment has just ended at row $i_b$. The next segment must begin somewhere at or
after $i_b$, and where exactly depends on how the table recorded the junction
between the two branches it separates. Published tables do this in one of three
ways, and the algorithm tries each in turn, taking the first that works,
\begin{equation}
i_a' = \text{first of } i_b,\ i_b+1,\ i_b+2 \text{ for which } i_b(i_a') \text{ is defined}.
\label{eq:next_start}
\end{equation}
The first case, $i_a' = i_b$, is a table that wrote a row exactly at the
junction. That row lies on both branches, since the branches meet there, and it
is fitted by both segments. The second case, $i_a' = i_b+1$, is a table with no
row at the junction. The last row on the lower branch is $i_b$ and the next row
is already on the upper branch, so the segments share nothing and the junction
falls in the gap between them. The third case, $i_a' = i_b+2$, is a table whose
row $i_b+1$ fits neither branch. Some tables, when interpolated across a kink,
produce a row whose slope is a blend of the two sides. Such a row is skipped
with a warning naming it, since it records a defect of the tabulation rather
than a feature of the matter. Nothing in Eq.~\eqref{eq:next_start} inspects the
shape of the data. Whichever start admits a passing segment is the convention
the table used, and trying three candidates is cheaper than deducing which.

\vspace{0.2cm} \noindent \textit{The loop.} The complete segmentation is then

\begin{enumerate}
\item Set $i_a$ to the first row of the table.
\item Compute $i_b(i_a)$ from Eq.~\eqref{eq:reach}.
If undefined, refuse the table.
Otherwise record the segment $[i_a, i_b]$.
\item If $i_b$ is the last row, stop.
\item Set $i_a$ to the first of $i_b$, $i_b+1$, $i_b+2$ admitting a segment, refusing the table if none does, and return to step 2.
\end{enumerate}

\vspace{0.2cm} \noindent \textit{Domains and fits.} Each segment carries two
ranges. Its \textit{fit} is the rows $[i_a, i_b]$ that passed, the only rows
entering the least-squares problem of Eq.~\eqref{eq:chi2}. Its \textit{domain}
is where the polynomial is evaluated, from $u_{i_a}$ to the first row of the
next segment, so the domains tile the table. When the next segment starts later,
the domain reaches slightly past the last fitted row, and a density in that gap
is then served by its own branch carried forward, not by the neighbor
extrapolated backward across the junction. The map of Eq.~\eqref{eq:affine_map}
therefore runs over the full domain, while $\chi^2$ runs over the fitted rows
alone.

\subsection{Step 4: classifying the boundaries}

The boundary between segments $q$ and $q+1$ is placed at the first row of
segment $q+1$, the first row known to lie on the new branch. When the segments
share a junction row this is that row. When the junction fell between rows it is
the first point demonstrably on the upper branch.

Both neighboring fits are evaluated at this point, each supplying the
logarithmic slope $\Gamma$ through Eq.~\eqref{eq:jacobian} with $n=1$. Since the
boundary is the right end of the lower segment's domain and the left end of the
upper's, each polynomial is read at an endpoint of its own domain and neither is
extrapolated. The rule stated in the main text is then
\begin{equation}
|\Delta\Gamma| \equiv |\Gamma_{\mathrm{upper}} - \Gamma_{\mathrm{lower}}| \ge 10^{-3} \quad\Longrightarrow\quad \text{interface},
\label{eq:interface_rule}
\end{equation}
and the boundary is a knot otherwise. For the tables used in this work every
boundary is a knot.

\subsection{Step 5: from the fit to the radial derivatives}

The fit delivers, on each segment, $w(u)$ and its derivatives $w_1$, $w_2$,
$w_3$ in closed form. The source terms require instead $d\varepsilon/dR$,
$d^2\varepsilon/dR^2$ and $d^3\varepsilon/dR^3$ at every node of the TOV grid.
The route between them passes through three exact conversions, from
$(u,w)$-derivatives to $\varepsilon$-derivatives of $p$, from those to
$p$-derivatives of $\varepsilon$, and from those to radial derivatives through
the structure equations. At no stage is any tabulated or profiled quantity
differentiated numerically.

\vspace{0.2cm} \noindent \textit{Derivatives of the EoS.} Since
$\varepsilon\,d/d\varepsilon = d/du$ and $p = e^{w(u)}$, repeated
differentiation gives
\begin{align}
c_s^2 \equiv \frac{dp}{d\varepsilon} &= \frac{p}{\varepsilon}\,w_1, \nonumber\\
\frac{d^2p}{d\varepsilon^2} &= \frac{p}{\varepsilon^2}\,q, \qquad q \equiv w_1^2 - w_1 + w_2, \label{eq:eos_derivatives}\\
\frac{d^3p}{d\varepsilon^3} &= \frac{p}{\varepsilon^3}\Big[(w_1-2)\,q + 2w_1w_2 - w_2 + w_3\Big]. \nonumber
\end{align}
We abbreviate $s\equiv c_s^2$, $s_1\equiv ds/d\varepsilon = d^2p/d\varepsilon^2$
and $s_2\equiv d^2s/d\varepsilon^2 = d^3p/d\varepsilon^3$. The first radial
derivative $d\varepsilon/dR$ reads $s$ alone, the second reads $s$ and $s_1$,
and the third reads all three. The fitted polynomial is therefore differentiated
exactly three times and no more.

\vspace{0.2cm} \noindent \textit{Inverting the barotropic relation.} Since $d/dp
= s^{-1}\,d/d\varepsilon$, the derivatives of $\varepsilon$ with respect to $p$
follow from $\varepsilon_p\equiv d\varepsilon/dp = 1/s$ by repeated
differentiation,
\begin{equation}
\varepsilon_p = \frac{1}{s}, \qquad
\varepsilon_{pp} = -\frac{s_1}{s^3}, \qquad
\varepsilon_{ppp} = \frac{3s_1^2 - s\,s_2}{s^5},
\label{eq:eps_p_derivatives}
\end{equation}
each line following from the one before through $ds/dp = s_1/s$.

\vspace{0.2cm} \noindent \textit{Radial derivatives of the pressure.} These come
from the structure equations. Writing
\begin{equation}
A \equiv \varepsilon + p, \qquad B \equiv M + 4\pi R^3 p, \qquad D \equiv R(R-2M),
\label{eq:ABD}
\end{equation}
the TOV and mass equations read
\begin{equation}
\frac{\mathrm{d}p}{\mathrm{d}R} = -\frac{AB}{D}, \qquad \frac{\mathrm{d}M}{\mathrm{d}R} = 4\pi R^2\varepsilon.
\label{eq:tov_compact}
\end{equation}
Differentiating once, with
\begin{align}
\frac{\mathrm{d}A}{\mathrm{d}R} &= \frac{\mathrm{d}\varepsilon}{\mathrm{d}R} + \frac{\mathrm{d}p}{\mathrm{d}R}, \nonumber\\
\frac{\mathrm{d}B}{\mathrm{d}R} &= \frac{\mathrm{d}M}{\mathrm{d}R} + 4\pi R^2\Bigl(3p + R\frac{\mathrm{d}p}{\mathrm{d}R}\Bigr), \label{eq:ABD_prime}\\
\frac{\mathrm{d}D}{\mathrm{d}R} &= 2\Bigl(R - M - R\frac{\mathrm{d}M}{\mathrm{d}R}\Bigr), \nonumber
\end{align}
gives
\begin{equation}
\frac{\mathrm{d}^2p}{\mathrm{d}R^2} = -\frac{\tfrac{\mathrm{d}A}{\mathrm{d}R}B + A\tfrac{\mathrm{d}B}{\mathrm{d}R}}{D} + \frac{AB\,\tfrac{\mathrm{d}D}{\mathrm{d}R}}{D^2}.
\label{eq:p_second}
\end{equation}
Differentiating again, with
\begin{align}
\frac{\mathrm{d}^2A}{\mathrm{d}R^2} &= \frac{\mathrm{d}^2\varepsilon}{\mathrm{d}R^2} + \frac{\mathrm{d}^2p}{\mathrm{d}R^2}, \nonumber\\
\frac{\mathrm{d}^2M}{\mathrm{d}R^2} &= 4\pi R\,\Bigl(2\varepsilon + R\frac{\mathrm{d}\varepsilon}{\mathrm{d}R}\Bigr), \nonumber\\
\frac{\mathrm{d}^2B}{\mathrm{d}R^2} &= \frac{\mathrm{d}^2M}{\mathrm{d}R^2} + 24\pi Rp + 24\pi R^2 \tfrac{\mathrm{d}p}{\mathrm{d}R} + 4\pi R^3 \tfrac{\mathrm{d}^2p}{\mathrm{d}R^2}, \label{eq:ABD_double_prime}\\
\frac{\mathrm{d}^2D}{\mathrm{d}R^2} &= 2\Bigl(1 - 2\tfrac{\mathrm{d}M}{\mathrm{d}R} - R\tfrac{\mathrm{d}^2M}{\mathrm{d}R^2}\Bigr), \nonumber
\end{align}
gives
\begin{align}
\frac{\mathrm{d}^3p}{\mathrm{d}R^3} = &-\frac{\tfrac{\mathrm{d}^2A}{\mathrm{d}R^2}B + 2\tfrac{\mathrm{d}A}{\mathrm{d}R}\tfrac{\mathrm{d}B}{\mathrm{d}R} + A\tfrac{\mathrm{d}^2B}{\mathrm{d}R^2}}{D} \nonumber\\
&+ \frac{2\Bigl(\tfrac{\mathrm{d}A}{\mathrm{d}R}B + A\tfrac{\mathrm{d}B}{\mathrm{d}R}\Bigr)\tfrac{\mathrm{d}D}{\mathrm{d}R} + AB\,\tfrac{\mathrm{d}^2D}{\mathrm{d}R^2}}{D^2} - \frac{2AB\,\bigl(\tfrac{\mathrm{d}D}{\mathrm{d}R}\bigr)^2}{D^3}.
\label{eq:p_third}
\end{align}

\vspace{0.2cm} \noindent \textit{Assembly and closure.} The energy density
depends on radius only through the pressure, so $\varepsilon(p(R))$ is
differentiated by the chain rule,
\begin{align}
\frac{d\varepsilon}{dR} &= \varepsilon_p\, dp/dR, \nonumber\\
\frac{d^2\varepsilon}{dR^2} &= \varepsilon_{pp}\, \left(\frac{dp}{dR}\right)^2 + \varepsilon_p\, \frac{d^2p}{dR^2}, \label{eq:eps_radial}\\
\frac{d^3\varepsilon}{dR^3} &= \varepsilon_{ppp}\, \left(\frac{dp}{dR}\right)^3 + 3\,\varepsilon_{pp}\, \frac{dp}{dR} \, \frac{d^2p}{dR^2} + \varepsilon_p\, \frac{d^3 p}{dR^3}. \nonumber
\end{align}
Equations~\eqref{eq:ABD_prime}--\eqref{eq:eps_radial} contain radial derivatives
of $\varepsilon$ on their right-hand sides, but only of lower order than the
quantity under construction. This is where the order reduction closes. Starting
from $dp/dR$, which needs the background alone, the derivatives of $p$ and
$\varepsilon$ follow in alternation, each step using only quantities already
computed. Each radial derivative of $\varepsilon$ reduces to a closed form in
$R$, $M$, $p$, $\varepsilon$, the fitted triple $(s,s_1,s_2)$, and derivatives
of lower order. Nothing in the construction is differentiated numerically. The
only numerical operations involved are the least-squares solve of Step 2 and the
integration of the background itself.

The three derivatives are evaluated at each node of the TOV grid, reading $R$,
$M$, $p$ and $\varepsilon$ from the solved background. Products appearing in the
source terms, such as $\varepsilon\,d^2\varepsilon/dR^2$ or
$(d\varepsilon/dR)^2$, are therefore mutually consistent by construction. The
same triple $(s,s_1,s_2)$ evaluated at $\varepsilon_c$ feeds the near-center
asymptotic expansions, so the boundary conditions and the interior solutions
they initialize rest on the same representation of the matter.

Below the lowest tabulated row the polynomial of the outermost segment is
extrapolated rather than clamped at the last tabulated value. The layer involved
is the outermost few centimeters of the star.

\begin{widetext}

\clearpage


\section{Perturbation-Function Profiles}
\label{app:profiles}

This appendix collects the complete set of perturbation-function profiles for
one configuration, so that the surface matching examined in
Sec.~\ref{sec:matching} can be seen to hold at every order and mode. The
configuration is the SLy star of TOV mass $M_*=1.5\,M_\odot$ from
Table~\ref{table:configurations}, with central energy density $\varepsilon_c =
598.880\,\mathrm{MeV\,fm^{-3}}$ and TOV radius $R_*=11.632\,\mathrm{km}$,
rotating at $f_*=716\,\mathrm{Hz}$. The profiles run from the background at
zeroth order through the odd- and even-parity functions at seventh, together
with the pressure and energy density. In every panel the interior numerical
solution, the exterior analytic solution and the near-center asymptotic
expansion are shown against the radial coordinate $R$, with the stellar interior
shaded up to $R_*$.

\begin{figure*}[h]
\centering
\includegraphics[width=\textwidth]{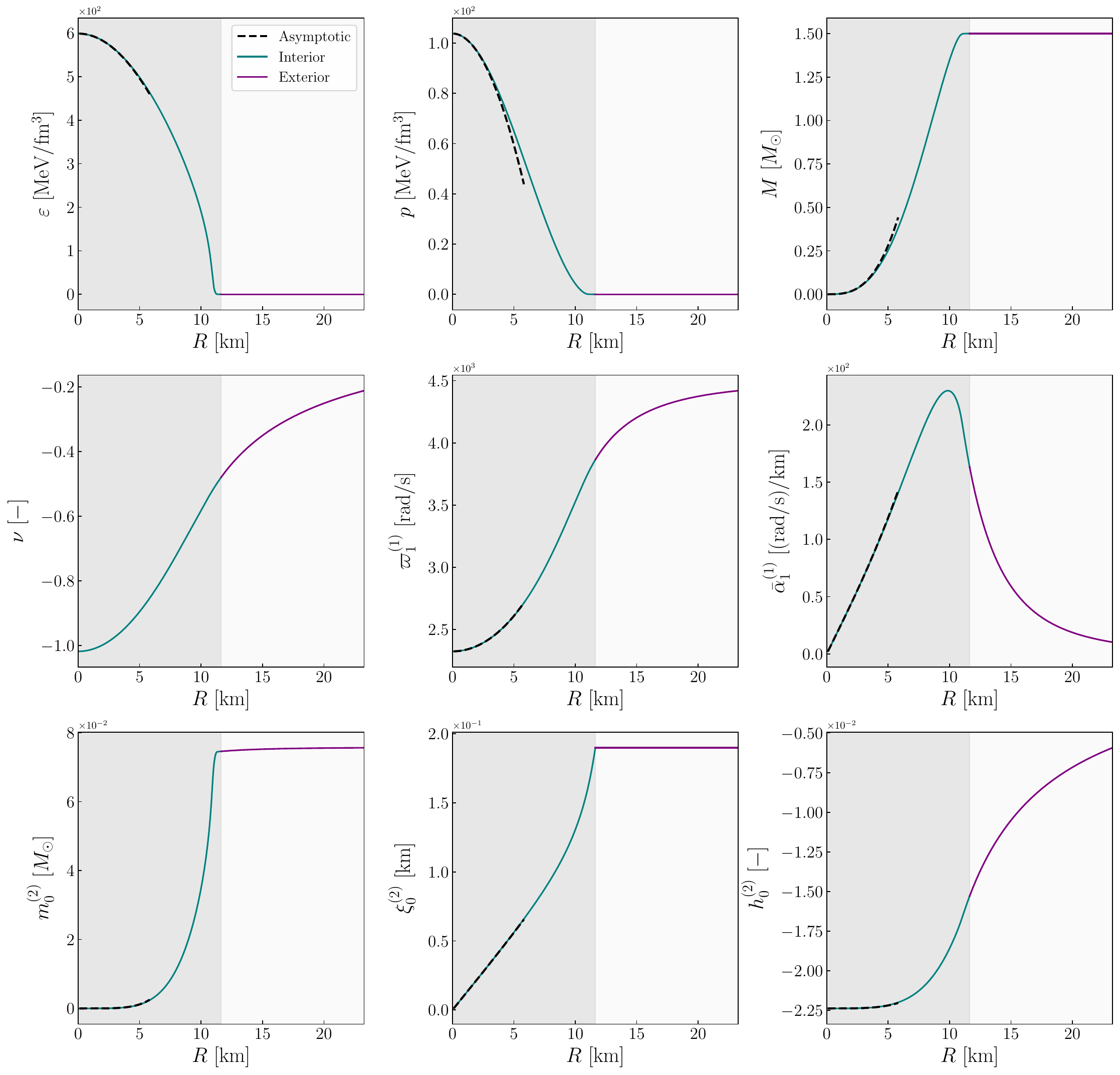}
\caption{Radial profiles of the background and of the lowest-order perturbations
for the $1.5\,M_\odot$ SLy star ($\varepsilon_c =
598.880\,\mathrm{MeV\,fm^{-3}}$, $R_{*}=11.632$~km), rotating at $f_{*}=716$~Hz.
\textit{Top row:} the background energy density, pressure, mass function and
metric potential. \textit{Middle:} the first-order pair $\varpi^{(1)}_{1}$ and
$\bar{\alpha}^{(1)}_{1}$. \textit{Bottom:} the second-order monopole
$m^{(2)}_{0}$, $\xi^{(2)}_{0}$ and $h^{(2)}_{0}$.}
  \label{fig:profiles1}
\end{figure*}
\begin{figure*}[p]
  \centering
  \includegraphics[width=\textwidth]{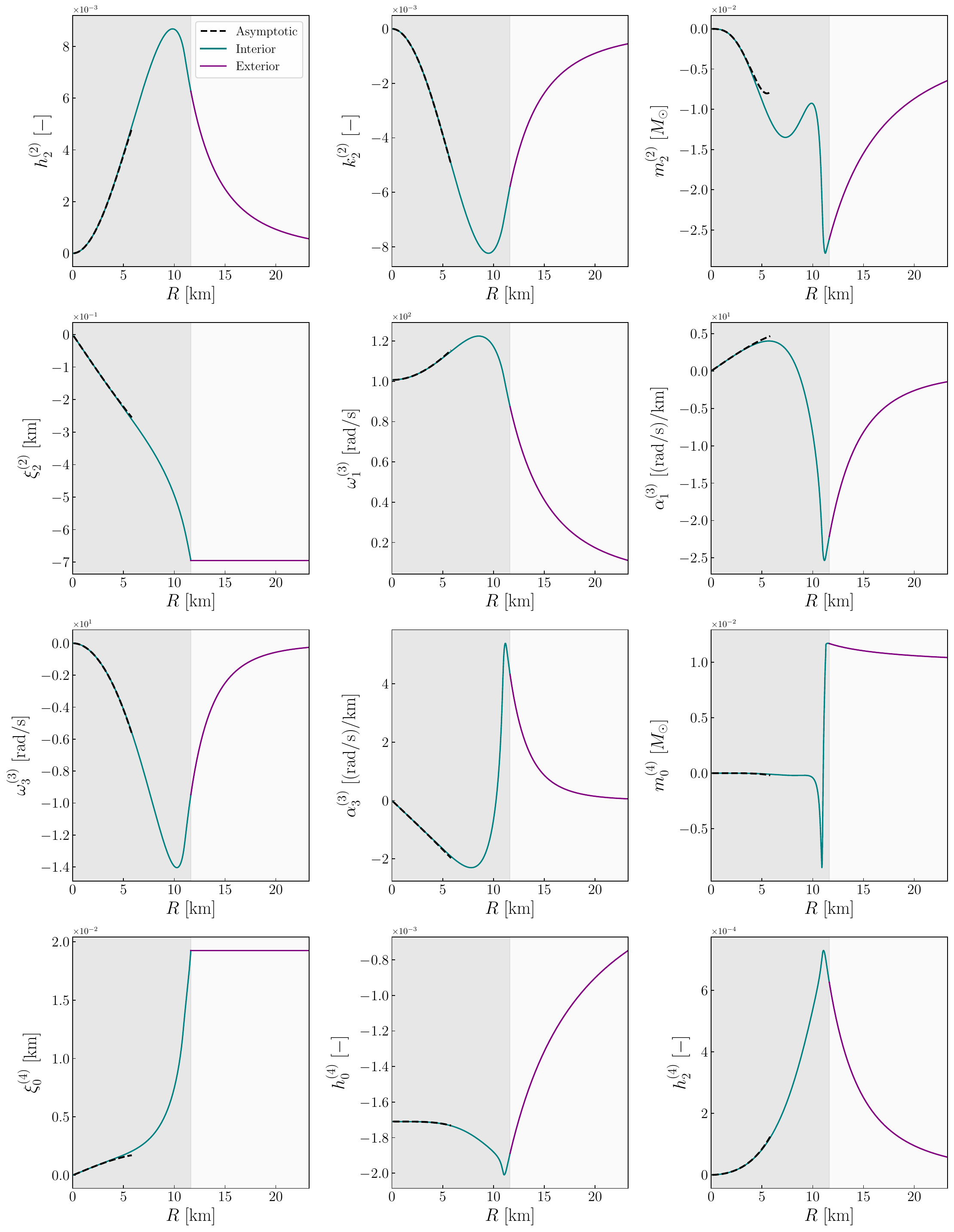}
\caption{As Fig.~\ref{fig:profiles1}, continuing with the second-order
quadrupole, the third-order odd sector at $\ell=1$ and $\ell=3$, and the first
of the fourth-order functions.}
\label{fig:profiles2}
\end{figure*}
\begin{figure*}[p]
  \centering
  \includegraphics[width=\textwidth]{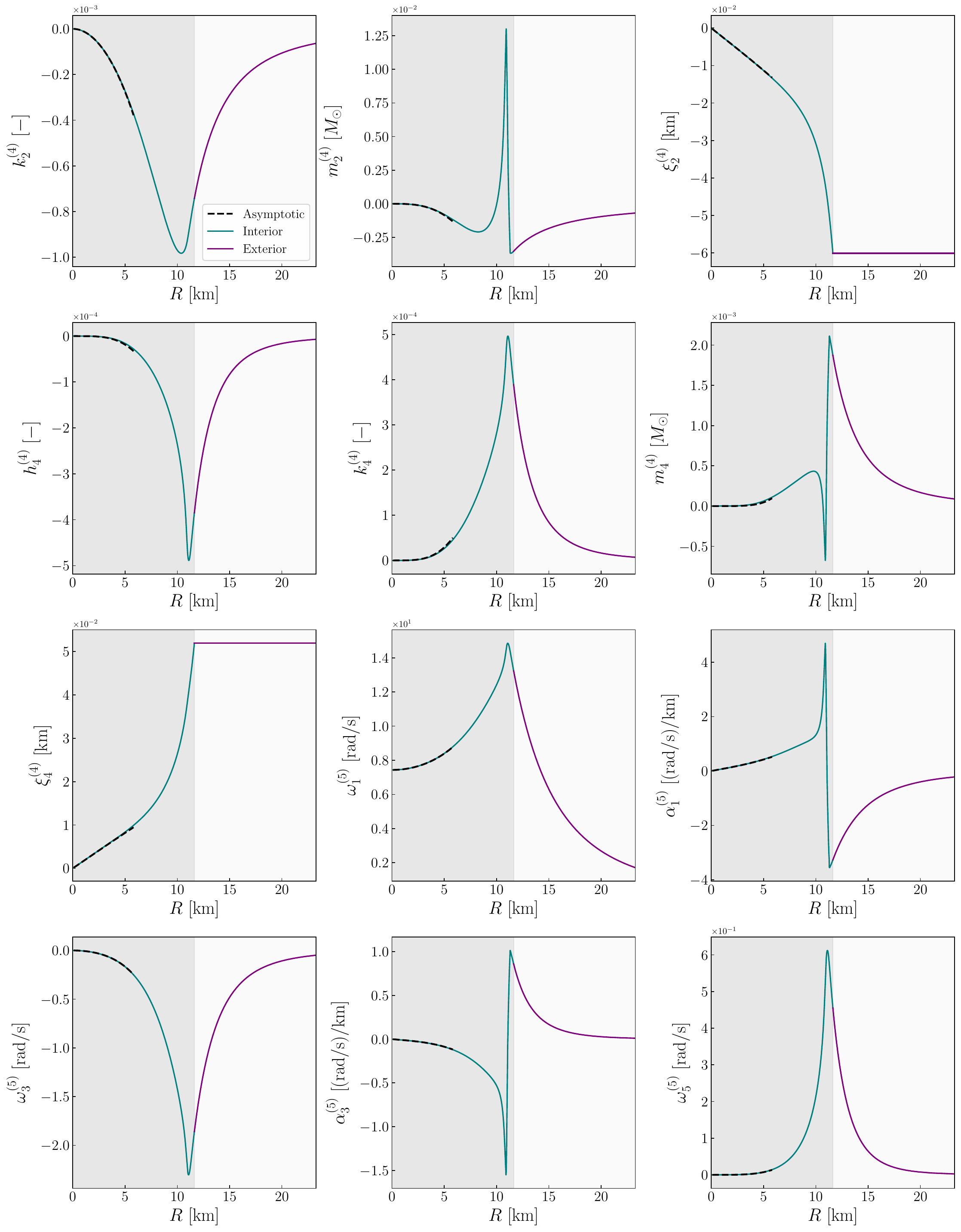}
\caption{As Fig.~\ref{fig:profiles1}, for the remaining fourth-order functions
at $\ell=2$ and $\ell=4$ and the fifth-order odd sector.}
\label{fig:profiles3}
\end{figure*}
\begin{figure*}[p]
  \centering
  \includegraphics[width=\textwidth]{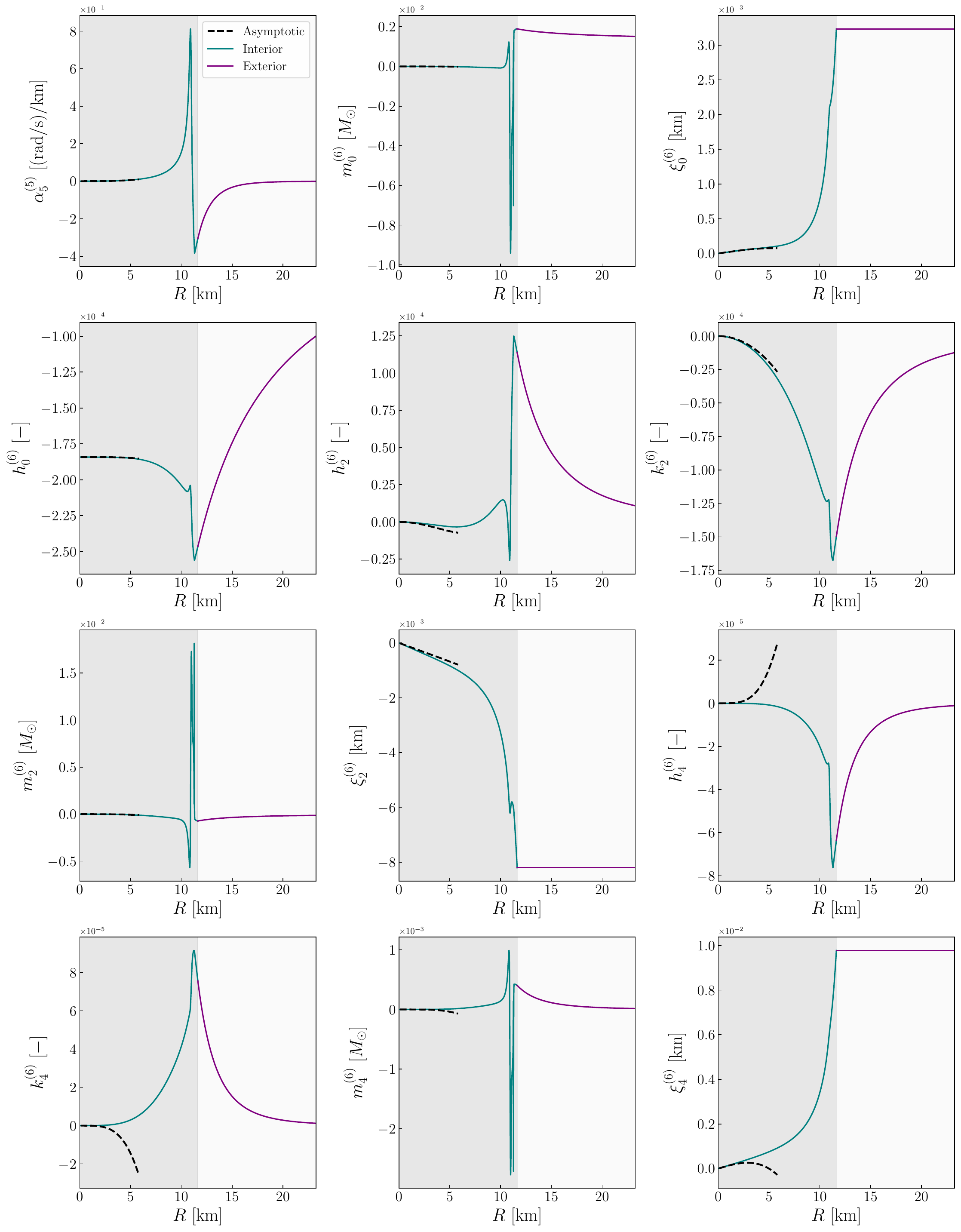}
\caption{As Fig.~\ref{fig:profiles1}, for the sixth-order monopole and the
sixth-order even sector at $\ell=2$ and $\ell=4$.}
\label{fig:profiles4}
\end{figure*}
\begin{figure*}[p]
  \centering
  \includegraphics[width=\textwidth]{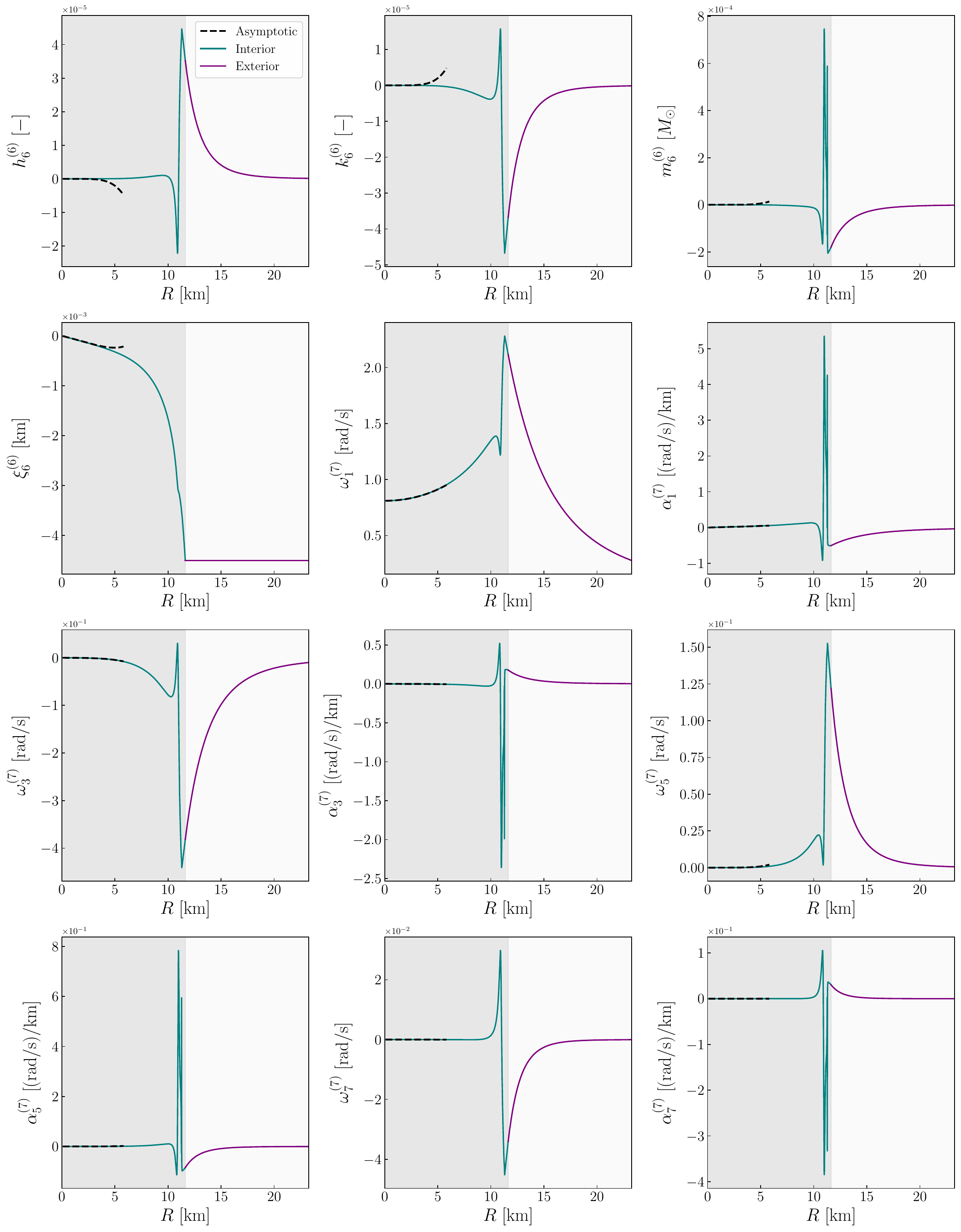}
\caption{As Fig.~\ref{fig:profiles1}, closing with the sixth-order $\ell=6$
functions and the seventh-order odd sector at $\ell=1,3,5$ and $7$.}
\label{fig:profiles5}
\end{figure*}
\clearpage
\end{widetext}

\begin{widetext}


\section{Multipole Moments and Their Spin-Induced Corrections}
\label{app:truncation}

\begin{table*}[h!]
\setlength{\tabcolsep}{1.0pt}
\renewcommand{\arraystretch}{1.0}
\begin{tabular}{l l rrr rrr rrr}
\toprule
 &  & \multicolumn{9}{c}{\textbf{SLy}} \\
\cmidrule(lr){3-11}
 &  & \multicolumn{3}{c}{$M_{*} = 1.0\,M_\odot$} & \multicolumn{3}{c}{$M_{*} =1.5\,M_\odot$} & \multicolumn{3}{c}{$M_{*} =2.0\,M_\odot$} \\
\cmidrule(lr){3-5}\cmidrule(lr){6-8}\cmidrule(lr){9-11}
$P_{\ell}$ & $P^{(n)}_{\ell}$ & Correction & $\mathcal{T}^{(n)}_{\ell}$ & $\mathcal{R}^{(n)}_{\ell}$ & Correction & $\mathcal{T}^{(n)}_{\ell}$ & $\mathcal{R}^{(n)}_{\ell}$ & Correction & $\mathcal{T}^{(n)}_{\ell}$ & $\mathcal{R}^{(n)}_{\ell}$ \\
\midrule
$M_{0}$ $[M_\odot]$ & $M_{*}$ & $1.000{\times}10^{0}$ & 90.19 & -- & $1.500{\times}10^{0}$ & 94.54 & -- & $2.000{\times}10^{0}$ & 97.79 & -- \\
 & $M^{(2)}_{0}$ & $8.761{\times}10^{-2}$ & 7.90 & 8.76 & $7.578{\times}10^{-2}$ & 4.78 & 5.05 & $4.221{\times}10^{-2}$ & 2.06 & 2.11 \\
 & $M^{(4)}_{0}$ & $1.741{\times}10^{-2}$ & 1.57 & 1.60 & $9.575{\times}10^{-3}$ & 0.60 & 0.61 & $2.747{\times}10^{-3}$ & 0.13 & 0.13 \\
 & $M^{(6)}_{0}$ & $3.737{\times}10^{-3}$ & 0.34 & 0.34 & $1.284{\times}10^{-3}$ & 0.08 & 0.08 & $1.877{\times}10^{-4}$ & 0.01 & $0.01$ \\
\cmidrule(lr){2-11}
 & \textbf{Total} & $\mathbf{1.109{\times}10^{0}}$ & \textbf{100.00} &  & $\mathbf{1.587{\times}10^{0}}$ & \textbf{100.00} &  & $\mathbf{2.045{\times}10^{0}}$ & \textbf{100.00} &  \\
\midrule
$S_{1}$ $[M_\odot^{2}]$ & $S^{(1)}_{1}$ & $4.437{\times}10^{-1}$ & 76.01 & -- & $7.629{\times}10^{-1}$ & 85.82 & -- & $1.022{\times}10^{0}$ & 94.19 & -- \\
 & $S^{(3)}_{1}$ & $1.063{\times}10^{-1}$ & 18.21 & 23.95 & $1.069{\times}10^{-1}$ & 12.02 & 14.01 & $5.834{\times}10^{-2}$ & 5.38 & 5.71 \\
 & $S^{(5)}_{1}$ & $2.679{\times}10^{-2}$ & 4.59 & 4.87 & $1.648{\times}10^{-2}$ & 1.85 & 1.89 & $4.313{\times}10^{-3}$ & 0.40 & 0.40 \\
 & $S^{(7)}_{1}$ & $6.993{\times}10^{-3}$ & 1.20 & 1.21 & $2.685{\times}10^{-3}$ & 0.30 & 0.30 & $3.587{\times}10^{-4}$ & 0.03 & 0.03 \\
\cmidrule(lr){2-11}
 & \textbf{Total} & $\mathbf{5.837{\times}10^{-1}}$ & \textbf{100.00} &  & $\mathbf{8.889{\times}10^{-1}}$ & \textbf{100.00} &  & $\mathbf{1.085{\times}10^{0}}$ & \textbf{100.00} &  \\
\midrule
$M_{2}$ $[M_\odot^{3}]$ & $M^{(2)}_{2}$ & $-1.847{\times}10^{0}$ & 81.21 & -- & $-1.682{\times}10^{0}$ & 88.81 & -- & $-9.727{\times}10^{-1}$ & 94.36 & -- \\
 & $M^{(4)}_{2}$ & $-3.327{\times}10^{-1}$ & 14.63 & 18.01 & $-1.780{\times}10^{-1}$ & 9.40 & 10.58 & $-5.268{\times}10^{-2}$ & 5.11 & 5.42 \\
 & $M^{(6)}_{2}$ & $-9.472{\times}10^{-2}$ & 4.16 & 4.35 & $-3.401{\times}10^{-2}$ & 1.80 & 1.83 & $-5.423{\times}10^{-3}$ & 0.53 & 0.53 \\
\cmidrule(lr){2-11}
 & \textbf{Total} & $\mathbf{-2.275{\times}10^{0}}$ & \textbf{100.00} &  & $\mathbf{-1.894{\times}10^{0}}$ & \textbf{100.00} &  & $\mathbf{-1.031{\times}10^{0}}$ & \textbf{100.00} &  \\
\midrule
$S_{3}$ $[M_\odot^{4}]$ & $S^{(3)}_{3}$ & $-1.729{\times}10^{0}$ & 69.67 & -- & $-1.660{\times}10^{0}$ & 81.79 & -- & $-7.765{\times}10^{-1}$ & 92.00 & -- \\
 & $S^{(5)}_{3}$ & $-5.688{\times}10^{-1}$ & 22.92 & 32.90 & $-3.077{\times}10^{-1}$ & 15.16 & 18.54 & $-6.109{\times}10^{-2}$ & 7.24 & 7.87 \\
 & $S^{(7)}_{3}$ & $-1.839{\times}10^{-1}$ & 7.41 & 8.00 & $-6.177{\times}10^{-2}$ & 3.04 & 3.14 & $-6.395{\times}10^{-3}$ & 0.76 & 0.76 \\
\cmidrule(lr){2-11}
 & \textbf{Total} & $\mathbf{-2.481{\times}10^{0}}$ & \textbf{100.00} &  & $\mathbf{-2.029{\times}10^{0}}$ & \textbf{100.00} &  & $\mathbf{-8.440{\times}10^{-1}}$ & \textbf{100.00} &  \\
\midrule
$M_{4}$ $[M_\odot^{5}]$ & $M^{(4)}_{4}$ & $8.291{\times}10^{0}$ & 78.62 & -- & $4.426{\times}10^{0}$ & 86.87 & -- & $9.140{\times}10^{-1}$ & 93.42 & -- \\
 & $M^{(6)}_{4}$ & $2.255{\times}10^{0}$ & 21.38 & 27.20 & $6.686{\times}10^{-1}$ & 13.13 & 15.11 & $6.437{\times}10^{-2}$ & 6.58 & 7.04 \\
\cmidrule(lr){2-11}
 & \textbf{Total} & $\mathbf{1.055{\times}10^{1}}$ & \textbf{100.00} &  & $\mathbf{5.094{\times}10^{0}}$ & \textbf{100.00} &  & $\mathbf{9.784{\times}10^{-1}}$ & \textbf{100.00} &  \\
\midrule
$S_{5}$ $[M_\odot^{6}]$ & $S^{(5)}_{5}$ & $9.920{\times}10^{0}$ & 70.33 & -- & $5.424{\times}10^{0}$ & 81.30 & -- & $8.685{\times}10^{-1}$ & 91.46 & -- \\
 & $S^{(7)}_{5}$ & $4.185{\times}10^{0}$ & 29.67 & 42.19 & $1.248{\times}10^{0}$ & 18.70 & 23.01 & $8.107{\times}10^{-2}$ & 8.54 & 9.33 \\
\cmidrule(lr){2-11}
 & \textbf{Total} & $\mathbf{1.411{\times}10^{1}}$ & \textbf{100.00} &  & $\mathbf{6.672{\times}10^{0}}$ & \textbf{100.00} &  & $\mathbf{9.495{\times}10^{-1}}$ & \textbf{100.00} &  \\
\midrule
$M_{6}$ $[M_\odot^{7}]$ & $M^{(6)}_{6}$ & $-5.215{\times}10^{1}$ & 100.00 & -- & $-1.631{\times}10^{1}$ & 100.00 & -- & $-1.182{\times}10^{0}$ & 100.00 & -- \\
\midrule
$S_{7}$ $[M_\odot^{8}]$ & $S^{(7)}_{7}$ & $-7.057{\times}10^{1}$ & 100.00 & -- & $-2.227{\times}10^{1}$ & 100.00 & -- & $-1.238{\times}10^{0}$ & 100.00 & -- \\
\bottomrule
\end{tabular}
\caption{Multipole moments of three SLy stars spinning at $f_{*}=716$~Hz:
$1.0\,M_\odot$: $\varepsilon_c=411.44$ MeV/fm$^3$, $R_{*}=11.848$~km;
$1.5\,M_\odot$: $\varepsilon_c=598.88$ MeV/fm$^3$, $R_{*}=11.632$~km;
$2.0\,M_\odot$: $\varepsilon_c=1148.58$ MeV/fm$^3$, $R_{*}=10.621$~km. Each
multipole is a series in the spin and one row is one term of that series. The
correction is $P^{(n)}_{\ell}\Omega^{n}$ with $\Omega=2\pi f_{*}$ the angular
frequency, $\mathcal{T}^{(n)}_{\ell}$ is that term as a fraction of its
multipole's total, and $\mathcal{R}^{(n)}_{\ell}$ is the term divided by the
cumulative value through the previous order. Truncating just above the row,
$\mathcal{R}^{(n)}_{\ell}$ estimates the error committed from the first
neglected term. The leading term of each multipole has no entry, having nothing
before it. $\mathcal{R}^{(n)}_{\ell}$ falls down every block, which is what
convergence looks like, and it falls faster for the heavier star. The largest
value in this table is 42.19\%, at the light end. Units used are $G=c=1$, so a
multipole of index $\ell$ carries $M_\odot^{\ell+1}$, as listed beside it.}
\end{table*}

\clearpage

\begin{table*}[h!]
\setlength{\tabcolsep}{1.0pt}
\renewcommand{\arraystretch}{1.0}
\begin{tabular}{l l rrr rrr rrr}
\toprule
 &  & \multicolumn{9}{c}{\textbf{BSk21}} \\
\cmidrule(lr){3-11}
 &  & \multicolumn{3}{c}{$1.0\,M_\odot$} & \multicolumn{3}{c}{$1.5\,M_\odot$} & \multicolumn{3}{c}{$2.0\,M_\odot$} \\
\cmidrule(lr){3-5}\cmidrule(lr){6-8}\cmidrule(lr){9-11}
$P_{\ell}$ & $P^{(n)}_{\ell}$ & Correction & $\mathcal{T}^{(n)}_{\ell}$ & $\mathcal{R}^{(n)}_{\ell}$ & Correction & $\mathcal{T}^{(n)}_{\ell}$ & $\mathcal{R}^{(n)}_{\ell}$ & Correction & $\mathcal{T}^{(n)}_{\ell}$ & $\mathcal{R}^{(n)}_{\ell}$ \\
\midrule
$M_{0}$ $[M_\odot]$ & $M_{*}$ & $1.000{\times}10^{0}$ & 87.51 & -- & $1.500{\times}10^{0}$ & 92.18 & -- & $2.000{\times}10^{0}$ & 95.62 & -- \\
 & $M^{(2)}_{0}$ & $1.096{\times}10^{-1}$ & 9.59 & 10.96 & $1.062{\times}10^{-1}$ & 6.53 & 7.08 & $8.157{\times}10^{-2}$ & 3.90 & 4.08 \\
 & $M^{(4)}_{0}$ & $2.662{\times}10^{-2}$ & 2.33 & 2.40 & $1.804{\times}10^{-2}$ & 1.11 & 1.12 & $9.079{\times}10^{-3}$ & 0.43 & 0.44 \\
 & $M^{(6)}_{0}$ & $6.525{\times}10^{-3}$ & 0.57 & 0.57 & $3.111{\times}10^{-3}$ & 0.19 & 0.19 & $9.537{\times}10^{-4}$ & 0.05 & 0.05 \\
\cmidrule(lr){2-11}
 & \textbf{Total} & $\mathbf{1.143{\times}10^{0}}$ & \textbf{100.00} &  & $\mathbf{1.627{\times}10^{0}}$ & \textbf{100.00} &  & $\mathbf{2.092{\times}10^{0}}$ & \textbf{100.00} &  \\
\midrule
$S_{1}$ $[M_\odot^{2}]$ & $S^{(1)}_{1}$ & $4.984{\times}10^{-1}$ & 70.63 & -- & $8.908{\times}10^{-1}$ & 80.39 & -- & $1.294{\times}10^{0}$ & 88.58 & -- \\
 & $S^{(3)}_{1}$ & $1.479{\times}10^{-1}$ & 20.96 & 29.67 & $1.734{\times}10^{-1}$ & 15.65 & 19.47 & $1.456{\times}10^{-1}$ & 9.96 & 11.24 \\
 & $S^{(5)}_{1}$ & $4.560{\times}10^{-2}$ & 6.46 & 7.06 & $3.626{\times}10^{-2}$ & 3.27 & 3.41 & $1.897{\times}10^{-2}$ & 1.30 & 1.32 \\
 & $S^{(7)}_{1}$ & $1.377{\times}10^{-2}$ & 1.95 & 1.99 & $7.652{\times}10^{-3}$ & 0.69 & 0.70 & $2.414{\times}10^{-3}$ & 0.17 & 0.17 \\
\cmidrule(lr){2-11}
 & \textbf{Total} & $\mathbf{7.056{\times}10^{-1}}$ & \textbf{100.00} &  & $\mathbf{1.108{\times}10^{0}}$ & \textbf{100.00} &  & $\mathbf{1.461{\times}10^{0}}$ & \textbf{100.00} &  \\
\midrule
$M_{2}$ $[M_\odot^{3}]$ & $M^{(2)}_{2}$ & $-2.609{\times}10^{0}$ & 77.11 & -- & $-2.784{\times}10^{0}$ & 84.83 & -- & $-2.261{\times}10^{0}$ & 90.34 & -- \\
 & $M^{(4)}_{2}$ & $-5.770{\times}10^{-1}$ & 17.06 & 22.12 & $-3.983{\times}10^{-1}$ & 12.14 & 14.31 & $-2.060{\times}10^{-1}$ & 8.23 & 9.11 \\
 & $M^{(6)}_{2}$ & $-1.972{\times}10^{-1}$ & 5.83 & 6.19 & $-9.949{\times}10^{-2}$ & 3.03 & 3.13 & $-3.573{\times}10^{-2}$ & 1.43 & 1.45 \\
\cmidrule(lr){2-11}
 & \textbf{Total} & $\mathbf{-3.383{\times}10^{0}}$ & \textbf{100.00} &  & $\mathbf{-3.281{\times}10^{0}}$ & \textbf{100.00} &  & $\mathbf{-2.503{\times}10^{0}}$ & \textbf{100.00} &  \\
\midrule
$S_{3}$ $[M_\odot^{4}]$ & $S^{(3)}_{3}$ & $-2.731{\times}10^{0}$ & 64.25 & -- & $-3.267{\times}10^{0}$ & 75.61 & -- & $-2.569{\times}10^{0}$ & 85.45 & -- \\
 & $S^{(5)}_{3}$ & $-1.103{\times}10^{0}$ & 25.95 & 40.39 & $-8.352{\times}10^{-1}$ & 19.33 & 25.57 & $-3.763{\times}10^{-1}$ & 12.52 & 14.65 \\
 & $S^{(7)}_{3}$ & $-4.167{\times}10^{-1}$ & 9.80 & 10.87 & $-2.184{\times}10^{-1}$ & 5.05 & 5.32 & $-6.115{\times}10^{-2}$ & 2.03 & 2.08 \\
\cmidrule(lr){2-11}
 & \textbf{Total} & $\mathbf{-4.251{\times}10^{0}}$ & \textbf{100.00} &  & $\mathbf{-4.320{\times}10^{0}}$ & \textbf{100.00} &  & $\mathbf{-3.006{\times}10^{0}}$ & \textbf{100.00} &  \\
\midrule
$M_{4}$ $[M_\odot^{5}]$ & $M^{(4)}_{4}$ & $1.613{\times}10^{1}$ & 75.12 & -- & $1.203{\times}10^{1}$ & 82.98 & -- & $5.498{\times}10^{0}$ & 89.19 & -- \\
 & $M^{(6)}_{4}$ & $5.340{\times}10^{0}$ & 24.88 & 33.11 & $2.468{\times}10^{0}$ & 17.02 & 20.51 & $6.667{\times}10^{-1}$ & 10.81 & 12.13 \\
\cmidrule(lr){2-11}
 & \textbf{Total} & $\mathbf{2.147{\times}10^{1}}$ & \textbf{100.00} &  & $\mathbf{1.450{\times}10^{1}}$ & \textbf{100.00} &  & $\mathbf{6.165{\times}10^{0}}$ & \textbf{100.00} &  \\
\midrule
$S_{5}$ $[M_\odot^{6}]$ & $S^{(5)}_{5}$ & $2.151{\times}10^{1}$ & 66.54 & -- & $1.757{\times}10^{1}$ & 76.18 & -- & $7.556{\times}10^{0}$ & 85.52 & -- \\
 & $S^{(7)}_{5}$ & $1.082{\times}10^{1}$ & 33.46 & 50.28 & $5.495{\times}10^{0}$ & 23.82 & 31.26 & $1.279{\times}10^{0}$ & 14.48 & 16.93 \\
\cmidrule(lr){2-11}
 & \textbf{Total} & $\mathbf{3.233{\times}10^{1}}$ & \textbf{100.00} &  & $\mathbf{2.307{\times}10^{1}}$ & \textbf{100.00} &  & $\mathbf{8.835{\times}10^{0}}$ & \textbf{100.00} &  \\
\midrule
$M_{6}$ $[M_\odot^{7}]$ & $M^{(6)}_{6}$ & $-1.373{\times}10^{2}$ & 100.00 & -- & $-7.174{\times}10^{1}$ & 100.00 & -- & $-1.840{\times}10^{1}$ & 100.00 & -- \\
\midrule
$S_{7}$ $[M_\odot^{8}]$ & $S^{(7)}_{7}$ & $-2.021{\times}10^{2}$ & 100.00 & -- & $-1.155{\times}10^{2}$ & 100.00 & -- & $-2.687{\times}10^{1}$ & 100.00 & -- \\
\bottomrule
\end{tabular}
\caption{Multipole moments of three BSk21 stars spinning at $f_{*}=716$~Hz:
$1.0\,M_\odot$: $\varepsilon_c=323.46$ MeV/fm$^3$, $R_{*}=12.467$~km;
$1.5\,M_\odot$: $\varepsilon_c=435.37$ MeV/fm$^3$, $R_{*}=12.593$~km;
$2.0\,M_\odot$: $\varepsilon_c=644.66$ MeV/fm$^3$, $R_{*}=12.295$~km. Each
multipole is a series in the spin and one row is one term of that series. The
correction is $P^{(n)}_{\ell}\Omega^{n}$ with $\Omega=2\pi f_{*}$ the angular
frequency, $\mathcal{T}^{(n)}_{\ell}$ is that term as a fraction of its
multipole's total, and $\mathcal{R}^{(n)}_{\ell}$ is the term divided by the
cumulative value through the previous order. Truncating just above the row,
$\mathcal{R}^{(n)}_{\ell}$ estimates the error committed from the first
neglected term. The leading term of each multipole has no entry, having nothing
before it. $\mathcal{R}^{(n)}_{\ell}$ falls down every block, which is what
convergence looks like, and it falls faster for the heavier star; the largest
value in this table is 50.28\%, at the light end. Units used are $G=c=1$, so a
multipole of index $\ell$ carries $M_\odot^{\ell+1}$, as listed beside it.}
\end{table*}

\clearpage


\section{Exterior Integration Constants}
\label{app:exterior_constants}

This appendix collects the rescaled exterior integration constants
$\tilde{C}^{(n)}_{\ell,\mathsf{ext}}$ along both stellar sequences of
Sec.~\ref{sec:sequences}, plotted against the central energy density
$\varepsilon_c$. Each sequence runs from $0.7\,M_\odot$ to the maximum mass the
EoS supports, $2.048\,M_\odot$ for SLy and $2.274\,M_\odot$ for BSk21, with
every star rotating at $f_*=716\,\mathrm{Hz}$. The constants are shown for every
order and mode from first through seventh, since the moments are built from them
and their smoothness in $\varepsilon_c$ is what guarantees the smoothness of the
moments themselves.

\vspace{0.5cm}

\begin{figure*}[htbp]
\centering
\includegraphics[width=1.0\textwidth]{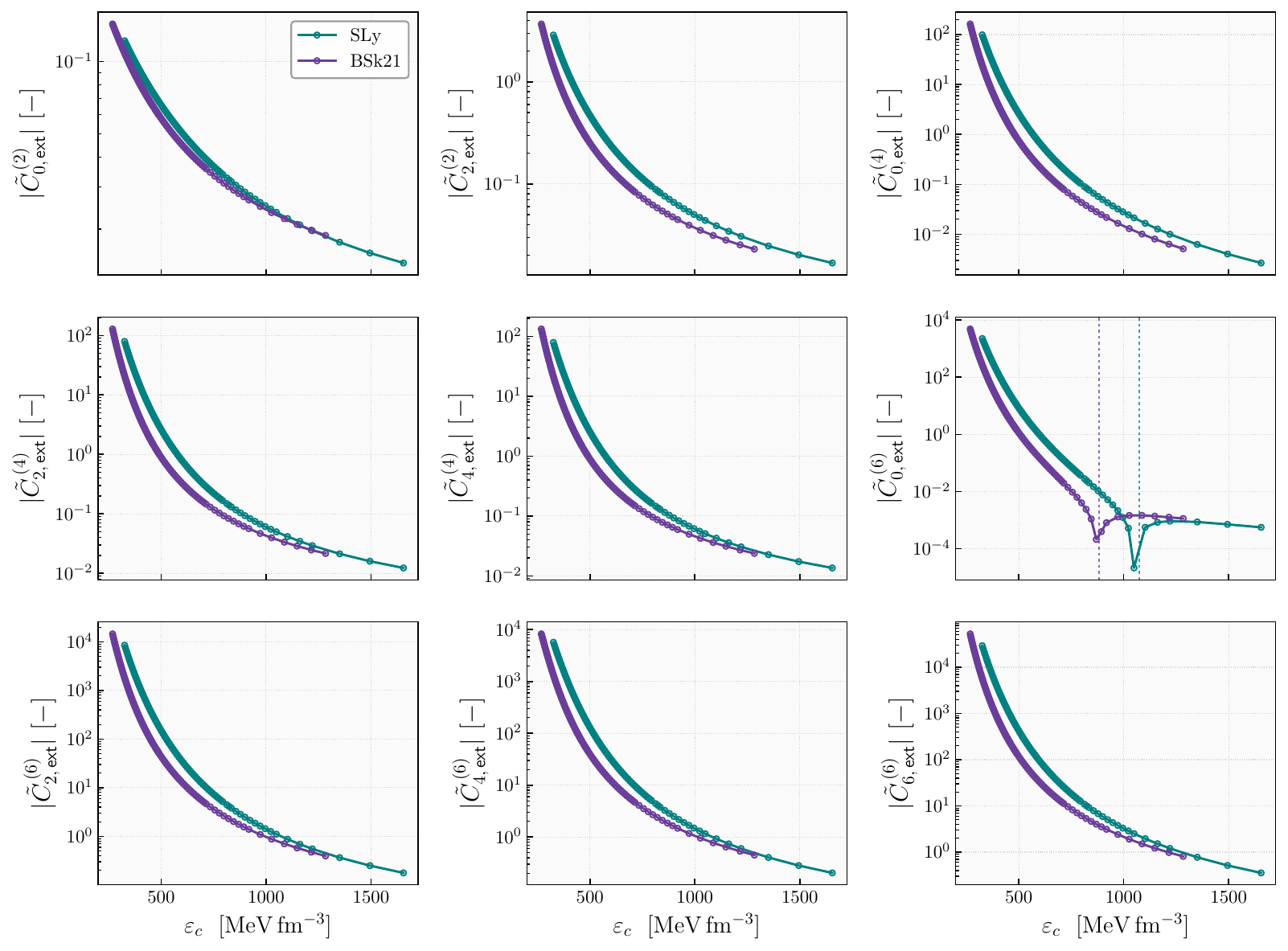}
\caption{The even-order exterior constants
$\tilde{C}^{(n)}_{\ell,\mathsf{ext}}$, orders $n=2$, $4$ and $6$, against the
central energy density. Both sequences are shown, SLy in teal and BSk21 in
purple. The vertical axis is logarithmic and carries the magnitude, so a
constant of either sign can be read on it. Every constant here is single-signed
along both sequences except $\tilde{C}^{(6)}_{0,\mathsf{ext}}$, which passes
through zero near $\varepsilon_c \simeq 1076$ and $882\,\mathrm{MeV\,fm^{-3}}$
for SLy and BSk21 respectively. The dashed verticals mark those crossings. The
constant itself is smooth there, as smooth as anywhere else on the branch. What
the plot shows is the magnitude diving toward the axis and returning, which is
what a sign change looks like under $|\cdot|$, and not a feature of the
constant.}
\label{fig:extconst_even}
\end{figure*}
\clearpage

\begin{figure*}[htbp]
    \centering
    \includegraphics[width=1.0\textwidth]{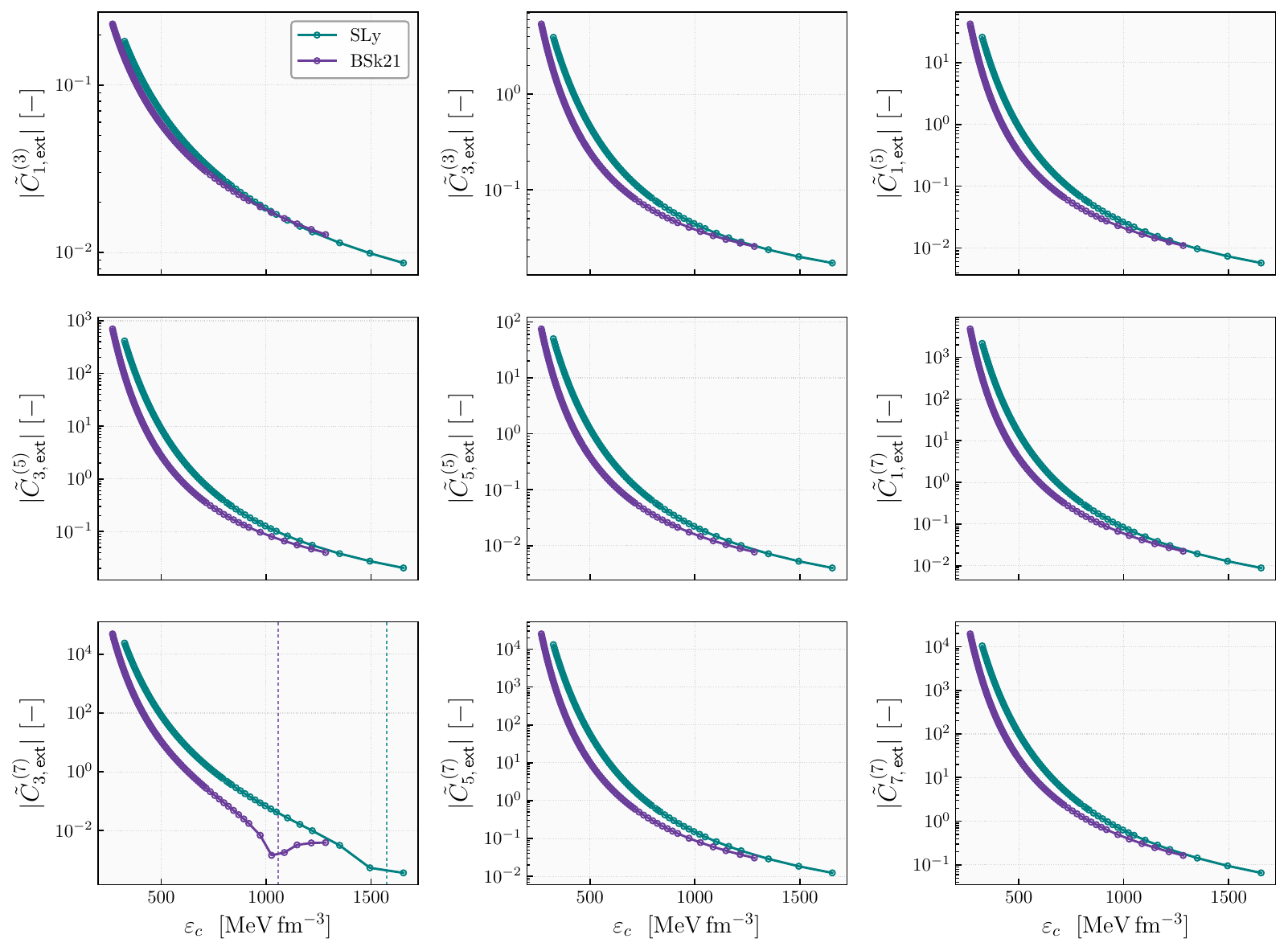}
\caption{As Fig.~\ref{fig:extconst_even}, for the odd-order exterior constants,
orders $n=3$, $5$ and $7$. Here the single constant that changes sign is
$\tilde{C}^{(7)}_{3,\mathsf{ext}}$, near $\varepsilon_c \simeq 1575$ and
$1057\,\mathrm{MeV\,fm^{-3}}$, marked in the same way. All other constants keep
one sign along both sequences.}
\label{fig:extconst_odd}
\end{figure*}

\end{widetext}

\end{document}